\documentclass[final,1p]{elsarticle}
\pdfoutput=1
\usepackage{amsmath}
\usepackage{amssymb}
\usepackage{mathrsfs}
\usepackage{lineno}
\usepackage{threeparttable}
\usepackage{pdflscape}
\usepackage{multirow}
\usepackage{soul}
\usepackage{verbatim}
\usepackage{stmaryrd}
\usepackage{mathtools}
\usepackage{array}
\newcolumntype{P}[1]{>{\centering\arraybackslash}p{#1}}
\newcolumntype{M}[1]{>{\centering\arraybackslash}m{#1}}

\usepackage{etoolbox}
\usepackage{color,hyperref}
\definecolor{darkblue}{rgb}{0.0,0.0,0.3}
\hypersetup{colorlinks,breaklinks,
            linkcolor=darkblue,urlcolor=darkblue,
            anchorcolor=darkblue,citecolor=darkblue}

\makeatletter
\def\@mkboth#1#2{}
\newlength\appendixwidth
\preto\appendix{\addtocontents{toc}{\protect\patchl@section}}
\newcommand{\patchl@section}{%
  \settowidth{\appendixwidth}{\textbf{Appendix }}%
  \addtolength{\appendixwidth}{1.5em}%
  \patchcmd{\l@section}{1.5em}{\appendixwidth}{}{\ddt}%
}
\makeatother

\journal{Progress in Materials Science}

\begin{document}

\begin{frontmatter}

\title{Grain-Boundary Kinetics: A Unified Approach\\
\textsf{\small{~\\Dedicated to the Memory of Prof. Dr. Lasar S. Shvindlerman, 1935-2018}}}

\author[mainaddress]{Jian Han}
\author[mainaddress]{Spencer L. Thomas}
\author[mainaddress,secondaryaddress]{David J. Srolovitz\corref{correspondingauthor}}

\address[mainaddress]{Department of Materials Science and Engineering, University of Pennsylvania, Philadelphia, Pennsylvania 19104, USA}
\address[secondaryaddress]{Department of Mechanical Engineering and Applied Mechanics, University of Pennsylvania, Philadelphia, Pennsylvania 19104, USA}

\cortext[correspondingauthor]{Corresponding author}
\ead{srol@seas.upenn.edu}

\begin{abstract}
Grain boundaries (GBs) are central defects for describing polycrystalline materials, and playing major role in a wide-range of physical properties of polycrystals. 
Control over GB kinetics provides effective means to tailor polycrystal properties through material processing. 
While many approaches describe different GB kinetic phenomena, this review provides a unifying concept for a wide range of GB kinetic behavior.
Our approach rests on a disconnection description of GB kinetics.
Disconnections are topological line defects constrained to crystalline interfaces with both step and dislocation character. 
These characteristics can be completely specified by GB bicrystallography and the macroscopic degrees of freedom of GBs.
GB thermal fluctuations, GB migration and the ability of GBs to absorb/emit other defects from/into the delimiting grains can be modeled via the nucleation, propagation and reaction of disconnections in the GB. 
We review the fundamentals of bicrystallography and its relationship to disconnections and ultimately to the kinetic behavior of GBs.
We then relate disconnection dynamics and GB kinetics to microstructural evolution.
While this review of the GB kinetics literature is not exhaustive, we review much of the foundational literature and draw comparisons from a wide swath of the extant experimental, simulation, and theoretical GB kinetics literature.
\end{abstract}


\end{frontmatter}
\tableofcontents

%\linenumbers

%%%%%%%%%%%%%%%%%%%%%%%%%%%%%%%%%%%%%%%%%%%%%%%%%%%%%%%
\section{Introduction}
\label{sec1}

Most natural and technological crystalline materials are polycrystalline; i.e., they are a space-filling aggregates of polyhedral single-crystalline grains of different crystallographic orientations. 
Grain boundaries (GBs) are the interfaces between pairs of contiguous grains. 
Many of the physical properties of polycrystals are determined or, at least, strongly influenced by the properties and spatial arrangement of the GBs (in addition, of course, to the properties of the single crystals that constitute the grains). 
Such properties include mechanical strength, ductility, creep rate, fatigue strength, radiation damage resistance,  diffusivity, susceptibility to corrosion, thermal and electrical resistivity, the shape of the magnetic hysteresis loop, magnetoresistance, critical current density for superconductivity, etc.~\cite{SuttonBalluffi} 
Therefore,  one  effective approach to tailor  material properties is through the exploitation of GB structure-properties-processing relationships to control the spatial, chemical, and crystallographic distribution of GBs (i.e., ``GB engineering''~\cite{watanabe1999control}). 

Like for materials in general, the study of GBs can be heuristically divided into thermodynamics and kinetics. 
Understanding equilibrium GB structure and energetics is in the bailiwick of thermodynamics; these and related issues have been extensively studied through electron microscopy observations (e.g. Refs.~\cite{Duscher04,yu2017segregation}), atomistic simulations (e.g. Refs.~\cite{bulatov2014grain,van2015multiscale,han2017grain}), and through a wide range of additional experimental and modeling approaches. 
However, GBs in polycrystals are rarely in equilibrium; they move in response to capillarity and other forces, interactions within the network of GBs, and interactions with defects from within the interiors of the grains.
The evolution of the GB network and structure of the GBs in a polycrystal is the subject of GB kinetics. 
GB kinetics are of considerable practical interest since they determine the evolution of microstructure, how a material evolves during (e.g., thermomechanical) processing, and their affect on the behavior of the bulk through absorption/emission of defects and/or solute from/into the grains.

GB kinetics studies have usually focused on either the ideal, isolated GB (bicrystal) case or  within the context of polycrystalline microstructures.  We refer to the bicrystal case as ``tame'' GBs and the polycrystalline case as GBs in the ``wild''.
In the tame case (isolated GBs), most GB kinetics studies have focused on the following phenomena:

\begin{itemize}
\item[(i)] \textit{Changes in GB composition}, including the kinetic process of GB diffusion~\cite{balluffi1982grain,suzuki2004diffusion,suzuki2005atomic}, GB segregation~\cite{raabe2014grain,lejvcek2017interfacial} and GB complexion/phase transitions~\cite{cantwell2014grain};  

\item[(ii)] \textit{Changes of GB profiles}, including the kinetic process of GB thermal fluctuations~\cite{trautt2006interface,karma2012relationship} and the roughening transition~\cite{yoon2005roughening}; 

\item[(iii)] \textit{Changes in GB position}, i.e., GB migration in response to driving forces~\cite{gottstein2009grain}; 

\item[(iv)] \textit{Changes in the shear across the GB}, i.e., the relative position of the two grains meeting at a GB (or GB sliding) in response to driving forces~\cite{langdon2006grain};  

\item[(v)] \textit{Changes in GB structure}; occurring in all kinetic processes through some form of atomic rearrangement which may often be described as some form of defect motion within the GB (i.e., conservative kinetics),  absorption/emission of defects into/out of the GB (non-conservative kinetics), and via GB structural (phase) transitions (e.g. Ref.~\cite{frolov2016phase}). 
\end{itemize} 

Since GBs in the wild are constrained by their surroundings (each grain is constrained by all of its surrounding grains) and by the presence of additional, geometrically necessary defects, such as triple junctions and quadruple points. 
The kinetics of tame GBs do not necessarily reflect the kinetics of the same GBs in the wild.
Hence, knowledge of GB kinetics from bicrystals may not provide an accurate description of microstructure evolution in polycrystals.  
For GBs in the wild (i.e., in polycrystals), most GB kinetics studies have focused on the following phenomena:

\begin{itemize}
\item[(i)] \textit{Evolution of the GB network}. 
Driven by the reduction of total energy of a polycrystal either by changing GB character or, more commonly, the decrease in  total GB area -- i.e., capillarity-driven grain growth.
Such capilllarity-driven grain growth is a classical problem in materials science~\cite{hillert1965theory,macpherson2007neumann,holm2010grain,grest1990abnormal,sonnweber2012kinetics}.
GB network evolution can also be driven by a wide range of other forces, such as those associated with crystal defect annihilation~\cite{rollett2004recrystallization,seita2013selective}, surface energy~\cite{thompson1990grain}, etc.
Recent studies have also found that GB migration can also be driven by stress (e.g., Ref.~\cite{rupert2009experimental}). 

\item[(ii)] \textit{Evolution of the  grains}. 
The evolution of the GBs and the evolution of their bounding grains are intrinsically linked.
Since stresses may drive GB motion, that motion necessarily leads to changes in the stress fields within the grains~\cite{thomas2017reconciling}.
The shear strain that accompanies GB migration may lead to grain rotation~\cite{harris1998grain,thomas2017reconciling,dake2016direct}.
The creation/annihilation of lattice dislocations and twins at the GB modify the plastic strain within the grains~\cite{thomas2016twins,rottmann2017experimental}. 
The absorption or emission of solute at the GB during GB migration changes the distribution of solute within the grain~\cite{pond1997diffusive,hirth2013interface}.
\end{itemize}

Several models of GB kinetics have been proposed in order to understand, describe and predict the evolution of GBs and GB networks (polycrystalline microstructure).
The simplest GB models consider the GB as a continuum surface, with an associated excess energy per unit area (i.e., surface tension). 
Such models can be used to predict the evolution of the GB shape~\cite{herringsurface}, GB migration~\cite{mullins1956two} and capillarity-driven grain growth~\cite{mason2015geometric}. 
However, by this model GB migration cannot be driven by stress and there is no description for the states of grains in a polycrystal.  
At the opposite extreme are models that are based upon the atomic structure of GBs, such as the polyhedral unit model~\cite{ashby1978structure} and the structural unit model~\cite{sutton1989structural,han2017grain}.  
In principle,  atomistic models can be used to study all GB kinetic behavior. 
However, since the fundamental parameters of such models are the coordinates of all of the atoms; the extraordinarily large  parameter space required to fully specify such models makes them extraordinarily difficult to apply to understand/predict GB kinetics. 
In practice, such models may be implemented in the framework of, for example, molecular dynamics (MD) simulations. 
Application of such models to study, for example, GB diffusion at low temperature is problematic because of the time-scale limitation in such simulations. Similarly, the temporal evolution of typical polycrystalline microstructures is nearly impossible today because neither the time nor length scales are compatible with MD simulations (in all except very fine nanocrystalline cases).

It is because of these limitations that we focus here largely on mesoscopic/defect-level descriptions which ignore many aspects of the atomistic details of GBs yet incorporate a much more microscopic approach than continuum descriptions. 
Our approach to GB character-sensitive GB kinetics is built upon the bicrystallography and dynamics of disconnections in GBs. 
The disconnection concept was initially proposed in the 1970s by Bollmann~\cite{bollmann1970general}, Ashby~\cite{ashby1972boundary} and Hirth and Balluffi~\cite{hirth1973grain}. 
Disconnections have both step and dislocation character and are constrained to crystalline interfaces. 
Both the step and dislocation characters are tied to the bicrystallography (the  symmetries of both crystals meeting at the interface).
The step character is responsible for GB migration and the dislocation character for GB sliding. 
Many other types of GB kinetic behavior can also be described on the basis of disconnection motion. 
Hence, the disconnection model provides a unifying approach to diverse kinetic phenomena associated with GBs (a notable exception is GB diffusion). 

In this paper, we review a wide range of GB kinetics problems within a disconnection framework.  
We note that, while we examine a wide cross-section of the experimental, simulation, and theoretical literature associated with GB kinetic phenomena, we have not aspired to providing an exhaustive review of the entire literature. 
Rather, our literature review focuses largely on foundational works and exemplars of the discussed kinetic phenomena. 
In Section~\ref{sec2} we provide a relatively succinct review of the fundamentals of bicrystallography, focusing more on the main concepts and results rather than mathematical generality. 
Bicrystallography describes the macroscopic geometry and symmetry of a bicrystal that contains a GB. 
We then discuss how disconnections naturally arise from the translational symmetry of the bicrystal. 
In Section~\ref{disconnection_model}, we  present a practical approach for the enumeration of disconnection modes as a function of the bicrystallography and propose an energetic description of each disconnection mode. 
Section~\ref{sec4} focuses on thermal equilibrium GB phenomena, such as GB roughening and intrinsic GB mobility. 
In Section~\ref{sec5}, we  examine  GB kinetics driven by external forces, including stress, bulk energy density difference across a GB, and their combination in situations that do not require interactions with defects from within the grains (i.e., conservative GB kinetics). 
Triple junction motion and microstructure evolution will also be considered within this disconnection framework. 
Section~\ref{sec6} focuses on non-conservative GB kinetics, including interactions between GBs and point defects and between GBs and lattice dislocations. 
We conclude with a discussion of the advantages and limitations of the disconnection-based approach to GB kinetics.

%%%%%%%%%%%%%%%%%%%%%%%%%%%%%%%%%%%%%%%%%%%%%%%%%%%%%%%
\section{Bicrystallography: refresher}
\label{sec2}

\subsection{CSL/DSC lattice}
\label{CSL_DSC}

\begin{figure}[!t]
\begin{center}
\scalebox{0.3}{\includegraphics{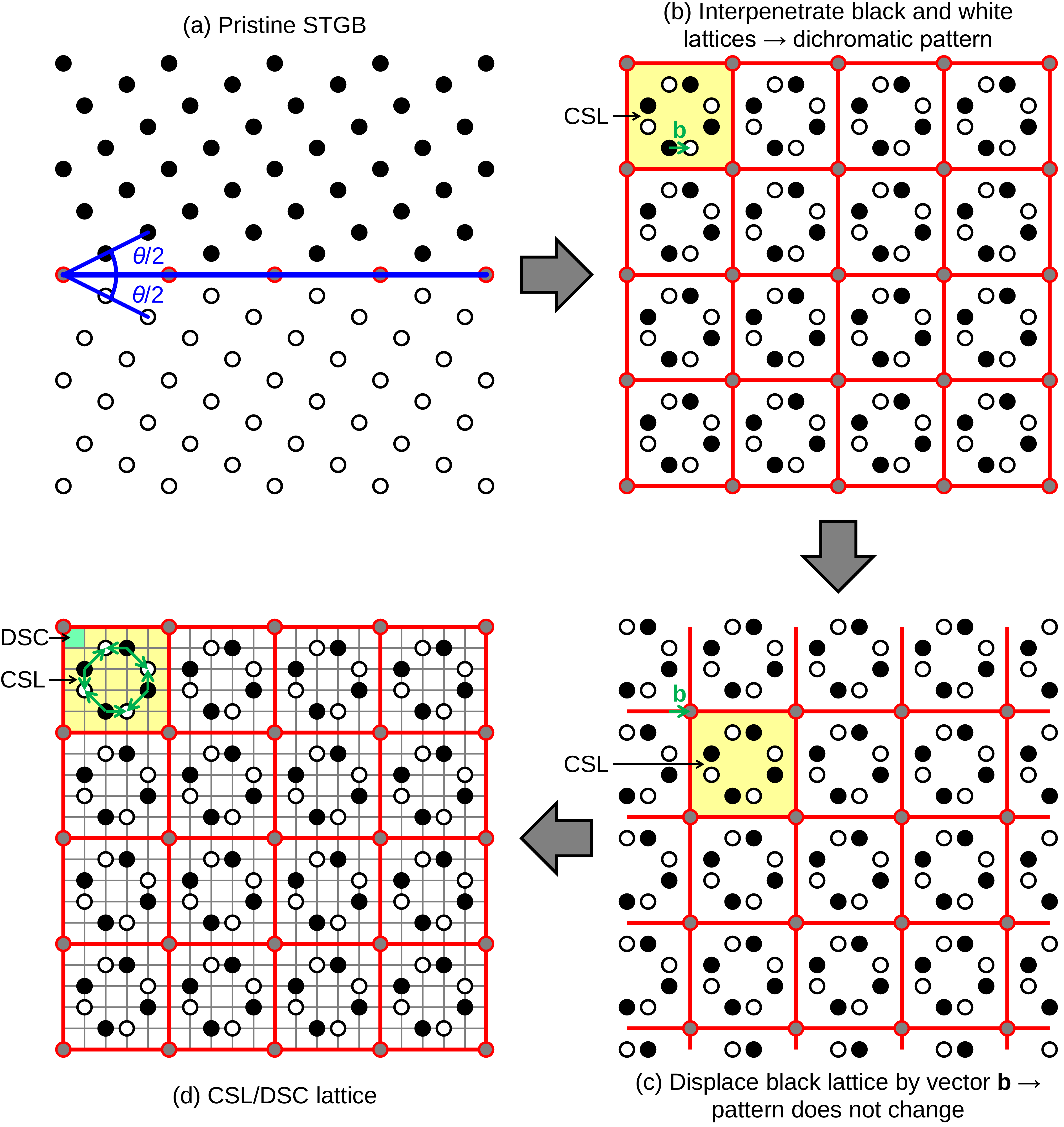}}
\caption{(a) A SC bicrystal which contains a $\Sigma 5$ $[100]$ STGB; the upper and the lower grains are colored black and white, respectively.  
(b) A dichromatic pattern formed by extending the black and the white lattices throughout the whole space; the lattice points which belong to both the black and the white lattices, called coincidence sites, are colored gray.
The coincidence sites form a lattice, called the coincidence-site lattice (CSL). The CSL is framed by the red lines; a CSL unit cell is shadowed yellow. 
(c) The entire black lattice is displaced (with respect to the white lattice) by the green vector $\mathbf{b}$, plotted in (b) and (c). 
By comparing the dichromatic pattern in (b) and that in (c), we note that the pattern does not change, but the CSL is shifted as an entity. 
(d) The same pattern as (b) with the addition of the gray lattice, called the DSC lattice; a DSC unit cell is shadowed green. 
A relative displacement by any DSC lattice vector, such as the green vectors, does not change the dichromatic pattern. } 
\label{DSC}
\end{center}
\end{figure}

A GB is the interface between two crystalline grains with identical phase but different orientation. 
Ideally, a GB can be studied in a bicrystal configuration, as illustrated in Fig.~\ref{DSC}a. 
The GB in the bicrystal is formed by joining two half-infinite grains with different orientation; the upper grain is colored black and the lower one is colored white. 
This GB is infinitely large and flat at mesoscale. 
The geometry of the GB in a bicrystal can be described by five macroscopic degrees of freedom (DOFs). 
Three of the macroscopic DOFs determine the crystallographic relation between two grains (i.e., misorientation); they are related to rotation axis $\mathbf{o}$ (2 DOFs) and rotation angle $\theta$ (1 DOF). 
The other two of the macroscopic DOFs determine the GB plane, which divides the black and white grains; they are related to the GB plane normal $\mathbf{n}$ (2 DOFs). 
In this section, we focus on the former three DOFs, which describe the misorientation. 
We will see that the misorientation relation between two grains features the existence of a coincidence-site lattice (CSL) and DSC lattice (DSC stands for ``displacement shift complete''~\cite{bollmann2012crystal} or ``displacements which are symmetry conserving''~\cite{pond1979symmetry}). 
For convenience and clearance, we will explain all the concepts based on the bicrystal as shown in Fig.~\ref{DSC}a, which contains a $\Sigma 5$ $[100]$ $(012)$ symmetric tilt GB (STGB) in a simple cubic (SC) material. 

In order to study the effect of misorientation, we first extend the black lattice and the white lattice throughout the entire space such that they form a dichromatic pattern (see Fig.~\ref{DSC}b). 
From the dichromatic pattern, we see that there are some sites where the black lattice points are coincident with the white lattice points; these are called coincidence sites and are colored gray. 
The set of coincidence sites constitutes a new lattice, which is framed by the red lines in Fig.~\ref{DSC}b. 
This new lattice is called the coincidence-site lattice (CSL); one primitive unit cell of CSL is shaded yellow in Fig.~\ref{DSC}b. 
We notice that, in the case shown in Fig.~\ref{DSC}, the size of a unit cell of CSL is five times that of crystal lattice (the black or white lattice). 
Therefore, the reciprocal coincidence site density is $\Sigma = 5$. 
Since the coincidence sites must belong to the subset of the crystal lattice points, $\Sigma$ is always an integer and $\Sigma \ge 1$. 
$\Sigma 1$ corresponds to the case of perfect crystal without a GB. 

In Fig.~\ref{DSC}b, $\mathbf{b}$ is a vector connecting one black lattice point to its nearest white lattice point. 
If we fix the white lattice and displace the entire black lattice by the vector $\mathbf{b}$, we will transform the dichromatic pattern in Fig.~\ref{DSC}b to the pattern in Fig.~\ref{DSC}c. 
By comparing the patterns in Figs.~\ref{DSC}b and c, we find that, after the displacement of the black lattice respect to the white lattice, the dichromatic pattern does not change, but the pattern is shifted (comparing the positions of the unit cell of CSL which are shaded yellow in Figs.~\ref{DSC}b and c). 
Therefore, the vector $\mathbf{b}$ represents a displacement which conserves the dichromatic pattern. 
Such vector is not unique; any vector that connects one black lattice point to one white lattice point will lead to the same result (i.e., conserve the pattern). 
All these vectors constitute a new lattice, called the DSC lattice, which is framed out by the gray and red lines in Fig.~\ref{DSC}d. 
The displacement of the black lattice with respect to the white lattice by any DSC lattice vector does not change the dichromatic pattern. 
The existence of the CSL implies the translational symmetry of the entire bicrystal while the existence of the DSC lattice implies the translational symmetry associated with the relative displacement between two grains. 

\subsection{Grain-boundary disconnection}
\label{GB_disconnection}

For a fixed misorientation (3 DOFs), a GB can be constructed by choosing a plane (2 DOFs) and then removing the black lattice below the plane and the white lattice above the plane. 
In this section, we will see that bicrystallography (including macroscopic geometry of a GB and bicrystal symmetry) implies the existence of line defects in the GB. 

\subsubsection{A special case: dislocation for a $\Sigma 1$ grain boundary}
\begin{figure}[!t]
\begin{center}
\scalebox{0.3}{\includegraphics{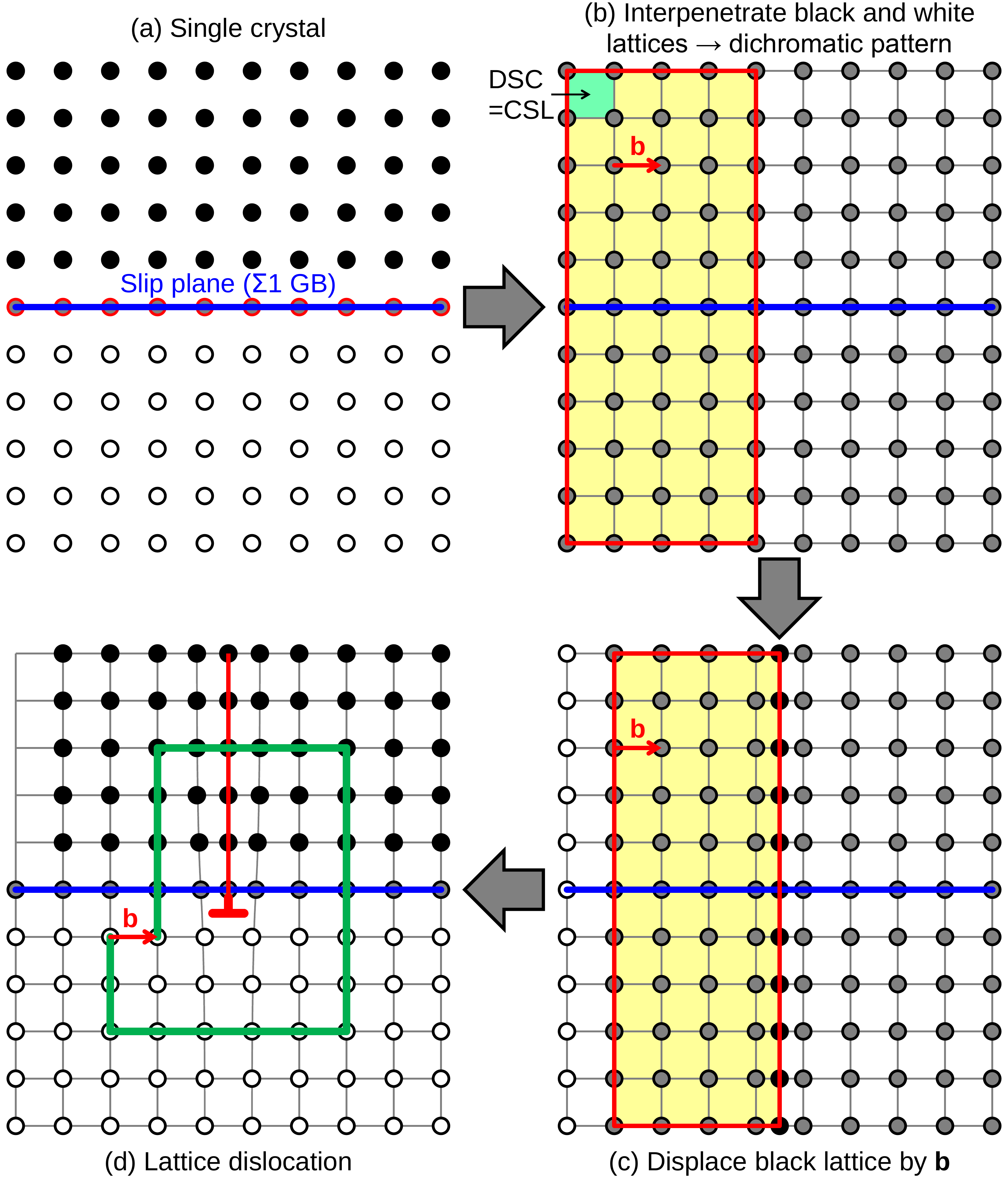}}
\caption{The formation of a lattice dislocation in a SC crystal by the operation of translational symmetry.
(a) A single crystal containing a slip plane denoted by the blue line (this is equivalent to a $\Sigma 1$ $[100]$ GB); the lattices above and below the slip plane are colored black and white, respectively. 
(b) A dichromatic pattern formed by extending the black and white lattices throughout the whole space; since all of the black and white lattice points coincide, all of the lattice points are colored gray and belong to the CSL. 
In this special case, the CSL is identical to the DSC lattice. 
(c) Based on (b), the black lattice in the yellow region is displaced (with respect to the white lattice) by the red vector $\mathbf{b}$. 
(d) A lattice dislocation is formed by removing the black lattice below the slip plane and the white lattice above the slip plane. 
The green line denotes the Burgers circuit and the closure failure of this circuit defines the Burgers vector, which is denoted by the red vector $\mathbf{b}$.  } 
\label{lattice_dislocation}
\end{center}
\end{figure}

It is instructive to first consider how a line defect (i.e., lattice dislocation) can be constructed in a perfect crystal. 
Figure~\ref{lattice_dislocation}a shows a perfect crystal. 
We can choose an arbitrary slip plane and color the lattice above the plane black and the lattice below the plane white.  
This perfect crystal can then be equivalently viewed as a bicrystal containing a $\Sigma 1$ GB ($\theta = 0$).

We can extend the black lattice and the white lattice throughout the entire space to form a dichromatic pattern (see Fig.~\ref{lattice_dislocation}b). 
For this special bicrystal, all the lattice points are located at the coincidence sites (colored gray). 
The crystal lattice, CSL, and DSC lattice are all identical; and the DSC lattice vector $\mathbf{b}$ is exactly the same as the translation vector of the perfect crystal or Burgers vector of lattice dislocation. 
From the structure in Fig.~\ref{lattice_dislocation}b, if we displace the black lattice in the yellow region with respect to the white lattice by the DSC lattice vector $\mathbf{b}$, we will obtain the structure in Fig.~\ref{lattice_dislocation}c. 
We find that, after such displacement, the pattern in the yellow region does not change (since the DSC lattice vector conserves the pattern). 
Based on the structure in Fig.~\ref{lattice_dislocation}c, if we remove the black lattice below the slip plane and the white lattice above the slip plane, we will obtain the structure in Fig.~\ref{lattice_dislocation}d. 
We find that, after the operation proposed above, we have introduced an edge dislocation with Burgers vector $\mathbf{b}$ into the slip plane, which is verified by the Burgers circuit in Fig.~\ref{lattice_dislocation}d. 
This special example for the $\Sigma 1$ GB shows that translational symmetry implies the existence of lattice dislocations. 

\subsubsection{Disconnection for a CSL grain boundary}
\label{disconnection_for_CSL}
\begin{figure}[!t]
\begin{center}
\scalebox{0.27}{\includegraphics{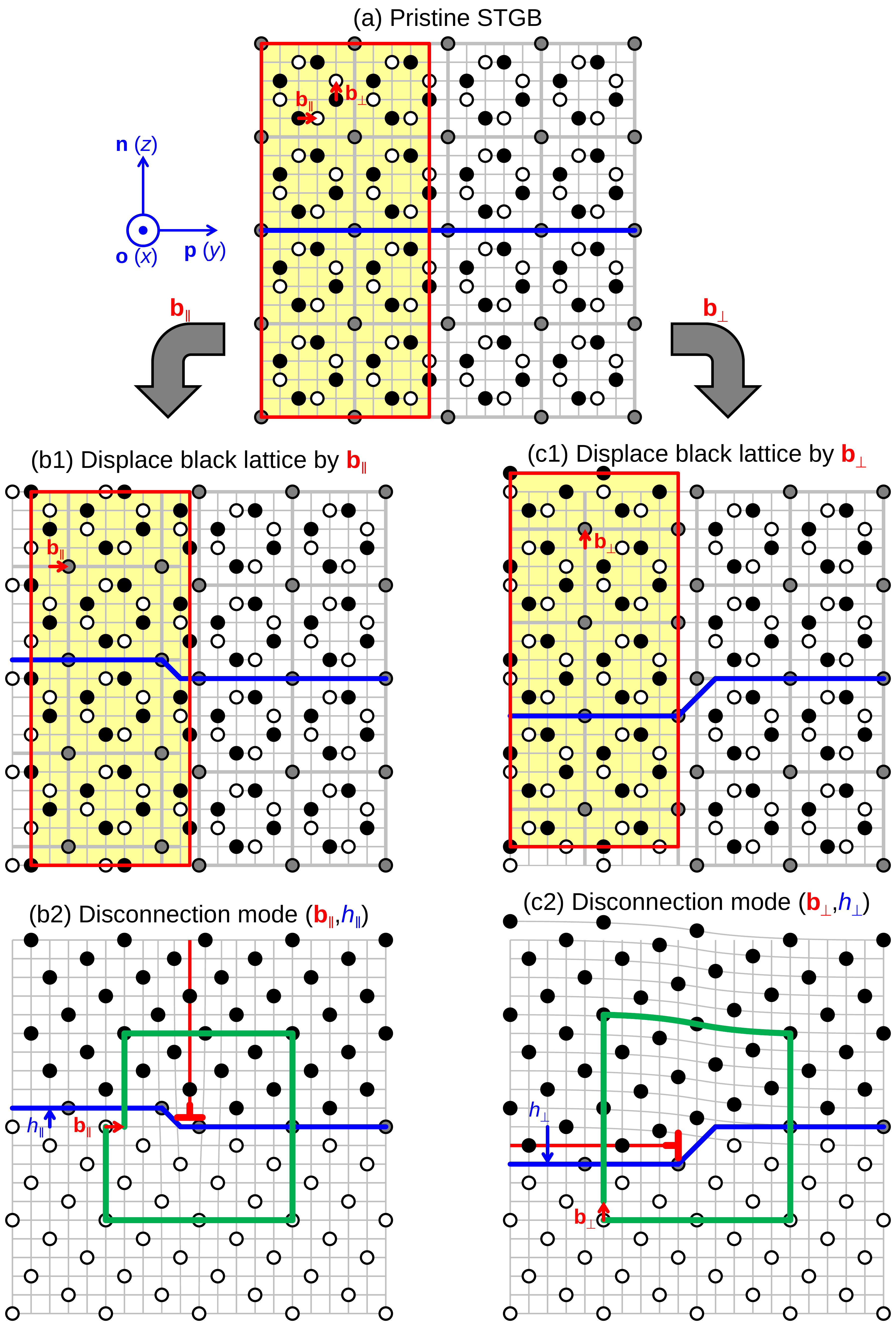}}
\caption{Formation of a disconnection in a $\Sigma 5$ $[100]$ STGB by the operation of translational symmetry related to the DSC lattice. 
(a) The same structure as that in Fig.~\ref{DSC}d. 
(b1) Based on (a), displace the black lattice in the yellow region with respect to the white lattice by the red vector parallel to the GB plane $\mathbf{b}_\parallel$; the blue line is the new GB plane after such displacement. 
(b2) A disconnection $(\mathbf{b}_\parallel, h_\parallel)$ is formed by removing the black lattice below the GB plane and the white lattice above the GB plane in (b1). 
(c1) Based on (a), the black lattice in the yellow region is displaced (with respect to the white lattice) by the red vector perpendicular to the GB plane $\mathbf{b}_\perp$; the blue line is the new GB plane after such a displacement. 
(c2) A disconnection $(\mathbf{b}_\perp, h_\perp)$ is formed by removing the black lattice below the GB plane and the white lattice above the GB plane in (c1). 
The green lines in (b2) and (c2) denote the Burgers circuit.  } 
\label{disconnection}
\end{center}
\end{figure}

Following a procedure similar to the one proposed above, we will show that translational symmetry implies the existence of line defects in a CSL GB (rather than a perfect crystal). 
Again, we will take the example of a $\Sigma 5$ $[100]$ $(012)$ STGB in a SC material, as shown in Fig.~\ref{DSC}a. 
After extending the black lattice and the white lattice throughout the entire space, we will obtain a dichromatic pattern as shown in Fig.~\ref{disconnection}a. 
If we displace the black lattice in the yellow region with respect to the white lattice by the DSC lattice vector $\mathbf{b}_\parallel$, we will obtain the structure in Fig.~\ref{disconnection}b1. 
After such displacement, the pattern in the yellow region does not change. 
However, unlike the special case of the $\Sigma 1$ GB (perfect crystal), the pattern in the yellow region is shifted. 
In order to keep the GB structure in the yellow region exactly same as that in the right unshaded region (i.e., intersecting the coincidence sites), we have to choose the new GB plane in the yellow region as shown in Fig.~\ref{disconnection}b1. 
The new GB plane in the yellow region is located higher than the GB plane in the unshaded region; hence, there is a step in the middle of the GB after the relative displacement in the yellow region. 
Finally, if we remove the black lattice below the GB plane and the white lattice above the GB plane, we will obtain the structure in Fig.~\ref{disconnection}b2. 
After the operation mentioned above, we have introduced an edge dislocation with Burgers vector $\mathbf{b}_\parallel$ and, at the same time, a GB step with step height $h_\parallel$ into the GB. 
This line defect in a GB, called a disconnection, is characterized by both Burgers vector and step height $(\mathbf{b}_\parallel, h_\parallel)$. 
The concept of GB disconnections was proposed and developed in the literature~\cite{bollmann1970general,hirth1973grain,balluffi1982csl,hirth2006disconnections,hirth2007spacing,king1980effects}. 

Previously, we chose a DSC vector $\mathbf{b}_\parallel$ parallel to the GB plane. 
We could also have chosen the DSC vector to be perpendicular to the GB plane (e.g. $\mathbf{b}_\perp$ as shown in Fig.~\ref{disconnection}a). 
If we displace the black lattice in the yellow region by $\mathbf{b}_\perp$ (see Fig.~\ref{disconnection}c1), we will obtain the structure in Fig.~\ref{disconnection}c2, where the disconnection is featured by $(\mathbf{b}_\perp, h_\perp)$. 
Since the Burgers vector $\mathbf{b}_\perp$ is perpendicular to the GB plane, we have to introduce an extra half plane lying along the GB plane (see Fig.~\ref{disconnection}c2). 

Although the discussion above is for a GB with particular geometry, it can be easily extended to the case of any CSL GB. 
Now, several generalizations will be made below. 

\begin{figure}[!t]
\begin{center}
\scalebox{0.32}{\includegraphics{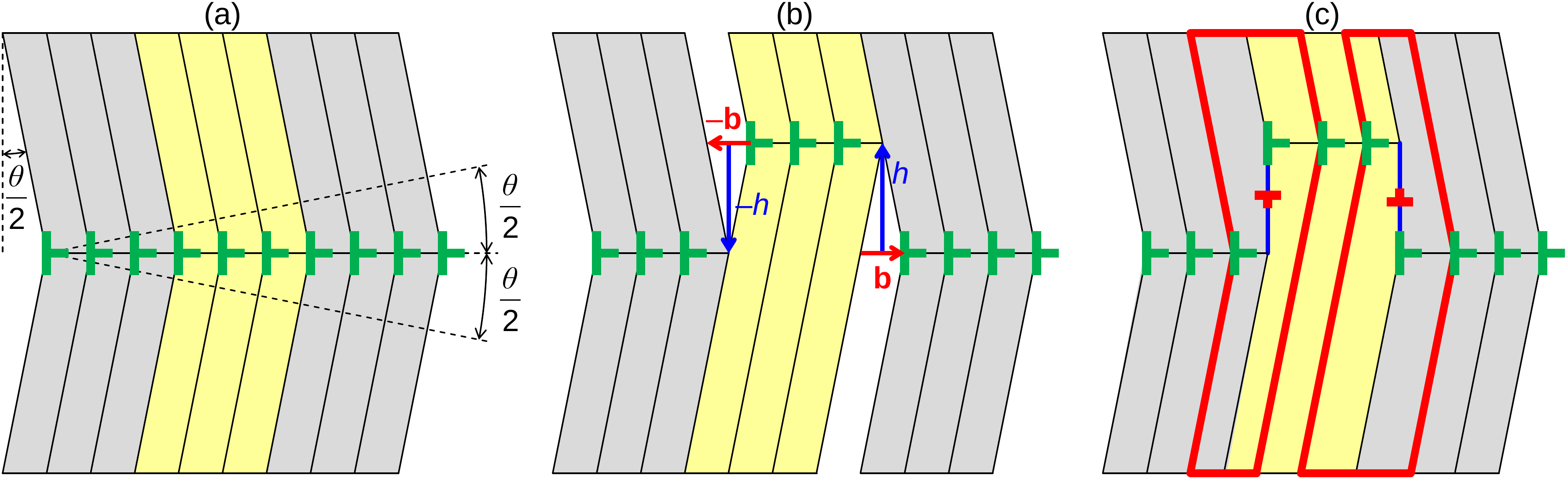}}
\caption{The formation of a pair of disconnections in a low-angle STGB. 
(a) A low-angle STGB composed of an array of lattice dislocations. 
(b) The free migration of a segment of GB by the glide of the lattice dislocations in the yellow region must lead to the cracks between the yellow region and the gray regions. 
The resulting step heights (right and left) are $h$ and $-h$  and the openings of the cracks are $\mathbf{b}$ and $-\mathbf{b}$, respectively. 
(c) The cracks are closed by straining the yellow and gray regions. 
The strain state (indicated by the red outlines) suggests that there are Burgers vectors $\mathbf{b}$ and $-\mathbf{b}$ located at the two GB steps.  } 
\label{wedge}
\end{center}
\end{figure}

\begin{itemize}
\item[(i)] The disconnection depicted in Fig.~\ref{disconnection} is of edge character; but this is not necessary. 
The disconnection line is not necessarily parallel to the rotation axis $\mathbf{o}$; it can be in any direction and even arbitrarily curved. 
For example, we can choose a plane parallel to the paper and displace the black lattice behind the plane with respect to the white lattice by $\mathbf{b}_\parallel$. 
In this way, we will obtain the disconnection $(\mathbf{b}_\parallel, h_\parallel)$ but of screw character. 

\item[(ii)] Low-angle GBs are associated with large $\Sigma$ values. 
Generally (but not universally), the size of the CSL unit cell is scaled by $\Sigma^{1/2}$ and the size of the DSC lattice unit cell is scaled by $\Sigma^{-1/2}$ for tilt GBs. 
Hence, low-angle GB disconnections are usually characterized by small Burgers vectors and large step heights. 
We will show that this is consistent with the analysis based on the GB dislocation model. 
The structure of a low-angle STGB is an array of edge lattice dislocations (different from disconnections)~\cite{read1950dislocation}, as illustrated in Fig.~\ref{wedge}a. 
The migration of the low-angle STGB is realized by the glide of the GB dislocations. 
As shown in Fig.~\ref{wedge}b, if the GB in the yellow region migrates upwards via the GB dislocation gliding, we will have to introduce wedge-shaped cracks between the yellow region and the gray regions in order to keep the bicrystal stress-free. 
Finally, as shown in Fig.~\ref{wedge}c, when we bring the yellow region and the gray regions together to eliminate the cracks, such an operation will introduce dislocations. 
The dislocations associated with the GB steps are disconnections. 
Since the Burgers vector of the dislocation is very small compared with the lattice constant, it is unlikely that we will see any line defect associated with the disconnection in the low-angle GB. 
Therefore, the disconnection exists for the low-angle GB in the sense of a dislocation-like elastic field (corresponding to the Burgers vector) and discontinuity of the GB plane (corresponding to the step height). 
 
\item[(iii)] The discussion about disconnections also applies to twist GBs (TwGBs).  
Based on the structure in Fig.~\ref{DSC}d, if we choose a plane parallel to the paper as the GB plane, remove the black lattice behind the plane and white lattice in front of the plane, we will obtain a TwGB. Since the relative displacement of the black lattice with respect to the white lattice by either $\mathbf{b}_\parallel$ or $\mathbf{b}_\perp$ does not require the change of GB plane, the disconnection in a TwGB could be $(\mathbf{b}_\parallel, 0)$ or $(\mathbf{b}_\perp, 0)$. 

\item[(iv)] The discussion about disconnections also applies to asymmetric tilt GBs (ATGBs). 
For a non-faceted ATGB, the disconnection is same as that in the STGB with the same misorientation. 
For a faceted ATGB, disconnections can be considered for each facet; each facet may correspond to a STGB or a non-faceted ATGB. 

\item[(v)] In general, CSL does not exist for any misorientation. 
If CSL does not exist for a misorientation, disconnections cannot be well defined for the GB with this particular misorientation. 
However, for any non-CSL misorientation, we can always find a CSL misorientation which is very close (in sense of the indices of $\mathbf{o}$ and the value of $\theta$) to the non-CSL one. 
The structure of any non-CSL GB can be viewed as the structure of the close-in-misorientation CSL GB plus non-periodically, sparsely distributed secondary GB dislocations~\cite{sutton1988irrational}. 
Then, the disconnection in this non-CSL GB is the same as that in the CSL GB. 
The possible approach to find the close-in-misorientation CSL GB for a non-CSL GB is to follow the Brandon criterion, i.e., to permit the tolerance of misorientation angle $\Delta \theta < \Delta \theta_0 \Sigma^{-1/2}$~\cite{brandon1966structure}. 
\end{itemize}

\subsection{Disconnection modes}
\label{multiple_modes}

\subsubsection{Non-uniqueness of step heights associated with each Burgers vector}
\begin{figure}[!t]
\begin{center}
\scalebox{0.3}{\includegraphics{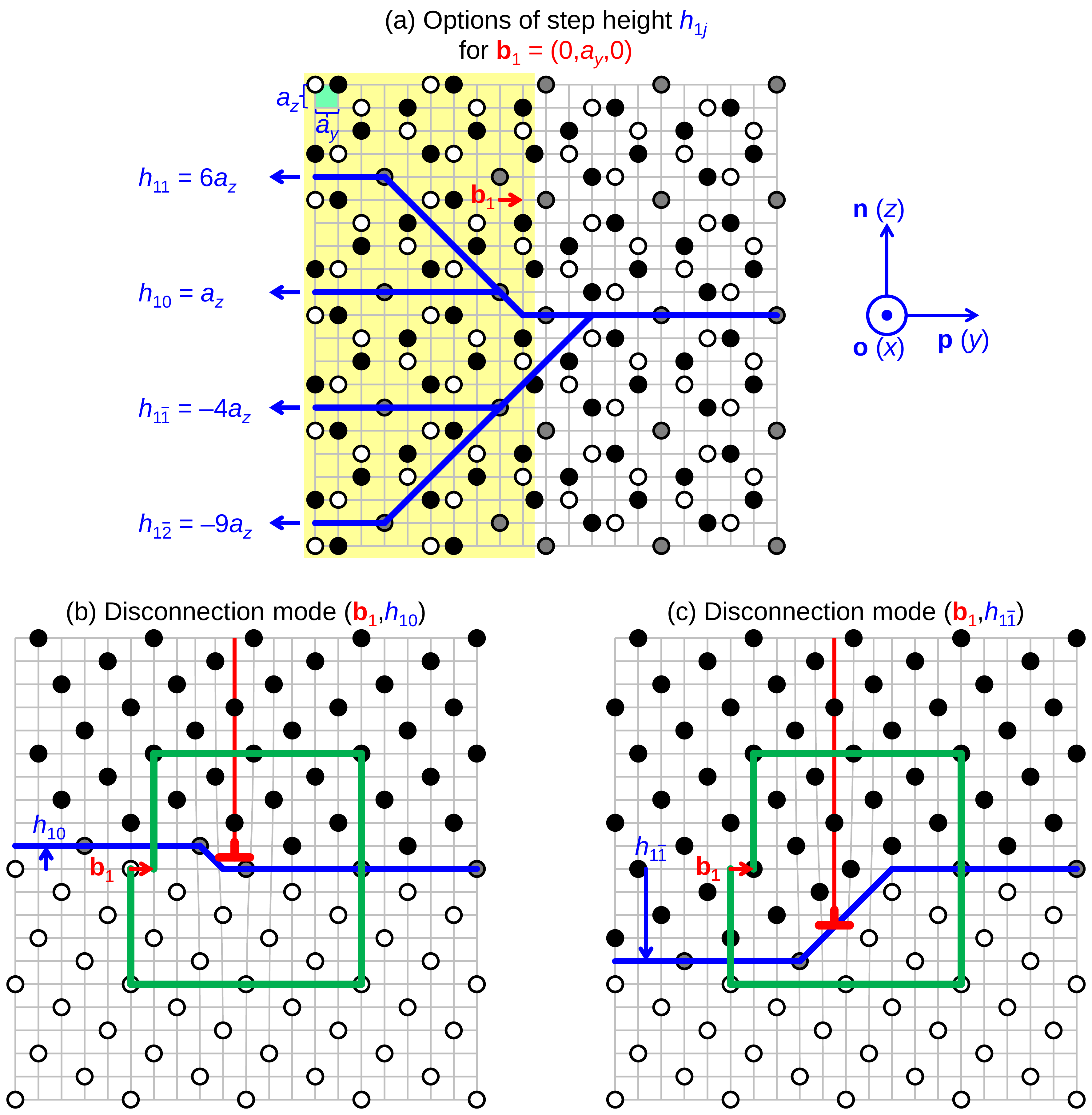}}
\caption{(a) Based on the DSC lattice of a $\Sigma 5$ $[100]$ STGB in a SC crystal, where $a_y$ and $a_z$ are the sizes of the DSC unit cell, the displacement of the black lattice (with respect to the white lattice) in the yellow region by the red vector $\mathbf{b}_1 = (0,a_y,0)$ leads to a disconnection with the Burgers vector $\mathbf{b}_1$ in this GB. 
This disconnection is associated with a step height. 
There are multiple possible step heights corresponding to a disconnection with Burgers vector $\mathbf{b}_1$; the step height can be $h_{1j} = (1+5j)a_z$ (where $j$ is integer). 
(b) One possible step height $h_{10} = a_z$, corresponding to the disconnection mode $(\mathbf{b}_1,h_{10})$. 
(c) Another possible step height is $h_{1\bar{1}} = -4a_z$, corresponding to the disconnection mode $(\mathbf{b}_1,h_{1\bar{1}})$. } 
\label{multiple_step}
\end{center}
\end{figure}

In Fig.~\ref{disconnection}b1, we find that, after the relative displacement of the black lattice with respect to the white lattice in the yellow region, in order to keep the GB structure in the yellow region the same as that in the unshaded region, we have to change the position of the GB plane in the yellow region. 
However, the choice of the new position of GB plane in the yellow region is not unique. 
Since different GB position in the yellow region corresponds to different step height, the step height is not unique associated with this particular relative displacement. 
As indicated in Fig.~\ref{multiple_step}a, corresponding to the relative displacement by the DSC lattice vector $\mathbf{b}_1$, the new position of GB plane in the yellow region could be $h_{11}$, $h_{10}$, $h_{1\bar{1}}$, $h_{1\bar{2}}$, etc. 
For this particular bicrystallography and Burgers vector $\mathbf{b}_1 = (0,1,0)a_y$, the step height can in general be expressed as $h_{1j} = (1+5j)a_z$ ($j$ is integer). 
For example, if we choose the GB plane located at $h_{10}$ (i.e., the step height is $h_{10}$), then, after removing the black lattice below the plane and the white lattice above the plane, we will obtain the disconnection $(\mathbf{b}_1, h_{10})$ as shown in Fig.~\ref{multiple_step}b. 
If we choose the GB plane located at $h_{1\bar{1}}$, we will obtain the disconnection $(\mathbf{b}_1, h_{1\bar{1}})$ as shown in Fig.~\ref{multiple_step}c. 

The step height can be positive or negative, implying that, with the disconnection gliding in the direction of $\mathbf{p}$ (from left to right in Fig.~\ref{multiple_step}), the GB migrates in the direction of $\mathbf{n}$ (upwards) or $-\mathbf{n}$ (downwards). 

\subsubsection{Multiple choices of Burgers vector}
We have considered all of the possible step heights associated with the Burgers vector $\mathbf{b}_1$ in the last section (see Fig.~\ref{multiple_step}). 
However, the choice of the Burgers vector is also not unique. 
Any DSC lattice vector can be taken as the Burgers vector. 
For example, in Fig.~\ref{disconnection}, both $\mathbf{b}_\parallel$ and $\mathbf{b}_\perp$ could be the Burgers vector. 
The Burgers vector can be zero and, in this special case, the corresponding disconnection is a pure step (without dislocation character). 
The Burgers vector of a lattice dislocation also corresponds to a DSC lattice vector. 
However, in the special case of the Burgers vector of a lattice dislocation, it is not necessary to introduce a GB step (or the step height is zero). 

Now, we know that, for any bicrystallography, there is a set of disconnection modes $\{(\mathbf{b}_n, h_{nj})\}$. 
In principle, the number of the disconnection modes is infinite. 
However, only the modes characterized by small Burgers vectors and step heights may occur in reality since they are associated with low energy. 

\subsection{Experimental evidence}
There is strong experimental evidence that directly demonstrates the existence of disconnections in GBs. 
For example, Merkle et al. clearly observed the presence of disconnections in thermally equilibrated GBs in Al and Au by high-resolution transmission electron microscopy~\cite{merkle2002thermally,merkle2004situ}.
Rajabzadeh et al. observed different types of disconnections (corresponding to different disconnection modes) in a STGB in a Cu bicrystal~\cite{rajabzadeh2014role}. 
Bowers et al. also observed disconnections in an irrational (incommensurate) tilt GB in a Au bicrystal via scanning transmission electron microscopy~\cite{bowers2016step}. 
Radetic et al. found that there were disconnections in the GB of a shrinking embedded cylindrical grain in Al and Au thin films~\cite{radetic2012mechanism,dahmen2016high}. 
Not limited to metals, disconnections were also observed in the GBs in ceramics, such as alumina~\cite{heuer2015disconnection,heuer2016band} and SrTiO$_3$~\cite{lee2009kinetic,sternlicht2015mechanism}. 
Disconnections were also inferred from earlier observations via transmission electron microscopy~\cite{baluffi1972electron,kegg1973grain,dingley1979interaction}. 

\subsection{Beyond bicrystallography: effect of microscopic degrees of freedom}
\label{microscopicDOFs}
\begin{figure}[!t]
\begin{center}
\scalebox{0.264}{\includegraphics{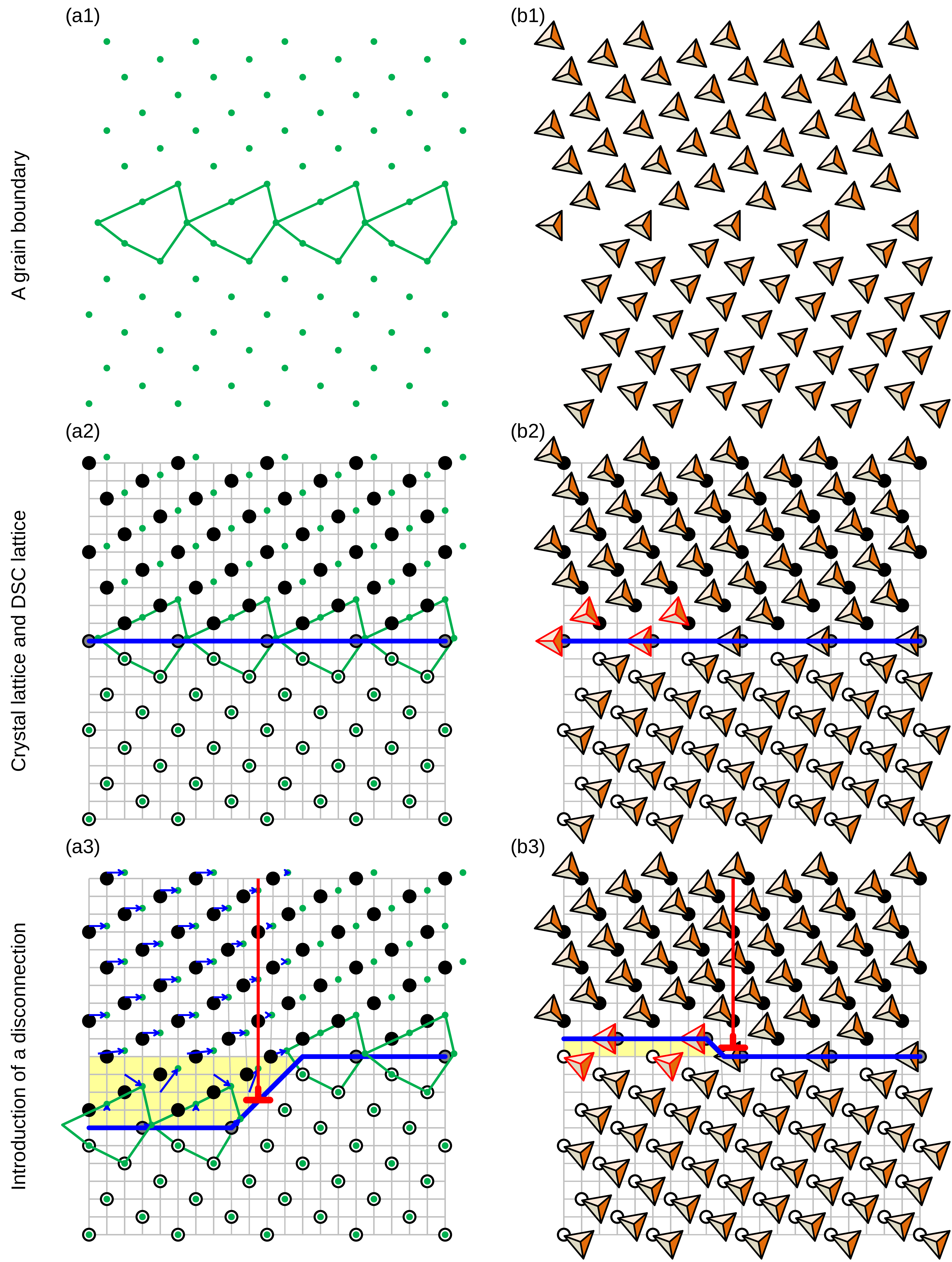}}
\caption{(a1) The atomic structure of a $\Sigma 5$ STGB in a real monatomic SC crystal; there is a relative shift of the upper grain with respect to the lower one and local atomic relaxation in the GB. 
(a2) The black and the white lattices are defined above and below the GB such that the two lattices coincide at the GB. 
To enforce coincidence at the GB, the atoms in the white lattice are located at lattice points while atoms in the black lattice are shifted with respect to the lattice points. 
(a3) The formation of a disconnection (corresponding to the mode $(\mathbf{b}_1, h_{1\bar{1}})$ in Fig.~\ref{multiple_step}c) in the GB structure in (a2). 
The blue arrows indicate the displacement of the atoms, pointing from the initial positions to the final positions.
(b1) The atomic structure of a $\Sigma 5$ STGB in a real complex SC crystal; the triangles denote the bases (motifs).
(b2) The black and the white lattices, to which the bases in (b1) are attached, are defined above and below the GB. 
(b3) The formation of a disconnection (corresponding to the mode $(\mathbf{b}_1, h_{10})$ in Fig.~\ref{multiple_step}b) in the GB structure in (b2). 
The red triangles are the bases undergoing shift and rotation during the formation of the disconnection.   } 
\label{complex}
\end{center}
\end{figure}

It should be emphasized that the discussion in the previous sections is based on bicrystallography (e.g. the bicrystal lattice shown in Fig.~\ref{DSC}a) rather than detailed atomic structure. 
Bicrystallography is uniquely determined by the five macroscopic DOFs; but the atomic structure of a GB is also influenced by the microscopic DOFs (except for macroscopic DOFs). 
The atomic structure of a $\Sigma 5$ $[100]$ $(012)$ STGB in a SC crystal may look like Fig.~\ref{complex}a1, which is different from the lattice structure shown in Fig.~\ref{DSC}a. 
In general, the atomic structure of the upper grain cannot exactly coincide with that of the lower grain at the GB; there could be a relative displacement of the upper grain with respect to the lower grain, atomic relaxation, and even a change of atomic fraction (or free volume) in the GB region -- all of these arise from the microscopic DOFs~\cite{SuttonBalluffi,han2016grain}. 
The atomic structure of a GB in a bicrystal can be viewed as the bicrystal lattice convoluted by a properly defined atomic basis in two grains, as illustrated in Fig.~\ref{complex}a2. 
In the case of Fig.~\ref{complex}a2, each atom is located at one white lattice point in the lower grain; in the upper grain, each atom is offset to one black lattice point by a fixed vector. 
We can construct a disconnection based on the bicrystal lattice, following the procedure shown in Fig.~\ref{disconnection}. 
Since each atom displaces with the corresponding lattice point, we will have the atomic structure of a disconnection as shown in Fig.~\ref{complex}a3.  
The Burgers vector and step height of this disconnection are only determined by the lattice structure (related to the macroscopic DOFs) and are independent of the detailed atomic structure (related to the microscopic DOFs). 

The formation of a disconnection (transformation from the structure in Fig.~\ref{complex}a2 to that in Fig.~\ref{complex}a3) requires atomic shuffling in the transformation region (see the blue arrows in the yellow shaded region in Fig.~\ref{complex}a3). 
The atomic shuffling can be partitioned into three types of contributions: (i) the shift of lattice points such that, in the transformation region, the lattice orientation changes from the orientation of the black lattice to that of the white lattice, (ii) the vector connecting each lattice point to the corresponding basis atom, and (iii) atomic relaxation. 
The first contribution is determined by the macroscopic DOFs, the second one is determined by the microscopic DOFs, and the third one is uncontrolled by the geometry (determined by the interaction among atoms). 

Following the same idea, we can also consider the GB disconnection in a complicated crystal structure, such as Fig.~\ref{complex}b1. 
Again, the disconnection in this GB is only related to the bicrystal lattice which is the same as that shown in Fig.~\ref{complex}a2. 
But the atomic shuffling during the glide of disconnection along the GB will, in addition, involve the rotation of the bases, as shown in Fig.~\ref{complex}b3. 

In summary, the discussion about CSL, DSC lattice and disconnection modes in the previous sections can be equally applied to the real GBs in materials with various crystal structure. 
The feature of atomic bases does not influence CSL, DSC lattice and disconnection modes, but is related to the way of atomic shuffling during the formation of disconnections. 

%%%%%%%%%%%%%%%%%%%%%%%%%%%%%%%%%%%%%%%%%%%%%%%%%%%%%%%
\section{Grain-boundary disconnection model}
\label{disconnection_model}

\subsection{Enumeration of disconnection modes}
\label{enumeration}
We already know that a disconnection mode is determined by a DSC Burgers vector and step height $(\mathbf{b}, h)$; for a particular bicrystallography (i.e., fixed macroscopic DOFs), there is a list of disconnection modes $\{(\mathbf{b}_n, h_{nj})\}$. 
Since the disconnection is the basic line defect in GBs, the character, energy, and dynamics of disconnections are related to many GB behaviors. 
In this section, we will determine a general approach to predict the characters of all possible disconnection modes based on a particular bicrystallography. 

\begin{figure}[!t]
\begin{center}
\scalebox{0.3}{\includegraphics{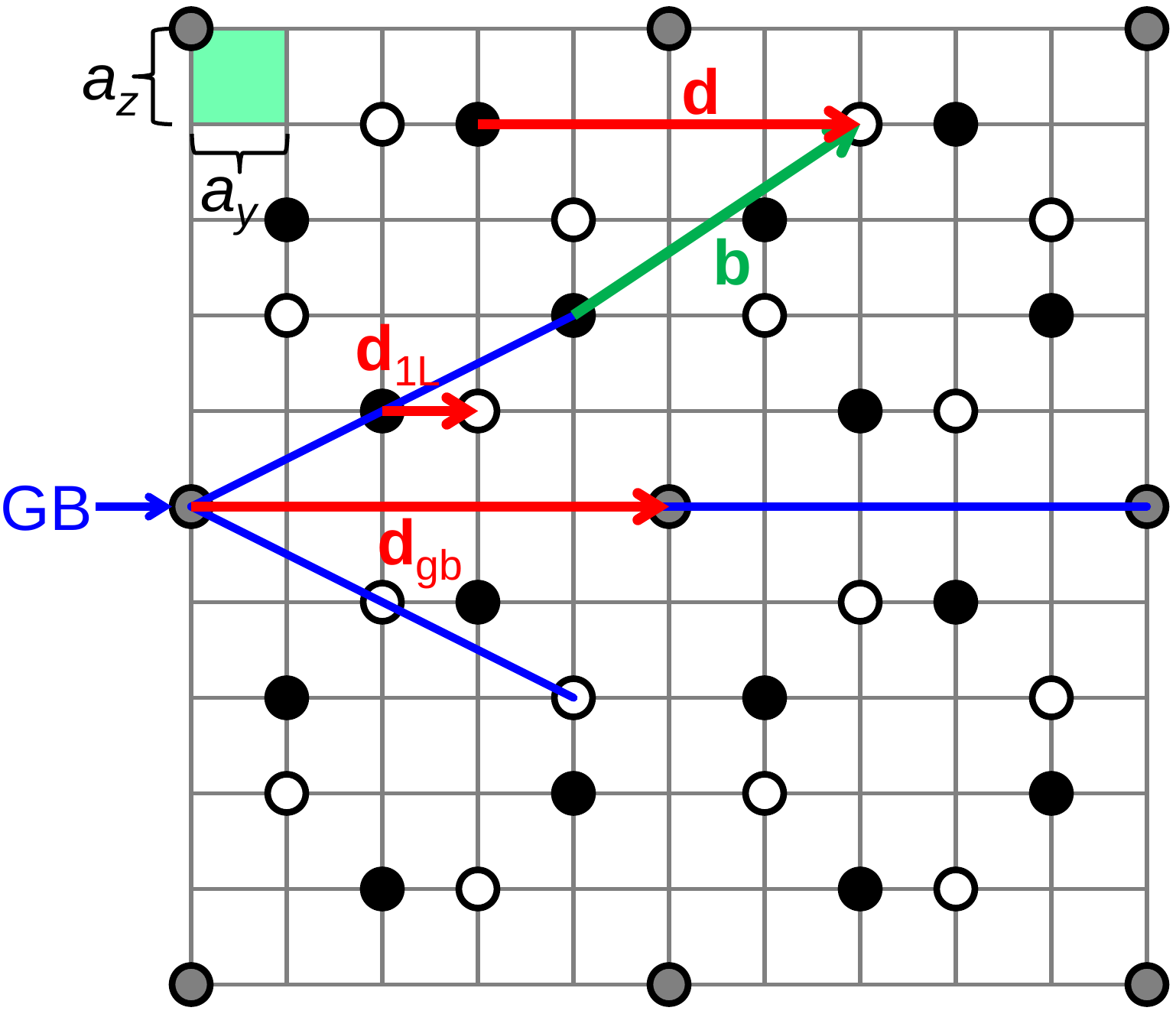}}
\caption{Dichromatic pattern of a $\Sigma 5$ $[100]$ STGB in a SC crystal. 
The gray lines denote the DSC lattice. 
$a_y$ and $a_z$ are the dimensions of the DSC unit cell. 
The green vector $\mathbf{b}$ is a DSC lattice vector; when a disconnection is formed by this vector, $\mathbf{b}$ will be the Burgers vector. 
$\mathbf{d}$, $\mathbf{d}_{\text{1L}}$ and $\mathbf{d}_{\text{gb}}$ represent the displacements of the black lattice with respect to the white lattice parallel to the GB. 
$\mathbf{d}_{\text{gb}}$ is the minimum displacement that preserves the CSL; $\mathbf{d}_{\text{1L}}$ is the displacement which can lead to one-layer upward shift of the GB plane; $\mathbf{d}$ is the displacement that results in the same dichromatic pattern as $\mathbf{b}$ does.   } 
\label{Diophantine}
\end{center}
\end{figure}

King and Smith~\cite{king1980effects} have provided a general formula to calculate all the possible step heights for a particular Burgers vector $\mathbf{b}$. 
Referred to the dichromatic pattern and DSC lattice shown in Fig.~\ref{Diophantine}, 
\begin{equation}
\mathbf{d} = \mathbf{d}_{\text{1L}} \tilde{h} + \mathbf{d}_{\text{gb}} N. 
\label{d}
\end{equation}
$\tilde{h} \equiv h/a_z$ is the step height scaled by the interlayer spacing in the $z$-direction; $\tilde{h}$ is the quantity to be solved. 
$\mathbf{d}$, $\mathbf{d}_{\text{1L}}$ and $\mathbf{d}_{\text{gb}}$ represent the displacements of the black lattice with respect to the white lattice parallel to the projection of $\mathbf{b}$ to the GB plane [i.e., in the direction of $(\mathbf{I}-\mathbf{n}\otimes\mathbf{n})\mathbf{b}$]; the definition of these displacements is given below. 
$\mathbf{d}_{\text{gb}}$ is the minimum displacement that preserves CSL; $\mathbf{d}_{\text{1L}}$ is the displacement which can induce a one-layer upward shift of the GB plane; $\mathbf{d}$ is the displacement that results in the same dichromatic pattern as $\mathbf{b}$ does. 
$N$ is any integer such that $\tilde{h}$ is also an integer. 
Equation~\eqref{d} can be understood as following. 
The left-hand side of Eq.~\eqref{d} is the displacement required by the introduction of $\mathbf{b}$.  
In order to produce the same displacement, we need to implement the displacement $\mathbf{d}_{\text{1L}}$ by $\tilde{h}$ times. 
Since any displacement by $\mathbf{d}_{\text{gb}}N$ does not change CSL (and, thus, the GB position), 
the term $\mathbf{d}_{\text{gb}}N$ leads to multiple solutions of $\tilde{h}$. 

In order to solve Eq.~\eqref{d} for $\tilde{h}$, we firstly simplify this equation. 
Since all the vectors in the equation are parallel, we ignore the vector direction ($\mathbf{d}\rightarrow d$, $\mathbf{d}_{\text{1L}}\rightarrow d_{\text{1L}}$, and $\mathbf{d}_{\text{gb}}\rightarrow d_{\text{gb}}$). 
All the displacements can be scaled by the interlayer spacing in the $y$-direction such that the dimensionless displacements are integers ($\tilde{d} = d/a_y$, $\tilde{d}_{\text{1L}} = d_{\text{1L}}/a_y$, and $\tilde{d}_{\text{gb}} = d_{\text{gb}}/a_y$). 
$\tilde{d}$ can be decomposed into the contribution from the component of $\mathbf{b}$ parallel to the GB plane, $\tilde{b}_{\parallel} = |(\mathbf{I}-\mathbf{n}\otimes\mathbf{n})\mathbf{b}|/a_y$, and the contribution from the component perpendicular to the GB plane, $\tilde{b}_{\perp} = \mathbf{b}\cdot\mathbf{n}/a_z$. 
$\tilde{b}_{\parallel}$ contributes to $\tilde{d}$ directly; while $\tilde{b}_{\perp}$ contributes to $\tilde{d}$ by $\tilde{b}_{\perp}\tilde{d}_{\text{u}}$, where the factor $\tilde{d}_{\text{u}}$ denotes the equivalent displacement parallel to the boundary plane (scaled by $a_y$) when $\tilde{b}_{\perp} = 1$. 
Hence, after the simplification mentioned above, Eq.~\eqref{d} becomes 
\begin{equation}
\tilde{d}
= \tilde{b}_{\parallel} + \tilde{b}_{\perp} \tilde{d}_{\text{u}}
= \tilde{d}_{\text{1L}} \tilde{h} + \tilde{d}_{\text{gb}} N, 
\label{tilded}
\end{equation}
where all quantities are integers. 
Eq.~\eqref{tilded} is a linear Diophantine equation in $\tilde{h}$ and $N$. 
After scaling $\tilde{d}$, $\tilde{d}_{\text{1L}}$ and $\tilde{d}_{\text{gb}}$ by $\text{gcd}[\tilde{d}_{\text{gb}}, \tilde{d}_{\text{1L}}]$ ($\text{gcd}[A,B]$ denotes the greatest common divisor of $A$ and $B$), we can obtain one solution to Eq.~\eqref{tilded}: 
\begin{equation}
\tilde{h}_0 
= \pm \text{mod}\left[
|\tilde{d}|
\tilde{d}_{\text{1L}}^{\Phi(\tilde{d}_{\text{gb}})-1}, 
\tilde{d}_{\text{gb}}
\right], 
\label{h0}
\end{equation}
where $\text{mod}[A,B]$ denotes $A$ modulo $B$, and $\Phi(n)$ is Euler's totient function which returns the number of positive integers that are less than and prime to $n$ ($n$ is an integer). 
We take the positive sign in Eq.~\eqref{h0} when $\tilde{d}>0$ and the negative sign when $\tilde{d}<0$. 
We also know that the difference between two neighboring solutions is 
\begin{equation}
\Delta\tilde{h} = \tilde{d}_{\text{gb}}. 
\label{Deltah}
\end{equation} 
Therefore, the complete set of the solutions to Eq.~\eqref{tilded} is 
\begin{equation}
\tilde{h}_j = \tilde{h}_0 + j\Delta\tilde{h}, 
\end{equation}
where $j$ is any integer. 

In principle, we can obtain the step heights $\{h_j\}$ for all possible $\mathbf{b}$ following the above algorithm. 
However, it is not necessary to calculate the step heights for all possible $\mathbf{b}$; we only need to know the step heights for the bases of the DSC lattice. 
The step heights of the other $\mathbf{b}$ are just combinations of them. 
If the bases of the DSC lattice are $\mathbf{b}^{(1)}$, $\mathbf{b}^{(2)}$ and $\mathbf{b}^{(3)}$, then any DSC Burgers vector can by represented by 
\begin{equation}
\mathbf{b}_n 
\equiv \mathbf{b}_{(q_1,q_2,q_3)}
= q_1 \mathbf{b}^{(1)} + q_2 \mathbf{b}^{(2)} + q_3 \mathbf{b}^{(3)}, 
\label{anyb}
\end{equation}
where $q_1$, $q_2$ and $q_3$ are integers; $n$ is an integer which is a compact notation for $(q_1, q_2, q_3)$. 
Since the step height is additive, the step height for the $\mathbf{b}_n$ in Eq.~\eqref{anyb} can be obtained by 
\begin{equation}
\tilde{h}_{nj} 
= q_1 \tilde{h}_0^{(1)} + q_2 \tilde{h}_0^{(2)} + q_3 \tilde{h}_0^{(3)} + j \Delta\tilde{h}, 
\label{anyh}
\end{equation}
where $\tilde{h}_0^{(1)}$, $\tilde{h}_0^{(2)}$ and $\tilde{h}_0^{(3)}$ are the step heights for $\mathbf{b}^{(1)}$, $\mathbf{b}^{(2)}$ and $\mathbf{b}^{(3)}$, respectively, and $\Delta\tilde{h}$ is the step spacing for $\mathbf{b}^{(1)}$ (as with $\mathbf{b}^{(2)}$ and $\mathbf{b}^{(3)}$). 
In summary, in order to obtain a complete list of the disconnection modes for a GB with particular geometry, we firstly need to figure out the bases of the DSC lattice $\mathbf{b}^{(1)}$, $\mathbf{b}^{(2)}$ and $\mathbf{b}^{(3)}$. 
Then, calculate their step heights $\tilde{h}_0^{(1)}$, $\tilde{h}_0^{(2)}$ and $\tilde{h}_0^{(3)}$ by Eq.~\eqref{h0} and the step spacing $\Delta\tilde{h}$ by Eq.~\eqref{Deltah}.  
Finally, the disconnection modes are ${(\mathbf{b}_n, h_{nj})}$, where $\mathbf{b}_n$ is constructed by Eq.~\eqref{anyb} and the corresponding $h_{nj}$ is obtained by Eq.~\eqref{anyh}. 

The practical algorithm for determining the disconnection modes for various geometries of STGBs in FCC crystals is provided in \ref{APPalgorithm}. 

\subsection{Connection between disconnections and grain-boundary kinetics}
We have established an approach to list all possible disconnection modes for any bicrystallography in the previous section. 
Next, we consider what we can do with such a list of disconnection modes. 
It has been observed experimentally that the formation and dynamics of various disconnections played an important role in different GB kinetic behaviors. 
For example, disconnections exist along thermally equilibrated GBs~\cite{merkle2002thermally,merkle2004situ}. 
The motion of disconnections was observed along the GB of a shrinking embedded cylindrical grain, suggesting that capillarity driven GB migration is facilitated by the motion of disconnections~\cite{radetic2012mechanism,dahmen2016high}. 
The motion of disconnections was also observed along the GBs in a polycrystal during a stress induced grain growth process, indicating that stress driven GB migration is closely related to the motion of disconnections~\cite{rajabzadeh2014role,rajabzadeh2013evidence}. 
Disconnection motion was also observed along the GB during GB sliding at elevated temperature~\cite{hosseinian2016quantifying}. 
The interaction between lattice dislocations and GBs were shown to involve the reaction and motion of disconnections~\cite{bollmann1972pseudo,dingley1979interaction,wise1988tem,cheng2008atomistic,shen1988dislocation,sangid2011energy,kacher2012quasi}. 
The ultimate aim of the disconnection model is to connect the set of disconnection modes to the understanding of all of the above GB kinetic behaviors. 
In order to reach this aim, we still need a model to describe the energetics of each disconnection mode, as follows in the next section. 

\subsection{Energetics of a disconnection mode}

\subsubsection{2D model}
\begin{figure}[!t]
\begin{center}
\scalebox{0.24}{\includegraphics{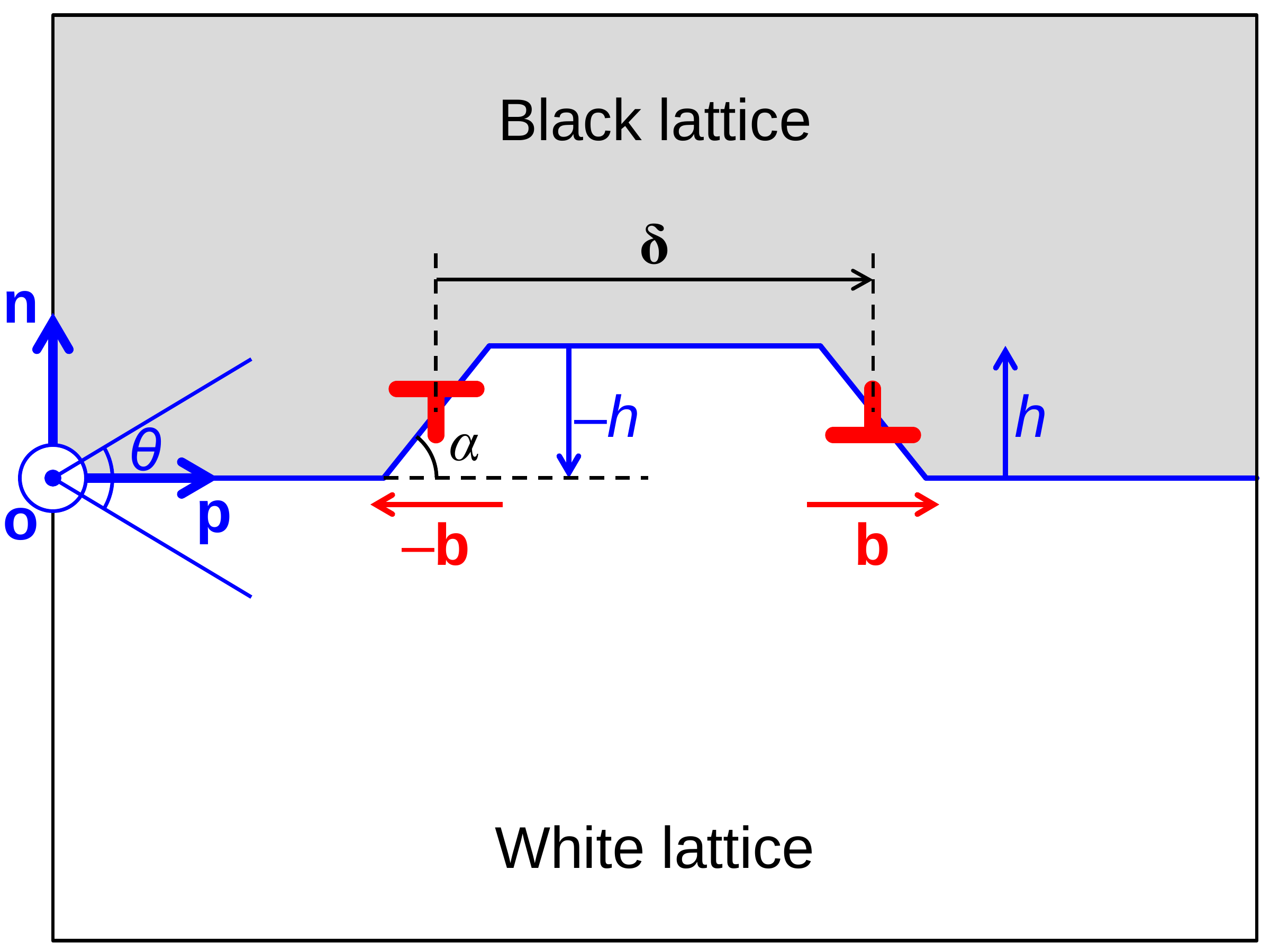}}
\caption{A bicrystal model containing a pair of disconnections.  } 
\label{energetics}
\end{center}
\end{figure}

The energy (per unit length) for the formation of a pair of disconnections,
$(\mathbf{b},h)$ and $(-\mathbf{b},-h)$ (since the analysis below is valid for any disconnection mode, the subscripts ``$n$'' and ``$nj$'' are omitted), 
as shown in Fig.~\ref{energetics}, 
can be partitioned as~\cite{HirthLothe,khater2012disconnection,mackain2017atomic} 
\begin{equation}
E 
= 2E_{\text{step}} + 2E_{\text{core}} + E_{\text{int}}, 
\label{Etot}
\end{equation}
where $E_{\text{step}}$ is the energy of each step, $E_{\text{core}}$ is the core energy of each disconnection, and $E_{\text{int}}$ is the elastic interaction energy between the disconnections. 
The energy of each step can be expressed as 
\begin{equation}
E_{\text{step}}
= \mathit{\Gamma}_\text{s}|h|, 
\label{Ese}
\end{equation}
where $\mathit{\Gamma}_\text{s}$ is the excess energy per area due to the introduction of the step. 
Based on the configuration shown in Fig.~\ref{energetics}, $\mathit{\Gamma}_\text{s}$ can be approximately estimated as~\cite{king1980effects}  
\begin{equation}
\mathit{\Gamma}_\text{s} 
= (\gamma_{\text{s}} - \gamma \cos{\Theta_\text{s}})/\sin{\Theta_\text{s}}, 
\end{equation}
where $\gamma$ is the interfacial energy of the reference boundary (assumed disconnection-free), $\gamma_{\text{s}}$ is the interfacial energy of the boundary that constitutes the step, and $\Theta_\text{s}$ is the inclination angle between the pristine boundary plane and the step plane. 
This is similar to the proposal by King~\cite{king1981properties}, except that in the excess energy due to the increase in GB area as suggested by Eq.~\eqref{Ese}, there are elastic interaction and core energies due to the dislocation dipoles  located at the step junctions. 
The relaxation of a GB structure usually results in relative displacement of two grains. 
The relative displacements for the pristine and the step boundaries usually differ from each other. 
The mismatch of the relative displacements at each step junction between the pristine and the step boundaries necessarily results in a dislocation with Burgers vector equal to the difference between the relative displacements. 
The elastic interaction energy between the two dislocations at the junctions varies logarithmically with $|h|$ while the core energy is independent of $h$. 
Here we ignore the energy attributed to the dislocation dipoles at the step junctions. 
Next, we consider the remaining terms in Eq.~\eqref{Etot}, i.e., $2E_\text{core}+E_\text{int}$. 
The elastic interaction energy between the disconnections can be expressed as~\cite{nabarro1952mathematical,HirthLothe}
\begin{equation}
E_{\text{int}}
= 2K b^2 \left[
\mathsf{f}(\alpha) \ln{\frac{\delta}{\delta_\text{c}}}
+ \frac{(\hat{\mathbf{b}}\cdot \mathbf{n})^2}{1-\nu}
\right], 
\label{Eint}
\end{equation}
where $K\equiv \mu/4\pi$ ($\mu$ is shear modulus), $\delta$ is the separation between the disconnections, $\hat{\mathbf{b}} \equiv \mathbf{b}/b$, $\delta_\text{c}$ is the disconnection core size, and 
\begin{equation}
\mathsf{f}(\alpha) \equiv \cos^2{\alpha} + \frac{\sin^2{\alpha}}{1-\nu} 
\end{equation}
($\nu$ is Poisson ratio and $\alpha \in [0,90^{\circ}]$ is the angle between $\mathbf{b}$ and $\mathbf{o}$). 
The determination of the core size $\delta_\text{c}$ and the core energy $E_{\text{core}}$ requires atomistic simulations. 
We can include the core energy by replacing $\delta_\text{c}$ in Eq.~\eqref{Eint} with the effective core size $\delta_0$~\cite{HirthLothe}. 
Therefore, the total energy for the formation of a pair of disconnections can be formulated as 
\begin{equation}
E
= 2\mathit{\Gamma}_\text{s}|h| + 2Kb^2\left[\mathsf{f}(\alpha)\ln\frac{\delta}{\delta_0} + \frac{(\hat{\mathbf{b}}\cdot \mathbf{n})^2}{1-\nu}\right]. 
\label{Etot2}
\end{equation}

\subsubsection{Effect of periodic boundary condition}
\label{2D_pbc}
In most simulations based on a bicrystal configuration, periodic boundary condition is applied along the $x$- and $y$-axes (e.g. Refs.~\cite{rajabzadeh2013elementary,mackain2017atomic}). 
We particularly focus on the glissile disconnections which have the Burgers vector parallel to the GB plane; thus, we ignore the last term in the square brackets of Eq.~\eqref{Etot2}. 
The energy of a pair of disconnections, with periodic boundary condition applied, is 
\begin{equation}
E 
= 2\mathit{\Gamma}_\text{s}|h| + 2Kb^2\mathsf{f}(\alpha) \ln\left|\frac{\sin(\pi\delta/L_y)}{\sin(\pi\delta_0/L_y)}\right|,  
\label{Etot_pbc}
\end{equation}
where $L_y$ is the period of the simulation supercell along the $y$-axis. 
Obviously, as $L_y \to \infty$, Eq.~\eqref{Etot_pbc} becomes Eq.~\eqref{Etot2}. 
The critical separation of the disconnections is $\delta^* = L_y/2$, and the energy barrier is 
\begin{equation}
E^*
= 2\mathit{\Gamma}_\text{s}|h| - 2Kb^2\mathsf{f}(\alpha) \ln\sin\frac{\pi\delta_0}{L_y}
\approx 2\mathit{\Gamma}_\text{s}|h| + 2Kb^2\mathsf{f}(\alpha) \ln\frac{L_y}{\pi\delta_0},  
\label{Etot_pbc_star}
\end{equation}
where the approximation is valid when $\delta_0/L_y \ll 1$. 

\subsubsection{$(2+\varepsilon)$D model}
\label{2plusD_pbc}
\begin{figure}[!t]
\begin{center}
\scalebox{0.35}{\includegraphics{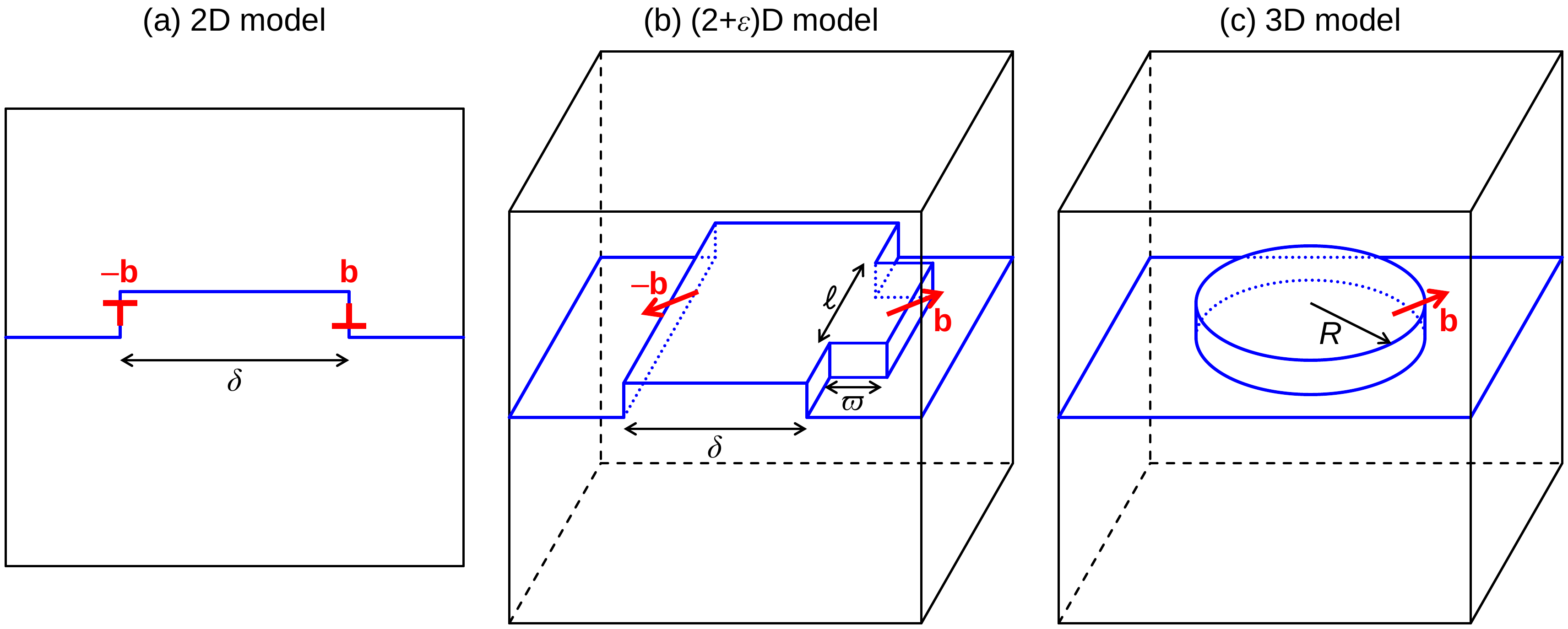}}
\caption{(a) A pair of disconnections on a 1D GB in a 2D bicrystal. 
(b) A pair of disconnections on a 2D GB in a 3D bicrystal. 
The disconnection moves by the nucleation of a pair of kinks. 
(c) A disconnection loop on a 2D GB in a 3D bicrystal. } 
\label{nDmodel}
\end{center}
\end{figure}

In the above 2D model, each disconnection is supposed to glide as a straight line. 
However, in reality, the glide of a disconnection is through the nucleation of a pair of kinks and the propogation of each kink along the disconnection line. 
The model explicitly including the mechanism of kink nucleation and propogation is called $(2+\varepsilon)$D model, as illustrated in Fig.~\ref{nDmodel}b.  
The attractive force between a pair of kinks is~\cite{HirthLothe} 
\begin{equation}
F_\text{k-int}
= \frac{1}{2} K \mathsf{f}_1(\alpha) b^2 \frac{\varpi^2}{\ell^2}, 
\label{Fkint}
\end{equation}
where $\ell$ is the separation between two kinks, $\varpi$ is the size of each kink, and 
\begin{equation}
\mathsf{f}_1(\alpha)
\equiv \frac{1+\nu}{1-\nu}\cos^2\alpha + \frac{1-2\nu}{1-\nu}\sin^2\alpha.  
\end{equation}
Different from the long-range interaction between two disconnections, which varies with the separation as $\delta^{-1}$, the interaction between two kinks varies with the separation as $\ell^{-2}$, which is of short-range. 
If periodic boundary condition is applied along the $x$-axis, Eq.~\eqref{Fkint} will become
\begin{equation}
F_\text{k-int}
= \frac{1}{2} K \mathsf{f}_1(\alpha) b^2 \varpi^2
\left[\frac{1}{\ell^2} - \frac{1}{(L_x-\ell)^2}\right], 
\end{equation}
Then, the elastic interaction energy is 
\begin{equation}
E_\text{k-int}
= \int_0^\ell F_\text{k-int}(\ell') \text{d}\ell'
= -\frac{1}{2} K b^2 \mathsf{f}_1(\alpha)  \frac{\varpi^2L_x}{\ell(L_x-\ell)}. 
\end{equation}
The energy for the formation of a pair of kinks is 
\begin{equation}
E_\text{k}
= 2\varpi\left\{ \mathit{\Gamma}_\text{k}|h| + Kb^2\left[\mathsf{f}_2(\alpha)\ln\frac{2\varpi}{e\delta_0} - \mathsf{f}(\alpha)\right] \right\}
-\frac{1}{2} K b^2 \mathsf{f}_1(\alpha) \frac{\varpi^2L_x}{\ell(L_x-\ell)}, 
\label{Ek}
\end{equation}
where $\mathit{\Gamma}_\text{k}$ is the excess energy per area located at the kink and 
\begin{equation}
\mathsf{f}_2(\alpha)
\equiv \frac{\cos^2\alpha}{1-\nu} + \sin^2\alpha. 
\end{equation}
The second term in the curly brackets in Eq.~\eqref{Ek} represents the self-energy of a single kink associated with dislocation character. 
The critical separation of the kinks is $\ell^* = L_x/2$, and the energy barrier is $E_\text{k}^* = E_\text{k}(\ell = L_x/2)$. 

The total energy for the formation of a pair of kinks on one of the pair of disconnections is 
\begin{align}
E_{(2+\varepsilon)\text{D}} 
&= 2L_x\left[ \mathit{\Gamma}_\text{s}|h| + Kb^2\mathsf{f}(\alpha) \ln\left|\frac{\sin(\pi\delta/L_y)}{\sin(\pi\delta_0/L_y)}\right| \right]
\nonumber \\
&+ 2\varpi\left\{ \mathit{\Gamma}_\text{k}|h| + Kb^2\left[\mathsf{f}_2(\alpha)\ln\frac{2\varpi}{e\delta_0} - \mathsf{f}(\alpha)\right] \right\}
\nonumber \\
&- \frac{1}{2} K b^2 \mathsf{f}_1(\alpha) \frac{\varpi^3}{(\delta-\mathcal{N}\varpi)[(\mathcal{N}+1)\varpi-\delta]}, 
\label{E2eD}
\end{align}
where we have applied the relation $\ell/L_x = (\delta-\mathcal{N}\varpi)/\varpi$ ($\mathcal{N}$ is the integer part of $\delta/\varpi$). 
The energy barrier is 
\begin{equation}
E_{(2+\varepsilon)\text{D}}^* 
= L_x E_\text{2D}^* + E_\text{k}^*
= A|\tilde{h}| + B \tilde{b}^2, 
\label{E2+star}
\end{equation}
where $A$ and $B$ are parameters. 
As $\varpi/L_x\ll 1$, $A = 2 (\mathit{\Gamma}_\text{s}+\mathit{\Gamma}_\text{k}\varpi/L_x) L_xa_z \approx 2\mathit{\Gamma}_\text{s}L_xa_z$. 
As $\varpi/L_x\ll 1$ and $\delta_0/L_y\ll 1$, $B \approx 2L_xK a_y^2 \mathsf{f} \ln(L_y/\pi\delta_0)$. 
Parameter $B$ measures the importance of Burgers vector contributing to the disconnection energy in comparison with step height; in trend (not exactly), $B \propto \Sigma^{-1}$ for tilt GBs. 
This implies that, for low-$\Sigma$ tilt GBs (e.g. coherent twin boundaries), the character of Burgers vector is important and the disconnection mode associated with small $\mathbf{b}$ is favorable in nucleation while, for high-$\Sigma$ tilt GBs, the charactor of GB step dominates the disconnection energy and the mode with small $h$ is favorable in nucleation. 

\subsubsection{3D model}
\label{3D_pbc}
In practice, disconnection is nucleated in an infinite flat GB in form of disconnection loop~\cite{wan2013shear,hadian2016atomistic} rather than a pair of parallel disconnections. 
The 3D model which describes disconnection mechanism as the nucleation and growth of disconnection loop is illustrated in Fig.~\ref{nDmodel}c. 
The energy of a circular disconnection loop as a function of radius $\delta$ can be written as 
\begin{equation}
E_\text{3D}
= 2\pi \delta \left[\bar{\mathit{\Gamma}}_\text{s}|h|
+ \frac{2-\nu}{2(1-\nu)} K b^2 \left(\ln\frac{8\delta}{\delta_0} - 2\right) \right], 
\end{equation}
where $\bar{\mathit{\Gamma}}_\text{s}$ is the average excess step energy over the disconnection loop. 

Although there is no simple analytical solution to the disconnection loop energy when periodic boundary condition is applied along the $x$-and $y$-axes, it is still reasonable to assume that a critical radius $\delta^*$ exists and the energy barrier has the approximated form of Eq.~\eqref{E2+star}. 
Therefore, no matter which type of model is employed, we always assume that the energy barrier for the formation of the disconnection $(\mathbf{b}_n, h_{nj})$ has the form: 
\begin{equation}
E^*_{nj} = A|\tilde{h}_{nj}| + B \tilde{b}_n^2, 
\label{Estar_nj}
\end{equation}
where $A$ and $B$ are parameters independent of disconnection mode.

\subsubsection{Evidence from atomistic simulations}
\begin{figure}[!t]
\begin{center}
\scalebox{0.29}{\includegraphics{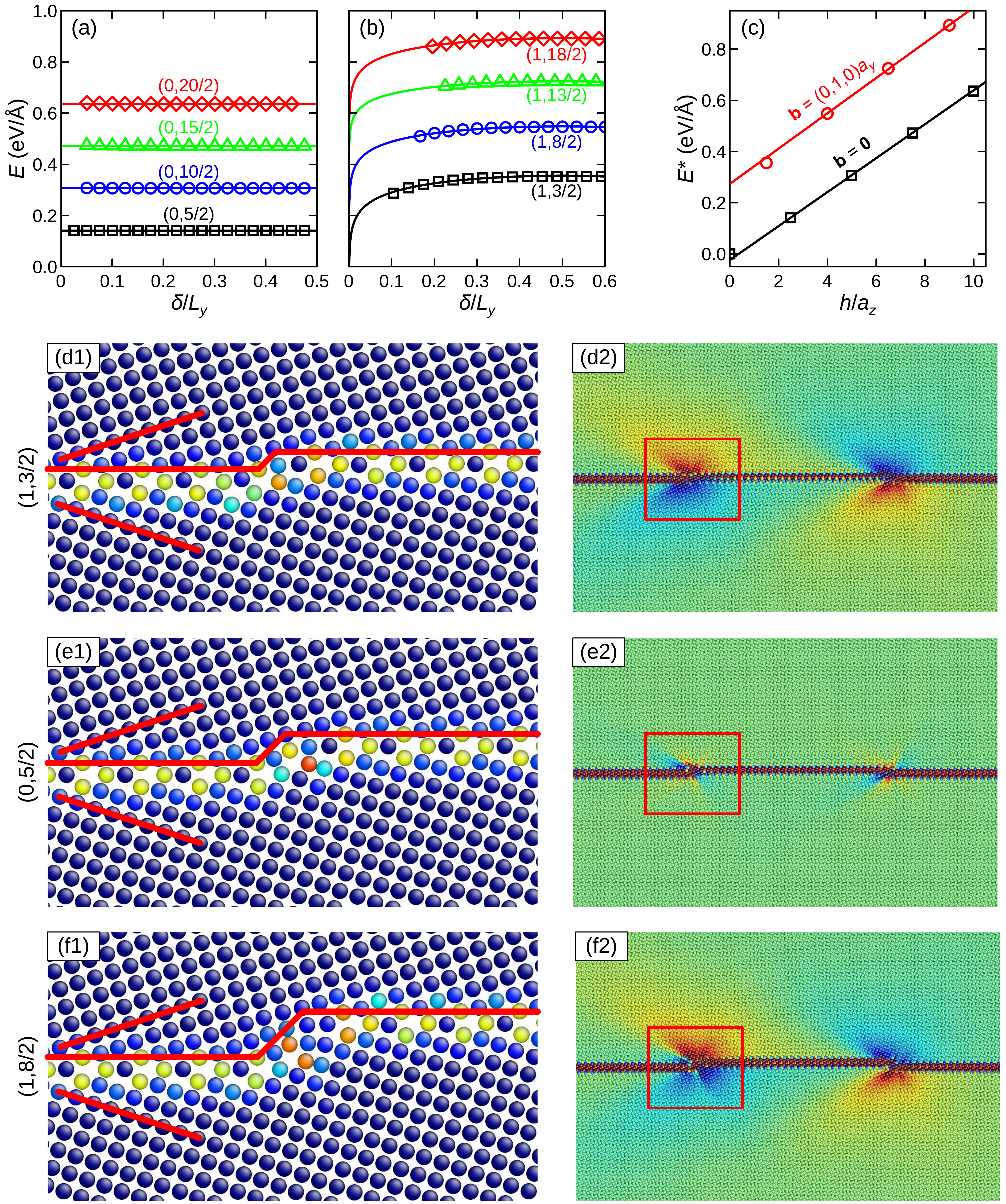}}
\caption{Molecular statics simulation results for a pair of disconnections on a $\Sigma 5$ $[100]$ $(013)$ STGB in FCC Cu. 
Energy vs. separation of a pair of disconnections (a) with zero Burgers vector and various step heights 
and (b) with Burgers vector $\mathbf{b}=(0,a_y,0)$ and various step heights in the 2D model shown in Fig.~\ref{nDmodel}a. 
This data is fitted by Eq.~\eqref{Etot_pbc}. 
(c) The energy barrier $E^*$ vs. step height. 
(d1), (e1) and (f1) show a single disconnection with the mode $(a_y, \frac{3}{2}a_z)$, $(0, \frac{5}{2}a_z)$ and $(a_y, 4a_z)$, respectively. 
The atoms are colored by the potential energy; the red lines are drawn to show the step height. 
(d2), (e2) and (f2) show a pair of disconnections corresponding to (d1), (e1) and (f1), respectively. 
The atoms are colored by the stress $\sigma_{yy}$ (see the coordinate system in Fig.~\ref{disconnection}).  } 
\label{MS_disconnection}
\end{center}
\end{figure}

The formulation of energy of each disconnection mode, derived in the previous section, has been checked by atomistic simulations. 
Based on the 2D model (Fig.~\ref{nDmodel}a) with periodic boundary condition, we calculated the formation energy for a pair of disconnections as a function of the disconnection separation $\delta$ in a $\Sigma 5$ $[100]$ $(013)$ STGB by the molecular statics method. 
The interaction between atoms is modeled by Mishin's embedded-atom-method (EAM) potential for Cu~\cite{mishin2001structural}. 
For this GB (with particular macroscopic DOFs), via the enumeration approach proposed in Section~\ref{enumeration}, we know that the list of disconnection modes with $\hat{\mathbf{b}}= \mathbf{p}$ can be expressed as $\{(b_n=na_y, h_{nj}=(\frac{3}{2}n+\frac{5}{2}j)a_z)\}$. 
Figure~\ref{MS_disconnection}d1 show the relaxed atomic structure for a disconnection $(b_1,h_{10})$. 
Clearly, we saw the character of a GB step and, from the stress field $\sigma_{yy}$ shown in Fig.~\ref{MS_disconnection}d2, we can confirm that there is a finite Burger vector (of an edge dislocation) located with this GB step. 
Figure~\ref{MS_disconnection}e1 show the relaxed structure for a disconnection $(b_0,h_{01})$, which is a pure step. 
From the stress field shown in Fig.~\ref{MS_disconnection}e2, we found that there was no obvious feature of dislocation. 
Figures~\ref{MS_disconnection}f1 and f2 show the result for a disconnection $(b_1,h_{11})$, which is characterized by both finite Burgers vector and step height. 

Figure~\ref{MS_disconnection}a shows the energy vs. separation for the modes $(b_0, h_{0j})$ ($j = 1,2,3,4$). 
These modes correspond to pure steps with different step height, so the energy is independent of the separation. 
We fitted the calculated nucleation barrier $E^*$ (i.e., the constant value of $E$) vs. $h$ by a linear relation and obtained the black line in Fig.~\ref{MS_disconnection}c. 
The black line has negligible intercept, which is consistent with Eq.~\eqref{Etot_pbc_star} (with $b = 0$).  
The slope of the black line gives the value of $2\mathit{\Gamma}_\text{s}\alpha_z$. 
Figure~\ref{MS_disconnection}b shows the results for the modes $(b_1, h_{1j})$ ($j = 0,1,2,3$). 
These modes are associated with identical finite Burgers vector, so the function $E(\delta)$ shows logarithmic dependence. 
By fitting the calculated $E^*(h)$ to a linear relation, we obtained the red line in Fig.~\ref{MS_disconnection}c. 
If Eq.~\eqref{Etot_pbc_star} (and, thus, Eq.~\eqref{Etot_pbc}) is valid, the slope of the red line will be the same as that of the black line; we find that this is approximately true. 
Such observation demonstrates that the simple partition of the total energy into the contributions due to Burgers vector and step height [suggested by Eq.~\eqref{Etot}] is an acceptable approximation. 

\begin{figure}[!t]
\begin{center}
\scalebox{0.33}{\includegraphics{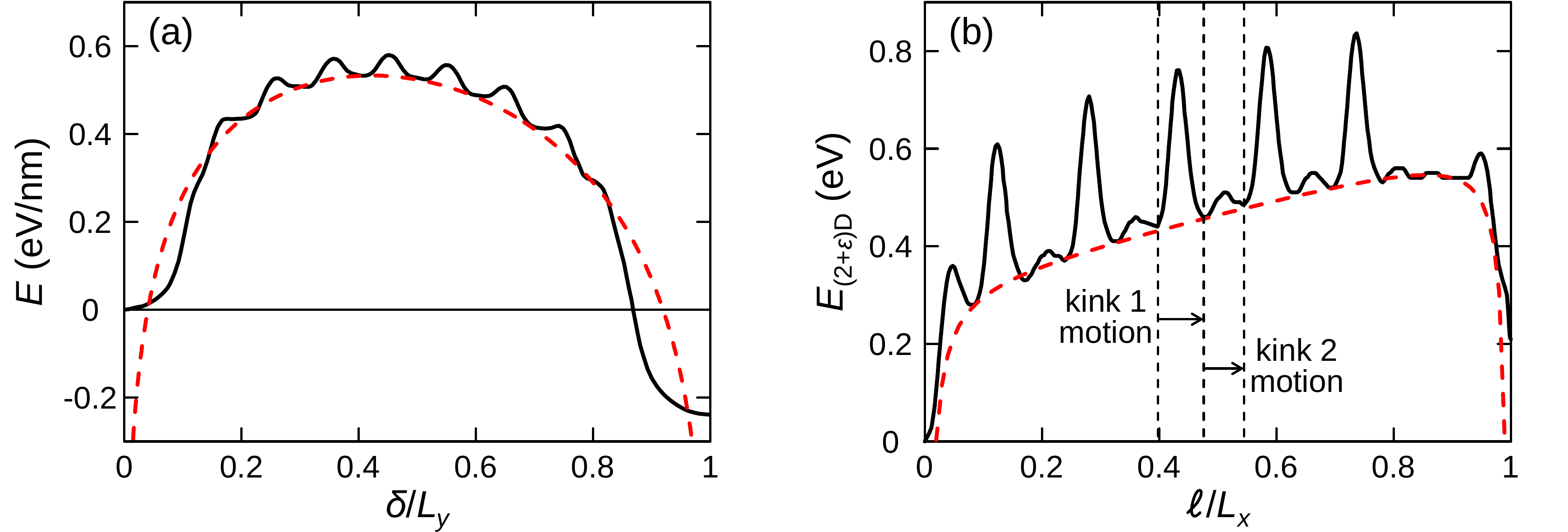}}
\caption{(a) Energy vs. separation (black solid line) of a pair of disconnections obtained by NEB calculation and fitted curve (red dotted line) according to Eq.~\eqref{Etot_pbc}. 
(a) is a replot in Fig.~4b in Ref.~\cite{rajabzadeh2013elementary} (N. Combe, American Physical Society 2013).
(b) Energy vs. separation (black solid line) of a pair of kinks along a disconnection obtained by NEB calculation and fitted curve (red dotted line) according to Eq.~\eqref{E2eD}. 
The higher barrier corresponds to the energy barrier for the motion of one kink and the lower barrier corresponds to the energy barrier for the motion of the other kink. 
(b) is a replot in Fig.~6 in Ref.~\cite{combe2016disconnections} (N. Combe, American Physical Society 2016). } 
\label{barrier}
\end{center}
\end{figure}

Rajabzadeh et al. also calculated the disconnection pair energy as a function of separation for a $\Sigma 13$ $[100]$ $(023)$ STGB in Cu based on a 2D model with periodic boundary condition~\cite{rajabzadeh2013elementary}. 
The disconnection mode under their consideration is $(2a_y, -5a_z)$. 
They found the barrier for glide of the disconnections by nudged elastic band (NEB) method~\cite{henkelman2000climbing}. 
Figure~\ref{barrier}a shows the result. 
Except for the small gliding barriers along the $E(\delta)$ curve, the data can be well fitted by Eq.~\eqref{Etot_pbc} plus a term representing the work done by the applied shear stress which is linear with $\delta$ (since, in this simulation, shear stress was applied to drive the glide of disconnections). 
The gliding barriers are attributed to the atomic shuffling mechanism mentioned in Section~\ref{microscopicDOFs} (in analogy to the Peierls barrier for the glide of a lattice dislocation~\cite{HirthLothe}). 

Explicitly considering kink propagation as the mechanism of disconnection motion, Combe et al. calculated the energy as a function of separation between two kinks along a disconnection for a $\Sigma 17$ $[100]$ $(014)$ STGB in Cu by NEB method based on a $(2+\varepsilon)$D model	 (Fig.~\ref{nDmodel}b)~\cite{combe2016disconnections}. 
Figure~\ref{barrier}b shows the result. 
$\ell/L_x = 0$ and $1$ correspond to two metastable configurations without kink (corresponding to two neighboring local minima in Fig.~\ref{barrier}a). 
The energies at $\ell/L_x = 0$ and $1$ are not identical since the separation between two disconnections $\delta$ is changing. 
Each higher peak along the $E_{(2+\varepsilon)\text{D}}(\ell)$ curve is the barrier for the motion of one kink; and each lower peak is the barrier for the motion of the other kink. 
It is expected that, if the kink mechanism is considered, the barrier between the metastable states (two neighboring local minia in Fig.~\ref{barrier}a) will be lower than that obtained from the 2D model.

%%%%%%%%%%%%%%%%%%%%%%%%%%%%%%%%%%%%%%%%%%%%%%%%%%%%%%%
\section{Thermal equilibrium of grain boundaries}
\label{sec4}

In this section, we will establish a unified description of thermally equilibrated GBs from the disconnection model. 
Two properties associated with thermal equilibrium of GBs will be reviewed based on this unified description; these properties are GB roughness and intrinsic GB mobility. 

\subsection{Description of a thermally equilibrated grain boundary}
\label{description}
Earlier understanding of thermally equilibrated GB structure was built in analogy to free surfaces. 
The structure of a thermally equilibrated free surface is usually described by terrace-ledge-kink (TLK) model~\cite{kossel1927theorie,burton1951growth,mullins1963microscopic}. 
In this model, a free surface at finite temperature features a distribution of steps on the surface and kinks along each step. 
Then, the evolution of the surface profile is carried out by the motion of the surface steps and kinks. 
The TLK model was simply extended to describe the thermally equilibrated GB structure by Gleiter~\cite{gleiter1969mechanism}. 
However, unlike the case of a free surface, the character of defects in a GB (or interface in a crystalline material) should be constrained by the bicrystallography. 
From Section~\ref{GB_disconnection}, we know that the line defect in a GB is not necessarily a step, as described in Gleiter's model, but a disconnection. 
Unlike a step on a free surface, a disconnection is characterized by not only step height but a Burgers vector as well. 
Therefore, based on the disconnection model, a GB at finite temperature features a distribution of disconnections (combined steps and DSC dislocations) on the GB and kinks along each disconnection; accordingly, the GB profile evolves via the motion of the disconnections and kinks. 

A thermally equilibrated GB can, in general, be described by a height function $z(x,y,t)$ [i.e., the height at the position $(x,y)$ at time $t$] and, at the same time, a relative displacement function $\mathbf{u}(x,y,t)$ [i.e., the relative displacement of the upper grain with respect to the lower one, referring to a bicrystal configuration such as Fig.~\ref{DSC}a, at the position $(x,y)$ at time $t$]. 

\begin{figure}[!t]
\begin{center}
\scalebox{0.4}{\includegraphics{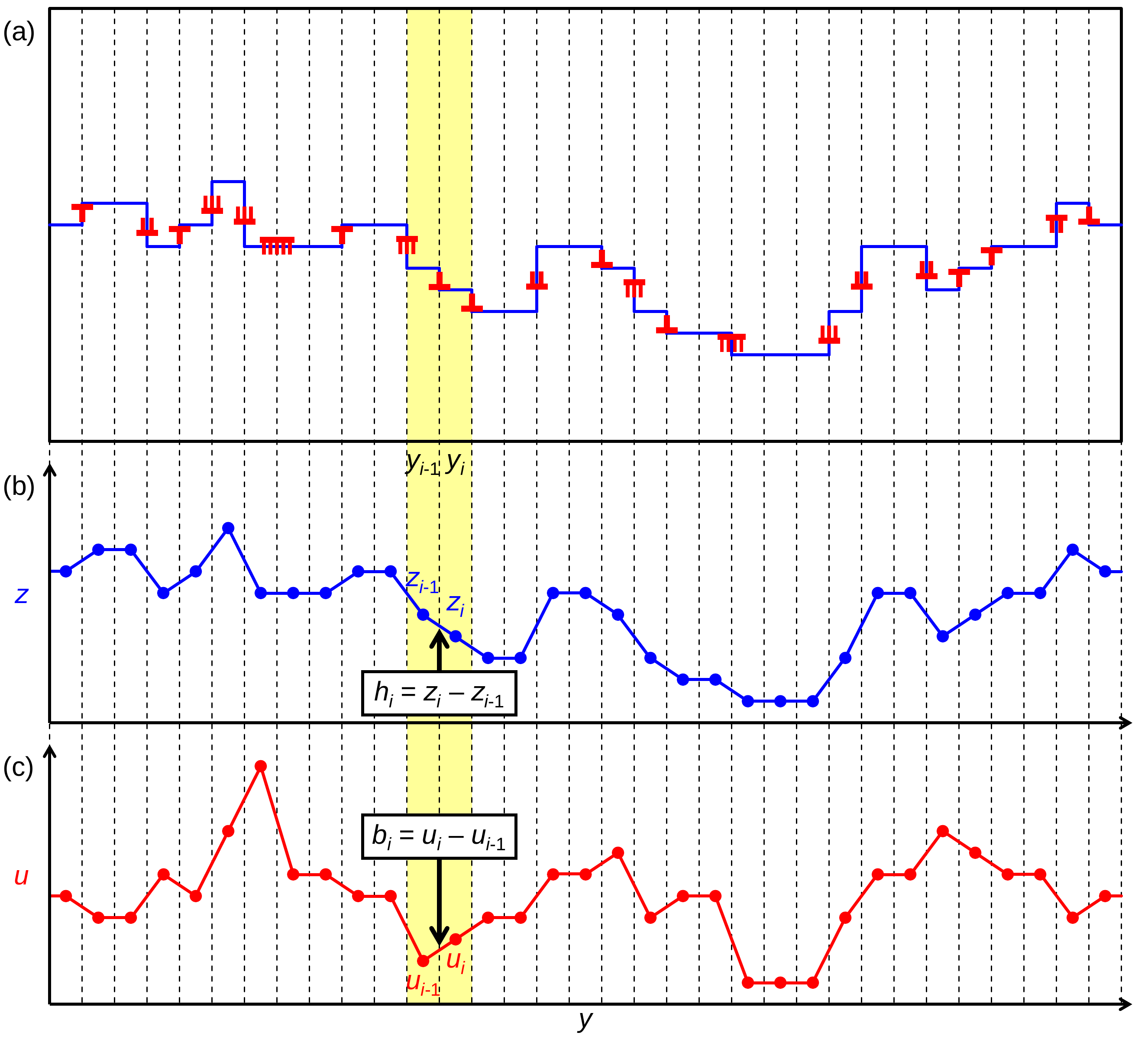}}
\caption{(a) Discrete description of a flat 1D GB lying along the $y$-axis in a 2D bicrystal, where the discretization is denoted by the dashed lines. The blue line represents the GB profile and the red symbols denote the dislocation character associated with the disconnections (the number of the vertical lines on each red symbol denotes the magnitude of the Burgers vector). Each disconnection is located at one dashed line. 
(b) The discrete GB height function $z(y)$, which is consistent with (a). The height of the $i$-th GB segment is represented by $z_i = z(y_i)$ and the step height of the $i$-th disconnection (between the $(i-1)$-th and the $i$-th GB segments) is $h_i = z_i - z_{i-1}$. 
(c) The discrete spatial distribution of relative displacement $u(y)$, which is consistent with (a). The relative displacement of the $i$-th GB segment is represented by $u_i = u(y_i)$ and the Burgers vector of the $i$-th disconnection is $b_i = u_i-u_{i-1}$. } 
\label{GBdescription}
\end{center}
\end{figure}

At microscopic scale, the height and relative displacement functions are not continuous in space. 
A discrete structural model of a GB is schematically depicted in Fig.~\ref{GBdescription}; it shows a snapshot of the GB at finite temperature in a 2D bicrystal [this GB is one-dimensional and described by $z(y)$ and $u(y)$]. 
The entire GB can be divided into pieces (by the dashed line as shown in Fig.~\ref{GBdescription}a) with the size $\Delta y$; $\Delta y$ is not smaller than twice the disconnection core size (corresponding to the highest resolution); the position of the $i$-th piece of the GB is denoted by $y_i$. 
For this GB configuration, the discretized $z(y)$ and $u(y)$ functions are shown in Figs.~\ref{GBdescription}b and \ref{GBdescription}c, respectively. 
Each disconnection is located between two neighboring pieces (if the states of two pieces are different). 
The step height of the disconnection between the $(i-1)$-th and the $i$-th pieces is $h_i = z_i-z_{i-1}$ [$z_i \equiv z(y_i)$]; the Burgers vector of the disconnection between the $(i-1)$-th and the $i$-th pieces is $b_i = u_i-u_{i-1}$ [$u_i \equiv u(y_i)$]. 

At mesoscopic scale, the height and the relative displacement functions can be treated as continuous functions in space. 
For a 1D GB in 2D bicrystal, both $z(y)$ and $u(y)$ become continuous curves. 
Then, the step height density is $\varrho^h(y') = \lim_{\Delta y \to 0}h_i/\Delta y =  (\text{d}z/\text{d}y)_{y'}$ ($y'$ is within the interval between $y_{i-1}$ and $y_i$). 
The Burgers vector density is $\varrho^b(y') = \lim_{\Delta y \to 0}b_i/\Delta y =  (\text{d}u/\text{d}y)_{y'}$. 
In general, for a 2D GB in 3D bicrystal, the step height tensor (of 1st order) is defined as 
\begin{equation}
\boldsymbol{\varrho}^h
= \left(\begin{array}{c}
\partial z/\partial y \\ \partial z/\partial x
\end{array}\right),  
\label{varrhoh}
\end{equation}
where $\varrho^h_m$ ($m$ stands for $x$ or $y$) is the net step height of disconnections threading through a line element of unit length lying in the boundary plane and perpendicular to the $m$-axis. 
The Burgers vector tensor (of 2nd order) is defined as 
\begin{equation}
\boldsymbol{\varrho}^b
= \left(\begin{array}{cc}
\partial u_x/\partial y	& \partial u_x/\partial x \\ 
\partial u_y/\partial y	& \partial u_y/\partial x
\end{array}\right),   
\label{varrhob}
\end{equation}
where $\varrho^b_{mn}$ ($m, n = x, y$) is the net Burgers vector in the $m$-direction of disconnections threading through a line element of unit length lying in the boundary plane and perpendicular to the $n$-axis. 
For example, if the only nonzero component of the relative displacement $\mathbf{u}$ is $u_y$ and the line element is perpendicular to the $x$-axis, then the Burgers vector density can be obtained by $\boldsymbol{\varrho}^b \mathbf{x} = (0, \partial u_y/\partial y)$, which reduces to the case of 1D GB in 2D bicrystal. 

In much of the literature, $z(x,y,t)$ alone is used to describe and analyze a thermally equilibrated GB. 
This is true of studies based on a solid-on-solid model~\cite{swendsen1977monte,hasenbusch1992cluster} or Ising model~\cite{hasenbusch1991cluster}. 
Only in a few cases, both $z(x,y,t)$ and $u(x,y,t)$ are explicitly considered. 
For example, Karma et al. analyticallly studied the equilibrium fluctuation of 1D GBs (in 2D bicrystal) by explicitly including the effect of both the height function and the relative displacement function and their coupling~\cite{karma2012relationship}. 
Through molecular dynamics (MD) simulation, they showed that, for most GBs in this study, the spectrum of GB fluctucation obeys $k^{-1}$ ($k$ is wave vector) behavior predicted by the model based on $z(y)$ and $u(y)$, rather than $k^{-2}$ behavior derived from the model only based on $z(y)$. 
Such observation demonstrates that it is not sufficent to describe thermally equilibrated GBs by height function alone (as suggested by TLK model or Gleiter's model); both height function and relative displacement function should be explicitly considered (consistent with the disconnection model). 

In the next sections, we will review two properties associated with thermal equilibrium of GBs: GB roughening and intrinsic GB mobility, using the novel model based on both height function and relative displacement function.

\subsection{Grain-boundary roughening}
\label{GB_roughening}

GB roughening usually refers to two distinct but related types of phenomenon. 
The first type is associated with asymmetric tilt GBs (ATGBs) which undergo decomposition into two sets of facets at low temperature. 
These faceted ATGBs will become flat either as temperature increases or as composition is changed. 
This is called a defaceting transition; it is also called a roughening transition in some literature~\cite{hsieh1989observations}.
This phenomenon was extensively observed in experiments for ATGBs in metals such as 
Al~\cite{hsieh1989observations}, 
Au~\cite{hsieh1989observations,goodhew1978low}, 
Ag~\cite{koo2001dependence}, 
Ni~\cite{lee2000grain}, 
Cu (Bi)~\cite{ference1988observation} and
stainless steel~\cite{choi2001temperature},
as well as ceramics such as 
Al$_2$O$_3$ (MgO)~\cite{park2000effects},  
BaTiO$_3$ (H$_2$)~\cite{lee2000grain} and
SrTiO$_3$~\cite{lee2003faceting}
(here, the elements in the brackets are impurities). 

The second type is in analogy to the roughening of singular surfaces. 
At low temperature, a singular surface is flat at atomic scale, since it corresponds to a local minimum of the surface energy vs. inclination of the surface plane. 
As temperature increases, the singular surface will become roughened at atomic scale (roughening refers to the enhancement of fluctuations of the surface profile); this phenomenon is also called a roughening transition. 
The roughening transition of surfaces was initially proposed by Burton, Cabrera and Frank~\cite{burton1949crystal,burton1951growth} based on an Ising model and observed in many experiments~\cite{yang1989high,robinson1991surface,kazantsev2017thermodynamic}. 
Similar to the case of singular surfaces, a singular GB is flat at atomic scale at low temperature. 
Here, a singular GB is one for which the energy is a local minimum with respect to the inclination angle of the GB plane for a fixed misorientation. 
As temperature increases, the singular GB will be roughened at atomic scale. 
The roughening transition of singular GBs was theoretically proposed for low-angle GBs by Rottman~\cite{rottman1986roughening,rottman1986thermal}. 
However, the direct experimental observation of such a phenomenon is very limited. 
Recently, Rajak et al. observed roughening of a $\Sigma 5$ $[100]$ $(013)$ STGB in SrTiO$_3$ in the experiment (measurement of the GB roughness was performed at the long facets of a $\Sigma 5$ $[100]$ ATGB)~\cite{rajak2016indication}.

The first type of phenomenon mentioned above is rigorously a defaceting transition rather than roughening transition. 
It is a ``non-congruent'' transition according to Cahn~\cite{cahn1982transitions,SuttonBalluffi}, since this transition involves change of inclination angles (the facets correspond to different inclination angles to that of the flat GB). 
The second type of phenomenon is rigorously a roughening transition, which is our interest in this paper. 
It is classified as a ``congruent'' transition according to Cahn~\cite{cahn1982transitions,SuttonBalluffi}, since the inclination angle does not change during this process. 
However, the roughening and defaceting transitions are related. 
One explanation for the defaceting transition is that, as temperature increases, the roughening transition occurring at the facets will eliminate the cusps along the GB energy vs. inclination and, thus, the flat GB will be associated with lower free energy than faceted one~\cite{rajak2016indication}. 
This was verified by the simultaneous occurrence of roughening transition and defaceting transition for both surfaces and GBs~\cite{robinson1991surface,hsieh1989observations,lee2007high}. 
In the next sections, we will focus on the roughening transition (i.e., the second type of phenomenon discussed above) rather than the defaceting transition (which is considered to originate from the roughening transition). 

The GB roughening transition will be reviewed based on the framework of the disconnection model. 
A thermally equilibrated GB is characterized by a distribution of disconnections and the GB profile is described by both the height function $z(x,y,t)$ and the relative displacement function $\mathbf{u}(x,y,t)$. 
Accordingly, the GB roughness is defined as 
\begin{equation}
w_z
= \sqrt{\left\langle \overline{z^2} - {\bar{z}}^2 \right\rangle_t}, 
\end{equation}
where $\bar{X} \equiv \int X(x,y,t) \mathrm{d}x\mathrm{d}y/\mathcal{A}$ (i.e., average over area) and $\langle X\rangle_t \equiv \int X(x,y,t) \mathrm{d}t/\mathcal{T}$ (i.e., average over time) for any quantity $X$ ($\mathcal{A}$ is total GB area and $\mathcal{T}$ is total time). 
Unlike the classical model of thermally equilibrated GBs which only relies on the height function $z(x,y,t)$ (such as the TKL model and Gleiter's model), based on the disconnection model, there is another kind of ``roughness'' which relates to the thermal vibration of one grain with respect to the other in any direction parallel to the GB plane (imagine that one grain is ``floating'' on the other); we simply call this ``roughness'' as GB $u$-roughness. 
The GB $u$-roughness is defined as 
\begin{equation}
w_u
= \sqrt{\left\langle \overline{\mathbf{u}^2} - {\bar{\mathbf{u}}}^2 \right\rangle_t}. 
\end{equation}
The $u$-roughness $w_u$ is usually of no practical interest because (i) it is to some extent prohibited in a polycrystal environment and (ii) it is hardly measured in experiments even in bicrystal systems. 
However, it may be of interest for a theoretical understanding since it represents a hidden DOF which may influence the other DOFs, which is of practical interest and can be measured. 

Below, we will see how the disconnection model helps to understand some features of GB roughening. 

\subsubsection{The simplest approach: Boltzmann statistics}
\label{simplest_Boltzmann}
First of all, we keep the model as simple as possible in order to gain a general idea about the implications of the disconnection model to GB roughening. 
We construct a toy model, which is a 2D bicrystal containing a CSL STGB with periodic boundary condition applied in any direction parallel to the GB plane. 
We consider that the GB profile is formed by the nucleation of disconnections of different modes (and, thus, different step heights). 
At low temperature, the GB is free of disconnections and, thus, atomically flat. 
As temperature increases, disconnections of different modes are nucleated such that the GB becomes roughened. 
The period of the bicrystal (which is artificially applied) is assumed to mimic the correlation length, beyond which disconnection pairs can be nucleated independently. 
In each period, the energy barrier for the nucleation of a particular disconnection mode $E_{nj}^*$ has the form of Eq.~\eqref{Estar_nj}; this form is reasonable for the 2D model (Section~\ref{2D_pbc}), $(2+\varepsilon)$D model (Section~\ref{2plusD_pbc}) and even 3D model (Section~\ref{3D_pbc}). 
The probability for the nucleation of the disconnection $(b_n, h_{nj})$ is proportional to the Boltzmann factor $\exp(-L_x E_{nj}^*/k_\text{B}T)$, where $E_{nj}^*= A|\tilde{h}_{nj}|+B\tilde{b}_n^2$ is the energy barrier for this disconnection mode in the 2D bicrystal model and is obtained by Eq.~\eqref{Etot_pbc_star}. 

In this toy model, the GB roughness can be estimated as 
\begin{equation}
w_z
= a_z\sqrt{\langle \tilde{h}^2 \rangle - \langle \tilde{h} \rangle^2}, 
\label{wz_Boltzmann}
\end{equation}
where the operation $\langle X\rangle$ represents ensemble average (weighted by the Boltzmann factor) of the quantity $X$ and is calculated by
\begin{equation}
\langle X\rangle 
= \frac{\displaystyle{ \sum_{n=-\infty}^\infty \sum_{j=-\infty}^\infty X_{nj} \exp\left(-\frac{A|\tilde{h}_{nj}|+B\tilde{b}_n^2}{k_\text{B}T/L_x}\right) }}
{\displaystyle{ \sum_{n=-\infty}^\infty \sum_{j=-\infty}^\infty \exp\left(-\frac{A|\tilde{h}_{nj}|+B\tilde{b}_n^2}{k_\text{B}T/L_x}\right) }}, 
\end{equation}
where the disconnection mode $(b_n, h_{nj})$ can be obtained by the enumeration method proposed in Section~\ref{enumeration}. 
Similarly, the GB $u$-roughness can be estimated as 
\begin{equation}
w_u
= a_y\sqrt{\langle \tilde{b}^2 \rangle - \langle \tilde{b} \rangle^2}.  
\label{wu_Boltzmann}
\end{equation}

\begin{figure}[!t]
\begin{center}
\scalebox{0.32}{\includegraphics{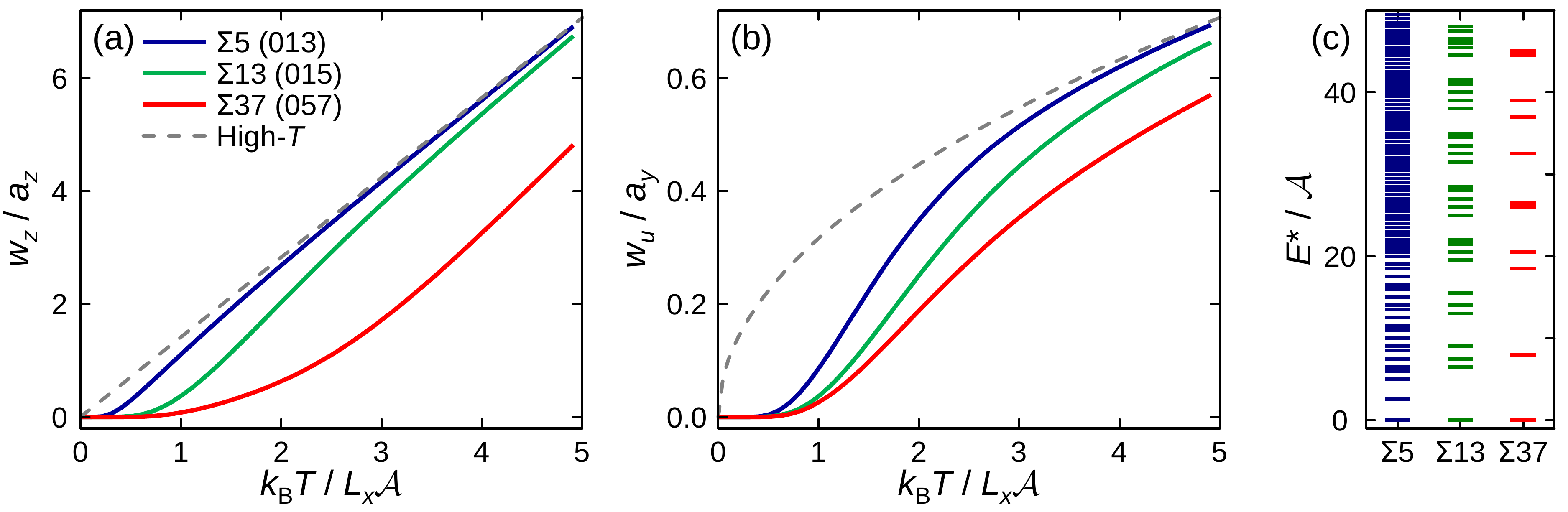}}
\caption{(a) and (b) GB roughness and $u$-roughness, respectively, as functions of temperature, obtained by the Boltzmann statistics of all disconnection modes for three STGBs in an FCC crystal. 
(c) The energy spectra for the three STGBs.  } 
\label{roughness_Boltzmann}
\end{center}
\end{figure}

We estimated the GB roughness and $u$-roughness according to Eq.~\eqref{wz_Boltzmann} and Eq.~\eqref{wu_Boltzmann}, respectively, for three different GB geometries: $\Sigma 5$ $[100]$ $(013)$, $\Sigma 13$ $[100]$ $(015)$ and $\Sigma 37$ $[100]$ $(057)$ STGBs in FCC crystals. 
The results are shown in Fig.~\ref{roughness_Boltzmann}. 
It can be derived that, as $T\to \infty$, $w_z/a_z \to \sqrt{2}T$ (gray dashed line in Fig.~\ref{roughness_Boltzmann}a) and $w_u/a_y \to \sqrt{A/2B}T^{1/2}$ (gray dashed line in Fig.~\ref{roughness_Boltzmann}b). 
We find that, at low temperature, the GB roughness $w_z$ clearly deviates from linear dependence on temperature. 
First, such deviation is primarily caused by the discrete energy spectrum of the disconnection modes. 
At low-temperature limit, the GB fluctuation is only dominated by the limited number of low-energy disconnection modes. 
However, at the high-temperature limit, the energy difference between different disconnection modes is not important and the behavior converges to the case for which the energy spectrum is a continuum. 
Second, we see that the deviation from the linear relation is GB-geometry dependent, since the input $\{(b_n, h_{nj})\}$ varies with GB geometry (or bicrystallography). 
This can be understood in terms of the discretization of the energy spectrum as shown in Fig.~\ref{roughness_Boltzmann}c. 
When the gap between the energy levels is large, the deviation of roughness from the high-temperature limit will be large, such as the case of $\Sigma 37$ GB; inversely, when the energy gap is small, the deviation will be small, such as the case of $\Sigma 5$ GB. 

Note that the toy model only provides a basic understanding of GB roughening based on the disconnection model. 
However, the assumptions adopted in this model are not necessarily reasonable. 
For example, the thermal fluctuation of the GB is assumed to be dominated by homogeneous nucleation of disconnection pairs. 
However, in reality, it may be carried out by the motion of disconnections or, equivalently, heterogeneous nucleation of disconnections (as discussed in Section~\ref{description}). 
In addition, the artificially applied periodic boundary condition is assumed to reflect the effect of correlation length; however, the correlation length should be temperature dependent and even diverge if the roughening transition exists as a phase transition~\cite{swendsen1977monte}. 
All of these problems can be solved by a more complicated model (i.e., a 1D lattice model) which will be discussed in the next section. 
Although the 1D lattice model is closer to the description of realistic thermally equilibrated GB than the simplest model mentioned above, it cannot be analytically solved and must be solved by Monte Carlo simulation. 

\subsubsection{Monte Carlo simulation}
\label{roughening_MC_simulation}
Now, we consider a 1D GB in a 2D bicrystal as depicted in Fig.~\ref{GBdescription}a. 
Periodic boundary condition is applied along the GB with period $L_y$. 
The 1D GB is discretized into $N$ cells (or segments) such that the size of each cell is $\Delta y = L_y/N$. 
The state of the $i$-th segment is denoted by $(z_i, u_i)$ ($i=1,\cdots,N$), where $z_i$ is the height of this GB segment and $u_i$ is the relative displacement of the upper grain with respect to the lower grain at this GB segment. 
When a disconnection characterized by the mode $(b,h)$ glides through this GB segment, the state of this segment will become $(z_i',u_i') = (z_i+h, u_i+b)$. 
As discussed in Section~\ref{multiple_modes}, for a particular bicrystallography, the disconnection can have any mode belonging to the list $\{(b_n, h_{nj})\}$. 
Therefore, the change of state $(b,h)$ is chosen from the list $\{(b_n, h_{nj})\}$. 

We then consider the energy difference caused by the change of state at the $i$-th GB segment. 
The energy change consists of three terms: 
\begin{equation}
\Delta E
= \Delta E_\text{step} + \Delta E_\text{core} + W_\tau.  
\label{energy_change}
\end{equation}
$\Delta E_\text{step}$ is the change of step energy: 
\begin{equation}
\Delta E_\text{step}
= \mathit{\Gamma}_\text{s} \left(
|z_i' - z_{i-1}| + |z_{i+1} - z_i'| -|z_i - z_{i-1}| - |z_{i+1} - z_i|
\right);  
\label{energy_change_step}
\end{equation}
$\Delta E_\text{core}$ is the change of disconnection core energy:
\begin{equation}
\Delta E_\text{core}
= \zeta \mathcal{K} \left[
(u_i' - u_{i-1})^2 + (u_{i+1} - u_i')^2 - (u_i - u_{i-1})^2 - (u_{i+1} - u_i)^2
\right], 
\label{energy_change_core}
\end{equation}
where $\zeta$ is a constant and $\mathcal{K}\equiv K\mathsf{f}(\pi/2)$; $W_\tau$ is the work done by the stress from all the other disconnections on the GB and the externally applied stress: 
\begin{equation}
W_\tau
= -\tau_i b \Delta y, 
\label{energy_change_work}
\end{equation}
where $\tau_i$ is the total stress on the $i$-th GB segment. 
This stress can be obtained, in the case of periodic boundary conditions, as
\begin{equation}
\tau_i 
= \frac{2\pi\mathcal{K}}{L_y} \left\{
\sum_{l=1}^N (u_l-u_{l-1})\cot\left[\frac{\pi}{N}\left(l-\frac{1}{2}-i\right)\right]
- b\cot\left(\frac{\pi}{2N}\right)
\right\}
+ \tau_\text{a}, 
\label{tau_i}
\end{equation}
where $\tau_\text{a}$ is the shear stress externally applied. 
The derivation of Eq.~\eqref{tau_i} is given in \ref{APPderivation}. 
If only the term $\Delta E_\text{step}$ is included, this model will reduce to the classical solid-on-solid model (i.e., Ising model with the interaction in the $z$-direction extending to infinity)~\cite{temperley1952statistical,swendsen1977monte}. 
If only the term $\Delta E_\text{core}$ is included, this model will reduce to the classical discrete-Gaussian model~\cite{chui1976phase,swendsen1977monte}. 

Based on the 1D lattice model described by the configuration $\{(z_i, u_i)\}$, the change of state $(b,h) \in \{(b_n,h_{nj})\}$ determined for a particular bicrystallography, and the energy formula Eq.~\eqref{energy_change}, we can design a Monte Carlo (MC) algorithm to simulate the thermally equilibrated GB. 
The detailed MC algorithm is provided in \ref{APPderivation}. 

\begin{figure}[!t]
\begin{center}
\scalebox{0.26}{\includegraphics{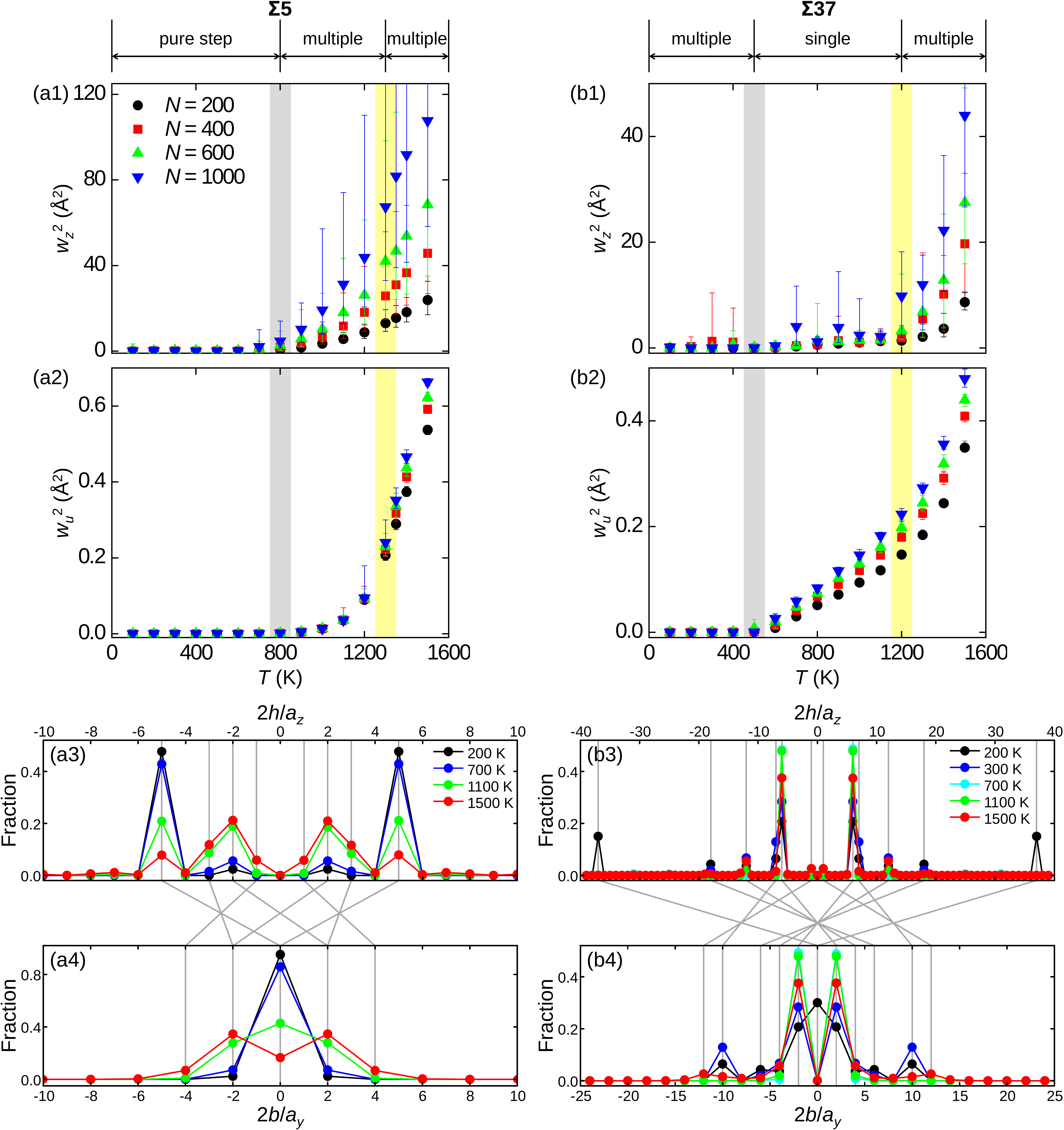}}
\caption{MC simulation results for the GB roughness and $u$-roughness of (a) $\Sigma 5$ $[100]$ $(013)$ and (b) $\Sigma 37$ $[100]$ $(057)$ STGBs in FCC crystals. 
(a1)/(b1) and (a2)/(b2) show the square of GB roughness $w_z^2$ and the square of $u$-roughness $w_u^2$ vs. temperature, respectively. 
Different temperature regimes are delimited according to the microscopic mechanism: ``pure step'', ``single'', and ``multiple'' stand for the mechanism associated with the activation of a pure step mode $(0,h)$, a single disconnection mode $(b,h)$ and multiple disconnection modes $\{(b_n,h_j)\}$. 
(a3)/(b3) and (a4)/(b4) show the fraction of step heights and the fraction of Burgers vectors, respectively. 
The step height in (a3) or (b3) and the Burgers vector in (a4) or (b4) which are connected by the gray line belong to one disconnection mode.  } 
\label{MC}
\end{center}
\end{figure}

We performed MC simulations on two different GB geometries: $\Sigma 5$ $[100]$ $(013)$ and $\Sigma 37$ $[100]$ $(057)$ STGBs in FCC crystals (taking EAM Ni as the model~\cite{foiles2006computation}). 
Figures~\ref{MC}a1, a2, b1 and b2 show simulation results for a square of GB roughness $w_z^2$ and a square of $u$-roughness $w_u^2$. 
There are two main observations. 
First, for both GBs, the GB roughness and $u$-roughness are nearly zero at low temperature and increases as temperature increases; such behavior is the GB (thermal) roughening transition. 
Second, the roughening behavior varies with different GB geometries. 
These two observations will be discussed below. 

First, does the roughening transition observed in the MC simulations correspond to a kind of phase transition or just a gradual increase in GB roughness? 
The roughening transition as a phase transition means that there exists a critical temperature $T_\text{r}$ such that, when $T<T_\text{r}$, the square of GB roughness $w_z^2$ converges to a constant with increase of the GB size (in one period) $N$; when $T>T_\text{r}$, $w_z^2$ diverges with $N$. 
For a 1D lattice model, it can be proved that, if only the short-range interaction [i.e., $\Delta E_\text{step}$ in Eq.~\eqref{energy_change}] is considered, $w_z^2$ will increase gradually with temperature and no phase transition will exist; however, when the long-range interaction [i.e., $W_\tau$ in Eq.~\eqref{energy_change}] is included, a phase transition will exist. 
The proof is provided in \ref{APPderivation}. 
Therefore, a necessary condition for $T_\text{r}$ is that, as $T>T_\text{r}$, $w_u^2$ must be finite (i.e., the Burgers vector plays the role). 
For the $\Sigma 5$ GB (see Figs.~\ref{MC}a1 and a2), we can find that, as $T > 800$~K (denoted by the vertical gray ribbon), $w_u^2$ becomes finite and $w_z^2$ diverges with $N$. 
Hence, $T_\text{r} \approx 800$~K for this GB. 
For the $\Sigma 37$ GB (see Figs.~\ref{MC}b1 and b2), as $T > 500$~K (vertical gray ribbon), $w_u^2$ becomes finite, but $w_z^2$ does not obviously diverge with $N$. 
Only when $T > 1200$~K (vertical yellow ribbon), $w_z^2$ diverges with $N$. 
Therefore, $T_\text{r} \approx 1200$~K for this GB. 

Second, we consider the GB geometry effect on the roughening behavior by comparing the cases of the $\Sigma 5$ GB and the $\Sigma 37$ GB. 
Since the MC simulations were based on the disconnection model, the geometry effect on the roughening behavior comes from the character of disconnection modes for a particular bicrystallography and the distribution of these modes during thermal equilibrium. 
Figures~\ref{MC}a3 and a4 plot the fraction of disconnections corresponding to each mode (characterized by step height $h$ and Burgers vector $b$) at various temperatures for the $\Sigma 5$ GB; and Figs.~\ref{MC}b3 and b4 for the $\Sigma 37$ GB. 
For the $\Sigma 5$ GB, there are three temperature regimes delimited by the vertical gray and yellow ribbons. 
In the $0<T<800~\text{K}$ ($T_\text{r}$) regime, the distribution is dominated by a pure step mode. 
In the $800~\text{K}<T<1300~\text{K}$ regime, multiple modes are activated, including a pure step mode as well as the modes with $b\ne 0$.  
In the $T>1300~\text{K}$ regime, multiple modes are activated and dominated by the modes $(b=\pm a_y, h = \mp a_z)$ and $(b=\pm a_y, h = \pm \frac{3}{2} a_z)$. 
Since the Burgers vector is large for the disconnection modes of this bicrystallography (in comparison with other higher-$\Sigma$ GBs), the disconnection mode with $b=0$ is favorable. 
Hence, the activation of the pure step mode dominates the GB roughening behavior (in the first and second regime). 
As an indicator of the mechanism dominated by pure step activation, the temperature dependence of $w_z^2$ shows a nearly parabolic trend (but with different coefficients at different regimes), which can be validated by Fig.~\ref{MC}a1. 
For the $\Sigma 37$ GB, there are also three temperature regimes delimited by the vertical gray and yellow ribbons. 
In the $0<T<500~\text{K}$ regime, multiple modes, including a pure step mode and the modes with $b\ne 0$, are activated. 
In the $500~\text{K}<T<1200~\text{K}$ ($T_\text{r}$) regime, the activated disconnections are dominated by a single mode $(b=\pm a_y, h = \mp 3a_z)$. 
In the $T>1200~\text{K}$ regime, multiple modes with $b\ne 0$ are activated. 
Since the Burgers vector is small for the disconnection modes of this bicrystallography, the disconnection mode with $b\ne0$ and small step height is favorable. 
In the second regime, which indicates a mechanism dominated by a single mode with $b\ne0$, $w_u^2$ is almost linear with temperature. 
In the third regime, which indicates a mechanism dominated by multiple modes with $b\ne0$, the temperature dependence of $w_u^2$ follows $T^\alpha$ ($\alpha>1$). 
These can be validated by Fig.~\ref{MC}b2. 

\subsubsection{Evidence from experiments and atomistic simulations}
\begin{figure}[!t]
\begin{center}
\scalebox{0.32}{\includegraphics{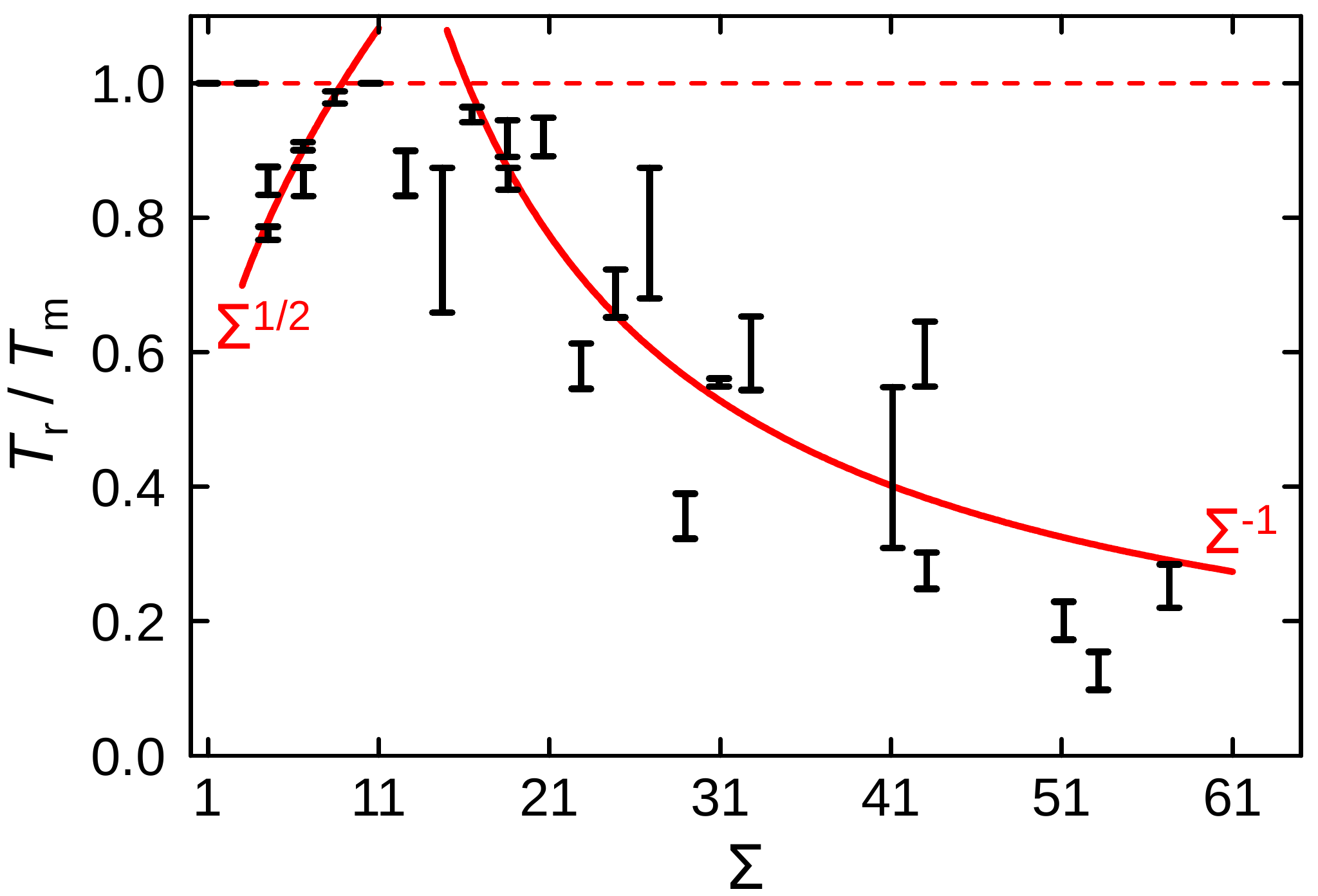}}
\caption{Experimental data for the roughening transition temperature vs. $\Sigma$ value for a series of GBs obtained by Shvindlerman and Straumal. 
The data (black bars) is reproduced from Ref.~\cite{shvindlerman1985regions} (L. S. Shvindlerman, Elsevier 1985). 
The left part of the data is fitted to the $\Sigma^{1/2}$ relation while the right part is fitted to the $\Sigma^{-1}$ relation.  } 
\label{Tr_vs_Sigma}
\end{center}
\end{figure}

There is no direct experimental evidence that proves the role of disconnection modes on GB roughening behavior. 
However, there are experimental measurements of a GB roughening transition temperature $T_\text{r}$ as a function of GB geometry. 
Based on the disconnection model, the influence of GB geometry is reflected in the character of the activated disconnection modes. 
Therefore, such experimental measurements can provide indirect evidence of the influence of the disconnection model. 
Shvindlerman and Straumal systematically measured $T_\text{r}$ as a function of reciprocal coincidence site density $\Sigma$ for a series of $[100]$ tilt and twist GBs in different materials	~\cite{shvindlerman1985regions}. 
Their experimental data is plotted in Fig.~\ref{Tr_vs_Sigma}. 
We can understand the trend of $T_\text{r}(\Sigma)$ based on the disconnection model. 
The roughening transition temperature is qualitatively scaled by the energy associated with the dominant disconnection mode; therefore, we consider the $\Sigma$-dependence of the energy barrier.  
From Eq.~\eqref{Estar_nj}, we can explicitly write the coefficients $A$ and $B$ as functions of $\Sigma$ for $[100]$ GBs: 
\begin{equation}
E_{nj}^*(\Sigma) 
= \left\{\begin{array}{ll}
A' \Sigma^{-1/2} |n\tilde{h}_0 + j\Sigma| + B' \Sigma^{-1} n^2, 	& \text{for tilt GBs} \\
A'|j| + B'\Sigma^{-1} n^2,										& \text{for twist GBs}
\end{array}\right., 
\label{Estar_Sgima}
\end{equation}
where $A'$ and $B'$ are coefficients which are insensitive to GB geometry for high-angle GBs. 
At the limit of small $\Sigma$ value, the coefficient before $n^2$ is large in comparison with the coefficient in the first term such that the pure step mode $(n=0, j=1)$ is favorable. 
The energy and, thus, $T_\text{r}$ are then scaled by $\Sigma^{1/2}$ for tilt GBs and $\Sigma^0$ for twist GBs. 
Indeed, from Fig.~\ref{Tr_vs_Sigma}, we find that $T_\text{r}/T_\text{m} \sim \Sigma^{1/2}$ at low temperature for a set of data for tilt GBs. 
For another set of twist GB data, $T_\text{r}/T_\text{m} \sim 1$ at low temperature, which indicates no roughening transition.  
At the limit of large $\Sigma$ value, the coefficient before $n^2$ is small such that the second term in Eq.~\eqref{Estar_Sgima} dominates. 
The energy and, thus, $T_\text{r}$ are expected to be scaled by $\Sigma^{-1}$ for both tilt and twist GBs. 
From Fig.~\ref{Tr_vs_Sigma}, we indeed see that $T_\text{r}/T_\text{m}$ decreases roughly as $\Sigma^{-1}$ at high temperature. 

\begin{figure}[!t]
\begin{center}
\scalebox{0.38}{\includegraphics{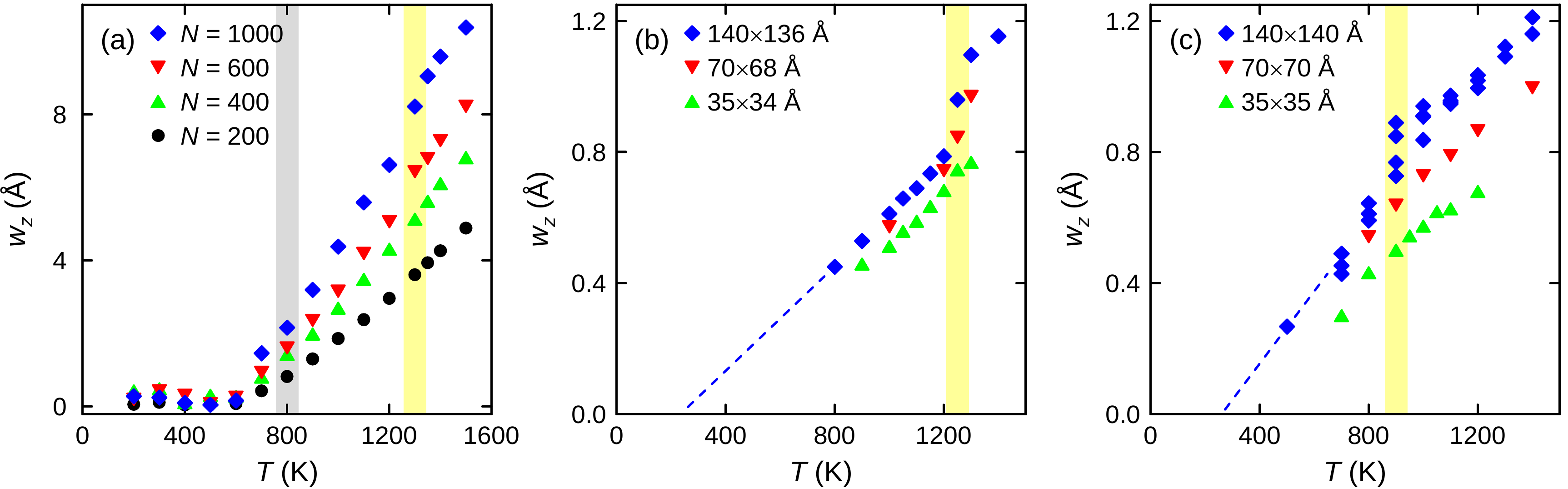}}
\caption{(a) MC simulation result for GB roughness vs. temperature for a $\Sigma 5$ $[100]$ $(013)$ STGB in an FCC material (EAM Ni). 
(b) and (c) show the MD simulation result for GB roughness for a $\Sigma 5$ $[100]$ $(013)$ STGB and $(001)/(034)$ ATGB respectively, obtained by Olmsted et al.; the data points are reproduced from Ref.~\cite{olmsted2007grain} (D. L. Olmsted, Elsevier 2007).
The dashed lines in (b) and (c) denote extrapolation of the data.  } 
\label{MD_comparison}
\end{center}
\end{figure}

Direct comparison can be made between the GB roughness obtained by the MC simulations in Section~\ref{roughening_MC_simulation} and that obtained by the MD simulations in literature. 
Olmsted et al. calculated the GB roughness for $\Sigma 5$ $[100]$ $(013)$ STGB and $(001)/(034)$ ATGB in EAM Ni by MD simulations~\cite{olmsted2007grain}. 
Figure~\ref{MD_comparison} shows the GB roughness $w_z$ as a function of temperature from the MC simulations and the MD simulations. 
First, the general trend is consistent that $w_z$ increases monotonically with temperature. 
At low temperature, $w_z$ is close to zero. 
The low-temperature trend is obtained by extrapolation for the MD data; it is always difficult to directly gain the low-temperature data by MD simulation because of slow dynamics and poor statistics. 
There is a discontinuous change in the slope of $w_z$ at a high temperature, denoted by $T_\text{d}$ and labeled by the vertical yellow ribbon in Fig.~\ref{MD_comparison}. 
The temperature $T_\text{d}$ is roughly consistent for the STGB from the MC result ($\sim 1300$~K) and from the MD result ($\sim 1250$~K). 
Olmsted et al. define $T_\text{d}$ as the ``roughening transition temperature'' because the GB mobility shows a discontinuous change at $T_\text{d}$~\cite{olmsted2007grain}. 
Note that this definition is different from the one mentioned in Section~\ref{roughening_MC_simulation} about $T_\text{r}$ (above which the square of GB roughness diverges with increase of the GB size $N$). 
With reference to Figs.~\ref{MC}a3 and a4, we find that the transition occurring at $T_\text{d}$ is caused by the change of dominant disconnection mode. 
When $T<T_\text{d}$, a pure step mode contributes to the roughening behavior, while, when $T>T_\text{d}$, roughening behavior is dominated by the modes with $b\ne 0$. 
The modes with $b\ne 0$ are characterized by finite GB sliding in the direction parallel to the GB plane and a smaller step height than the height of the pure step. 
Hence, the activation of the modes with $b\ne 0$ above $T_\text{d}$ lowers the slope of $w_z(T)$ (and GB mobility as a function of temperature) but increases the slope of $w_u(T)$. 
We also notice that the main distinction between the MC and MD results is the magnitude of the roughness. 
There are two possible reasons. 
(i) The size of the GB used in the MC simulation is much larger than that in the MD simulation. 
(ii) The MC simulation is performed based on 2D model while the MD simulation is based on a 3D model; it is likely that the change of dimension influences the roughening behavior (for example, the roughening transition exists as a phase transition in the 3D Ising model but not in the 2D Ising model~\cite{swendsen1977monte,chandler1987introduction}). 

\subsection{Intrinsic grain-boundary mobility}
\label{intrinsicGBmobility}

\subsubsection{What do we mean by intrinsic mobility? }
The second property associated with the thermal equilibrium of GBs is intrinsic GB mobility. 
The fundamental and classical kinetic equation for GB migration is~\cite{turnbull1951theory,SuttonBalluffi,balluffi2005kinetics} 
\begin{equation}
v_\perp = M_zF, 
\label{MF}
\end{equation}
where $v_\perp$ is the velocity of GB migration (normal to the GB plane), $F$ is the applied driving force, and the GB mobility $M_z$ is defined as the coefficient relating $v_\perp$ and $F$. 
This relation plays a central role in the study of microstructure evolution in polycrystalline materials~\cite{hillert1965theory}. 
Several constraints will be imposed, as below, on Eq.~\eqref{MF} in order to uniquely define the GB mobility $M_z$ such that $M_z$ is intrinsic in the sense that it only depends on material parameters, temperature and bicrystallography. 
(i) $M_z$ is defined in the limit of $F\to 0$; thus, $M_z$ does not depend on any external condition. 
Therefore, the intrinsic GB mobility is an equilibrium GB property. 
(ii) $M_z$ is defined in an infinitely large bicrystal configuration. 
(iii) The bicrystal contains an infinitely large, flat (at mesoscale) GB. 
The intrinsic GB mobility is of theoretical interest since it is well defined. 
It can also be used as a parameter to understand some complicated phenomena (such as grain growth) which are of practical interest. 

Based on a 2D bicrystal containing a 1D GB, the thermally equilibrated GB can be described by a height function $z(y,t)$ and a relative displacement function $u(y,t)$, as proposed in Section~\ref{description}. 
The intrinsic GB mobility can be evaluated by monitoring the evolution of $z(y,t)$. 
There are several ways to do so. 
The first way is to calculate $M_z$ directly according to the definition Eq.~\eqref{MF}, or the slope of the function $v_\perp(F)$ in the small $F$ regime. 
This method was widely used in MD simulations with different kinds of driving forces $F$ (such as stress, energy density jump across a GB, etc.)~\cite{zhang2004computer,zhang2005mobility,janssens2006computing,olmsted2009survey,zhou2011towards,song2012molecular,homer2014trends,rahman2014comprehensive,priedeman2017role}. 
However, the shortcoming of this method lies on the contradiction between the small driving force required by the definition and the large driving force required for the time scale of dynamics accessible by MD simulations. 
According to the fluctuation-dissipation theory~\cite{chandler1987introduction}, the intrinsic GB mobility can also be obtained from GB fluctuation. 
The second way to calculate $M_z$ is through the random walk of a GB under finite temperature. 
By tracing the (1D) trajectory of the average GB position $\bar{z}(t)$, we can first obtain the diffusion coefficient $D_z$ from the mean square displacement of the average GB position: 
\begin{equation}
\langle {\bar{z}}^2\rangle - \langle\bar{z}\rangle^2
= D_zt.  
\label{Dz}
\end{equation}
Then, the GB mobility can be measured as 
\begin{equation}
M_z
= \frac{D_z \mathcal{A}}{2k_\text{B}T}, 
\label{Mz}
\end{equation}
which is analogous to the Einstein relation for 1D diffusion of a Brownian particle. 
This relation was rigorously derived for the interface in the case of capillary fluctuation~\cite{trautt2006interface} and the case of GB migration-shear coupling~\cite{karma2012relationship}. 
This method was used to calculate the GB mobility in MD simulations~\cite{trautt2006interface,deng2011atomistic,rahman2014comprehensive,priedeman2017role} and experiments~\cite{skinner2010grain}. 
The third way is to extract the GB mobility $M_z$ from the GB fluctuation based on the capillary fluctuation wave (CFW), which is also derived from the fluctuation-dissipation theory. 
The GB profile can be expanded in a Fourier series: $z(\mathbf{r},t) = \sum_{\mathbf{k}} z_{\mathbf{k}}(t) e^{i\mathbf{k}\cdot\mathbf{r}}$, where $\mathbf{r} \equiv (x,y)$ and $\mathbf{k}\equiv (k_x,k_y)$. 
We then can have the equilibrium static fluctuation spectrum: 
\begin{equation}
\left\langle |z_{\mathbf{k}}(t)|^2 \right\rangle
= \frac{k_\text{B}T}{\mathcal{A}\mathit{\Gamma} k^2}
\text{ or }
\left\langle |z_{\mathbf{k}}(t)|^2 \right\rangle
= \frac{k_\text{B}T}{\mathcal{A}C\beta^2 k}, 
\label{A2}
\end{equation}
where $\langle \cdot\rangle$ denotes configurational average, $k \equiv |\mathbf{k}|$, $\mathit{\Gamma} \equiv \gamma + \gamma''$ is the GB stiffness (the prime denotes the derivative with respect to the orientation of the GB normal)~\cite{herringsurface}, $C$ is a combination of elastic constants, and $\beta$ is the shear-coupling factor, which is $b/h$ if the activated disconnection mode is $(b,h)$ (see Section~\ref{phenomenon_stress} for details). 
The first relation in Eq.~\eqref{A2} was derived for the case of capillary fluctuation (no shear coupling)~\cite{hoyt2010fluctuations,karma2012relationship}, while the second relation in Eq.~\eqref{A2} was derived for the case of shear-coupled GB migration~\cite{karma2012relationship}. 
The temporal correlation of the amplitude is 
\begin{equation}
\left\langle z_{\mathbf{k}}(t)z^*_{\mathbf{k}}(t') \right\rangle 
= \left\langle |z_{\mathbf{k}}(t)|^2 \right\rangle e^{-(t'-t)/t_\text{R}}, 
\end{equation}
where the relaxation time $t_\text{R}$ can be obtained by fitting this relation to the MD data.  
Then, based on the evaluation of $t_\text{R}$ and the relation Eq.~\eqref{A2}, the intrinsic GB mobility can be calculated by 
\begin{equation}
M_z = \frac{\left\langle |z_{\mathbf{k}}(t)|^2 \right\rangle \mathcal{A}}{t_\text{R} k_\text{B}T}. 
\end{equation}
The CFW method was initially used to calculate the mobility of liquid-solid interfaces by MD simulations~\cite{hoyt2002atomistic,monk2009determination}. 
This method was also used to extract $M_z$ for a $\Sigma 7$ $[111]$ $(11\bar{2})/(13,\bar{2},\bar{11})$ ATGB in EAM Ni by Foiles et al.~\cite{foiles2006computation} and a series of $[100]$ STGBs in EAM Cu by Karma et al.~\cite{karma2012relationship} 

The GB mobility calculated by each of the three methods listed above was intrinsic. 
As follows, we will consider how the intrinsic GB mobility can be qualitatively estimated based on the disconnection model and see the influence of the disconnection modes on the mobility. 

\subsubsection{The simplest approach: nucleation-controlled kinetics}
Just as in Section~\ref{simplest_Boltzmann}, we will construct a toy model that is kept as simple as possible in order to gain a general idea of the implications of the disconnection model on the intrinsic GB mobility. 
This toy model is associated with a 2D bicrystal containing a STGB with periodic boundary conditions applied (such as that shown in Fig.~\ref{DSC}a). 
The main assumption is that the GB migration is dominated by (homogeneous) nucleation of disconnection pairs. 
Such an assumption is made for two reasons. 
(i) The energy barrier for nucleation is much larger than the energy barrier for disconnection gliding along the GB, which is true according to the result of the atomistic simulation shown in Fig.~\ref{barrier}a. 
(ii) The propagation and reaction of disconnections is equivalent to heterogeneous nucleation of disconnection pairs occurring at the existing disconnections; the energy barrier for heterogeneous nucleation might be qualitatively approximated by that for homogeneous nucleation (with certain modifications). 
The disconnection pair can adopt one of the disconnection modes $\{(b_n, h_{nj})\}$ for a particular bicrystallography (determined by the algorithm provided in Section~\ref{enumeration}). 
In each period, the energy barrier for the nucleation of a disconnection pair of the mode $(b_n, h_{nj})$, $E_{nj}^*$, has the form of Eq.~\eqref{Estar_nj}. 
Then, according to the harmonic transition state theory~\cite{vineyard1957frequency}, the rate for the nucleation of this pair of disconnections $r_{nj}$ is proportional to $\exp(-L_x E_{nj}^*/k_\text{B}T)$. 
The fluctuation of the average GB position $\bar{z}$ can be viewed as a 1D random walk. 
The intrinsic GB mobility can be extracted from the simulation of such a 1D random walk based on the nucleation rate formula. 

\subsubsection{Kinetic Monte Carlo simulations}
\label{KMC_simulations}
We simulated the 1D random walk of a GB (in a 2D bicrystal under periodic boundary condition) by the kinetic Monte Carlo (KMC) method. 
The KMC simulations were performed for $\Sigma 5$ $[100]$ $(013)$, $\Sigma 13$ $[100]$ $(015)$, and $\Sigma 37$ $[100]$ $(057)$ STGBs in a FCC material. 
For each GB geometry, we have a list of disconnection modes $\{(b_m, h_m)\}$ [$m$ is the reduced notation for $(n,j)$]. 
For the mode list, we calculated the associated energy barriers $\{E^*_m\}$ and constructed the rate list $\{r_m\}$. 
The activity of this system is then $\mathcal{R}=\sum_m r_m$ and the probability for nucleating the $m$-th disconnection mode is $P_m = r_m/\mathcal{R}$. 
In each KMC step, the GB migrates by a disconnection mode, e.g. $(b_m, h_m)$, which is randomly chosen from the mode list according to its probability $P_m$. 
The (reduced) time for this KMC step is $\Delta \tilde{t} = \mathcal{R}^{-1}\ln \chi^{-1}$, where $\chi$ is a random number between $0$ and $1$. 
In this way, the GB position evolves as $\bar{z}(\tilde{t}+\Delta\tilde{t}) = \bar{z}(\tilde{t}) + h_m$ and the relative displacement of two grains evolves as $\bar{u}(\tilde{t}+\Delta\tilde{t}) = \bar{u}(\tilde{t}) + b_m$. 

\begin{figure}[!t]
\begin{center}
\scalebox{0.29}{\includegraphics{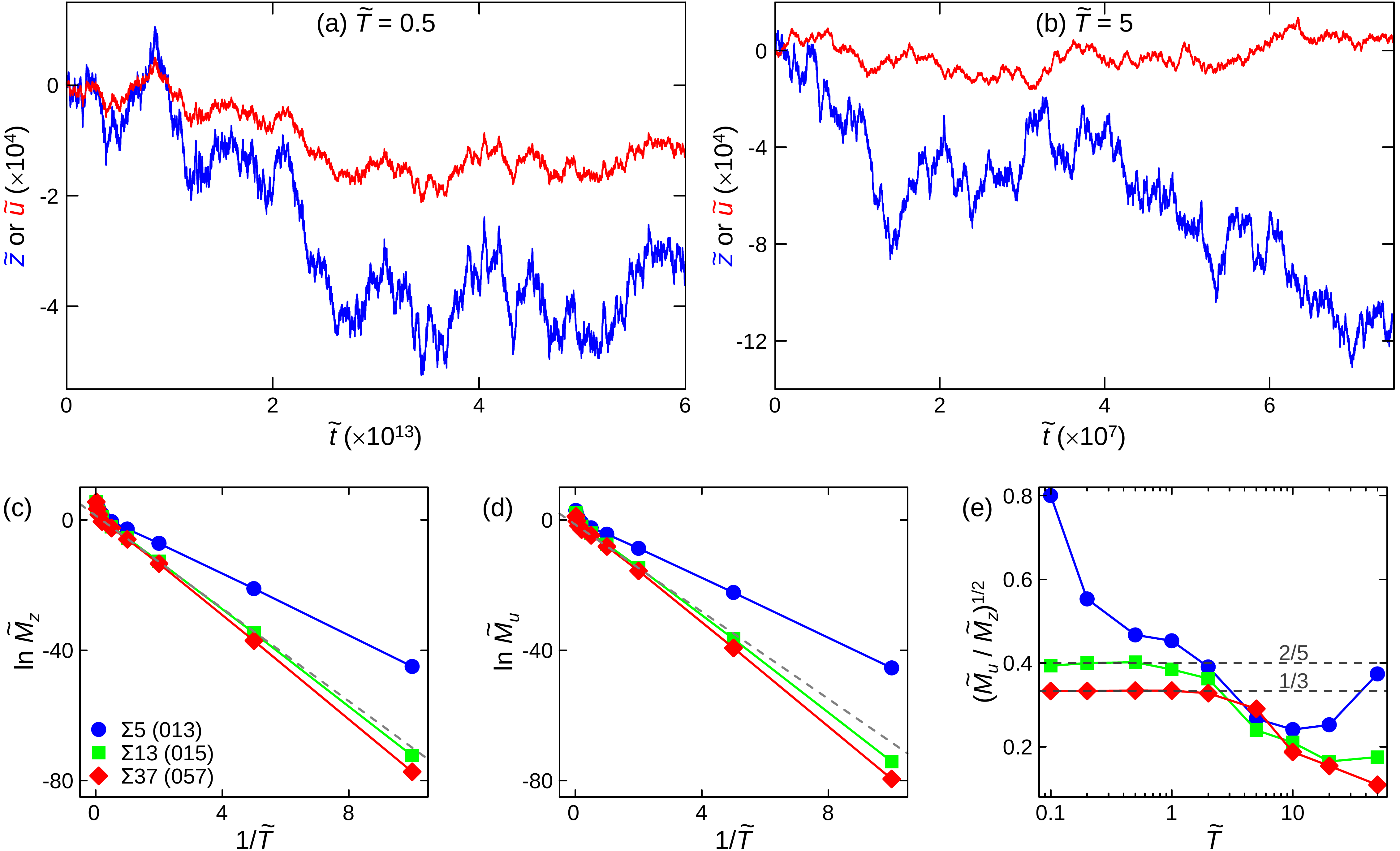}}
\caption{(a) and (b) Reduced GB height $\tilde{z} \equiv \bar{z}/\alpha_z$ (the blue curves) and reduced relative displacement $\tilde{u} \equiv \bar{u}/\alpha_y$ (red curves) vs. the reduced time ($\tilde{t} \equiv \nu t$) at two reduced temperatures: $\tilde{T}=0.5$ and $5$, respectively, obtained from KMC simulations for a $\Sigma 13$ $[100]$ $(015)$ STGB in an FCC material. 
(c) and (d) Semilogarithmic plots of the reduced intrinsic GB mobility $\tilde{M}_z \equiv [4\mathit{\Gamma}_\text{s}/(\omega L_y\alpha_z)] M_z$ and the reduced sliding coefficient $\tilde{M}_u \equiv [4\mathit{\Gamma}_\text{s}\alpha_z/(\omega L_y\alpha_y^2)] M_u$ vs. the inverse of reduced temperature for three GB geometries ($\omega$ is the attempt frequency). 
(e) shows the plots of $(\tilde{M}_u/\tilde{M}_z)^{1/2}$ vs. the reduced temperature. 
The dashed lines denote the ideal shear-coupling factors $\beta$ for the $\Sigma 13$ and $\Sigma 37$ STGBs. } 
\label{GBmobility2}
\end{center}
\end{figure}

From the KMC simulations, we obtained the trajectories $\bar{z}(\tilde{t})$ and $\bar{u}(\tilde{t})$; the examples for the $\Sigma 13$ $[100]$ $(015)$ STGB at two different temperatures: $\tilde{T} = 0.5$ and $5$ are shown in Figs.~\ref{GBmobility2}a and b, respectively [the reduced temperature is defined as $\tilde{T} = k_\text{B}T/(2\mathit{\Gamma}_\text{s}L_x a_z)$]. 
We find that, at low temperature (see Fig.~\ref{GBmobility2}a), the shape of $\bar{z}(\tilde{t})$ scales with that of $\bar{u}(\tilde{t})$; the scaling factor is $\beta = \bar{u}(\tilde{t})/\bar{z}(\tilde{t}) \approx 0.4$, which is called shear-coupling factor (see Section~\ref{theoretical_background} for details). 
According to the bicrystallography, the disconnection mode with the smallest nonzero Burgers vector and step height is $(b = a_0/\sqrt{26}, h = \frac{5}{2}a_0/\sqrt{26})$ and $b/h = 0.4$. 
This indicates that, at low temperature, this disconnection mode is activated and dominates the GB migration. 
As a result, GB migration and GB sliding are coupled at low temperature. 
At high temperature (see Fig.~\ref{GBmobility2}b), we find that $\langle \bar{u}(\tilde{t})/\bar{z}(\tilde{t})\rangle_t \approx 0$, implying that there is no correlation between GB migration and GB sliding.  

The intrinsic GB mobility $M_z$ can be calculated based on the random walk of the GB position $\bar{z}(\tilde{t})$ and according to Eqs.~\eqref{Dz} and \eqref{Mz}. 
The sliding coefficient $M_u$ can be defined similarly to $M_z$: $v_\parallel = M_u F$, where $v_\parallel$ is the sliding velocity of one grain with respect to the other. 
$M_u$ can be calculated in a similar manner to $M_z$, i.e., based on the random walk of one grain with respect to the other $\bar{u}(\tilde{t})$ and according to Eqs.~\eqref{Dz} and \eqref{Mz} but with ``$z$'' replaced by ``$u$''. 
The intrinsic GB mobilities and the sliding coefficients as functions of temperature obtained by the KMC simulations are shown in Figs.~\ref{GBmobility2}c and d, respectively. 
First, we notice that the temperature dependence of GB mobility $M_z$ and the sliding coefficient $M_u$ are different for different GB geometries. 
This results from the list of disconnection modes, which is determined for a particular GB geometry. 
Second, we find that, at low temperature, the $M_z$ value is highly related to the $M_u$ value for the $\Sigma 13$ and $\Sigma 37$ STGBs. 
If a single disconnection mode characterized by $b/h=\beta$ and $b\ne 0$ is activated at low temperature, we will have the relationship: $|\beta| = (M_u/M_z)^{1/2}$, which was derived by Karma et al.~\cite{karma2012relationship} 
As shown in Fig.~\ref{GBmobility2}e, $(M_u/M_z)^{1/2}\approx 0.4$ and $0.33$ for the $\Sigma 13$ and $\Sigma 37$ STGBs, respectively. 
The $(M_u/M_z)^{1/2}$ values are consistent with the values of $|\beta|$ associated with the disconnection mode characterized by the smallest nonzero Burgers vector and step height for these two GB geometries, i.e., $|\beta| = 2/5$ and $1/3$ for the $\Sigma 13$ and $\Sigma 37$ STGBs, respectively. 
However, at high temperature, the $(M_u/M_z)^{1/2}$ values deviate from the ideal values of $|\beta|$ determined by the GB geometry, because the coupling between GB migration and GB sliding becomes weak, as observed in Fig.~\ref{GBmobility2}b. 
The data of $\Sigma 5$ STGB does not satisfy the coupling relation. 
From the previous MC simulation of the thermal equilibrium GB (Section~\ref{roughening_MC_simulation}), we see that the GB migration is dominated by nucleation of pure steps (see Fig.~\ref{MC}a3 and a4) rather than a single disconnection mode with $b\ne 0$. 
Therefore, the $(M_u/M_z)^{1/2}$ value becomes lower than the ideal value of $|\beta|$ determined by the GB geometry (ideally $|\beta| = 1$ for the $\Sigma 5$ GB geometry). 
In addition, as shown in Fig.~\ref{GBmobility2}c, the plots of $\ln M_z$ vs. $1/T$ do not exactly follow a linear relationship, indicative of non-thermal (non-Arrhenius) behavior~\cite{cantwell2015anti}. 
The slopes (absolute value) of these plots become slightly smaller as temperature increases. 
Such non-thermal behavior seems counter-intuitive; usually, as temperature increases, the events with higher activation energies will occur more frequently, resulting in a larger slope of the semilogarithmic plot as temperature increases (such as GB diffusivity~\cite{hwang1978possible,suzuki2005atomic}). 
How can we understand the smaller slope at higher temperature for the intrinsic GB mobility? 
According to the disconnection model, GB motion involves the nucleation and gliding of disconnections which are characterized by both step heights and Burgers vectors. 
Therefore, GB motion should include two (three) DOFs in a 2D (3D) bicrystal: GB migration and GB sliding. 
As temperature increases, the high-energy disconnection modes, characterized by large Burgers vector but small step height, might be activated. 
If so, even if the event with high activation energy occurs at high temperature, it does not contribute much to GB migration (via step height), leading to lower slope of $\ln M_z$ vs. $1/T$. 
A similar explanation was given by Sch\"{o}nfelder et al. based on observations from their MD simulations of $[100]$ twist GBs~\cite{schonfelder2005comparative}. 

\subsubsection{Evidence from atomistic simulations}
\begin{figure}[!t]
\begin{center}
\scalebox{0.36}{\includegraphics{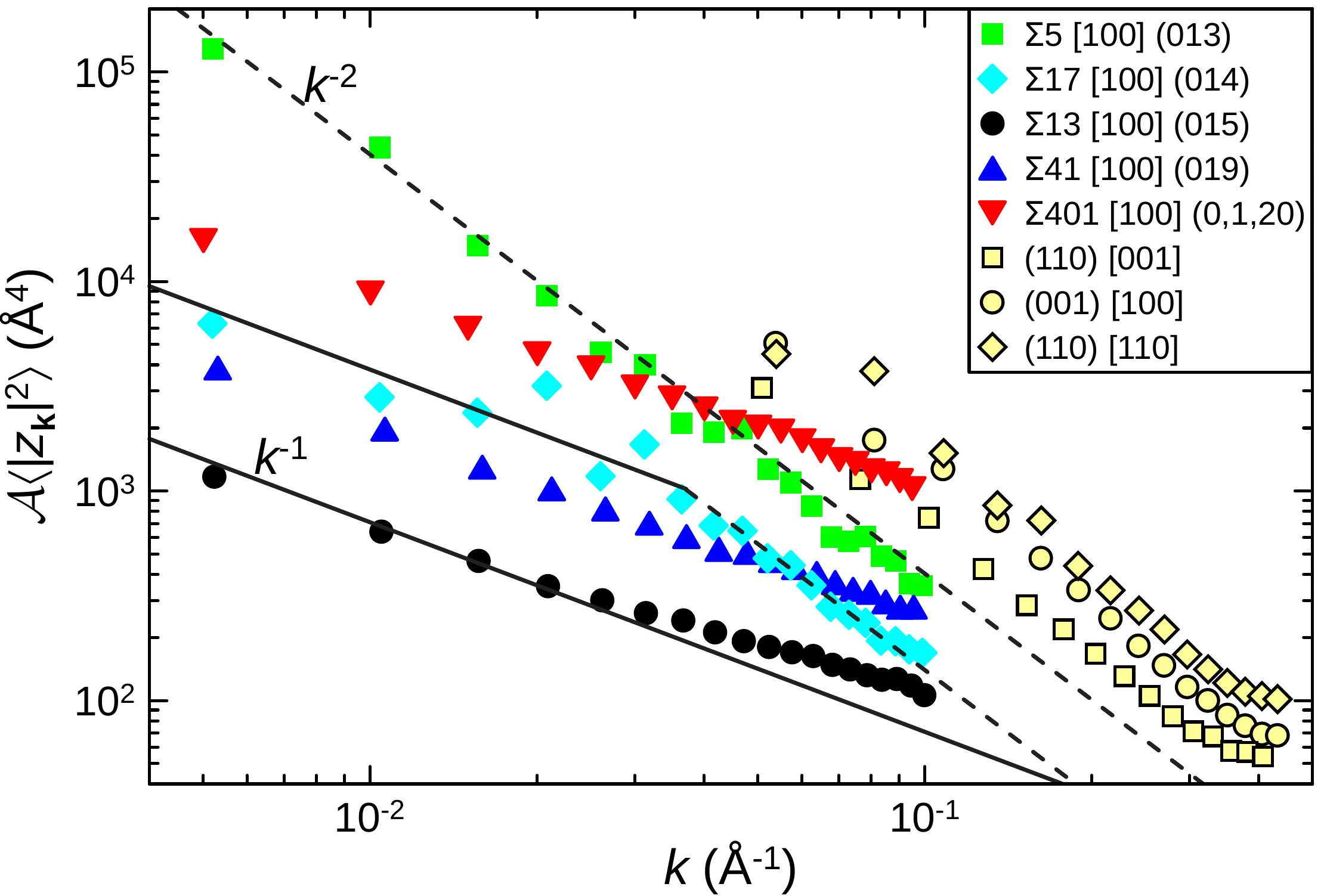}}
\caption{The power spectra of interface thermal fluctuation for $[100]$ STGBs (in EAM Cu) and liquid-solid interfaces (in EAM Ni), obtained by MD simulations. 
The figure is reproduced from Ref.~\cite{karma2012relationship} (A. Karma, American Physical Society 2012) and Ref.~\cite{hoyt2003atomistic} (J.J.Hoyt, Elsevier 2003).
The lines with slope $-1$ (solid) and slope $-2$ (dashed) are shown for reference. } 
\label{CFW_MD}
\end{center}
\end{figure}

We will check the disconnection model and the KMC simulations by investigating MD simulations and experiments in literature. 
The intrinsic GB mobility was calculated by MD simulations using the CFW method~\cite{foiles2006computation,karma2012relationship}. 
Karma et al. proved via simulations that the disconnections with $b/h = \beta \ne 0$ play a role in GB migration~\cite{karma2012relationship}. 
According to Eq.~\eqref{A2}, when both elastic energy (contributed by Burgers vector) and GB stiffness (contributed by step height) are taken into account, 
\begin{equation}
\left\langle |z_{\mathbf{k}}|^2 \right\rangle
\sim \frac{1}{k^2 + c k}, 
\label{A2_mix}
\end{equation}
where $c = C\beta^2/\mathit{\Gamma}$. 
If a pure step, rather than a disconnection with $\beta \ne 0$, dominates the GB migration (such as in the TKL or Gleiter's model), $c=0$ and $\langle |z_{\mathbf{k}}|^2 \rangle \sim k^{-2}$. 
However, if a single disconnection with $b\ne 0$ operates, $\langle |z_{\mathbf{k}}|^2 \rangle$ will approach the trend of $k^{-1}$. 
Figure~\ref{CFW_MD} summarizes the MD results of $\langle |z_{\mathbf{k}}|^2\rangle$ vs. $k$ for several $[100]$ STGBs~\cite{karma2012relationship} and liquid-solid interfaces~\cite{hoyt2003atomistic}. 
We find that the $\Sigma 5$ GB follows the trend of $k^{-2}$ (or $c\ll 1$), similar to the behavior of the liquid-solid interfaces; this indicates that nucleation of pure step dominates GB migration, which is consistent with the finding in the previous MC simulation (Figs.~\ref{MC}a3 and a4). 
The $\Sigma 17$ GB follows Eq.~\eqref{A2_mix}, indicative of comparable contribution from the mode of $\beta \ne 0$ and the pure step mode. 
The other GBs follow the trend of $k^{-1}$ (or $c\gg 1$), which is strong evidence for the existence of a single disconnection mode with $b\ne 0$ as the dominant mechanism (perfect shear coupling). 

The intrinsic GB mobility was also calculated by MD simulations using a driven-GB-migration method~\cite{zhang2004computer,zhang2005mobility,janssens2006computing,olmsted2009survey,zhou2011towards,song2012molecular,homer2014trends,rahman2014comprehensive} and the random-walk method~\cite{trautt2006interface,deng2011atomistic,rahman2014comprehensive}. 
Non-thermal behavior was observed in the semilogarithmic plots of GB mobility $M_z$ for many GBs. 
For example, the decrease of $\ln M_z$ vs. $1/T$ at high temperature was observed for 
$\Sigma 5$ $[100]$ $(0\bar{1}1)/(0\bar{7}1)$ ATGB in EAM Ni~\cite{mendelev2013comparison},
$\Sigma 5$ $[100]$ $(013)$ STGB in EAM Ni~\cite{olmsted2007grain}, 
non-CSL $[100]$ ATGBs in EAM Ni~\cite{zhang2007atomic,zhang2009grain}, 
$\Sigma 651$ $[112]$ $(11,13,\overline{12})$ STGB in EAM Al~\cite{rahman2014comprehensive}, 
and a series of $[001]$ twist GBs in Lennard-Jones Cu~\cite{schonfelder2005comparative}. 
The mobilities of many GBs in the study by Homer et al. show non-thermal (or non-Arrhenius) behavior~\cite{homer2014trends}. 
There are three types of explanation to the phenomenon of non-thermal temperature dependence of intrinsic GB mobility observed in these MD simulations. 
\begin{itemize}
\item[(i)] \textit{Structural phase transition}. It has been recognized that, in general, a GB with fixed macroscopic DOFs could have multiple metastable structures~\cite{han2016grain,han2017grain}. 
As temperature increases, the most stable structure may change from one to another~\cite{Frolov13}. 
It is probable that different GB structures are associated with different intrinsic GB mobilities. 
If the most stable GB structure at high temperature corresponds to lower GB mobility than that at low temperature, the GB mobility will decrease with the increase of temperature. 

\item[(ii)] \textit{``Roughening'' transition}. As temperature increases, GB structure may become roughened. 
This is called a ``roughening'' transition in some literature (which is different from the definition in Section~\ref{GB_roughening}).  
The transition temperature corresponds to $T_\text{d}$ (rather than $T_\text{r}$) in Section~\ref{GB_roughening}. 
Olmsted et al. found that $T_\text{d}$ is consistent with the temperature above which the GB mobility decreases~\cite{olmsted2007grain}. 
Hence, such ``roughening'' transition is considered to be a possible reason for the non-thermal behavior~\cite{olmsted2007grain,homer2014trends}. 

\item[(iii)] \textit{Glass transition}. If a GB has many metastable GB structures (associated with the variation of microscopic DOFs), the GB may undergo a glass transition as temperature increases. 
Zhang et al. found that the relationship between GB mobility $M_z$ and temperature followed Vogel-Fulcher law~\cite{vogel1921law,fulcher1925analysis} (rather than Arrhenius law), which is signature of glass-like behavior~\cite{zhang2009grain}. 
If Vogel-Fulcher law is followed, the slope of $\ln M_z$ vs. $1/T$ will gradually decrease with the increase of temperature (without sharp transition). 

\item[(iv)] \textit{Phonon drag}. It is well-known that dislocation mobility decreases with the increase of temperature because of phonon drag (or thermal damping)~\cite{gorman1969mobility,HirthLothe}. 
For this reason, the mobility of a low-angle GB comprised of an array of lattice dislocations is expected to decrease with the increase of temperature. 
For the migration of high-angle GBs, facilitated by disconnection motion, a similar temperature dependence of GB mobility may also apply because of phonon drag acting on the motion of disconnections. 
Hence, phonon drag (or thermal damping) is considered as another possible reason for the non-thermal behavior~\cite{olmsted2009survey,homer2014trends,priedeman2017role}. 
\end{itemize}
Only the non-thermal behavior due to (ii) ``roughening'' transition is taken into account in the KMC simulations based on the disconnection model proposed in Section~\ref{KMC_simulations}, because the transition at $T_\text{d}$ can be captured by the disconnection model and indeed appears in the previous MC simulations mentioned in Section~\ref{roughening_MC_simulation} (see Fig.~\ref{MD_comparison}). 
The other possible causes of non-thermal behavior cannot be included in the KMC simulation based on the disconnection model. 
However, the reasons [other than (iii) glass transition~\cite{zhang2009grain}] provided for explaining the non-thermal behavior observed in the MD simulations have not been proven from the atomic mechanism of GB motion. 
Manifold, complicated temperature dependence of the intrinsic GB mobility for various GB geometries is still a problem waiting for a satisfactory explanation and direct evidence~\cite{homer2014trends}.

%%%%%%%%%%%%%%%%%%%%%%%%%%%%%%%%%%%%%%%%%%%%%%%%%%%%%%%
\section{Conservative grain-boundary kinetics}
\label{sec5}

GB kinetics in a single-component system may be divided into two types: conservative and non-conservative~\cite{SuttonBalluffi,balluffi2005kinetics}. 
We define conservative kinetics as processes which only involve the GB itself (i.e., no interaction with other defects). 
To investigate a conservative kinetic process, a GB can be studied as a closed system. 
On the contrary, we define the non-conservative kinetics as the process which involves interaction between the GB and the defects in the neighboring grains, such as point defects (self-interstitial atoms and vacancies)  and lattice dislocations. 
Therefore, in our definition, the so-called conservation refers to the invariance of both atomic number and total Burgers vector along the GB. 

\subsection{Conservative kinetics: driven grain-boundary migration}

A GB can migrate without absorption/emission of point defects or lattice dislocations from/into the neighboring grains; this is a typical conservative process. 
GB migration can occur only if the total free energy of the whole system decreases through this process. 
Therefore, the driving force for the GB migration can be expressed as 
\begin{equation}
F(x,y) = - \frac{\delta E_\text{tot}[z(x,y)]}{\delta z(x,y)}, 
\label{F1}
\end{equation}
where $z(x,y)$ is the GB height function describing a curved plane in 3D space and $E_\text{tot}$ is the total free energy of the whole system (due to the excess energy associated with the GB structure) which is a functional of the GB morphology. 

In general, the driving force is defined locally (point by point) according to Eq.~\eqref{F1}. 
However, this definition can be largely simplified for a planar GB in a bicrystal system. 
The driving force for GB migration of a planar GB can be expressed as 
\begin{equation}
F = - \frac{\text{d} E_\text{tot}}{\text{d} \bar{z}}.  
\label{F2}
\end{equation}
Here, $E_\text{tot}$ is the total free energy of the bicrystal system per unit area. 
In most studies of driven GB migration, a bicrystal system was used such that complications due to GB morphology are avoided. 

The driving force for the conservative GB migration can be divided into two types: stress and energy density jump across a GB. 
We will review the studies of driven GB migration in the next two sections from these two types of driving forces. 
We will show how the effect of these two types of driving forces can be understood based on the disconnection model.

\subsection{Stress-driven grain-boundary migration}

\subsubsection{Phenomenon of stress-driven grain-boundary migration}
\label{phenomenon_stress}
\begin{figure}[!t]
\begin{center}
\scalebox{0.36}{\includegraphics{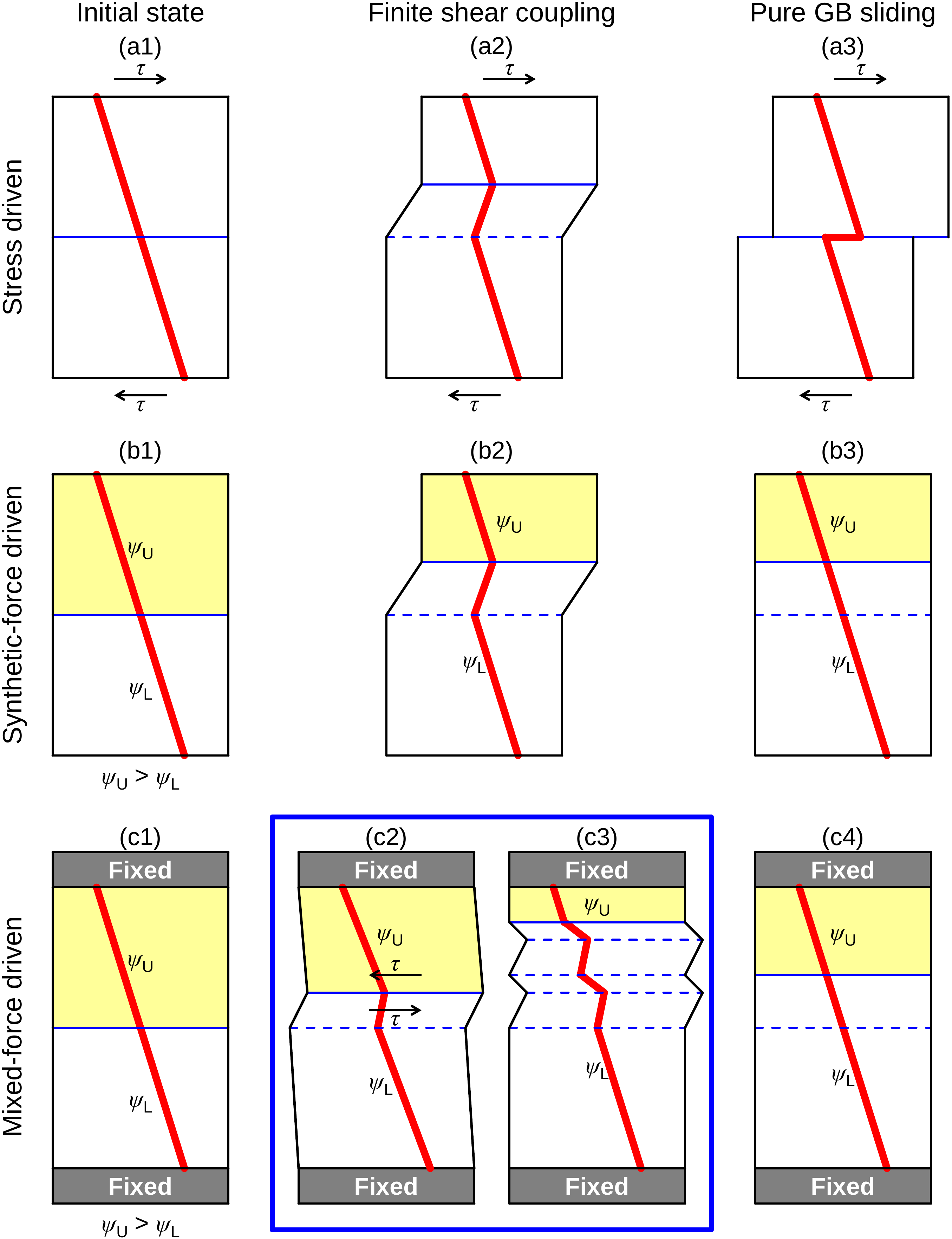}}
\caption{Schematics of GB migration in a bicrystal driven by shear stress ($\tau$), energy jump (i.e., the energy density difference $\psi_\text{U}-\psi_\text{L}$) with two free surfaces, and energy jump with two fixed surfaces, respectively. 
In each figure, the blue solid line denotes the current GB position, the blue dashed line denotes the initial GB position, and the red line denotes the fiducial line. 
(a1), (b1) and (c1) show the initial bicrystal states. 
(a2), (b2) and (c2) or (c3) show the final states for the case of finite shear-coupling factor. (c2) depicts the scenario of GB migration stagnation, accompanied by the accumulation of internal stress. (c3) depicts continuous GB migration via repeated switch between two coupling modes.
(a3), (b3) and (c4) show the final states for the case of pure GB sliding. } 
\label{coupling_schematic}
\end{center}
\end{figure}

The phenomenon of stress-driven GB migration in a bicrystal system is shown schematically in Figs.~\ref{coupling_schematic}a and b. 
The red lines denote the fiducial lines (i.e., the markers attached to the material). 
When the shear stress is applied parallel to the GB plane, the GB migrates upwards. 
From the change of the fiducial line, we notice that plastic shear deformation accompanies the GB migration. 
This is stress-driven GB migration. 

\begin{figure}[!t]
\begin{center}
\scalebox{0.38}{\includegraphics{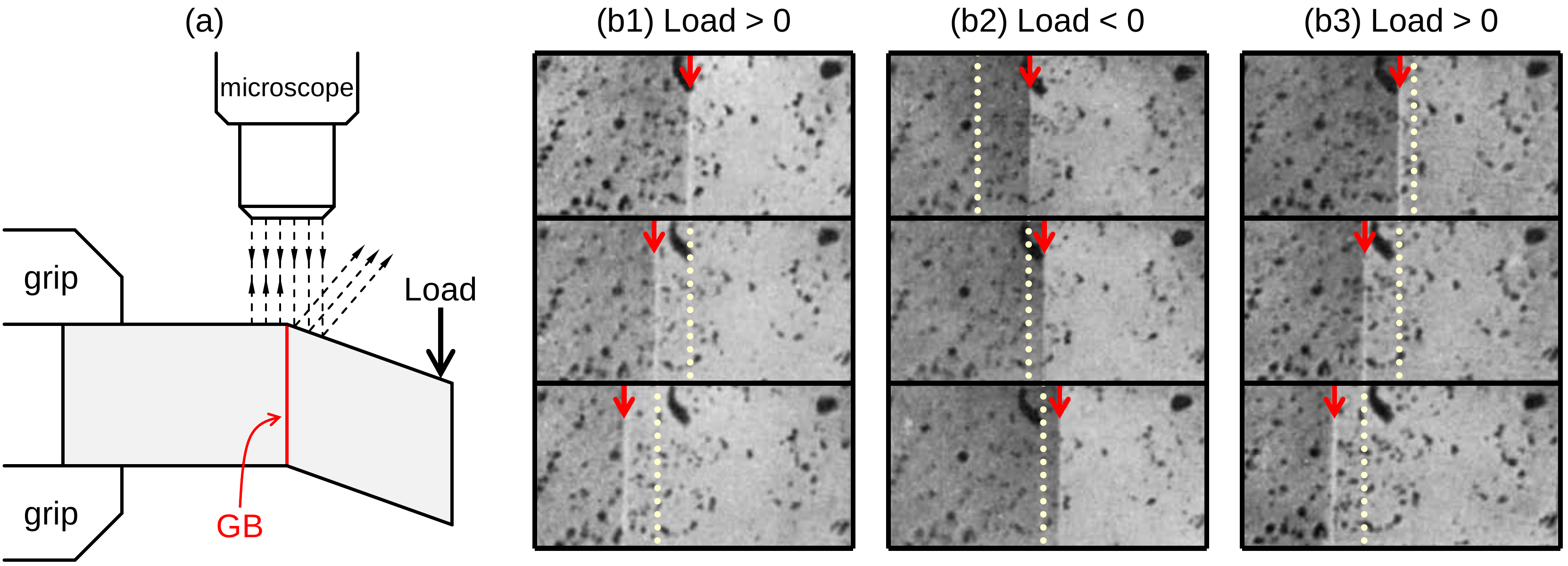}}
\caption{(a) Illustration of the experimental setup used to measure GB migration driven by an applied load. 
The GB position is detected by the brightness difference under illumination. 
(b1), (b2) and (b3) show a series of photographs of GB migration under positive load, negative load, and positive load again. 
The red arrows indicate the current GB position; the dotted lines label GB positions from previous photographs. 
The figures are reproduced from Ref.~\cite{li1953stress} (C. H. Li, Elsevier 1953).  } 
\label{washburn}
\end{center}
\end{figure}

GB migration driven by an applied shear stress has already been demonstrated in many experiments. 
The first experimental observation of stress-driven GB migration was made by Washburn et al.~\cite{washburn1952experiments,li1953stress} 
They prepared a pure Zn bicrystal sample containing a low-angle tilt boundary with tilt angle $\theta = 2^\circ$. 
This bicrystal was then loaded as a cantilever beam, as illustrated in Fig.~\ref{washburn}a. 
Observation was made using an optical microscope under oblique illumination such that the GB position could be determined by the brightness difference across the GB. 
They found that, as shown in Fig.~\ref{washburn}b, when the load was positive, the GB migrated to the left; when the load was negative, the GB migrated to the right. 
This process was reversible. 
This observation demonstrated that shear stress is the driving force for the GB migration. 

Similar stress-driven GB migration was widely observed in later experiments on bicrystals, including cases of both low- and high-angle STGBs and mixed tilt-twist GBs. 
Literature examples are listed below: Bainbridge et al. (STGB with $\theta = 2^\circ$ in Zn~\cite{bainbridge1954recent}), 
Fukutomi et al. ($[112]$, $[110]$ and $[100]$ STGBs with $\theta=3^\circ$-$17^\circ$ in Al~\cite{fukutomi1981stress}), 
Winning et al. ($[112]$ and $[111]$ STGBs with $\theta=3^\circ$-$35^\circ$ in Al~\cite{winning2001stress}; $[100]$ STGBs with $\theta=3^\circ$-$34^\circ$ in Al~\cite{winning2003motion}), 
Gorkaya et al. ($[100]$ STGBs with $\theta=5.6^\circ$-$84.2^\circ$ in Al~\cite{gorkaya2009stress}), 
Yoshida et al. ($[110]$ with $\theta=20^\circ$, $70.5^\circ$ and $129.5^\circ$ STGBs in ZrO$_2$~\cite{yoshida2004high}), 
Gorkaya et al. ($[100]$ mixed GBs with $\theta=18.2^\circ$ and a $20^\circ$ twist component in Al~\cite{gorkaya2011concurrent}), 
etc.  

\begin{figure}[!t]
\begin{center}
\scalebox{0.4}{\includegraphics{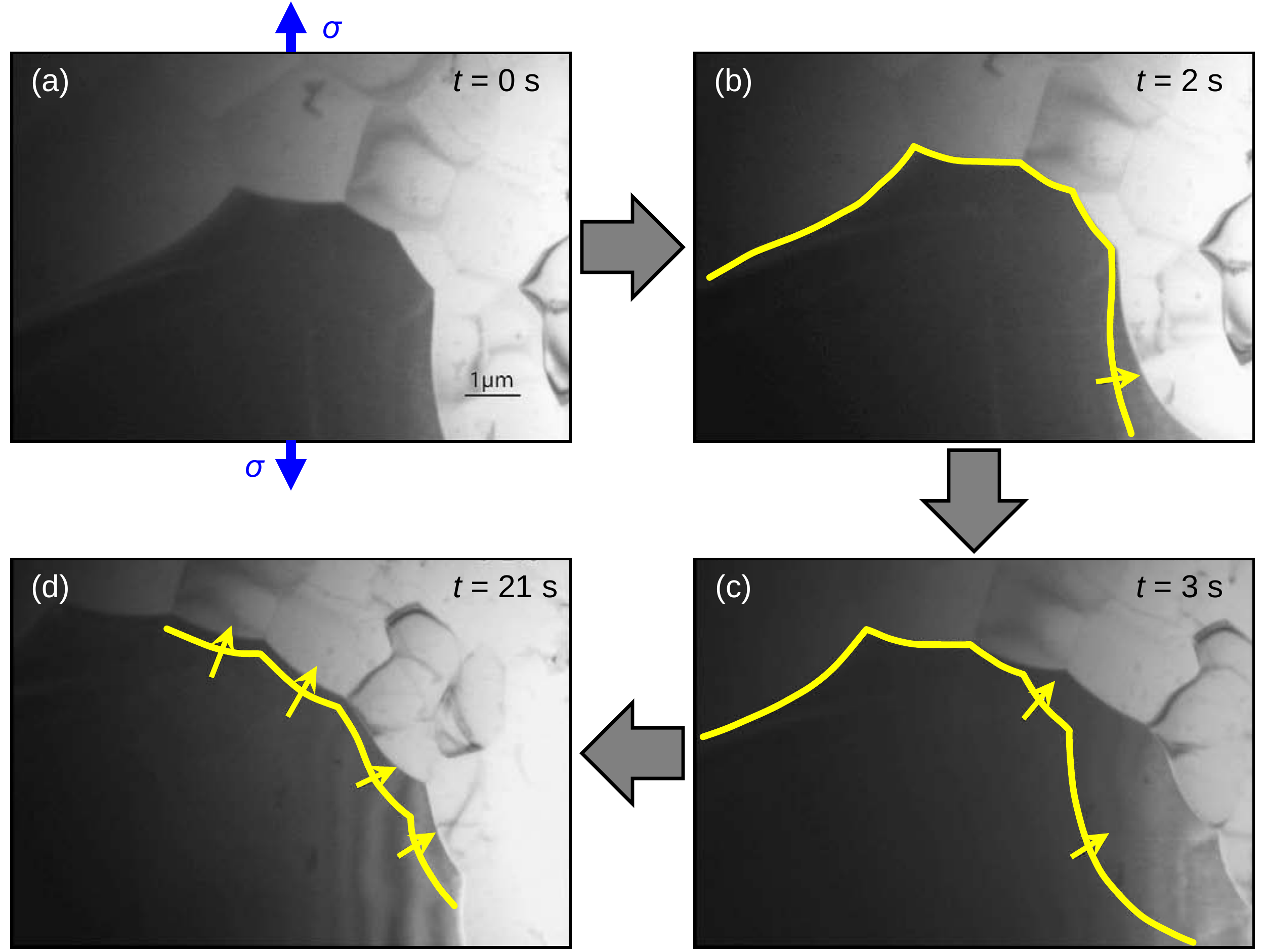}}
\caption{(a-d) Grain growth under applied stress (tension in the vertical direction) at $350^\circ$C, obtained from the in situ experiments performed by Mompiou, Caillard and Legros~\cite{mompiou2009grain}.
The yellow lines indicate previous GB positions and the arrows indicate the direction of GB migration.
The figures are reproduced from Ref.~\cite{mompiou2009grain} (F. Mompiou, Elsevier 2009).  } 
\label{mompiou}
\end{center}
\end{figure}

In a polycrystal system, stress-driven GB migration was also observed in experiments, although such migration process is inevitably influenced by other factors (such as the constraints applied by the triple junctions). 
For example, Rupert et al. found that the distribution of grain size is correlated with the distortional energy density~\cite{rupert2009experimental} in polycrystalline Al thin films. 
First, grain growth occurred when stress was applied on the film; second, the grains grow faster at the places where the distortional energy density is large. 
They concluded that the stress-driven GB migration influences the evolution of the microstructure of polycrystals. 
A similar result was also reported for the polycrystalline Pt thin films~\cite{sharon2011stress}. 
By means of in situ transmission electron microscopy (TEM), it was also observed that GB migration dynamics in polycrysals were sensitive to an applied stress. 
For example, Fig.~\ref{mompiou} shows the evolution of a GB (marked by the yellow lines) with a tensile stress applied in the vertical direction in an Al polycrystal, which was reported by Mompiou et al.~\cite{mompiou2009grain} 

\subsubsection{Theoretical background of literature on shear-coupling factor}
\label{theoretical_background}
How can we characterize the phenomenon of stress-driven GB migration quantitatively? 
Since the stress-driven GB migration is characterized by the simultaneous occurrence of GB sliding (defined as the relative displacement of the two neighboring grains) and GB migration, the stress-driven GB migration behavior can be characterized by a quantity, called the shear-coupling factor, which is generally defined as 
\begin{equation}
\beta_\ell(x,y,t) = \frac{\text{d}u_t}{\text{d}z_n}\bigg|_t (x,y)
\approx \frac{\text{d}u}{\text{d}z}\bigg|_t (x,y)
= \frac{\text{d}u/\text{d}t|_{x,y,t}}{\text{d}z/\text{d}t|_{x,y,t}}
\label{beta1}
\end{equation}
at an arbitrary position $(x,y)$ in a curved GB in the time interval from $t$ to $t+\text{d}t$, where $\text{d}z_n$ is the change of the normal component of the GB migration distance and $\text{d}u_t$ is the change of the tangential component of the relative displacement. 
In the approximated expression in Eq.~\eqref{beta1}, $\text{d}u$ is the relative displacement parallel to the $x$-$y$ plane and $\text{d}z$ is the GB migration distance normal to the $x$-$y$ plane; this approximation is valid when the gradient of the normal $\nabla \mathbf{n}$ is small everywhere along the GB. 

For a planar GB in a bicrystal system, the experimentally measurable shear-coupling factor (or ``effective'' shear-coupling factor), is defined as
\begin{equation}
\beta
= \frac{\langle\text{d}\bar{u}/\text{d}t\rangle_t}{\langle\text{d}\bar{z}/\text{d}t\rangle_t}
= \frac{v_\parallel}{v_\perp}, 
\label{beta2}
\end{equation}
where $v_\parallel$ and $v_\perp$ are the average velocities of GB sliding and GB migration, respectively, and $\langle\cdot\rangle_t$ denotes the average over time after steady-state GB migration is reached. 
In comparison with the last equation in Eq.~\eqref{beta1}, $\beta$ is $\beta_\ell$ with both the numerator and the denominator averaged over space and time. 
This definition is consistent with that used by Ashby~\cite{ashby1972boundary} and Cahn et al.~\cite{cahn2004unified,cahn2006coupling} 

\begin{figure}[!t]
\begin{center}
\scalebox{0.42}{\includegraphics{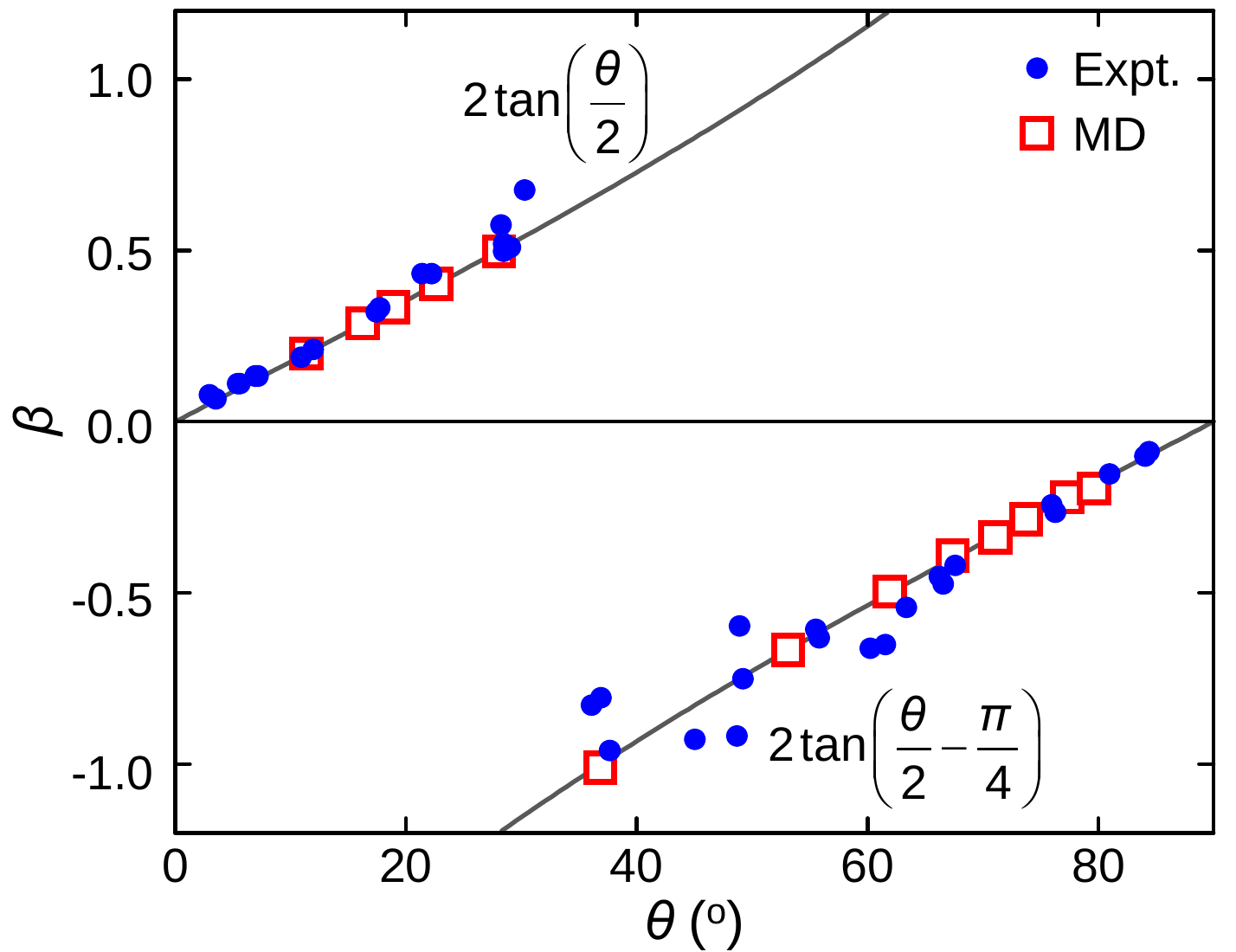}}
\caption{The shear-coupling factor $\beta$ vs. the misorientation angle for STGBs in FCC materials. 
The blue solid circles are the experimental data obtained for Al bicrystals~\cite{gorkaya2009stress} and the red hollow squares are the MD simulation data obtained for EAM Cu~\cite{cahn2006coupling}.
The dark gray lines are the theoretical prediction according to Eq.~\eqref{branch12}.  } 
\label{cahn_gorkaya}
\end{center}
\end{figure}

The values of $\beta$ were reported in many experiments and MD simulations for various bicrystal systems. 
The most interesting feature of $\beta$ is the geometry dependence. 
Cahn et al. performed MD simulations on a series of $[100]$ STGBs in Cu and showed that the value of $\beta$ vs. misorientation angle $\theta$ simply follows two branches, as shown in Fig.~\ref{cahn_gorkaya}. 
The two branches are described by the functions 
\begin{equation}
\beta(\theta) 
= \left\{\begin{array}{ll}
2\tan(\theta/2),	& 0 \le \theta \le \theta_\text{d} \\
2\tan[(\theta-90^\circ)/2], 	& \theta_\text{d} < \theta \le 90^\circ
\end{array}\right., 
\label{branch12}
\end{equation}
where the switch between two branches occurs at the angle $\theta_\text{d}$ (the exact value depends on the material). 
These simulation results have been validated by the experimental measurement performed by Gorkaya et al.~\cite{gorkaya2009stress}, as shown in Fig.~\ref{cahn_gorkaya}. 
However, in an in situ experiment on a polycrystalline Al sample, Mompiou et al.~\cite{mompiou2009grain} noticed that the trend of $\beta(\theta)$ can deviate from the two branches for some GBs; i.e., the measured value of $\beta$ can be lower than the value following the two branches. 
There is a dependence of $\beta$ on GB misorientation, but this dependence is not as simple as suggested by Eq.~\eqref{branch12} (for $[100]$ STGBs and similarly for other STGBs). 

\subsubsection{Prediction of shear-coupling factor}
\label{prediction_beta_stress}
How can we predict the value of $\beta$ based on a knowledge of GB geometry (bicrystallography)? 
A simple geometry dependence of $\beta$ was first proposed by Cahn et al.~\cite{cahn2006coupling,cahn2006duality}, which is exactly expressed as Eq.~\eqref{branch12} for $[100]$ STGBs in FCC materials. 
A similar geometry dependence was also suggested for the other STGBs by Homer et al.~\cite{homer2013phenomenology} 
Such a dependence can be rationalized based on the GB dislocation model~\cite{read1950dislocation,cahn2006coupling,cahn2006duality}. 
The idea is straightforward for low-angle tilt GBs. 
The structure of a low-angle STGB can be viewed as an array of edge lattice dislocations~\cite{read1950dislocation}. 
GB migration is associated with the glide of these lattice dislocations out of the initial GB plane; this process will inevitably induces shear of the entire system in the direction parallel to the GB plane. 
This is the origin of shear-coupled GB migration for low-angle tilt GBs. 
According to the geometry shown in Fig.~\ref{wedge}, we can immediately obtain the first branch in Eq.~\eqref{branch12}. 
The structure of a high-angle tilt GB cannot be viewed as an array of lattice dislocations. 
However, the GB dislocation content can be formally defined based on the Frank-Bilby equation~\cite{bilby1955types,frank1950resultant}. 
The shear-coupling factor for high-angle tilt GBs can still be predicted by Eq.~\eqref{branch12} following the same line of reasoning as for low-angle tilt GBs; the rigorous derivation based on the Frank-Bilby equation was provided by Cahn et al.~\cite{cahn2006coupling,cahn2006duality}  

However, as mentioned previously, the prediction based on the GB dislocation model is not always consistent with the experimental measurement; there are exceptions~\cite{mompiou2009grain}.  
The problem lies with the fact that the formal GB dislocation content for high-angle tilt GBs is not necessarily physically meaningful; the determination of GB dislocation content depends on the choice of reference and this reference is not unique~\cite{SuttonBalluffi}. 
The Frank-Bilby equation just reflects the bicrystallography rather than the real mechanism associated with the GB migration. 
Without relying on the (possibly unmeaningful) definition of GB dislocations, Mompiou et al. proposed another way to predict $\beta(\theta)$ directly based on the GB geometry~\cite{mompiou2010smig}, which they called the shear migration geometrical (SMIG) model. 
They considered the transformation from the structure of one grain to the structure of the other through a simple shear of a chosen unit cell and atomic shuffling inside this unit cell. 
For a GB with fixed macroscopic geometric DOFs, multiple choices of the unit cell result in multiple shear-coupling factors. 
Some of these shear-coupling factors can be smaller than the value predicted based on the GB dislocation model, which is consistent with the experimental observations. 
The problem with the SMIG model is that it only provides us a list of all possible $\beta$ values for each particular GB geometry. 
It does not tell us which $\beta$ value (or coupling mode) will occur under certain condition in practice. 

The issue with the GB dislocation model~\cite{cahn2006coupling,cahn2006duality} is that it is based on the unphysical GB migration mechanism (i.e., glide of the formal GB dislocation out of the GB plane). Therefore, the GB dislocation model may give the wrong prediction especially for high-angle GBs. 
The SMIG model is purely a geometric model such that it can give a list of all possible disconnection modes but cannot predict which one occurs in reality. 
Now, we will propose a new way to predict the geometry dependence of $\beta$ based on the correct GB migration mechanism. 

\begin{figure}[!t]
\begin{center}
\scalebox{0.38}{\includegraphics{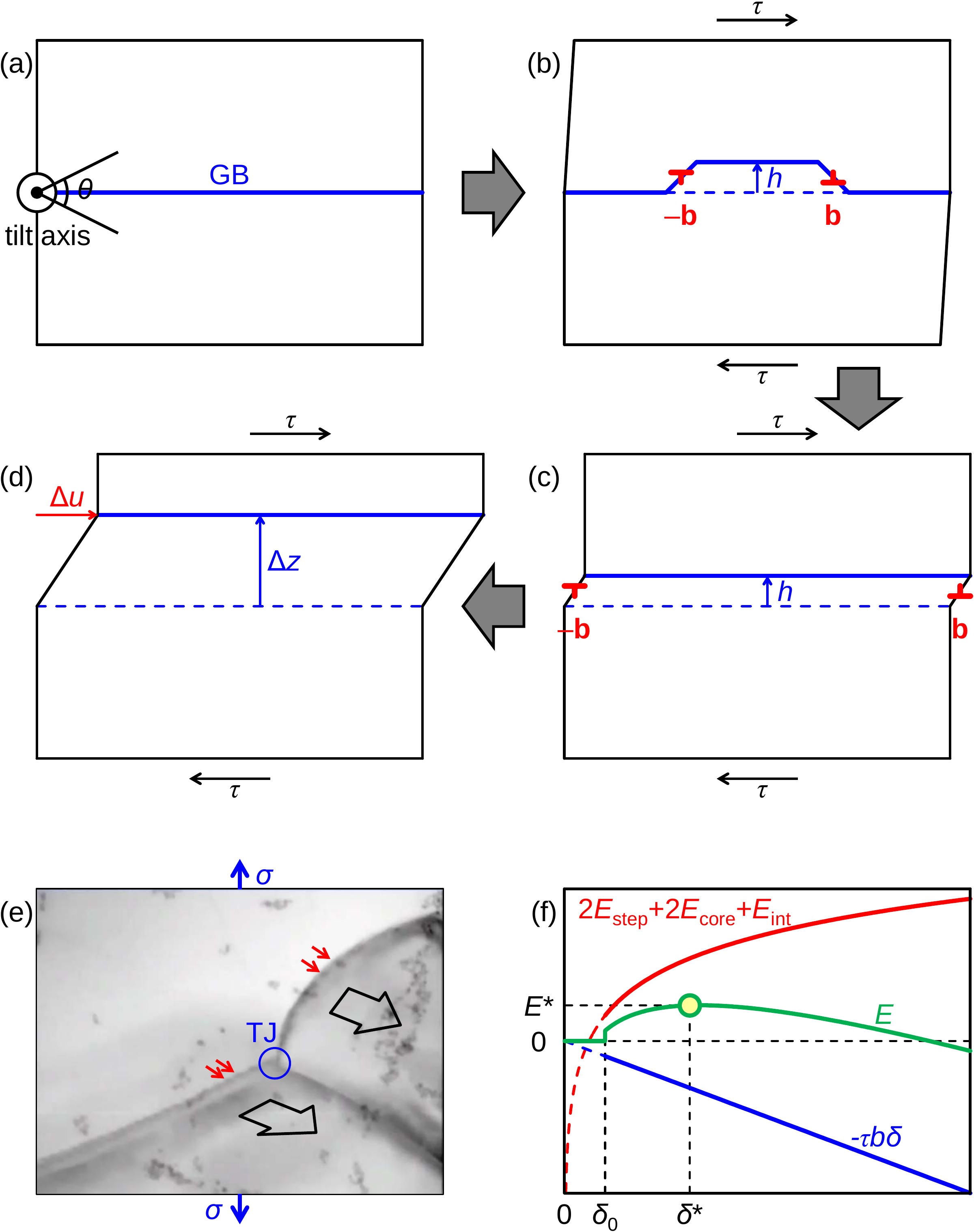}}
\caption{(a-d) Schematic of the microscopic mechanism of shear stress-driven GB migration in a bicrystal containing a tilt GB: (a) an initially flat GB; (b) nucleation of a pair of disconnections (denoted by the red ``$\perp$'' symbols) characterized by $(\mathbf{b},h)$ and $(-\mathbf{b},h)$; (c) separation of the two disconnections, which causes the GB to migrate by $h$; (d) the relative displacement $\Delta u$ and GB migration $\Delta z$ induced by repeated nucleation and separation of disconnection pairs. 
(e) Migration of GBs and TJ under the applied stress observed in the experiment by Rajabzadeh et al.~\cite{rajabzadeh2013evidence}. 
The black, hollow arrows denote the direction of GB migration. 
The red arrows indicate several moving disconnections along the GBs, which demonstrates that the GB migrates via the motion of disconnections. 
%This figure is reproduced from Ref.~\cite{rajabzadeh2013evidence}. 
(f) Schematic plot of Eq.~\eqref{E_W1} or \eqref{E_W2}. 
The red curve corresponds to the step energy, disconnection core energy, and elastic energy. The blue curve corresponds to the work done by the shear stress $\tau$ and the green curve corresponds to the total energy. 
In the total energy, there is a barrier $E^*$ at the critical disconnection separation $\delta^*$.  } 
\label{conservative}
\end{center}
\end{figure}

GB migration occurs through the nucleation and glide of the GB disconnections along a GB. 
This was proposed by Smith et al.~\cite{king1980effects,rae1980mechanisms}, Fukutomi et al.~\cite{fukutomi1985grain,fukutomi1991sliding} and Cahn et al.~\cite{cahn2006coupling} 
As illustrated in Figs.~\ref{conservative}a and b, starting from a clean (free of extrinsic defects), planar GB in a bicrystal, the applied shear stress $\tau$ triggers the nucleation of a pair of disconnections (with opposite sign of both Burgers vector and step height). 
Then, driven by the stress $\tau$, the two disconnections glide away from each other. 
Since it is likely that a disconnection will be associated with a step height, the glide of the disconnections will cause the GB to migrate upwards by an amount equal to the step height $h$ (see Fig.~\ref{conservative}c). 
When this nucleation and glide process occurs repeatedly, the shear strain will be established and, at the same time, the GB will migrate upwards continuously. 
In practice (for example, in a polycrystal), the nucleation of disconnections is probably not homogeneous; it is possible that the disconnections are nucleated at the triple junctions or free surface. 
Such mechanism has already been observed in the in situ experiments performed by Rajabzadeh et al.~\cite{rajabzadeh2013evidence}
As shown in Fig.~\ref{conservative}e, when the tensile stress was applied in the vertical direction, the GBs would migrate downwards. 
We can see that the GB migration was carried out through the motion of line defects along the GB, and the line defects in the GB are disconnections. 
This mechanism was also demonstrated in MD simulations~\cite{hyde2005atomistic,monk2006role,race2014role,race2015mechanisms}. 

As stress-driven GB migration occurs via the repeated nucleation and glide of disconnections, as illustrated in Fig.~\ref{conservative}, the shear-coupling factor can be written as 
\begin{equation}
\beta = b/h
\label{beta3}
\end{equation}
based on the microscopic mechanism, where $b$ and $h$ are the Burgers vector and the step height of the activated disconnections. 
Therefore, in order to predict the value of $\beta$, we only need to know what the active disconnection (or disconnection mode) is for each particular GB geometry. 

From Section~\ref{multiple_modes}, for any particular GB geometry, the disconnection mode, i.e., the $(b,h)$ pair, is not unique. 
We have already determined a way to list all the possible $(b,h)$ pairs for any GB geometry; the complete list is denoted as $\{(b_n, h_{nj})\}$. 
Then, according to Eq.~\eqref{beta3}, the complete list of the shear-coupling factors is $\{\beta_{nj}\equiv b_n/h_{nj}\}$; this list is consistent with the list predicted by the SMIG model. 
Now, from the list $\{\beta_{nj}\}$, we need to choose the one that really occurs. 
According to the disconnection pair nucleation mechanism (as illustrated in Fig.~\ref{conservative}), $\beta$ is one element in the list $\{\beta_{nj}\}$ which corresponds to the lowest nucleation barrier. 
Based on Eq.~\eqref{Etot}, the energy (per unit length) for the formation of a pair of disconnections under the applied stress $\boldsymbol{\sigma}$ can be written as 
\begin{equation}
E = 2E_\text{step} + 2E_\text{core} + E_\text{int} + W, 
\label{E_W1}
\end{equation}
where $W$ is the work done by $\boldsymbol{\sigma}$ and exerted on the disconnections. 
In general, based on the coordinate system in Fig.~\ref{energetics}, $W = [\mathbf{o}\times(\boldsymbol{\sigma}\mathbf{b})]\cdot \boldsymbol{\delta} = -\delta (\boldsymbol{\sigma}\mathbf{n})\cdot\mathbf{b}$. 
If a pure shear stress $\boldsymbol{\sigma} = \tau (\mathbf{p}\otimes\mathbf{n} + \mathbf{n}\otimes\mathbf{p})$ is applied, then $W = -\tau \mathbf{b}\cdot\boldsymbol{\delta}$. 
Since the stress-driven GB migration is conservative, $\mathbf{b}$ has no component perpendicular to the GB plane (otherwise, the motion of disconnection requires atomic diffusion). 
In addition, it is reasonable to consider case of $\mathbf{b} = b\mathbf{p}$ because the component parallel to $\mathbf{p}$ is much smaller than the component parallel to $\mathbf{o}$. Then, Eq.~\eqref{E_W1} becomes 
\begin{equation}
E
= 2\mathit{\Gamma}_\text{s}|h| + 2\mathcal{K}b^2 \ln\frac{\delta}{\delta_0} - \tau b\delta, 
\label{E_W2}
\end{equation}
where $\mathcal{K}\equiv K\mathsf{f}(\pi/2) = \mu/[4\pi(1-\nu)]$. 
Without the the work term (i.e., the last term) in Eq.~\eqref{E_W2}, the energy $E$ will always increase with the seperation of the disconnection pair (see the red line in Fig.~\ref{conservative}f), and thus the GB will never migrate. 
With the addition of the work term, $E$ will first increase with an increase of $\delta$ and then decrease with further increase of $\delta$ (see the blue line in Fig.~\ref{conservative}f). 
Hence, with the work term, there is a nucleation barrier, corresponding to the critical separation $\delta^*$ determined by the condition $(\text{d}E/\text{d}\delta)_{\delta^*} = 0$, i.e., 
\begin{equation}
\delta^* = \frac{2\mathcal{K}b^2}{\tau b}. 
\label{deltastar_stress}
\end{equation}
The barrier for the nucleation of a pair of disconnections characterized by $(b,h)$ can then be expressed as 
\begin{equation}
E^* = E(\delta^*) = 2\mathit{\Gamma}_\text{s}|h| + 2\mathcal{K}b^2 \ln\frac{2\mathcal{K}b^2}{e\tau b\delta_0}.  
\label{Estar_stress}
\end{equation}
For the special case where $b = 0$, the nucleation of such disconnections (i.e., pure steps) cannot be driven by the applied shear stress; accordingly, $\delta^* \to \infty$. 
Since there are multiple disconnection modes $\{(b_n, h_{nj})\}$, corresponding to each mode, we can obtain a particular nucleation barrier $E^*_{nj}$ based on Eq.~\eqref{Estar_stress}. 
The disconnection mode that really occurs is associated with the lowest nucleation barrier; this mode determines the shear-coupling factor $\beta$ for this particular GB geometry. 

The procedure to predict the shear-coupling factor $\beta$ for a particular GB geometry under the applied shear stress is summarized below. 
\begin{enumerate}
\item[(i)] According to the GB geometry, determine the complete list of disconnection modes $\{(b_n, h_{nj})\}$, using the algorithm proposed in Section~\ref{enumeration}. 
\item[(ii)] For each mode $(b_n, h_{nj})$, calculate the nucleation barrier $E^*_{nj}$, using Eq.~\eqref{Estar_stress}. 
\item[(iii)] Identify the mode corresponding to the lowest nucleation barrier, denoted as $(b_0, h_0)$. 
\item[(iv)] The shear-coupling factor for this GB geometry is $\beta = b_0/h_0$. 
\end{enumerate}

\subsubsection{Comparison with simulation and experimental results}
\begin{figure}[!t]
\begin{center}
\scalebox{0.39}{\includegraphics{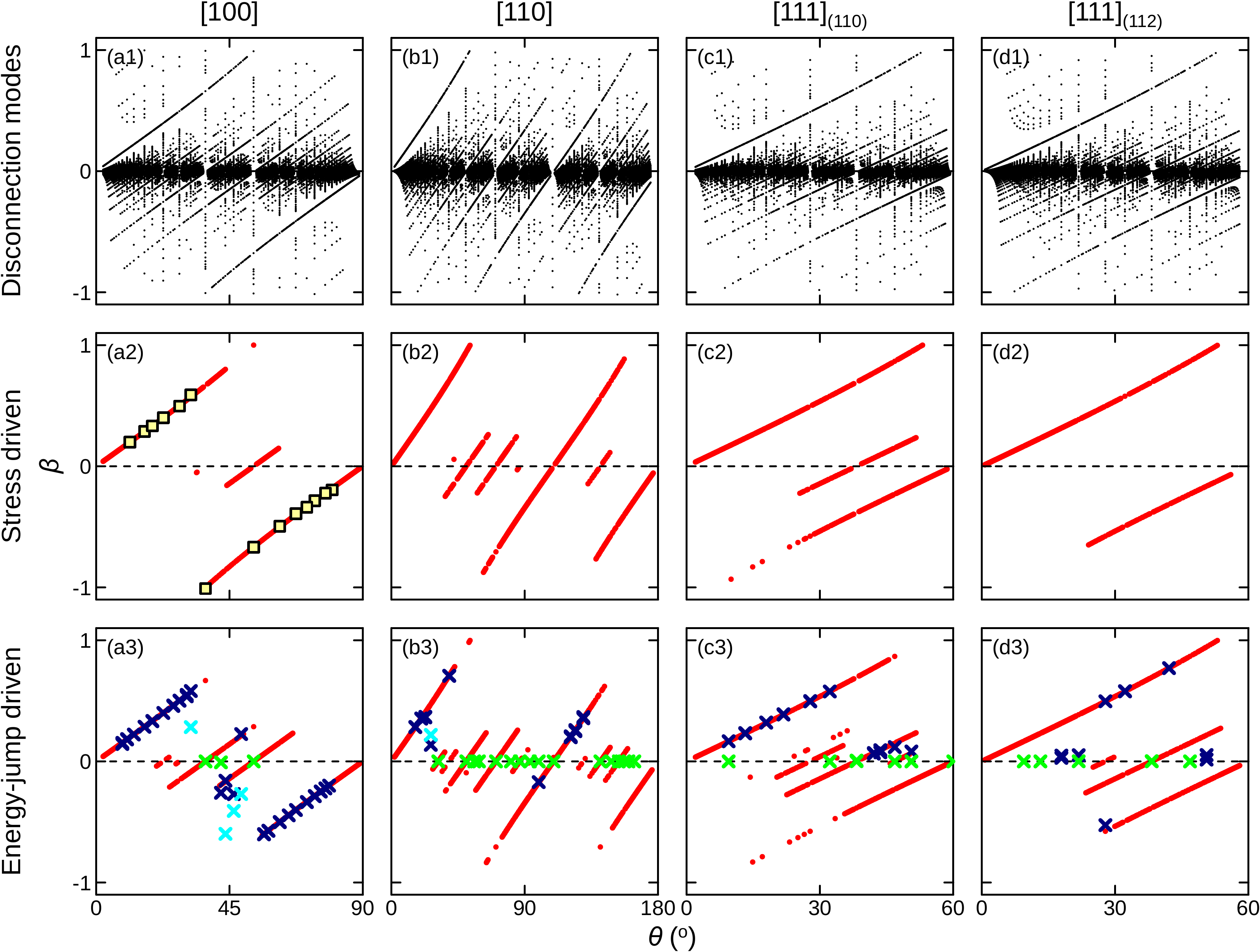}}
\caption{Shear-coupling factor $\beta$ vs. misorientation angle $\theta$ for (a) $[100]$, (b) $[110]$, (c) $[111]_{(110)}$ (the subscript denotes the median plane) and (d) $[111]_{(112)}$ STGBs in a FCC material.
In the top panels, the black dots represent all possible $\beta$ values (in the range of $|\beta|<1$), corresponding to all possible disconnection modes. 
In the middle panels, the red dots are the predicted $\beta$ values for the case of stress-driven GB migration.
The squares are the data obtained by MD simulations~\cite{cahn2006coupling}. 
In the bottom panels, the red dots are the predicted $\beta$ values for the case of energy-jump-driven GB migration. 
The colored crosses are the data obtained by MD simulations~\cite{homer2013phenomenology}. 
The green crosses are the data for the case where GB migration is not observed in the simulations; the blue crosses correspond to the case where the prediction is consistent with simulation; the light blue crosses correspond to where the prediction is inconsistent with simulation.   } 
\label{beta_predict}
\end{center}
\end{figure}

Following the procedure listed in the last section, we predicted the shear-coupling factor $\beta$ as a function of the misorientation angle $\theta$ for four series of GBs (i.e., $[100]$, $[110]$, $[111]_{(110)}$, and $[111]_{(112)}$ STGBs) in an FCC material (the material parameters are taken for Cu). 
First, following Step (i) in the last section, we obtained the $\{(b_n, h_{nj})\}$ list for each $\theta$ value. 
As shown in Figs.~\ref{beta_predict}a1, b1, c1 and d1, corresponding to each $\theta$ value, there are multiple values of shear-coupling factor (black dots), defined as $\{\beta_{nj}\equiv b_n/h_{nj}\}$. 
Then, following Step (ii) and (iii), for each $\theta$ value, there is only one disconnection mode associated with the lowest nucleation barrier. 
This mode gives the shear-coupling factor $\beta$ according to Step (iv). Figures~\ref{beta_predict}a2, b2, c2 and d2 show the $\beta$ values obtained via this procedure (red dots); in this figure, the shear-coupling factor is unique for each $\theta$ value. 
This is the prediction of the shear-coupling factor for the practical situation of stress-driven GB migration. 

In Fig.~\ref{beta_predict}a2, the $\beta$ values obtained by the MD simulations~\cite{cahn2006coupling} are shown as the squares. 
By comparison, we notice that the available simulation data is completely consistent with our prediction. 
However, the available simulation data is also consistent with the prediction based on the GB dislocation model, i.e., Eq.~\eqref{branch12}. 
Different from the prediction based on the GB dislocation model, the disconnection model predicts the existence of lower $\beta$ values for high-angle GBs (this is particularly obvious for the $[110]$ STGBs). 
In the experiments~\cite{mompiou2009grain}, it was indeed found that the measured $\beta$ values were lower than the prediction based on the GB dislocation model for some high-angle GBs. 
Therefore, the disconnection model provides better prediction of the geometry dependence of $\beta$ than the GB dislocation model. 

\subsubsection{Temperature dependence of shear coupling}
\label{temperature_stress}
Up to now, we do not consider the effect of temperature on the stress-driven GB migration; so far, the discussion of $\beta$ is rigorously true only at zero temperature. 
In this section, we will review work on the temperature dependence of shear coupling. 

The extreme case for illustrating the influence of temperature is ``pure'' GB sliding at high temperature (usually $T>0.4 T_\text{m}$). 
Here, we define ``pure'' GB sliding as the relative displacement of one grain with respect to the other parallel to the GB plane without GB migration for a microscopically planar GB. 
As schematically shown in Fig.~\ref{coupling_schematic}a3, pure GB sliding is demonstrated for a bicrystal system. 
From the change of the fiducial line with respect to the initial state, we notice that the plastic shear deformation is established but the GB position does not change. 
According to the definition from Eq.~\eqref{beta2}, $\beta \to \infty$ for pure GB sliding, since in this case $v_\parallel$ is finite but $v_\perp$ is zero. 
We know that, at zero temperature, $\beta$ has a finite value which depends on the GB geometry (such as the case shown in Fig.~\ref{coupling_schematic}a2). 
The general trend of the temperature dependence is that $\beta$ goes from a finite value to infinity as temperature increases. 

\begin{figure}[!t]
\begin{center}
\scalebox{0.32}{\includegraphics{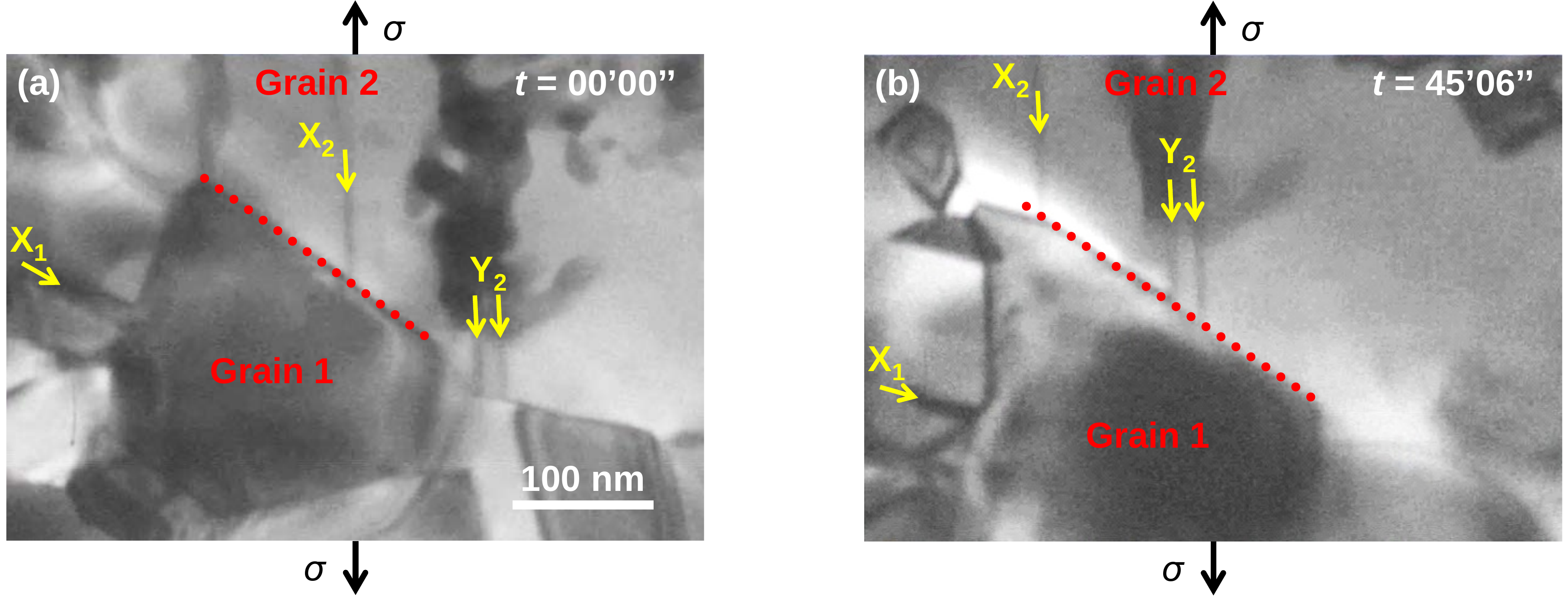}}
\caption{(a-b) GB sliding observed in the experiment performed by Hosseinian et al.~\cite{hosseinian2016quantifying} 
Sliding occurs between Grain 1 and Grain 2; the red dotted line indicates the position of the GB between these two grains. 
The yellow arrows (denoted by X$_1$, X$_2$ and Y$_2$) indicate stacking faults fixed in the grains. 
Taking X$_1$ as a reference, we find that Grain 2 moves with respect to Grain 1 but the GB does not apparently migrate. 
This figure is reproduced from Ref.~\cite{hosseinian2016quantifying} (O. N. Pierron, Royal Society of Chemistry 2016). } 
\label{GBsliding}
\end{center}
\end{figure}

Pure GB sliding has been observed in experiments on planar GBs. 
For example, Hosseinian et al. performed an in situ experiment and observed the occurrence of pure GB sliding in a Au film~\cite{hosseinian2016quantifying}. 
As shown in Fig.~\ref{GBsliding}, we focus on the behavior of the GB between Grain 1 and Grain 2. 
In Grain 2, X$_2$ and Y$_2$ denote the immobile stacking faults, which serve as markers. 
X$_1$ is another marker at another GB neighboring Grain 1. 
After the tensile stress is applied, the microstructure changes from Fig.~\ref{GBsliding}a to b. 
On one hand, by comparing the distance between the marker X$_1$ and the GB plane, we can find that the GB does not migrate during the loading process. 
On the other hand, via tracing the motion of the markers X$_2$ and Y$_2$ after the tensile stress is applied, we can notice the relative displacement of Grain 2 with respect to Grain 1. 
This observation demonstrates that, under an applied stress, pure GB sliding occurs. 

As to the understanding of pure GB sliding, there are many mechanisms proposed for the pure sliding of non-planar GBs~\cite{SuttonBalluffi,raj1971grain,ashby1972boundary}; however, there is still no satisfying explanation for the pure sliding of planar GBs. 
It is always suggested that GB sliding of a planar GB should be coupled with GB migration. 
This is naturally implied by the disconnection model. 
As mentioned previously, the disconnection nucleated under the applied shear stress is characterized by a particular $(b,h)$ pair ($h\ne 0$ since pure step is not energetically favorable in the stress-driven kinetics); thus, shear-coupled GB migration is always expected (with finite value of $\beta$, i.e., $b/h$). 
However, this is in conflict with observations of pure GB sliding (i.e., $\beta\to \infty$) in experiments~\cite{yoshida2004high,hosseinian2016quantifying,gorkaya2009stress} and MD simulations~\cite{van2001grain,cahn2006coupling,qi2007molecular}. 

Now, we confront whether it is possible to understand pure GB sliding based on the disconnection model. 
As mentioned in the previous section, for a given bicrystallography, there existing multiple disconnection modes. 
According to the disconnection model, the disconnection mode corresponds to the lowest nucleation barrier of disconnections. 
However, this is rigorously true only at zero temperature. 
At finite temperature, multiple disconnection modes (not only the one with the lowest nucleation barrier) can be activated with certain probabilities. 
As already demonstrated in Fig.~\ref{multiple_step}, even for the modes with the same Burgers vector, the step height can be positive or negative, implying that the same shear deformation can make the GB migrate upwards or downwards. 
Thus, on average, the shear-coupling factor becomes larger than that in the case where only one particular mode is activated. 
At the limit of high temperature, the nucleation probability for many disconnection modes may be very close such that the shear-coupling factor will get close to infinity -- this explains pure GB sliding in the framework of disconnection model. 

Based on this idea, we can estimate the effective shear-coupling factor as a function of temperature, denoted as $\bar{\beta}(T)$, in the sense of an ensemble average. 
The rate for the nucleation of a pair of disconnections $(b_n, h_{nj})$ is 
\begin{equation}
r_{nj} =  
\omega\exp\left(-\frac{E^*_{nj}}{k_\text{B}T/L_x}\right), 
\label{rnj}
\end{equation}
where $\omega$ is the attempt frequency and $E^*_{nj}$ is obtained by Eq.~\eqref{Estar_stress} with $b_n$ and $h_{nj}$ as the input. 
The nucleation and glide of this pair of disconnections leads to the relative displacements $b_n$ and the GB migration distance $h_{nj}$. 
The velocities of relative displacement and GB migration are then
\begin{equation}
v_\parallel = \sum_{n=-\infty}^\infty\sum_{j=-\infty}^\infty b_n r_{nj}
\text{, and }
v_\perp = \sum_{n=-\infty}^\infty\sum_{j=-\infty}^\infty h_{nj} r_{nj}, 
\label{vpara_vperp}
\end{equation}
respectively. 
Then, the effective shear-coupling factor can be obtained according to Eq.~\eqref{beta2}, which can be written explicitly as 
\begin{equation}
\bar{\beta}(T)
= \frac{\displaystyle{ \sum_{n=-\infty}^\infty\sum_{j=-\infty}^\infty b_n \exp\left(-\frac{E^*_{nj}}{k_\text{B}T/L_x}\right) }}
{\displaystyle{ \sum_{n=-\infty}^\infty\sum_{j=-\infty}^\infty h_{nj} \exp\left(-\frac{E^*_{nj}}{k_\text{B}T/L_x}\right) }}.  
\label{barbeta}
\end{equation}

\begin{figure}[!t]
\begin{center}
\scalebox{0.42}{\includegraphics{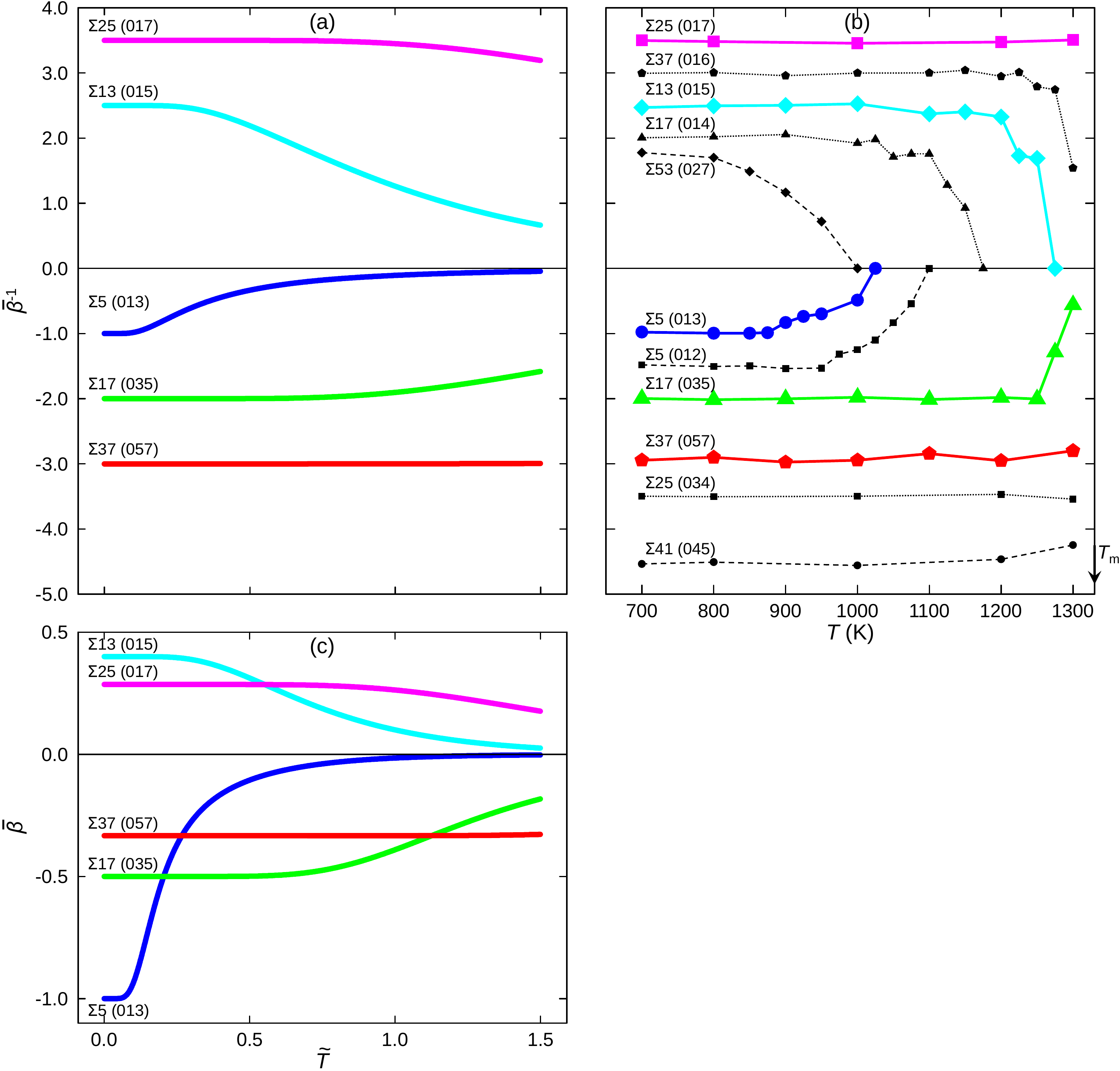}}
\caption{The temperature dependence of the effective shear-coupling factor $\beta$ for a series of $[100]$ STGBs in an FCC material. 
(a) shows the prediction in the case of GB migration under constant stress. 
(b) shows MD simulation results in the case of GB migration under constant shear rate; this is reproduced from Ref.~\cite{cahn2006coupling} (Y. Mishin, Elsevier 2006). 
(c) shows the prediction in the case of GB migration driven by energy-jump driving force. } 
\label{beta_T}
\end{center}
\end{figure}

Figure~\ref{beta_T} shows the temperature dependence of $\bar{\beta}^{-1}$ for five different GB geometries, i.e., $\Sigma 5$ $[100]$ $(013)$, $\Sigma 13$ $[100]$ $(015)$, $\Sigma 17$ $[100]$ $(035)$, $\Sigma 25$ $[100]$ $(017)$, and $\Sigma 37$ $[100]$ $(057)$ STGBs in FCC materials. 
Figure~\ref{beta_T}a shows the prediction from Eq.~\eqref{barbeta} while Fig.~\ref{beta_T}b shows the data obtained from MD simulations~\cite{cahn2006coupling}. 
The prediction is consistent in trend with the MD results; i.e., at low temperature $\bar{\beta}^{-1} \to \beta^{-1}$ while at high temperature $\bar{\beta}^{-1} \to 0$. 
We find that the prediction from Eq.~\eqref{barbeta} can reflect the correct geometry effect. 
For example, the ideal shear-coupled GB migration behavior for the $\Sigma 37$ STGB can persist to a very high temperature, which is consistent with the feature from the MD results. 
We also notice the discrepancy between the prediction and the MD results -- the transition of $\bar{\beta}^{-1}$ from the ideal zero-temperature value $\beta^{-1}$ to $0$ as temperature increases is different. 
The disconnection model predicts a gradual change of $\bar{\beta}^{-1}$ with temperature while the MD results show a relatively sharp transition. 
This discrepancy may result from the different loading condition. 
The prediction based on Eq.~\eqref{barbeta} is applicable for the condition of constant stress (note that there is a parameter $\tau$ in $E^*_{nj}$) while the MD simulation was performed under the condition of constant shear rate. 

%%%%%%%%%%%%%%%%%%%%%%%%%%%%%%%%%%%%%%%%
\subsection{Energy-jump-driven grain-boundary migration}
Other than shear stress, conservative GB migration can be driven by another type of driving force, i.e, energy density jump across a GB. 
Energy-jump-driven GB migration is demonstrated for a bicrystal simulation in Figs.~\ref{coupling_schematic}b1 and b2. 
If we artificially set the free energy density in the upper grain $\psi_\text{U}$ to be higher than that in the lower grain $\psi_\text{L}$, the GB will migrate upwards in order to lower the total free energy of the whole system. 
This phenomenon refers to energy-jump-driven GB migration, and the artificially applied free energy density jump across the GB is denoted by $\psi$ ($\equiv \psi_\text{U}-\psi_\text{L}$). 
Since, in general, GB migration couples with the shear deformation, the fiducial line will transform from the initial state to the state shown in Fig.~\ref{coupling_schematic}b2 after GB migration occurs. 

\subsubsection{Physical interpretation to energy-jump driving force}
The initial aim for the application of energy jump is to study GB migration by MD simulations  (also called synthetic driving force in MD simulations)~\cite{janssens2006computing,olmsted2009survey,homer2013phenomenology,hadian2016atomistic}. 
However, we will show below that the energy jump can serve as a model driving force, which represents a large class of real driving forces different from shear stress. 

\begin{itemize}
\item[(i)] \textit{Elastic anisotropy}. 
When stress or strain is applied on a bicrystal, both grains will be elastically deformed and elastic energy will be stored. 
Since the elastic modulus is anisotropic and the orientations of two grains are different, the elastic energy density in two grains is different under the same stress or strain. 
The difference in elastic energy density plays the role of an energy-jump driving force for GB migration. 
The GB migration due to elastic anisotropy was observed in MD simulations performed by Zhang et al.~\cite{zhang2004computer,zhang2009grain} 

\item[(ii)] \textit{Excess defect density}. 
If the density of defects (point defects or dislocations) in the upper grain is higher than that in the lower grain (ref. Fig.~\ref{coupling_schematic}b1), the GB will migrate upwards such that the defect density in the whole bicrystal system decreases. 
Each defect is associated with excess energy, so the difference in energy density is proportional to the difference in defect density between the two grains. 
This results in a driving force on the GB.  
Such a situation typically occurs in the process of recrystallization. 
During recrystallization, the newly nucleated, defect-free grains grow in the expense of the surrounding highly defected grains (e.g. Ref.~\cite{rollett2004recrystallization}); the total defect density in the whole system is lowered via GB migration. 
Another related process is irradiation-induced selective grain growth~\cite{haessner1970boundary,van1978ion,atwater1988interface,seita2013selective}. 
Seita et al.~\cite{seita2013selective} performed experiments in which they irradiated a polycrystalline Au film by ion beam and found that the grains with a channeling orientation relative to the direction of the ion beam grew at the expense of the remaining grains. 
Because the irradiation-induced point defect density in the grains with a channeling orientation is lower than that in the grains with other orientations, the difference in point defect density provides the driving force for the GB migration.  

\item[(iii)] \textit{Magnetic field}. 
When a magnetic field is uniformly applied on a bicrystal, there will be a difference in magnetic free energy density in two grains because the magnetic susceptibility is anisotropic. 
The difference in magnetic free energy density can, in effect, be modeled by the energy-jump driving force. 
The GB migration due to the anisotropy of magnetic susceptibility has been experimentally demonstrated in Bi~\cite{mullins1956magnetically,molodov1998true},  Zn~\cite{molodov2006grain,gunster2010magnetically}, Ti~\cite{molodov2004effect} and Zr~\cite{molodov2010observations}. 

\item[(iv)] \textit{Surface energy}. 
At the free surface of a polycrystalline sample, grains with different orientations exposed to the vacuum (or gas) are associated with the free surfaces with different surface planes and, thus, different surface energy. 
The GBs will tend to migrate such that the grains with lower surface energy grow while those with higher surface energy shrink. 
The difference in surface energy provides a driving force for GB migration. 
For example, minimization of surface energy is responsible for the $(111)$ texture in the FCC polycrystalline thin films~\cite{thompson1990grain}. 

\item[(v)] \textit{Capillary force}.
Perhaps the most important and well-known type of driving force during grain growth is capillary force.  
There is an excess energy in a polycrystal due to the existence of GBs (proportional to the total GB area). 
As a result, there is a pressure $\gamma \kappa$ (Gibbs-Thomson equation)~\cite{burke1952recrystallization} on each GB, where $\gamma$ is the GB energy (assumed to be isotropic) and $\kappa$ is the local mean curvature of the GB plane. 
In this sense, the capillary force $\gamma\kappa$ is analogous to the energy-jump force~\cite{mullins1956two,mullins1956magnetically}. 
\end{itemize}

\subsubsection{Prediction of shear-coupling factor}
\label{prediction_beta_synthetic_force}
Since a broad class of driving forces can be modeled by the energy-jump driving force (i.e., the free energy density difference across a GB $\psi$), we will consider how to predict the shear-coupling factor $\beta$ in the case of energy-jump-driven GB migration. 
As already discussed in Section~\ref{prediction_beta_stress}, GB migration mechanism is associated with the nucleation and glide of GB disconnections along the GB; this is true no matter which type of driving force is applied. 
As with the case of stress-driven GB migration, the value of $\beta$ can be determined by Eq.~\eqref{beta3} with $b$ and $h$ corresponding to the disconnection mode which has the lowest nucleation barrier. 

\begin{figure}[!t]
\begin{center}
\scalebox{0.38}{\includegraphics{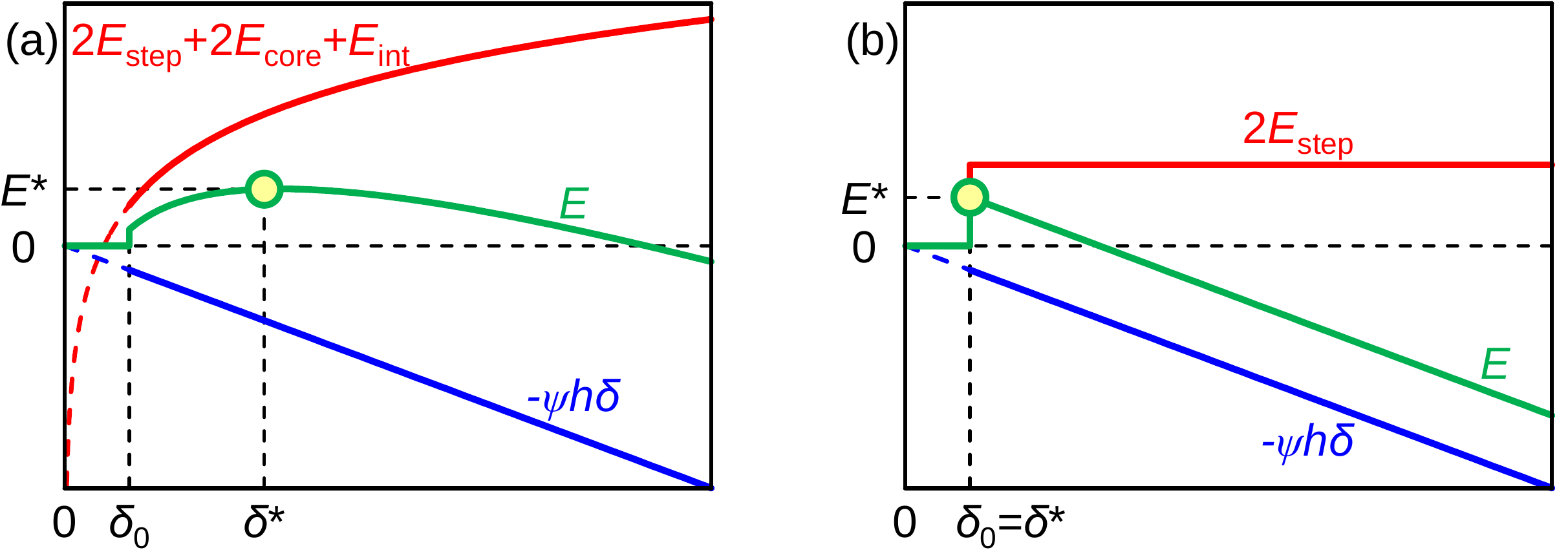}}
\caption{(a) Schematic plot of Eq.~\eqref{E_W1} or (equivalently) Eq.~\eqref{E_W3}. 
The red curve corresponds to the step energy, disconnection core energy, and elastic energy. The blue curve corresponds to the work done by the energy-jump driving force $\psi$. The green curve corresponds to the total energy. 
In the total energy, there is a barrier $E^*$ at the critical disconnection separation $\delta^*$. 
(b) The special case of (a), where $E_\text{core} = E_\text{int} = 0$ (i.e., this represents a pure-step disconnection mode).   } 
\label{synthetic_energy}
\end{center}
\end{figure}

The energy (per unit length) for the formation of a pair of disconnections under the applied energy jump $\psi$ can be expressed as Eq.~\eqref{E_W1}. 
However, unlike the case of stress-driven GB migration, the work done by the energy-jump driving force on the disconnections becomes $W = -\psi h\delta$. 
In this case, Eq.~\eqref{E_W1} becomes
\begin{equation}
E
= 2\mathit{\Gamma}_\text{s}|h| + 2\mathcal{K}b^2 \ln\frac{\delta}{\delta_0} - \psi h\delta.  
\label{E_W3}
\end{equation}
This yields the energy $E$ as a function of the separation of the disconnection pair $\delta$, as shown in Fig.~\ref{synthetic_energy}a. 
As a result, the critical separation is
\begin{equation}
\delta^* = \frac{2\mathcal{K}b^2}{\psi h}, 
\end{equation}
and the nucleation barrier is 
\begin{equation}
E^* = 2\mathit{\Gamma}_\text{s}|h| + 2\mathcal{K}b^2 \ln\frac{2\mathcal{K}b^2}{e \psi h\delta_0}.  
\label{Estar_synthetic_force}
\end{equation}
For the special case where $b = 0$ (i.e., pure steps), $\delta^* = \delta_0$ and $E^* = 2\mathit{\Gamma}_\text{s}|h|$. 
This is the formation energy of a pair of pure steps, as shown in Fig.~\ref{synthetic_energy}b. 
In comparison with Eq.~\eqref{deltastar_stress} and Eq.~\eqref{Estar_stress}, we simply replace $\tau b$ by $\psi h$. 
The value of $\beta$ can the be determined following the procedure proposed in the end of Section~\ref{prediction_beta_stress}, but with $E^*_{nj}$ evaluated by Eq.~\eqref{Estar_synthetic_force}. 

\subsubsection{Comparison with simulation and experimental results}
Following the approach proposed in the last section, we predicted the shear-coupling factor $\beta$ as a function of the misorientation angle $\theta$ for four series of GBs (i.e., $[100]$, $[110]$, $[111]_{(110)}$, and $[111]_{(112)}$ STGBs) in an FCC material (the material parameters are taken for Ni); the data are shown as the red dots in Figs.~\ref{beta_predict}a3, b3, c3 and d3. 
In this figure, the shear-coupling factor is unique for each $\theta$ value, corresponding to the disconnection mode associated with the lowest nucleation barrier under the applied energy-jump driving force. 

In Figs.~\ref{beta_predict}a3, b3, c3 and d3, the $\beta$ values obtained by the MD simulation~\cite{homer2013phenomenology} are shown as the crosses. 
The blue crosses correspond to the cases where GB migration was observed within the simulation time while the green crosses correspond to the cases where the GB was immobile (i.e., the green crosses are not useful data). 
Noting the discrepancy between the red dots in Fig.~\ref{beta_predict}a3 (b3, c3 or d3) and those in Fig.~\ref{beta_predict}a2 (b2, c2 or d2), we conclude that the $\beta$ value really depends on which type of driving force is applied. 
Qualitatively speaking, the shear stress tends to drive the disconnection mode with large $b$ and small $|h|$; on the contrary, the disconnection mode with small $b$ and large $|h|$ is favored by the energy-jump driving force. 
Therefore, for each fixed GB geometry, the $\beta$ value for the stress-driven GB migration is always larger or equal to that for the energy-jump-driven GB migration. 
By comparing the predicted $\beta$ values (red dots) and the useful MD data (blue crosses) in Fig.~\ref{beta_predict}, we find that they are almost consistent. 
The wrong predictions actually correspond to other disconnection modes; this may be attributed to the approximation implied in the energy expression Eq.~\eqref{Estar_synthetic_force} (for example, entropy is not considered although the MD simulations were always performed at finite temperature). 

In principle, when the energy jump is very small (i.e., $\psi \to 0$), the disconnection modes with zero $b$ and finite $h$ (i.e., pure step and $\beta = 0$) will be preferred. 
This is because, for the nucleation of a pair of pure steps, $E^* = 2\mathit{\Gamma}_\text{s}|h|$ which is always smaller than Eq.~\eqref{Estar_synthetic_force} as $\psi \to 0$. 
However, when the energy jump becomes large enough, it is possible to obtain the nonzero $\beta$ value, which is equal to that obtained in the case of stress-driven GB migration. 

\subsubsection{Temperature dependence of shear coupling}
\label{temperature_synthetic}
The prediction following the approach in Section~\ref{prediction_beta_synthetic_force} is, in principle, reasonable only at zero temperature.  
Similar to the discussion for the case of stress-driven GB migration in Section~\ref{temperature_stress}, the temperature dependence of the shear-coupling factor for the case of energy-jump-driven GB migration can be considered as an ensemble average over all possible disconnection modes. 

From Eq.~\eqref{barbeta}, we can estimate the shear-coupling factor $\bar{\beta}$ as a function of temperature; but, in the case of energy-jump-driven GB migration, $E^*_{nj}$ is obtained by Eq.~\eqref{Estar_synthetic_force}. 
Figure~\ref{beta_T}c shows the predictions of $\bar{\beta}(T)$ for five different GB geometries, i.e., $\Sigma 5$ $[100]$ $(013)$, $\Sigma 13$ $[100]$ $(015)$, $\Sigma 17$ $[100]$ $(035)$, $\Sigma 25$ $[100]$ $(017)$, and $\Sigma 37$ $[100]$ $(057)$ STGBs in FCC materials. 
In these predictions, the energy jump is set large enough that shear coupling occurs at zero temperature (if the energy jump is too small, pure steps will be activated at zero temperature and there will be no shear coupling, such as the case shown in Fig.~\ref{synthetic_energy}b). 
The results suggest that the magnitude of $\bar{\beta}$ decreases with temperature; this trend is opposite to that in the case of stress-driven GB migration (see Fig.~\ref{beta_T}a). 
At the limit of low temperature, $\bar{\beta}$ converges to a constant; this situation is schematically shown in Fig.~\ref{coupling_schematic}b2. 
At the limit of high temperature, multiple disconnection modes can be activated with close probability such that, with reference to Fig.~\ref{coupling_schematic}b3, the probability for the relative displacements of both grains to the left will be close to that for the relative displacement to the right, which implies that $\bar{\beta}\to 0$. 
Therefore, it is reasonable to expect that the magnitude of $\bar{\beta}$ decreases with temperature. 
In addition, depending on the GB geometry, the decrease of $|\bar{\beta}|$ with temperature can be very slow. 
For example, for the $\Sigma 37$ STGB, the ideal shear-coupling GB migration behavior can persist to a very high temperature. 
The $\beta$ value for the case of energy-jump-driven GB migration is usually smaller than that for the case of stress-driven GB migration (comparing Fig.~\ref{beta_T}a with b). 
The energy-jump driving force tends to drive the operation of the disconnection mode associated with large $h$ and thus small $\beta$, while the stress tends to drive the mode with large $b$ and thus large $\beta$. 

The prediction based on the ensemble average is consistent with the MD results obtained by Homer et al.~\cite{homer2013phenomenology}
From their MD simulations, the useful data (for the GBs which are mobile and show quantifiable, correlated shear coupling behavior) include the results for 156 STGBs in FCC Ni. 
Out of the 156 STGBs, they found that 56\% GBs exhibit decrease of $\bar{\beta}$ with increasing temperature; 35\% GBs exhibit temperature independence of $\bar{\beta}$; and 9\% GBs exhibit sharp change of disconnection modes (and, thus, $\bar{\beta}$) at a finite temperature. 
The trend of temperature dependence of $\bar{\beta}$ which we predicted by ensemble average of disconnection modes is consistent with most of these MD results, except that we cannot predict the sharp change of disconnection modes at a certain temperature (only observed in a small fraction of the MD results). 

%%%%%%%%%%%%%%%%%%%%%%%%%%%%%
\subsection{Mixed-force-driven grain-boundary migration}
\label{mixed_force_driven_gb_migration}
In most simuations of practical interest, GB migration is driven by multiple driving forces simultaneously. 
For example, in the case of grain growth under an applied magnetic field~\cite{molodov1998true,molodov2006grain,gunster2010magnetically,molodov2004effect,molodov2010observations}, the driving force is the mixture of the magnetic free energy density difference and capillary force. 
In the cases of recrystallization~\cite{rollett2004recrystallization} and irradiation-induced selective grain growth~\cite{seita2013selective}, the driving force is the mixture of the stored deformation energy density difference and capillary force. 
In the case of stress-driven grain growth in a polycrystal, the driving force is usually attributed to shear stress, capillary force, and the elastic energy density difference (due to elastic anisotropy); although the elastic energy density difference contributes much less than the other two driving forces and can be ignored. 
In any case, the mixed driving force can be simply expressed as the combination of shear stress and energy jump. 

\subsubsection{Molecular dynamics simulations}
\label{mixed_MD}
In order to model mixed-force-driven GB migration, we consider a constrained bicrystal system as illustrated in Fig.~\ref{coupling_schematic}c1. 
In this model system, we applied the energy density difference in the upper and lower grains $\psi = \psi_\text{U} - \psi_\text{L} >0$. 
If the top and bottom surfaces of this bicrystal are traction-free, then the GB will migrate upwards, which is exactly the case of energy-jump-driven GB migration (same as Fig.~\ref{coupling_schematic}b2). 
However, we may also adopt the boundary condition that the top and bottom surfaces are frozen (i.e., fixed-displacement boundary condition). 
We can use such a bicrystal with frozen surfaces to model the GB migration in a polycrystal; in this case, the GB is driven by capillary force (modeled by the energy-jump driving force), but the neighboring grains cannot displace freely since they are constrained by the other grains in a polycrystal environment (modeled by the fixed-displacement boundary condition). 
The problem, now, is to determine how the GB migrates in this constrained bicrystal system. 
This problem was solved by MD simulation for two particular GBs: $\Sigma 13$ $[111]_{(110)}$ $(3\bar{4}1)$ and $\Sigma 39$ $[111]_{(110)}$ $(5\bar{7}2)$ STGBs in EAM Ni. 

\begin{figure}[!t]
\begin{center}
\scalebox{0.32}{\includegraphics{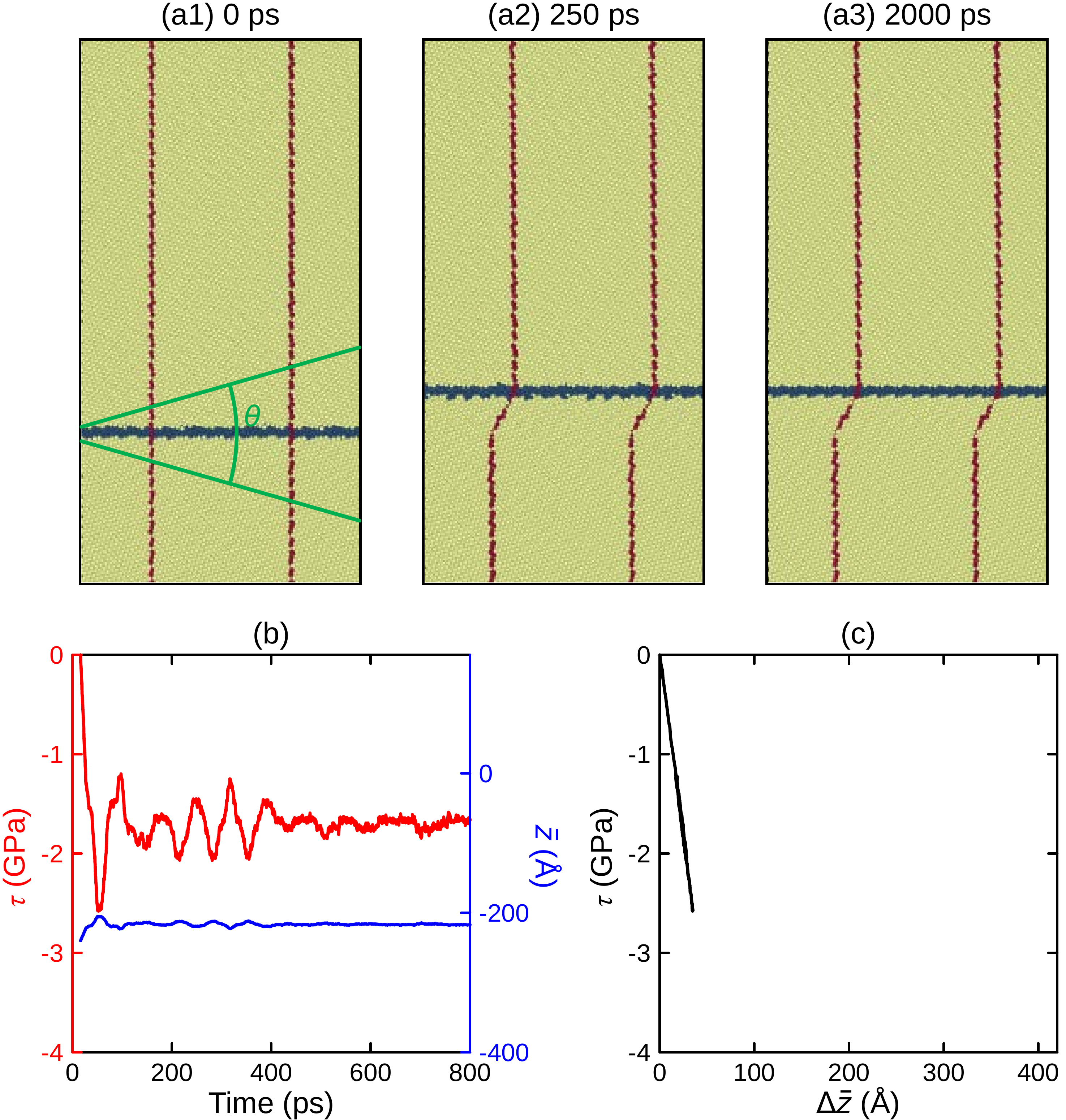}}
\caption{Stagnation of GB migration observed for a $\Sigma 39$ $[111]_{(110)}$ $(5\bar{7}2)$ STGB in EAM Ni by MD simulation. 
(a) The snapshots at different times, where the atoms at the GB are colored blue and some atoms are always colored red to serve as reference (i.e., a fiducial marker). 
(b) The time evolution of the established internal shear stress (red) and the GB position (blue). 
(c) The internal shear stress vs. GB migration distance. 
The figures are reproduced from Ref.~\cite{thomas2017reconciling} (D. J. Srolovitz, 2017).  } 
\label{S39_mixed}
\end{center}
\end{figure}

Figure~\ref{S39_mixed}a shows the simulation result for the $\Sigma 39$ STGB at 300~K, where the blue line traces the GB position and the dark red lines are the fiducial lines. 
We found that, initially, this GB migrated upwards since $\psi > 0$ (from Fig.~\ref{S39_mixed}a1 to a2), and then the GB migration stopped (from Fig.~\ref{S39_mixed}a2 to a3). 
The deformation of the fiducial lines indicates that plastic shear strain was established with the GB migration, which is a feature of shear coupling. 
The shear-coupling factor, obtained by the inverse of the slope of the fiducial lines in the deformed region, is $0.58$, which is very close to the predicted value based on the bicrystallography $1/\sqrt{3}$ corresponding to the disconnection mode $(b = a_0/\sqrt{26}, h = a_0\sqrt{3/26})$ (following the approach proposed in Section~\ref{enumeration}). 
Figure~\ref{S39_mixed}b plots the evolution of the GB position ($\bar{z}$) and the shear stress ($\tau$) in the system. 
The evolution of $\bar{z}$ is consistent with the observation in Fig.~\ref{S39_mixed}a. 
The evolution of $\tau$ shows that the shear stress is established with GB migration and, then, kept around a constant (nonzero) after the GB migration stopped. 
This clearly demonstrates the existence of shear coupling. 
Figure~\ref{S39_mixed}c plots the shear stress versus the GB migration distance, which shows linear relation. 

\begin{figure}[!t]
\begin{center}
\scalebox{0.32}{\includegraphics{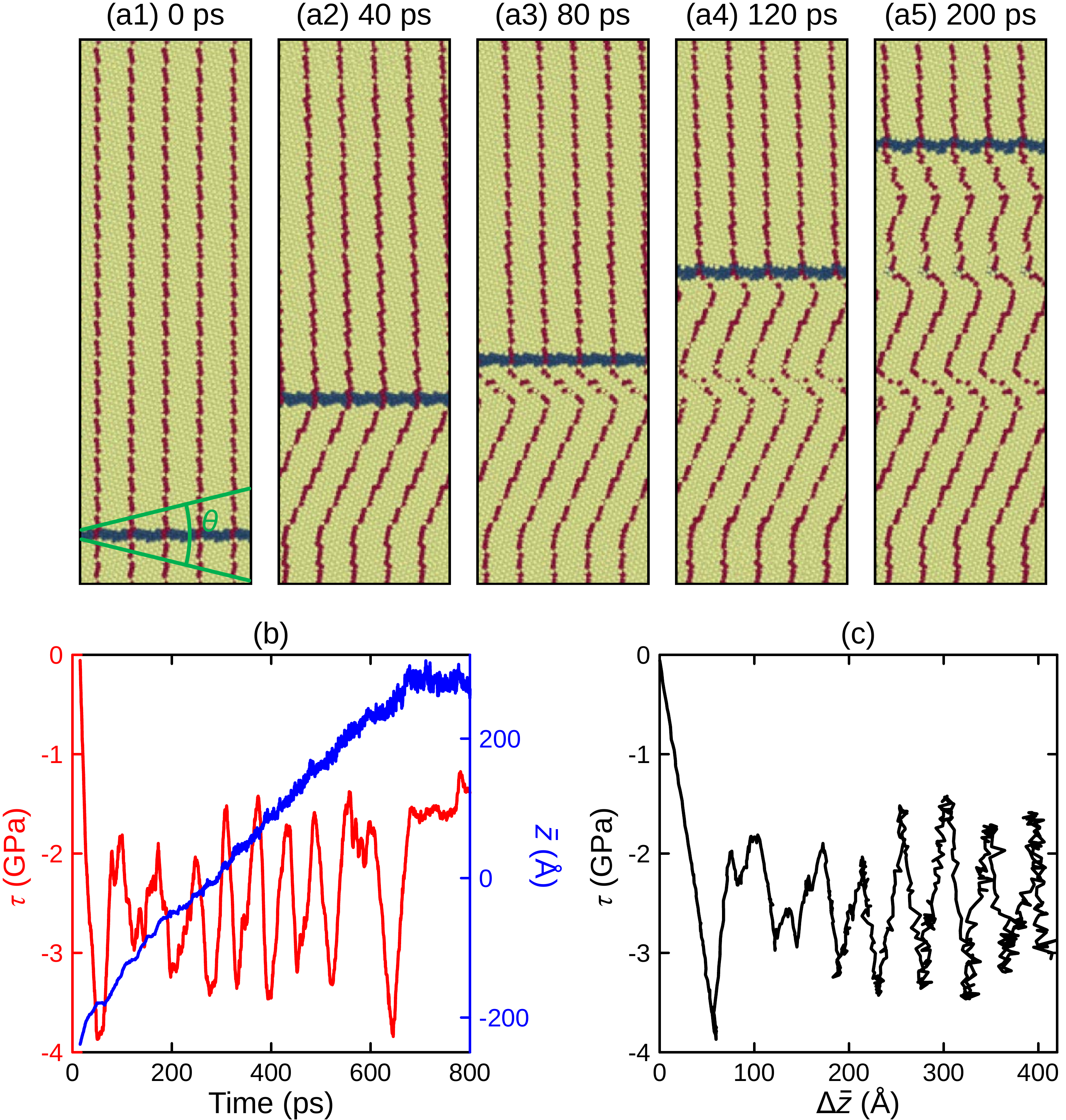}}
\caption{Continuous GB migration via repeated switching between different disconnection modes (with opposite sign of $\beta$) observed for a $\Sigma 13$ $[111]_{(110)}$ $(3\bar{4}1)$ STGBs in EAM Ni by MD simulation. 
(a) The snapshots at different times, where the atoms at the GB are colored blue and some atoms are always colored red to serve as reference (i.e., a fiducial marker). 
(b) The time evolution of the established internal shear stress (red) and the GB position (blue). 
(c) The internal shear stress vs. GB migration distance. 
The figures are reproduced from Ref.~\cite{thomas2017reconciling} (D. J. Srolovitz, 2017).   }
\label{S13_mixed}
\end{center}
\end{figure}

The simulation result for the $\Sigma 13$ STGB at 600~K, as shown in Fig.~\ref{S13_mixed}, is different from that for the $\Sigma 39$ STGB. 
For the $\Sigma 13$ STGB, the GB migrated initially upwards due to positive $\psi$ (from Fig.~\ref{S13_mixed}a1 to a2). 
By measuring the inverse slope of the fiducial lines, we obtained the value $0.45$ which is consistent with the predicted shear-coupling factor $2\sqrt{3}/7$ corresponding to the disconnection mode $(b = a_0\sqrt{3/26}, h = 7a_0/2\sqrt{26})$. 
Unlike the behavior of the $\Sigma 39$ STGB, the $\Sigma 13$ STGB never stopped migrating; with the GB migration, the slope of the fiducial lines is changed (from Fig.~\ref{S13_mixed}a2 and a3). 
The measured inverse slope for the fiducial line segment with opposite slope to the initial one is $-0.63$ which is close to the predicted shear-coupling factor $-1/\sqrt{3}$ corresponding to the disconnection mode $(b = a_0\sqrt{3/26}, h = -3a_0/\sqrt{26})$.
We found that the continuous GB migration was accompanied by repeatedly switching between two different shear-coupling factors (or disconnection modes) (from Figs.~\ref{S13_mixed}a3 and a5). 
Figure~\ref{S13_mixed}b plots the evolution of the GB position ($\bar{z}$) and the shear stress ($\tau$) in the system. 
The evolution of $\bar{z}$ is consistent with the observation in Fig.~\ref{S13_mixed}a. 
The evolution of $\tau$ shows oscillation about a nonzero value, indicating cyclic establishment and release of shear stress in the system. 
Figure~\ref{S13_mixed}c plots the shear stress versus the GB migration distance, which also shows the cyclic change of shear stress along with GB migration. 

The MD simulation results suggest that, for a constrained bicrystal (with the displacement of surfaces fixed) under the applied energy-jump driving force, 
(i) accompanied by GB migration, shear stress arises, 
(ii) GB migration may stagnate for some GBs (e.g. $\Sigma 39$ STGB), 
and (iii) GB migration may continue by repeatedly switching between two disconnection modes (characterized by different shear-coupling factors) for the other GBs. 
Now, how can we understand these observations based on the disconnection model? 

\subsubsection{Analytical disconnection model}
\begin{figure}[!t]
\begin{center}
\scalebox{0.36}{\includegraphics{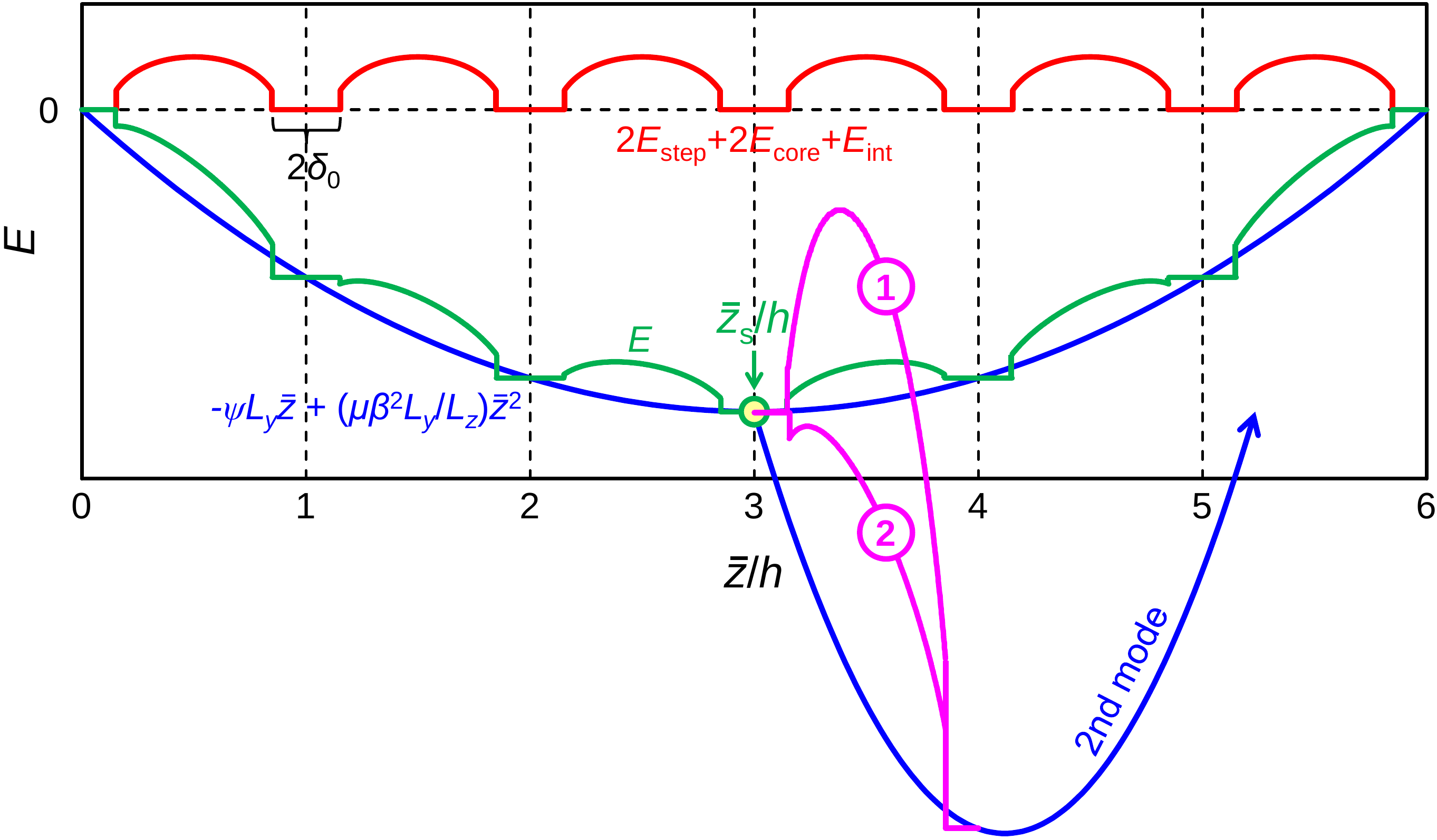}}
\caption{Schematic plot of Eq.~\eqref{mixed2}. 
The red curve corresponds to the step energy, disconnection core energy, and elastic energy. The blue curve corresponds to the work done by the energy-jump driving force and the induced shear stress. The green curve corresponds to the total energy. 
The blue curve labeled ``2nd mode'' corresponds to the work if the GB migrates by switching to the second disconnection mode (associated with a sign of $\beta$ opposite the first mode). 
The purple line labeled ``1'' indicates the case where the energy barrier is so high that the switch to the second mode is difficult and GB migration tends to stagnate. The purple line labeled ``2'' indicates the case where there is no energy barrier such that the switch to the second mode is natural and the GB can continue to migrate. } 
\label{mixed_energy}
\end{center}
\end{figure}

In the case of GB migration in a constrained bicrystal mentioned in the last section, the energy-jump driving force is applied and shear stress is established during this process. 
Hence, this is a typical case of mixed-force-driven GB migration. 
We consider an analytical model based on a bicrystal system with periodic boundary conditions applied along the $x$- and $y$-axes (i.e., GB plane). 
Modified based on Eq.~\eqref{Etot_pbc}, the energy (per unit length) for the formation of a pair of disconnections under both the energy density difference $\psi$ and the shear stress $\tau$ is 
\begin{equation}
E 
= 2\mathit{\Gamma}_\text{s}|h| + 2\mathcal{K}b^2 \ln\left|\frac{\sin(\pi\delta/L_y)}{\sin(\pi\delta_0/L_y)}\right| - \left(\psi h + \tau b \right)\delta.   
\label{mixed}
\end{equation}
In the case of bicrystal with fixed surfaces, shear stress is established with GB migration. 
Assuming isotropic linear elasticity, the induced shear stress is $\tau = -\mu \bar{u}/L_z = -\mu\beta \bar{z}/L_z$ (we take the initial GB position at $\bar{z} = 0$ such that $\bar{z}$ stands for the GB migration distance). 
The linear relation between $\tau$ and $\bar{z}$ is consistent with the MD simulation result (see Fig.~\ref{S39_mixed}c). 
Since periodic boundary condition is applied along the GB, the relation between the disconnection separation and the GB migration distance is $\delta/L_y = \bar{z}/h$. 
Equation~\eqref{mixed} then becomes 
\begin{equation}
E(\bar{z}) 
= \left( 2\mathit{\Gamma}_\text{s}|h| + 2\mathcal{K}b^2 \ln\left|\frac{\sin(\pi \bar{z}/h)}{\sin(\pi\delta_0/L_y)}\right| \right)
- \left(\psi L_y \bar{z} - \frac{\mu\beta^2 L_y}{L_z}\bar{z}^2 \right).   
\label{mixed2}
\end{equation}
The function $E(\bar{z})$ is schematically plotted as the green line in Fig.~\ref{mixed_energy}. 
We can find that the terms in the first set of parentheses of Eq.~\eqref{mixed2} correspond to the periodic corrugation along the $E(\bar{z})$ curve (see the red line in Fig.~\ref{mixed_energy}), which leads to many local minima . 
The terms in the second set of parentheses of Eq.~\eqref{mixed2} (i.e., the terms due to driving forces) contribute to the general shape of the $E(\bar{z})$ curve (see the blue line in Fig.~\ref{mixed_energy}), which leads to the global energy minimum at the position where the GB stagnates $\bar{z}_\text{s} = \psi L_z / (2\mu\beta^2)$ and the corresponding shear stress is $\tau_\text{s} = -\psi/(2\beta)$. 
The shape of the $E(\bar{z})$ curve suggests that, initially driven by $\psi$, the GB migrates from $\bar{z}=0$ to $\bar{z}_\text{s}$, and, at the same time, the shear stress is established; when the shear stress reaches $\tau_\text{s}$, the GB will stagnate at the position $\bar{z}_\text{s}$. 

In the analysis above, it is assumed that there is only one active disconnection mode (since $b$, $h$ and $\beta$ are constants). 
However, when the GB migrates to the position $\bar{z}$ by the operation of this disconnection mode, the GB will stagnate. 
At this time, in principle, the GB migration can continue by the operation of a different disconnection mode which has a Burgers vector of opposite sign to the first activated mode. 
For example, if the first mode is characterized by $(b_1>0, h_1>0)$, this mode will be activated by the applied energy jump $\psi > 0$ and finally stopped by the induced shear stress $\tau <0$. 
Then, the positive $\psi$ and negative $\tau$ may activate the second mode with $(b_2<0,h_2>0)$. 
If so, GB migration can continue by switching from the first mode to the second mode. 
As schematically shown in Fig.~\ref{mixed_energy}, when the first mode ceases to operate, the GB will have a chance to switch to the second mode to continue migrating. 
There will be two different cases: 
\begin{itemize}
\item[(i)] As shown in Fig.~\ref{mixed_energy}, if the nucleation barrier for the second mode is not high (see the magenta line 2) such that it is reduced by the driving forces (i.e., positive $\psi$ and negative $\tau$), then the GB migration will continue by switching to the second mode. 
By the operation of the second mode, the shear stress $\tau$ will gradually become positive and the second mode will cease to operate; at this time, the first mode will be activated again. 
GB migration can continue by cyclically switching between the first and the second modes; such behavior is consistent with the MD simulation result for the $\Sigma 13$ STGB (see Fig.~\ref{S13_mixed}). 

\item[(ii)] As shown in Fig.~\ref{mixed_energy}, if the nucleation barrier for the second mode is too high (see the magenta line 1), then the GB will stagnate (or, rigorously speaking, migrate very slowly); such behavior is consistent with the MD simulation result for the $\Sigma 39$ STGB (see Fig.~\ref{S39_mixed}). 
\end{itemize}
Therefore, the observation in the MD simulations mentioned in Section~\ref{mixed_MD} can be analytically understood based on the disconnection model. 

\subsubsection{Generalized kinetic equation for grain-boundary motion}
Now we consider a more general case of mixed-force-driven GB migration; this case is not limited to the constrained bicrystal system, and there are multiple disconnection modes available (rather than one or two modes, as discussed in the last section). 
No matter which boundary condition is applied, Eq.~\eqref{mixed} is applicable for each disconnection mode [but Eq.~\eqref{mixed2} is not]. 
When the driving forces are small, the nucleation barrier is approximately reached at the separation $\delta^* = L_y/2$ and the nucleation barrier is 
\begin{equation}
E^*
\approx \mathcal{Q} - \mathcal{Q}^\text{d}
= \left( 2\mathit{\Gamma}_\text{s}|h| - 2\mathcal{K}b^2 \ln\sin\frac{\pi\delta_0}{L_y} \right) - \left(\psi h + \tau b \right) \frac{L_y}{2},    
\label{mixed_star}
\end{equation}
where $\mathcal{Q}$ is the nucleation barrier with zero driving force and $\mathcal{Q}^\text{d}$ is the modification of the nucleation barrier due to the driving force. 
Equation~\eqref{mixed_star} applies to each disconnection mode. 
For the mode $(b_n, h_{nj})$, we can obtain $E^*_{nj}$, $\mathcal{Q}_{nj}$ and $\mathcal{Q}^\text{d}_{nj}$ with $b=b_n$ and $h=h_{nj}$ in Eq.~\eqref{mixed_star}. 
From Eq.~\eqref{rnj}, the nucleation rate for a pair of disconnections $(b_n, h_{nj})$ is 
\begin{equation}
r_{nj}
= \omega\exp\left(-\frac{\mathcal{Q}_{nj} - \mathcal{Q}^\text{d}_{nj}}{k_\text{B}T/L_x}\right)
\approx \omega e^{-L_x\mathcal{Q}_{nj}/k_\text{B}T} \left(1 + \frac{\mathcal{Q}^\text{d}_{nj}}{k_\text{B}T/L_x}\right), 
\label{rnj2}
\end{equation}
where the approximation (expansion to the first order) is valid when $\mathcal{Q}^\text{d}_{nj}\ll 1$ (or $\psi \ll 1$ and $\tau \ll 1$). 
Similar to Eq.~\eqref{vpara_vperp}, the velocity of GB sliding is  
\begin{align}
v_\parallel
&= \frac{\mathcal{A}\omega}{2k_\text{B}T} \left(
\tau \sum_{n,j} b_n^2 e^{-L_x\mathcal{Q}_{nj}/k_\text{B}T}
+ \psi \sum_{n,j} b_n h_{nj} e^{-L_x\mathcal{Q}_{nj}/k_\text{B}T}
\right)
\nonumber \\
&= M_{11} \tau + M_{12} \psi, 
\label{v_para}
\end{align}
and the velocity of GB migration is 
\begin{align}
v_\perp
&= \frac{\mathcal{A}\omega}{2k_\text{B}T} \left(
\tau \sum_{n,j} b_n h_{nj} e^{-L_x\mathcal{Q}_{nj}/k_\text{B}T}
+ \psi \sum_{n,j} h_{nj}^2 e^{-L_x\mathcal{Q}_{nj}/k_\text{B}T}
\right)
\nonumber \\
&= M_{21} \tau + M_{22} \psi. 
\label{v_perp}
\end{align}
Combining Eq.~\eqref{v_para} and Eq.~\eqref{v_perp}, we can obtain the generalized kinetic equation for GB motion: 
\begin{equation}
\mathbf{V} = \mathbf{M} \mathbf{F}, 
\label{generalized_kinetic}
\end{equation}
where $\mathbf{V} \equiv (v_\parallel, v_\perp)^T$, $\mathbf{M}\equiv [M_{ij}]$ ($i,j = 1,2$), and $\mathbf{F} \equiv (\tau, \psi)^T$. 
Unlike the classical kinetic equation for GB migration Eq.~\eqref{MF}, 
the generalized kinetic equation provides complete description of GB motion (not only GB migration but GB sliding as well). 

In Eq.~\eqref{generalized_kinetic}, $\mathbf{M}$ is a symmetric second-order tensor. 
The diagonal elements of $\mathbf{M}$ relate each velocity directly to their conjugate driving forces; 
the off-diagonal elements are responsible for the coupling effects. 
$\mathbf{M}$ depends on temperature. 
From Eq.~\eqref{v_para} and Eq.~\eqref{v_perp}, the terms in the summation included in $M_{11}$ or $M_{22}$ are always positive, so the diagonal elements are always positive.
But the terms in the summation included in $M_{12}$ ($=M_{21}$) can be positive or negative. 
When $T\to \infty$, $M_{12} \to 0$, implying that there is no coupling effect. 
When $T\to 0$, only the term associated with the smallest $Q_{nj}$ in the summation included in $M_{ij}$ will dominate the value of the summation. 
In summary, 
\begin{equation}
\mathbf{M}(T\to\infty)
= \frac{\mathcal{A}\omega}{2k_\text{B}T} 
\left(\begin{array}{cc}
\sum_{n,j}b_n^2 & 0 \\
0 & \sum_{n,j}h_{nj}^2 
\end{array}\right), 
\end{equation}
and
\begin{equation}
\mathbf{M}(T\to 0)
= \frac{\mathcal{A}\omega}{k_\text{B}T} 
\left(\begin{array}{cc}
b_0^2 & b_0h_0 \\
b_0h_0 & h_0^2 
\end{array}\right), 
\end{equation}
where we denote the disconnection mode with the smallest $Q_{nj}$ as $(b_0, h_0)$. 
As $T\to 0$, $(M_u/M_z)^{1/2} = (M_{11}/M_{22})^{1/2} = |b_0/h_0| = |\beta|$, which is consistent with the relation derived by Karma et al.~\cite{karma2012relationship}

\begin{figure}[!t]
\begin{center}
\scalebox{0.28}{\includegraphics{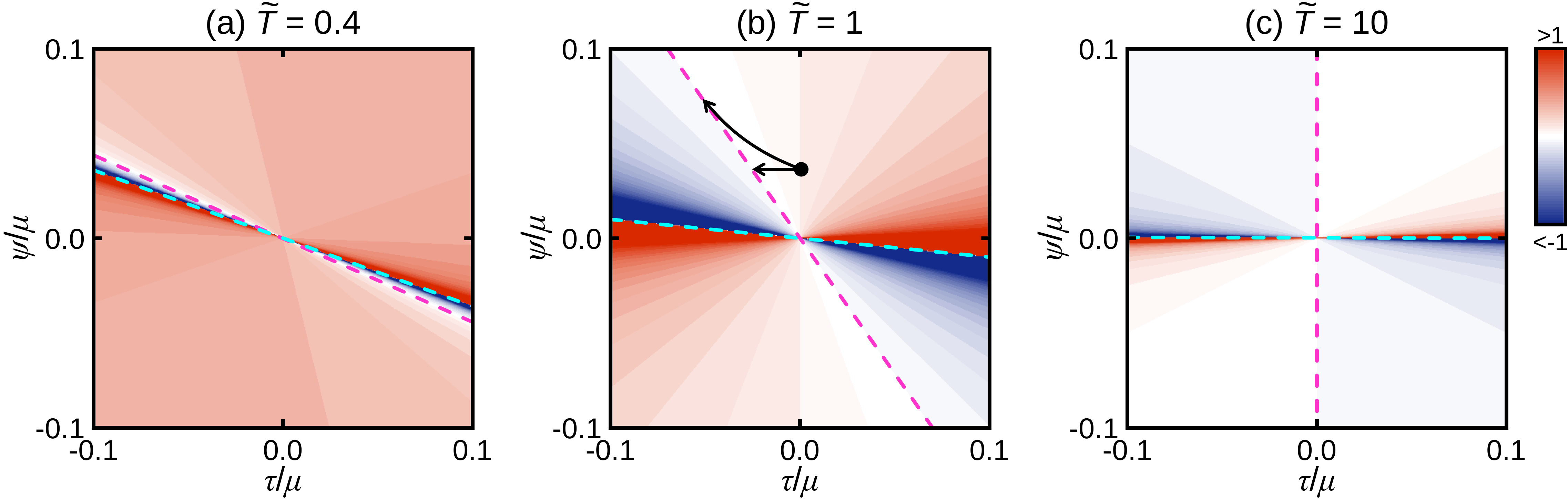}}
\caption{Contour plots of the effective shear-coupling factor $\bar{\beta}$ as a function of the applied shear stress $\tau$ and energy jump $\psi$ for a $\Sigma 13$ $[100]$ $(015)$ STGB in a FCC crystal at the (reduced) temperatures: (a) $\tilde{T}=0.4$, (b) $\tilde{T}=1$ and (c) $\tilde{T}=10$, according to Eq.~\eqref{beta_M}. 
The purple dashed line is the contour line of $\bar{\beta} = 0$ and the light blue dashed line is the contour line of $\bar{\beta}\to \pm \infty$. 
The horizontal arrow in (b) corresponds to the process of GB migration driven by energy jump with two fixed surfaces (i.e., the scenario depicted in Fig.~\ref{coupling_schematic}c). 
The curved arrow in (b) corresponds to the process of grain shrinkage (i.e., the scenario depicted in Fig.~\ref{grainshrink}a3). } 
\label{mixed_map}
\end{center}
\end{figure}

According to Eq.~\eqref{v_para} and Eq.~\eqref{v_perp}, the shear-coupling factor, in general case, is 
\begin{equation}
\bar{\beta}(T)
= \frac{M_{11} \tau + M_{12} \psi}{M_{12} \tau + M_{22} \psi}. 
\label{beta_M}
\end{equation}
In the case of stress-driven GB migration (where $\tau$ is finite but $\psi = 0$), $\bar{\beta} = M_{11}/M_{12}$, which is approximately equal to Eq.~\eqref{barbeta}. 
At the limit of $T\to 0$, $\bar{\beta} \to b_0/h_0$ which is associated with the behavior shown in Fig.~\ref{coupling_schematic}a2; at the limit of $T\to \infty$, $\bar{\beta} \to \infty$, which is associated with behavior as shown in Fig.~\ref{coupling_schematic}a3. 
In the case of energy-jump-driven GB migration (where $\tau=0$ but $\psi$ is finite), $\bar{\beta} = M_{12}/M_{22}$. 
As an example, Fig.~\ref{mixed_map} shows the $\bar{\beta}$ value as a function of $\tau$ and $\psi$ for a $\Sigma 13$ $[100]$ $(015)$ STGB for three different temperatures according to Eq.~\eqref{beta_M}. 
At low temperature (Fig.~\ref{mixed_map}a), $\bar{\beta} \approx 0.4$ almost throughout the whole $\tau$-$\psi$ phase space, which is consistent with the prediction shown in Figs.~\ref{beta_T}a and c. 
At the medium temperature (Fig.~\ref{mixed_map}b), the region where $\bar{\beta}<0$ becomes larger. 
The scenario discussed in Section~\ref{mixed_force_driven_gb_migration} corresponds to the arrow show in Fig.~\ref{mixed_map}b: under a constant applied energy-jump driving force, negative stress is established; the state in the $\tau$-$\psi$ phase space finally reaches the dashed purple line where $\bar{\beta}=0$ and, thus, no apparent shear coupling appears on average, as illustrated in Fig.~\ref{coupling_schematic}c3. 
At the high temperature (Fig.~\ref{mixed_map}c), $\bar{\beta} \approx 0$ in most region of the phase space, corresponding to the case shown in Fig.~\ref{coupling_schematic}b3. 
However, near the line where $\psi = 0$, $\bar{\beta} \to \pm\infty$, corresponding to the case shown in Fig.~\ref{coupling_schematic}a3. 

\subsection{Extension to asymmetric tilt/mixed grain boundaries}
\begin{figure}[!t]
\begin{center}
\scalebox{0.3}{\includegraphics{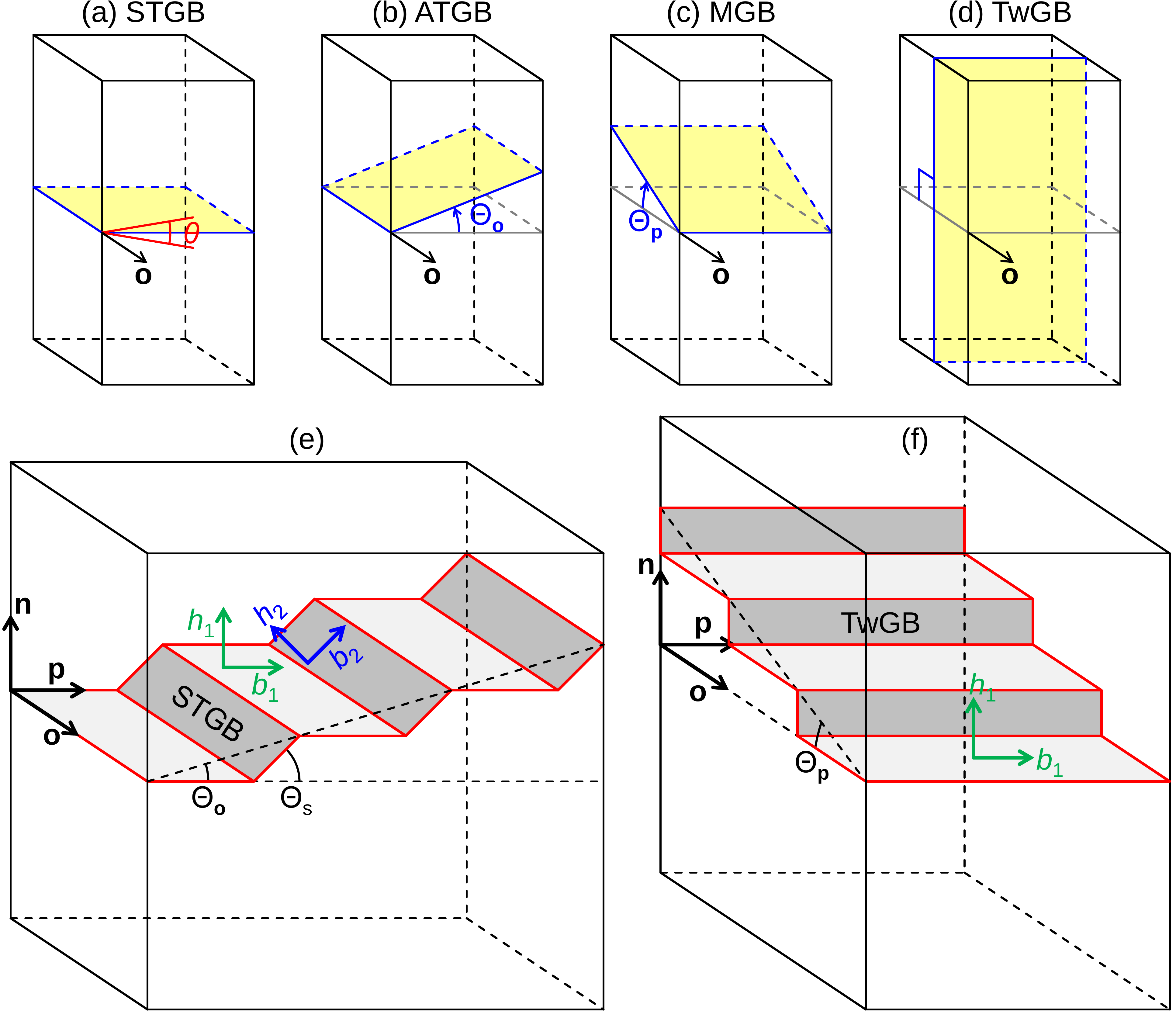}}
\caption{Macroscopic GB geometry for (a) a symmetric tilt GB, (b) an asymmetric tilt GB, (c) a mixed GB, and (d) a pure twist GB, where the $\mathbf{o}$-axis is the rotation (tilt) axis, $\theta$ is the rotation (misorientation) angle of two grains, $\Theta_\mathbf{o}$ is the inclination angle by which the GB plane rotates about the $\mathbf{o}$-axis, and $\Theta_\mathbf{p}$ is the inclination angle by which the GB plane rotates about the $\mathbf{p}$-axis. 
(e) shows a faceted structure of (b), where the horizontal facets are associated with the STGB with $\Theta_\mathbf{o}=0$ and the inclined facets are associated with the STGB with $\Theta_\mathbf{o}=\Theta_\text{s}$. 
$(b_1,h_1)$ and $(b_2,h_2)$ are the disconnection modes for the horizontal and the inclined facets, respectively. 
(f) shows a faceted structure of (c), where the horizontal facets are associated with the STGB with $\Theta_\mathbf{p}=0$ and the inclined facets are associated with the TwGB with $\Theta_\mathbf{p}=90^\circ$. 
$(b_1,h_1)$ is the disconnection mode for the horizontal facet. } 
\label{beta_mixed_geometry}
\end{center}
\end{figure}

In the above sections, we limit the discussion to the shear-coupling factor for STGBs just for convenience. 
In practice, however, many GBs in a polycrystal are not STGBs; they could be pure twist GBs (TwGBs), asymmetric tilt GBs (ATGBs) or, more generally, mixed tilt-twist GBs (MGBs)~\cite{randle2008five}. 
In this section, we will demonstrate the extension to the prediction of shear-coupling factor for TwGBs, ATGBs and MGBs. 
Any TwGB, ATGB or MGB can be constructed from a STGB with the same misorientation. 
Starting from a STGB with misorientation angle $\theta$ (see Fig.~\ref{beta_mixed_geometry}a), if we incline the GB plane around the tilt axis $\mathbf{o}$ by the angle $\Theta_\mathbf{o}$, we will obtain an ATGB, as shown in Fig.~\ref{beta_mixed_geometry}b.  
If we incline the GB plane around the axis $\mathbf{p}$ by the angle $\Theta_\mathbf{p}$, we will obtain a MGB, as shown in Fig.~\ref{beta_mixed_geometry}c. 
At the limit of $\Theta_\mathbf{p} = 90^\circ$, the MGB becomes a TwGB, as shown in Fig.~\ref{beta_mixed_geometry}d. 
Now, the problem is that, if we know the shear-coupling factor for the STGB with a particular misorientation $\beta(\theta)$, how do we predict the shear-coupling factor as a function of the inclination angles [i.e., $\beta(\theta; \Theta_\mathbf{o}, \Theta_\mathbf{p})$]? 

\subsubsection{Prediction of shear-coupling factor}
First, there is no shear coupling in TwGBs (see Fig.~\ref{beta_mixed_geometry}c with $\Theta_\mathbf{o}=90^\circ$). 
We take, for example, a SC bicrystal with a $\Sigma 5$ misorientation, as illustrated in Fig.~\ref{DSC}d. 
If we choose the horizontal plane (blue line) as the GB plane in Fig.~\ref{DSC}a and remove the black lattice below the plane and white lattice above the plane, we will obtain an STGB (exactly the case in Fig.~\ref{DSC}a). 
However, if we choose the plane with a normal perpendicular to the paper and remove the black lattice behind the plane and white lattice in front of the plane, we will obtain a TwGB. 
In the case of TwGB, the shortest nonzero DSC vectors (i.e., the green vectors in Fig.~\ref{DSC}d) are always parallel to the GB plane and associated with zero step height, corresponding to the disconnection mode $(\mathbf{b},0)$; the shortest nonzero step height is always associated with zero DSC vector, corresponding to the disconnection mode $(\mathbf{0},a_0)$. 
Therefore, GB sliding and GB migration are not coupled for TwGBs. 
If a TwGB is driven to slide by applied shear stress, then $\beta \to \infty$ (GB sliding as shown in Fig.~\ref{coupling_schematic}a3); if a TwGB is driven to migrate by applied energy jump, then $\beta = 0$ (as shown in Fig.~\ref{coupling_schematic}b3). 
The same conclusion can also be drawn from the analysis of GB dislocations for low-angle TwGBs~\cite{cahn2004unified}. 

Second, we consider the shear-coupling factor for ATGBs (see Fig.~\ref{beta_mixed_geometry}b). 
The structure of an ATGB is usually faceted, as illustrated in Fig.~\ref{beta_mixed_geometry}e. 
One facet plane corresponds to an STGB with $\Theta_\mathbf{o} = 0$; the other facet plane corresponds to another STGB with $\Theta_\mathbf{o} = \Theta_\text{s}$. 
For example, for a $\Sigma 5$ $[100]$ tilt GB in FCC material, one facet is $\Sigma 5$ $[100]$ $(013)$ STGB and the other is $\Sigma 5$ $[100]$ $(012)$ STGB, and $\Theta_\text{s} = 45^\circ$ (with reference to the former). 
Such a faceted structure is supported by observations from atomistic simulations and experiments~\cite{hasson1972theoretical,pond1977periodic,hsieh1989experimental,medlin2017defect,muschik1993energetic}. 
For example, Hasson et al. found that $\Sigma 5$ $[100]$ $(0,9,17)$ ATGB is decomposed into two facets characterized by $\Sigma 5$ $[100]$ $(013)$ and $\Sigma 5$ $[100]$ $(012)$ STGBs ($\Theta_\text{s} = 45^\circ$) after relaxation~\cite{hasson1972theoretical}. 
Such faceting was also observed in experiments by Medlin et al.~\cite{medlin2017defect}
Pond et al. also found that a $\Sigma 3$ $[110]$ $(\bar{5}51)$ ATGB is decomposed into two facets characterized by $\Sigma 3$ $[110]$ $(\bar{1}11)$ and $\Sigma 3$ $[110]$ $(\bar{1}12)$ STGBs ($\Theta_\text{s} = 90^\circ$)~\cite{pond1977periodic}; this is consistent with experimental observations by Muschik et al.~\cite{muschik1993energetic}
Let the shear-coupling factor for the facet with $\Theta_\mathbf{o}=0$ (i.e., one STGB) be $\beta_1$ and that for the other facet with $\Theta_\mathbf{o}=\Theta_\text{s}$ (i.e., another STGB) be $\beta_2$. 
The shear-coupling factor for the ATGB can then be obtained by the geometry shown in Fig.~\ref{beta_mixed_geometry}e: 
\begin{equation}
\beta(\Theta_\mathbf{o})
= \frac{\sin^2\Theta_\text{s}}
{\beta_1^{-1} \sin^2(\Theta_\text{s}-|\Theta_\mathbf{o}|) + \beta_2^{-1} \sin^2|\Theta_\mathbf{o}|}. 
\label{beta_ATGB}
\end{equation}
At the limit of $\Theta_\mathbf{o} = 0$, $\beta = \beta_1$; at the limit of $\Theta_\mathbf{o} = \Theta_\text{s}$, $\beta = \beta_2$. 
When $\beta_1$ and $\beta_2$ are of the same sign, $\beta$ will be finite. 
However, when $\beta_1$ and $\beta_2$ are of the opposite sign, there will be a special inclination angle $\Theta_\mathbf{o}$ such that the denominator in Eq.~\eqref{beta_ATGB} is zero and $\beta \to \pm \infty$. 
This corresponds to the case where two facets migrate in opposite directions such that there is no net migration for the ATGB as an entity. 
Note that, although not all ATGBs are faceted into two STGBs as suggested in Fig.~\ref{beta_mixed_geometry}e (e.g. $\Sigma 11$ $[110]$ ATGB in Ref.~\cite{hsieh1989experimental}), Eq.~\eqref{beta_ATGB} is still valid since it only relies on GB geometry (i.e., DSC lattice and GB plane) rather than GB structure. 
Basak et al. also constructed the functional form for $\beta(\Theta_\mathbf{o})$~\cite{basak2014two} which is different from Eq.~\eqref{beta_ATGB}; but their construction is empirical rather than physics-based. 

Third, we think of the shear-coupling factor for MGBs (see Fig.~\ref{beta_mixed_geometry}c). 
The structure of a MGB is also usually faceted, as illustrated in Fig.~\ref{beta_mixed_geometry}f, which is supported by simulations~\cite{humberson2017anti}. 
One facet plane corresponds to a STGB with $\Theta_\mathbf{p} = 0$; the other facet plane corresponds to a TwGB with $\Theta_\mathbf{p} = 90^\circ$. 
Since there is no shear coupling for TwGBs, the shear-coupling behavior of the MGB is simply that of the STGB which characterizes the facets resolved to the plane of the MGB, i.e., 
\begin{equation}
\beta(\Theta_\mathbf{p})
= \beta_1 \sec\Theta_\mathbf{p}. 
\label{beta_MGB}
\end{equation}
Although not all MGBs are faceted into one STGB and one TwGB as suggested in Fig.~\ref{beta_mixed_geometry}f (e.g. Ref.~\cite{hadian2016atomistic}), Eq.~\eqref{beta_MGB} is still valid. 

\begin{figure}[!t]
\begin{center}
\scalebox{0.29}{\includegraphics{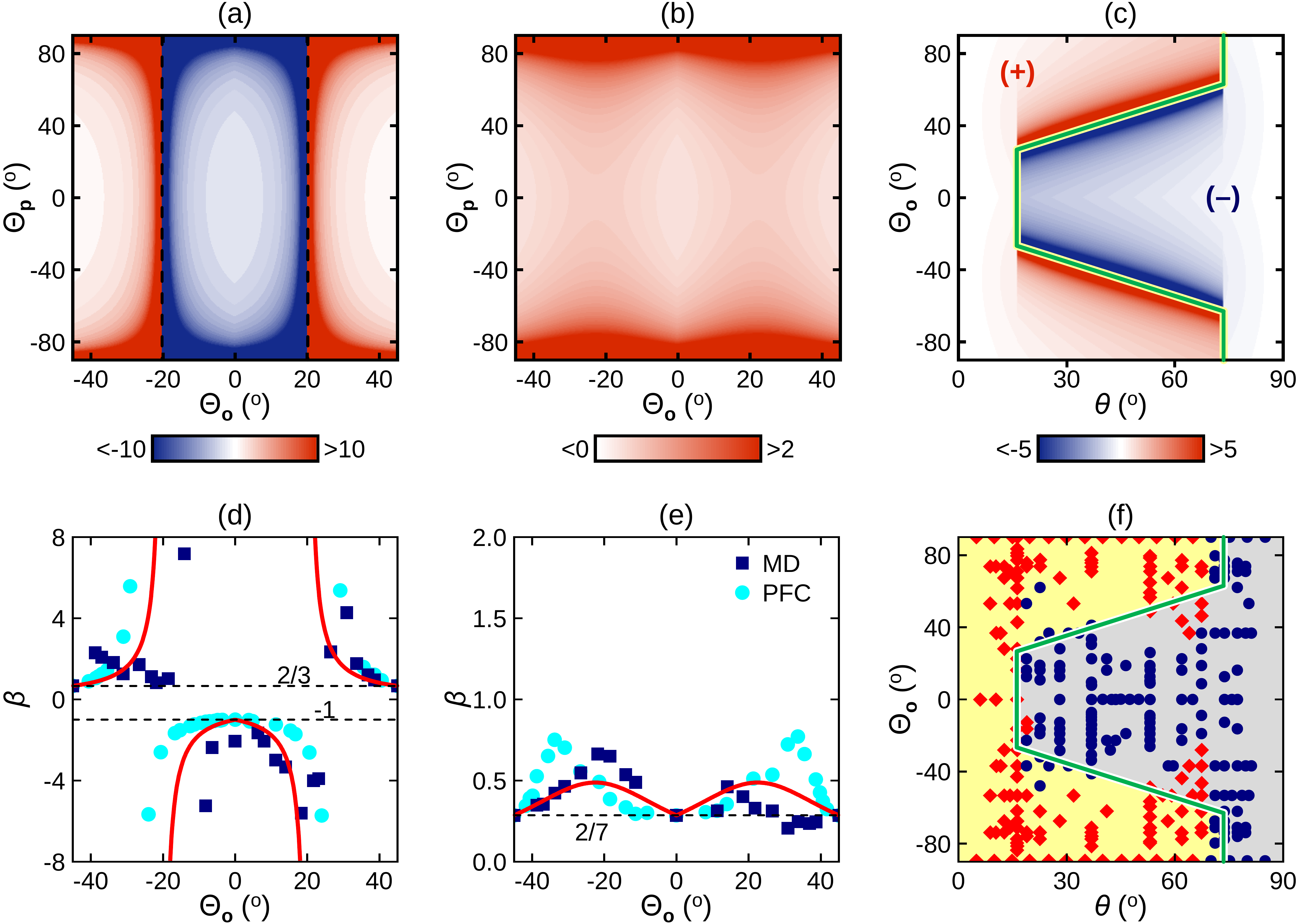}}
\caption{(a) and (b) show the predicted shear-coupling factor $\beta$ as a function of the inclination angles $\Theta_\mathbf{o}$ and $\Theta_\mathbf{p}$ (see the geometry in Fig.~\ref{beta_mixed_geometry}) for $\Sigma 5$ ($\theta = 36.9^\circ$) and $\Sigma 25$ ($\theta = 16.3^\circ$) misorientations, respectively. 
(c) shows the predicted shear-coupling factor as a function of the misorientation angle $\theta$ and the inclination angles $\Theta_\mathbf{o}$ for the case of tilt boundaries (i.e., $\Theta_\mathbf{p}=0$). The green solid line indicates a discontinuity in (and change in sign of) $\beta$.
(d) and (e) show comparisons between the predicted shear-coupling factor and the simulation (MD and PFC) data~\cite{trautt2012coupled} for $\Sigma 5$ and $\Sigma 25$ tilt boundaries (including STGBs and ATGBs), respectively. 
(f) shows the sign of the shear-coupling factor by varying $\theta$ and $\Theta_\mathbf{o}$ with $\Theta_\mathbf{p} = 0$ according to the PFC simulation results~\cite{trautt2012coupled}, where the red diamonds denote positive 
$\beta$ values, while the blue circles denote negative $\beta$ values. } 
\label{beta_mixed_predict}
\end{center}
\end{figure}

Finally, in the most general case where both $\Theta_\mathbf{o}$ and $\Theta_\mathbf{p}$ are nonzero, the shear-coupling factor can be estimated as 
\begin{equation}
\beta(\theta; \Theta_\mathbf{o}, \Theta_\mathbf{p})
= \frac{\sin^2\Theta_\text{s} \sec\Theta_\mathbf{p}}
{\beta_1^{-1}(\theta) \sin^2(\Theta_\text{s}-|\Theta_\mathbf{o}|) + \beta_2^{-1}(\theta) \sin^2|\Theta_\mathbf{o}|}. 
\label{beta_general}
\end{equation}
Figures~\ref{beta_mixed_predict}a and b show the shear-coupling factor $\beta$ in the space spanned by $\Theta_\mathbf{o}$ and $\Theta_\mathbf{p}$ for the ATGBs and MGBs originated from $\Sigma 5$ ($\theta = 36.9^\circ$) and $\Sigma 25$ ($\theta=16.3^\circ$) $[100]$ STGBs, respectively. 
Figure~\ref{beta_mixed_predict}c shows the prediction of $\beta$ values in the space spanned by the misorientation angle $\theta$ and the inclination angle $\Theta_\mathbf{o}$, where we assume that the misorientation dependence of $\beta$ follows Eq.~\eqref{branch12} with $\theta_\text{d} = 16.5^\circ$. 

\subsubsection{Comparison with simulations and experiments}
\label{ATGB}
The DFT simulation on a $\Sigma 5$ $[001]$ TwGB in Al performed by Molteni et al. shows that, driven by an applied shear deformation, GB sliding occurs without GB migration (i.e., $\beta \to \infty$)~\cite{molteni1996first,molteni1997sliding}. 
However, experiments~\cite{badirujjaman2006cyclic} and MD simulations~\cite{schonfelder2006atomistic} suggest that the migration of TwGB induced by the elastic energy density difference (i.e., energy-jump driving force) occurs without noticeable GB sliding (i.e., $\beta=0$). 
All of these observations support that there is no shear coupling for TwGBs. 

Trautt el al. obtained the shear-coupling factor for a series of ATGBs with various values of $\Theta_\mathbf{o}$ (but $\Theta_\mathbf{p}=0$) by MD and phase-field crystal (PFC) simulations~\cite{trautt2012coupled}. 
Figures~\ref{beta_mixed_predict}d and e plot the MD and PFC results for $\Sigma 5$ ($\theta = 36.9^\circ$) and $\Sigma 25$ ($\theta=16.3^\circ$) $[100]$ ATGBs, respectively. 
On the same figures, the prediction based on Eq.~\eqref{beta_ATGB} is also plotted as red lines, which is consistent in trend with the MD and PFC results. 
The PFC simulations also provide the sign of the $\beta$ values for varying $\theta$ and $\Theta_\mathbf{o}$ as shown in Fig.~\ref{beta_mixed_predict}f, which is consistent with our prediction shown in Fig.~\ref{beta_mixed_predict}c. 

Gorkaya et al. measured the shear-coupling factor $\beta$ for a $[100]$ $\theta = 18.2^\circ$ $\Theta_\mathbf{p} = 20^\circ$ MGB in Al~\cite{gorkaya2011concurrent}. 
They found the $\beta$ value was close to that of a $[100]$ $\theta = 18.2^\circ$ STGB; i.e., the shear coupling behavior is only determined by the tilt component of the MGB but unaffected by the twist component. 

We could not find a systematic collection of the shear-coupling factors for the general GBs with nonzero $\Theta_\mathbf{o}$ and $\Theta_\mathbf{p}$ in literature. 
Therefore, Eq.~\eqref{beta_general} is still waiting for validation.

\subsection{Continuum equation for grain-boundary migration}
\label{continuum_eqn_for_GB_migration}
Now, we consider a continuum description for conservative GB migration incorporating shear coupling and, in principle, all other types of driving forces. 
For a GB driven solely by capillarity or curvature  $z_{,yy}$ (see Fig.~\ref{continuum}a), the classical continuum equation for GB migration may be expressed as $z_{,t}(y,t) = M\mathit{\Gamma}z_{,yy}$, where $M$ is the GB mobility and $\mathit{\Gamma}$ is the GB stiffness~\cite{burke1952recrystallization,du2007properties}. 
However, we also know that GB migration can be driven by stress (shear-coupling). 
How does shear coupling modify the  continuum GB equation of motion? 
Such a continuum equation for GB migration that includes the shear-coupling effect was proposed by Zhang et al.~\cite{zhang2017equation}; we  review this proposal here. 
In Section~\ref{prediction_beta_stress} we demonstrated that GBs migrate by the nucleation and propagation of disconnections along the GBs; the disconnections are characterized by a Burgers vector $b$ and a step height $h$, corresponding to a shear-coupling factor $\beta = b/h$ (assuming that only a single disconnection mode operates). 
Based on this disconnection model, the GB migration velocity is 
\begin{equation}
z_{,t} = -v_\text{d}z_{,y} = -M_\text{d} F_\text{d}z_{,y}, 
\label{vdzy}
\end{equation}
where $v_\text{d}$ is the disconnection velocity, $z_{,y}$ is the slope of the GB profile (which corresponds to the signed disconnection density), $M_\text{d}$ is the disconnection glide mobility, and $F_\text{d}$ is the driving force on the disconnection (i.e., minus the variation of the energy of the system with respect to the displacement of the disconnection along the GB). 
There are two classes of driving forces: one  associated with a shear stress (i.e., the Peach-Koehler force) and the other with a energy-jump driving force.  
The driving force can be expressed as 
\begin{equation}
F_\text{d}
= \left[\left(\tau_\text{int}[z_{,y}]+\tau\right) b
+ \left(\psi - \mathit{\Gamma} z_{,yy}\right) h\right] z_{,y}/|z_{,y}|, 
\label{Fd}
\end{equation}
where $\tau$ is the externally applied shear stress resolved along the GB, $\psi$ is the energy density difference across the GB, and $\tau_\text{int}$ is the shear stress from all of the disconnections in the system resolved onto the GB (this is a functional of the step height density $z_{,y}$): 
\begin{equation}
\tau_\text{int}[z_{,y}]
= \mathcal{K}\int_{-\infty}^\infty \frac{\beta z_{,y}(y',t)}{y-y'}~\text{d}y'. 
\end{equation}
Combining Eq.~\eqref{vdzy} and Eq.~\eqref{Fd}, we obtain the continuum equation for GB migration: 
\begin{equation}
z_{,t}(y,t)
= - M_\text{d} \left[
\left(\tau_\text{int}[z_{,y}] + \tau\right) b + \left(\psi - \mathit{\Gamma} z_{,yy}\right) h
\right]
\left(|z_{,y}| + \eta\right). 
\label{continuum_eq}
\end{equation}
$\eta$ can be thought of as a disconnection source (or nucleation) term, without which a flat GB ($z_{,y} = 0$) would not be able to move (in contradiction to simulation observations of driven migration of a flat GB).
Zhang et al.~\cite{zhang2017equation} suggested replacing the  disconnection source with the assumption that  $\eta$ represents the temperature-dependent, thermal equilibrium step density: $\eta = (h/a_\text{b}) e^{-E_\text{d}/k_\text{B}T}$, where $a_\text{b}$ is the atomic spacing along the GB and $E_\text{d}$ is half the energy of the formation of a disconnection pair. 
With this equation of GB motion Eq.~\eqref{continuum_eq}, we can numerically  solve for the evolution of a GB profile under any type of or combination of driving forces given the appropriate  boundary conditions;  examples of this are shown in Section~\ref{GBmwTJ}. 

Another continuum description of GB structure, energy and migration behavior was proposed, based on the distribution and dynamics of intrinsic GB dislocations~\cite{Zhu-Xiang-2014,Zhang-Gu-Xiang-2017,Zhang-Xiang-2017}. 
This approach is mainly applicable to low-angle GBs, but may, in principle, be extended to high-angle GBs by considering secondary GB dislocations. 
Even for the case of low-angle GBs, we would emphasize that the description based on the motion of intrinsic GB dislocations is equivalent to a description based on disconnections (as illustrated in Fig.~\ref{wedge} and discussed in Section~\ref{disconnection_for_CSL}).

\subsection{Implications for grain growth/shrinkage}
\label{grain_growth}

Driven GB migration and the associated shear-coupling effect has been widely studied for individual flat GB usually in the bicrystal context. 
However, in the vast majority of applications, the materials of interest are polycrystals (rather than single or bi-crystals). 
Hence, it is important to examine the implications of shear coupling for the evolution of polycrystalline microstructure. 

It is often fruitful to model a polycrystalline microstructure as a GB network. 
The evolution of the microstructure can then be described as the evolution of the GB network~\cite{mason2015geometric}. 
Traditionally, the driving force for the evolution of the GB network (in a well-annealed sample and in the absence of an applied stress) is assumed to be surface tension (capillarity), meaning that the GB network tends to evolve in order to lower the total GB energy (and, thus, total GB area) in the entire polycrystalline system. 
For this driving force, the kinetic equation for GB migration, is usually expressed as $v_\perp = M\gamma \kappa$. 
This is the equation of motion that describes ideal soap froth coarsening, where only the surface tension plays the role. 
However, polycrystalline materials are not soap froths. 
First, unlike soap froths, crystalline materials are anisotropic; GB properties (such as GB energy, GB mobility, etc.) depend on the orientations of the neighboring grains (no such issue arises for the isotropic membranes of soap bubbles). 
Note that a small modification of this kinetic equation can account for anisotropic GB energy and mobility~\cite{abbruzzese1986theory}. 
Second, many experiments have demonstrated that GB migration may be driven by an applied stress~\cite{rupert2009experimental,mompiou2009grain,sharon2011stress} (such elastic stress-induced GB motion never occurs in a froth with liquid membranes). 
Third, grain rotation occurs during the evolution of polycrystalline microstructure~\cite{harris1998grain,shan2004grain}; grain rotation has no meaning for soap bubbles. 
Therefore, the classical model for capillarity-driven microstructure evolution fails to capture many important phenomena associated with the GB network evolution in polycrystals. 

As suggested in the previous sections, the main missing element in the classical model for the evolution of microstructure is related to GB bicrystallography and its inevitable shear-coupling effect. 
Since the shear-coupling factor $\beta$ is determined by the bicrystallography, much of the anisotropy of GB properties may naturally be included by virtue of the disconnection model. 
Second, since GB migration occurs by the motion of disconnections, GB migration can be driven by (internal and/or applied) stresses (even in a polycrystal system). 
Third, since GB migration is accompanied by shear deformation, grain rotation naturally occurs provided that the shear deformation occurs cooperatively (in the same sense) for the GBs enclosing individual grains. 

We now review our present understanding of GB migration in microstructure coarsening and especially how it is related to shear coupling. 

\subsubsection{A shrinking cylindrical grain}
\begin{figure}[!t]
\begin{center}
\scalebox{0.29}{\includegraphics{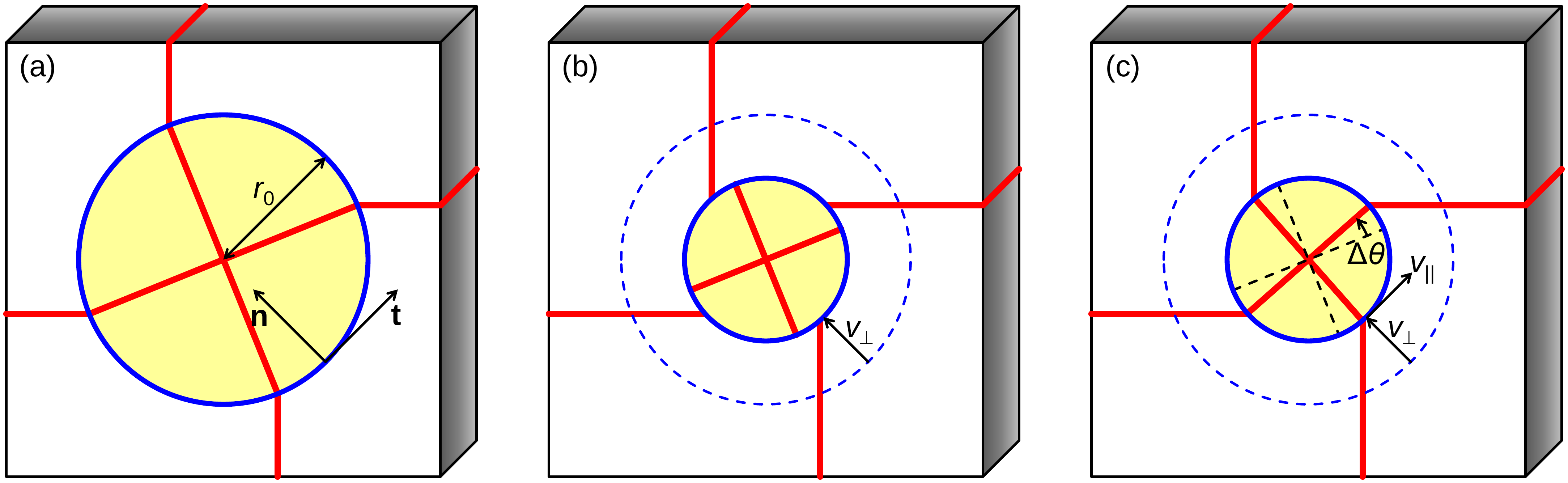}}
\caption{(a) A cylindrical grain (colored yellow) with initial radius $r_0$ embedded in a matrix of different misorientation. 
The blue solid line indicates the current GB position and the red lines are fiducial marks, attached to material points.  
(b) Grain shrinkage in the case of no coupling effect (pure GB sliding). 
The blue dashed line indicates the initial GB position. 
(c) Grain shrinkage when shear-coupled GB migration occurs. 
The black dashed line denotes the fiducial line if the cylindrical grain does not rotate. 
The comparison between the black dashed line and the red solid line in the cylindrical grain indicates grain rotation by the angle $\Delta\theta$.
This figure is reproduced from Ref.~\cite{cahn2004unified} (J. W. Cahn, Elsevier 2004). } 
\label{cylindrical_grain}
\end{center}
\end{figure}

A simplified model for studying the evolution of microstructure is a circular cylindrical grain embedded in a matrix, as illustrated in Fig.~\ref{cylindrical_grain}a. 
This is an example of capillarity-driven evolution without the complications of triple junctions and multiple grain orientations.
This may be a reasonable approximation to a grain surrounded by high-angle/high-energy GBs where all of the other GBs meeting it at triple junctions are low-angle/low-energy, such that  the effect of triple junctions is small. 
The circular shape of the embedded grain may, in principle, allow for easy grain rotation. 

Driven by the capillary force, this embedded grain tends to shrink in order to lower the total GB energy. 
According to the classical law of capillarity-driven grain growth (in the absence of shear coupling), the grain shrinkage velocity is 
\begin{equation}
-\dot{r} = v_\perp = M\gamma/r, 
\label{dr_classic}
\end{equation}
where $\dot{A} \equiv \text{d}A/\text{d}t$ for any quantity $A$, $r$ is the radius of the grain, and $v_\perp$ is the magnitude of the GB migration velocity in the direction normal to the local GB plane (we define the GB normal as pointing into the embedded grain). 
The solution to Eq.~\eqref{dr_classic} is 
\begin{equation}
r^2 = r_0^2 - 2M\gamma t,
\label{r2_1}
\end{equation}
which is simply the classical parabolic grain growth/shrinkage law. 

If there is a (constant) shear-coupling factor associated with the GB, then  grain shrinkage will be accompanied by grain rotation, as illustrated in Fig.~\ref{cylindrical_grain}c. 
Cahn et al. analyzed the shrinkage dynamics of the cylindrical grain in such situation~\cite{cahn2004unified}. 
The magnitudes of the GB migration velocity $v_\perp$ and GB sliding velocity $v_\parallel$ are (see Fig.~\ref{cylindrical_grain}c)
\begin{equation}
v_\perp = -\dot{r} 
\text{ and }
v_\parallel = r\dot{\theta}. 
\end{equation}
If these two velocities are coupled by the factor $\beta$, then 
\begin{equation}
\beta = \frac{v_\parallel}{v_\perp} = - r\frac{\text{d}\theta}{\text{d}r}.  
\label{coupling_relation}
\end{equation}
The driving force for  grain shrinkage is the capillary force (reduction of the total GB energy), i.e., 
\begin{equation}
F = \frac{1}{2\pi r} \frac{\text{d}(2\pi r\gamma)}{\text{d}r}
= (\gamma - \beta\gamma')/r, 
\end{equation}
where Eq.~\eqref{coupling_relation} is used, $\gamma$ is in general considered as a function of the misorientation angle $\theta$, and $\gamma' \equiv \text{d}\gamma/\text{d}\theta$. 
Thus, the kinetic equations for GB motion are
\begin{equation}
-\dot{r} = v_\perp = MF = M(\gamma - \beta\gamma')/r, 
\label{dr_couple}
\end{equation}
\begin{equation}
r\dot{\theta} = v_\parallel = \beta v_\perp = \beta M(\gamma - \beta\gamma')/r. 
\label{rdtheta_couple}
\end{equation}
For $\beta = 0$ (i.e., no shear coupling, e.g. at high temperature), Eq.~\eqref{dr_couple} exactly becomes Eq.~\eqref{dr_classic}, which gives the classical parabolic law; and from Eq.~\eqref{rdtheta_couple}, we will have $\dot{\theta} = 0$, indicating that the grain will not rotate; this is the scenario illustrated in Fig.~\ref{cylindrical_grain}b. 
In general, $\beta$, $M$, $\gamma$ and $\gamma'$ are all functions of $\theta$. 

We first consider the  low-angle GB case ($\theta \ll 1$) such that the GB is an array of lattice dislocations.
In this case, the shear-coupling factor  $\beta(\theta) \approx \theta$ [this is the small-$\theta$ branch in Eq.~\eqref{branch12}]. 
The GB mobility in this limit can be thought of as arising from the motion of its constituent lattice dislocations.  Hence, $M(\theta)= m_\text{b}b_\text{b}/\theta = M_0/\theta$, where $m_\text{b}$ and $b_\text{b}$ are the mobility and Burgers vector of these lattice dislocations and $M_0\equiv m_\text{b}b_\text{b}$~\cite{cahn2004unified,karma2012relationship}. 
The low-angle GB energy is $\gamma(\theta) = \gamma_0\theta (a - \ln\theta)$~\cite{read1950dislocation}. 
Thus, Eq.~\eqref{dr_couple} becomes 
\begin{equation}
-\dot{r} = M_0\gamma_0/r
\Rightarrow 
r^2 = r_0^2 - 2M_0\gamma_0 t.
\label{r2_t}
\end{equation} 
This is the classical result of Eq.~\eqref{r2_1} in form.
Inserting these low-angle GB assumptions into the shear rate expression, Eq.~\eqref{rdtheta_couple}, this implies a (nearly) constant rotation rate: 
\begin{equation}
\frac{\dot{\theta}}{\theta} = \frac{M_0\gamma_0}{r^2} = \frac{M_0\gamma_0}{r_0^2 - 2M_0\gamma_0 t}
\Rightarrow 
\frac{\theta}{\theta_0}
= \frac{1}{\sqrt{1 - (2M_0\gamma_0/r_0^2) t}} 
\approx 1 + \frac{M_0\gamma_0}{r_0^2} t.
\label{theta_t}
\end{equation}
We see that the inclusion of shear coupling does not modify the classical prediction of how fast a grain shrinks, yet it does explain grain rotation (which is not in the classical theory). 

MD simulations were performed to measure grain rotation during grain shrinkage~\cite{upmanyu2006simultaneous,trautt2012grain,barrales2016capillarity}. 
In particular, the MD results obtained by Trautt et al. show qualitative agreement with Eq.~\eqref{r2_t} and Eq.~\eqref{theta_t}~\cite{trautt2012grain}. 
However, in some MD simulations and experiments, the rotation of an embedded grain was not observed  during grain shrinkage~\cite{radetic2012mechanism,mompiou2012situ,barrales2016capillarity}. 

Several explanations of this apparent contradiction are possible. 
Trautt et al. found that, if the embedded grain rotated such that the misorientation angle reached $\theta_\text{d}$ [ref. Eq.~\eqref{branch12}, i.e., the angle where the $\beta$ value switches from one branch to the other], the grain would stop rotating at this angle (i.e., $\theta_\text{d}$ is a stationary misorientation)~\cite{trautt2012grain}. 
Trautt et al. also suggests that we should describe the rotation of the embedded grain based on a  $\beta$ which is the average of the shear-coupling factors around the entire grain~\cite{trautt2014capillary}. 
If the shear-coupling factor changes sign for different GB segment for the GBs around the grain,  the inclination-averaged $\beta$ will be close to zero and, thus, no grain rotation will be observed. 
This is consistent with the predictions of $\beta(\Theta_\mathbf{o})$ in Fig.~\ref{beta_mixed_predict}a where we see the sign of  $\beta$ switching with GB inclination.
However, if there were only single disconnection modes operating at different positions along the GB, then stresses would develop and this should stop GB motion altogether, which is not consistent with the simulation results.
Alternatively, as the grain shrinks, the stresses that develop will lead to cyclic switching between disconnection modes with positive and negative $\beta$ values (see Section~\ref{mixed_force_driven_gb_migration} and Fig.~\ref{coupling_schematic}c3), as illustrated in Fig.~\ref{cylindrical_grain}c. 
This will allow shrinking without rotation -- consistent with the simulations.
Examination of  Figs.~\ref{beta_mixed_predict}a and b suggest that the inclination-averaged value of $\beta$ may be zero or nonzero, indicating that for some grains rotation should occur while for others no rotation will be observed.

\subsubsection{Grain-boundary migration with triple junctions}
\label{GBmwTJ}
\begin{figure}[!t]
\begin{center}
\scalebox{0.36}{\includegraphics{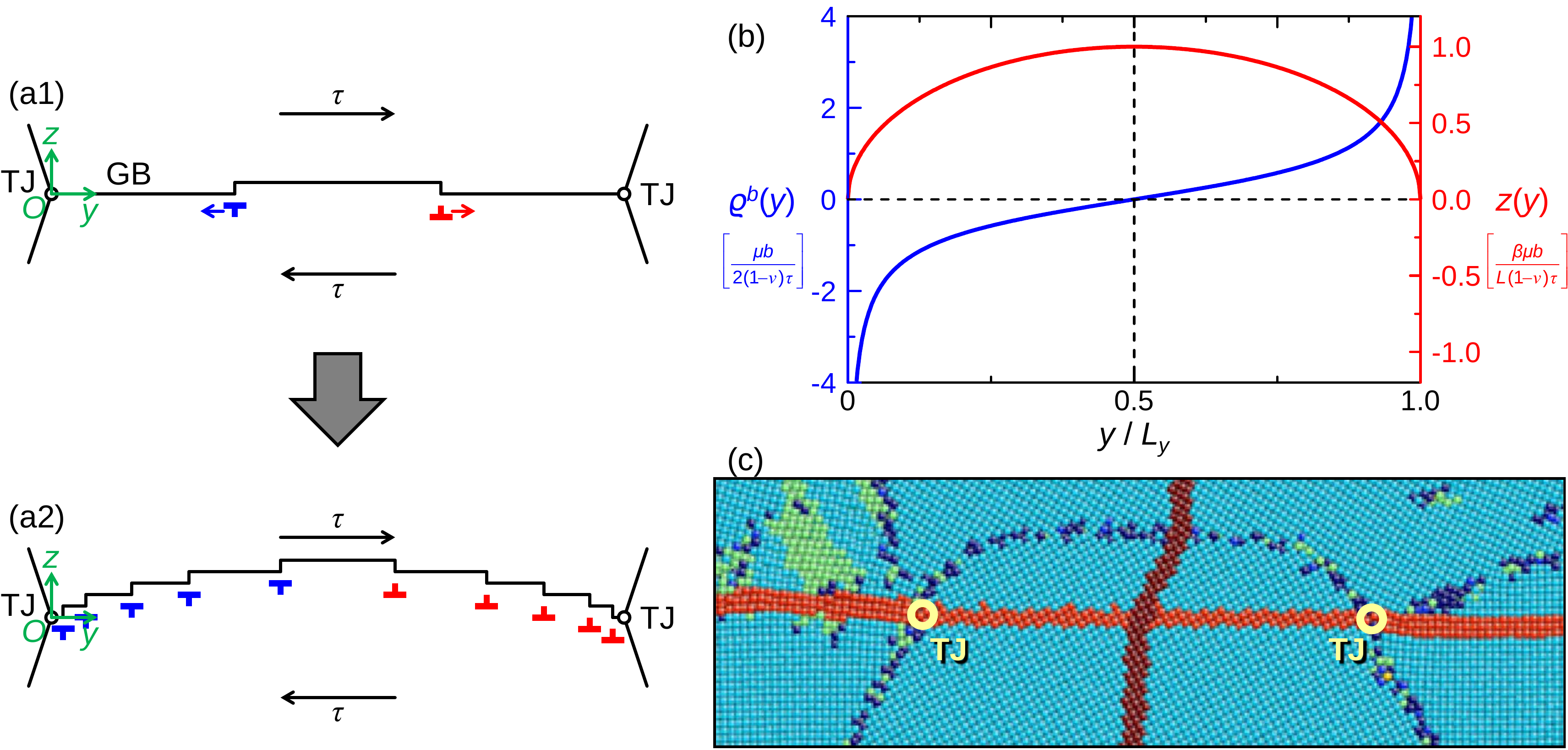}}
\caption{(a1) Model of a GB with two ends pinned by TJs. 
The applied shear stress $\tau$ induces the nucleation of a pair of disconnections on the GB. 
(a2) More and more disconnection pairs are nucleated and separated and pile up at the TJs. 
The equilibrium distribution of disconnections leads to the curved GB profile. 
(b) The plots of the distribution of Burgers vector density $\varrho^b(y)$ according to Eq.~\eqref{eqBurgers} (the blue curve) and the GB profile $z(y)$ according to Eq.~\eqref{eqGBprofile} (the red curve). 
(c) The snapshot from the MD simulation performed by Aramfard and Deng~\cite{aramfard2014influences}, where the atoms are colored according to the local lattice orientation, except that some atoms are artificially colored red to serve as a reference. }
%This figure is reproduced from Ref.~\cite{aramfard2014influences} (C. Deng, IOP Publishing 2014). } 
\label{TJaGB}
\end{center}
\end{figure}

In polycrystals, GBs are neither of infinite extent nor periodic (as often modeled in simulations), but are instead delimited by triple junctions (TJs) -- 1D defects along which three GBs meet~\cite{SuttonBalluffi,gottstein2010thermodynamics}. 
While our previous discussion considered a finite grain size (in the form of an embedded cylindrical grain), it did not discuss the effects of TJs on GB migration. 
Here, we consider the evolution of a GB profile associated with the effect of the delimiting TJs within the framework of the disconnection model. 
Figure~\ref{TJaGB}a1 shows a schematic illustration of a single 1D GB terminated by two TJs. 

We initially assume that the TJs are immobile and consider the equilibrium GB profile, i.e., $z(y)$, under an applied shear stress $\tau$. 
As mentioned in Section~\ref{prediction_beta_stress} and shown in Fig.~\ref{conservative}, the shear stress triggers repeated nucleation and separation of disconnection pairs. 
Since the TJs are immobile (i.e., the disconnection fluxes into/out of the TJs are zero), the nucleated disconnections pile up at the TJs, as shown in Fig.~\ref{TJaGB}a2. 
The experiments of Pond, Smith and Southerden~\cite{pond1978role} demonstrated that the extent of GB sliding decreases along the GB as it approaches an immobile TJ. 
Since  GB sliding occurs by disconnection motion, this finding supports the notion that disconnections  indeed pile up at  TJs. 
An analytical solution exists for the equilibrium Burgers vector density distribution associated with this classical double-ended pileup of lattice dislocations under an applied shear stress $\tau$~\cite{HirthLothe}:
\begin{equation}
\varrho^b(y)= \frac{2(1-\nu)\tau}{\mu b} \frac{y-L_y/2}{\sqrt{y(L_y-y)}}, 
\label{eqBurgers}
\end{equation}
which is based on the geometry and coordinate system shown in Fig.~\ref{TJaGB}a. This function is plotted as the blue curve in Fig.~\ref{TJaGB}b. 
In the disconnection model, a Burgers vector pileup creates a corresponding pileup of steps (constant $\beta$ case); hence, the stress that drives the Burgers vectors also drives a GB profile change.
From Eqs.~\eqref{varrhoh}, \eqref{varrhob} and \eqref{beta1}, we know that $\varrho^h = \text{d}z/\text{d}y$, $\varrho^b = -\text{d}u/\text{d}y$, and $\beta = \text{d}u/\text{d}z$. Hence, $\varrho^b = -\beta \varrho^h$. 
With this relationship, we obtain the equilibrium GB profile as 
\begin{equation}
z(y)
= \int_0^y \varrho^h(y') ~\text{d}y'
= - \frac{1}{\beta} \int_0^y \varrho^b(y') ~\text{d}y'
= \frac{2(1-\nu)\tau}{\beta\mu b} \sqrt{y(L_y -y)}, 
\label{eqGBprofile}
\end{equation}
which is plotted as the red curve in Fig.~\ref{TJaGB}b. 
A GB network geometry similar to that in Fig.~\ref{TJaGB}a1 was examined in the MD simulation of Aramfard and Deng~\cite{aramfard2014influences} and the resultant profile is shown in  Fig.~\ref{TJaGB}c.  
Comparison of this profile with our prediction (see the red curve in Fig.~\ref{TJaGB}b) shows excellent agreement. 
A similar equilibrium GB profile terminated by two TJs and under a shear stress was also simulated by Velasco, Van Swygenhoven and Brandl~\cite{velasco2011coupled}. 
In the discussion above, we neglected the capillary force due to the GB profile curvature. 
According to the solution to Eq.~\eqref{eqGBprofile}, the curvature goes to infinity near the TJs (as $y\to 0$ or $L_y$). 
Therefore, if the capillary force is taken into account, the curvature should be lowered near the TJs. 

\begin{figure}[!t]
\begin{center}
\scalebox{0.36}{\includegraphics{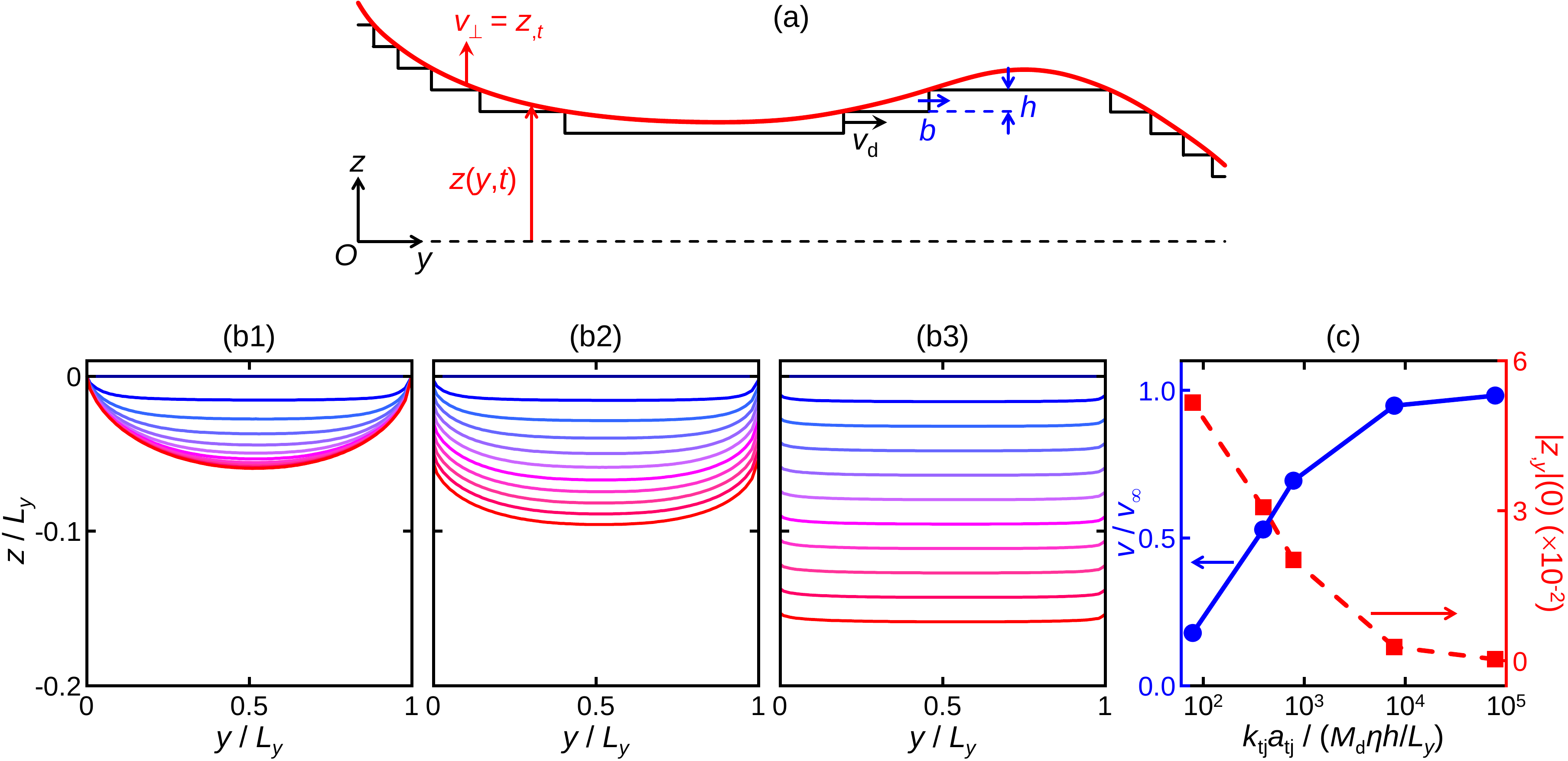}}
\caption{(a) A curved 1D GB represented by a continuum description (red) and by a discrete distribution of disconnections of mode $(\pm b, \pm h)$ (black). 
(b1), (b2) and (b3) The numerical results of the GB profile evolution under the applied shear stress $\tau = 5\times 10^{-2}\mu$ for $k_\text{tj}a_\text{tj}/(M_\text{d}\eta h/L_y) = 0$, $78.1$ and $781$, respectively. 
In each case, blue line color denotes an earlier time while red line color denotes a later time. 
These plots are reproduced from the supplemental material of Ref.~\cite{zhang2017equation} (L. Zhang, American Physical Society 2017). 
(c) The GB migration velocity (the solid blue curve) and the slope of the GB profile at the boundary (the dashed red curve) as functions of the kinetic parameter $k_\text{tj}a_\text{tj}$.   } 
\label{continuum}
\end{center}
\end{figure}

The time evolution of the GB profile (including both transient and equilibrium profiles) can be obtained by numerically solving Eq.~\eqref{continuum_eq} with the boundary condition $z_{,t}(0, t) = z_{,t}(L_y, t) = 0$; this equation includes the effect of capillarity $\mathit{\Gamma}z_{,yy}$. 
Figure~\ref{continuum}b1 shows the numerical results for stress-driven GB migration with the GB pinned by two TJs obtained by Zhang et al.~\cite{zhang2017equation}.
The equilibrium GB profile (red curve) is similar to that predicted from Eq.~\eqref{eqGBprofile}, but the difference is that the curvature near the TJs is finite. 
We note that when the disconnection source parameter $\eta$ is small, the evolving GB profile near $y=L_y/2$  has very small curvature. 
As the disconnections are rapidly pushed to the TJs by the shear stress and join the disconnection pileup at the TJs, the disconnection density remains small (which implies a near flat GB profile) around $y=L_y/2$. 
Zhang et al.~\cite{zhang2017equation} demonstrated that changing $\eta$ (e.g., by changing the temperature) changes the shape of the evolving GB profile but not its equilibrium profile. 

TJ-pinned GB migration corresponds to the scenario where the disconnections flowing along the GB cannot annihilate at the TJs. 
In this case, when the disconnections are driven to the TJs, the Burger vectors (carried by the disconnections) will inevitably accumulate at the TJs. 
The accumulation of Burgers vectors at the TJs will establish a back stress on the disconnections along the GB and eventually lead to a stationary GB profile. 
The TJ-pinned GB profile evolution, however, is not a good representation of the GB profile evolution in a polycrystalline microstructure. 
In reality, GBs are usually not pinned by TJs (otherwise, grain growth would never be observed). 
Rather, disconnections do not necessarily accumulate at the TJs because they may react and annihilate with  disconnections arriving at the TJ from the other GBs. 
Two extreme cases are expected. 
If the reaction rate at the TJ is zero, disconnection accumulation will occur and lead to a pinned TJs and the cessation of GB migration. 
If the reaction rate tends to infinity, the TJ motion will be a slave to the GB migration and the GB profile will remain flat as it migrates. 
We can consider the effect of mobile TJs by modifying the boundary condition at the TJ. 
The GB velocity at the TJ located at $y=0$ is 
\begin{equation}
z_{,t}(0,t) = J(0,t)  h = \varrho(0,t)  v_\text{d}(0,t) h = z_{,y}(0,t) v_\text{d}(0,t), 
\label{zti}
\end{equation}
where $J$ is the disconnection flux flowing towards the TJ and $\varrho$ is the disconnection density (the same boundary conditions applies to the TJ at $y = L_y$). 
We assume that disconnection reaction kinetics at the TJ is of first order, leading to a disconnection velocity at the TJ: $v_\text{d}(0,t) = k_\text{tj}a_\text{tj}$, where $k_\text{tj}$ is the reaction rate constant and $a_\text{tj}$ is a characteristic distance to the TJ within which the disconnection annihilation occurs spontaneously. 
The reaction rate constant can be expressed as $k_\text{tj} = \nu e^{-E_{\text{gb}\to\text{tj}}^*/k_\text{B}T}$, where $\nu$ is an attempt frequency and $E_{\text{gb}\to\text{tj}}^*$ is the energy barrier for a disconnection absorption into the TJ. 
Hence, the boundary condition accounting for the effect of finite TJ mobility is 
\begin{equation}
z_{,t}(y,t) = k_\text{tj}a_\text{tj} z_{,y}(y,t) 
\text{ for }
y = 0 \text{ and } L_y. 
\label{ztBC}
\end{equation}
Zhang et al.~\cite{zhang2017equation} solved Eq.~\eqref{continuum_eq} with this boundary condition; the numerical results of the GB profile evolution for $k_\text{tj}a_\text{tj}L_y/M_\text{d}\eta = 0$, $78.1$ and $781$ are shown in Figs.~\ref{continuum}b1, b2 and b3, respectively. 
From these results we can see that, following an initial transient, a steady-state GB profile is established. 
These steady-state GB profiles (i.e., the red curves in Fig.~\ref{continuum}b) can not be simply described by Eq.~\eqref{continuum_eq}. 
The steady-state GB velocity and the absolute value of the slope of the steady-state GB profile at $y=0$ as functions of the parameter $k_\text{tj}a_\text{tj}$ are shown in in Fig.~\ref{continuum}c. 
These results show that, with increasing $k_\text{tj}a_\text{tj}$, the GB and TJs move faster and the GB profile becomes increasingly flat. 
The results also show that for small $k_\text{tj}a_\text{tj}$ the GB slope at the TJ is large and as $k_\text{tj}a_\text{tj}\to \infty$ the slope goes to zero. 
This demonstrates that TJ drag, induced by disconnection reaction barriers, indeed changes the TJ angle from its equilibrium value ($0$ here), as discussed earlier by Shvindlerman and co-workers ~\cite{czubayko1998influence,protasova2001triple,mattissen2005drag,upmanyu1999triple,upmanyu2002molecular}.

\subsubsection{Triple junctions}
\label{TJconstraints}
In the previous section, the effects of the TJs were treated as a boundary condition applied to the terminations of a single GB. 
However, because the mechanism behind these TJ effect is related to the reactions among the disconnections coming from the three GBs meeting at the TJ, the TJ motion should depend on disconnection motion along all three GBs; this cannot be described by the simple boundary condition Eq.~\eqref{ztBC} (except in special cases). 
Here, we focus on  TJ motion and propose a more rigorous manner to treat TJ effect on  GB migration. 

TJs are not simply the terminations of GBs, but are unique defects in their own right; i.e., they have their own crystallography, thermodynamics, and kinetic properties~\cite{gottstein2010thermodynamics}.
For example, as alluded to above, both experiments and MD simulations have demonstrated that TJs can create drag on GB migration, especially at low temperature~\cite{czubayko1998influence,protasova2001triple,mattissen2005drag,upmanyu1999triple,upmanyu2002molecular}. 
This implies that TJs are not simply the geometric intersection of three GBs, which would imply that TJs have infinite mobility (i.e., they are in equilibrium with respect to the GBs).
Rather, TJs have their own dynamics that give rise to finite TJ mobilities.

\begin{figure}[!t]
\begin{center}
\scalebox{0.22}{\includegraphics{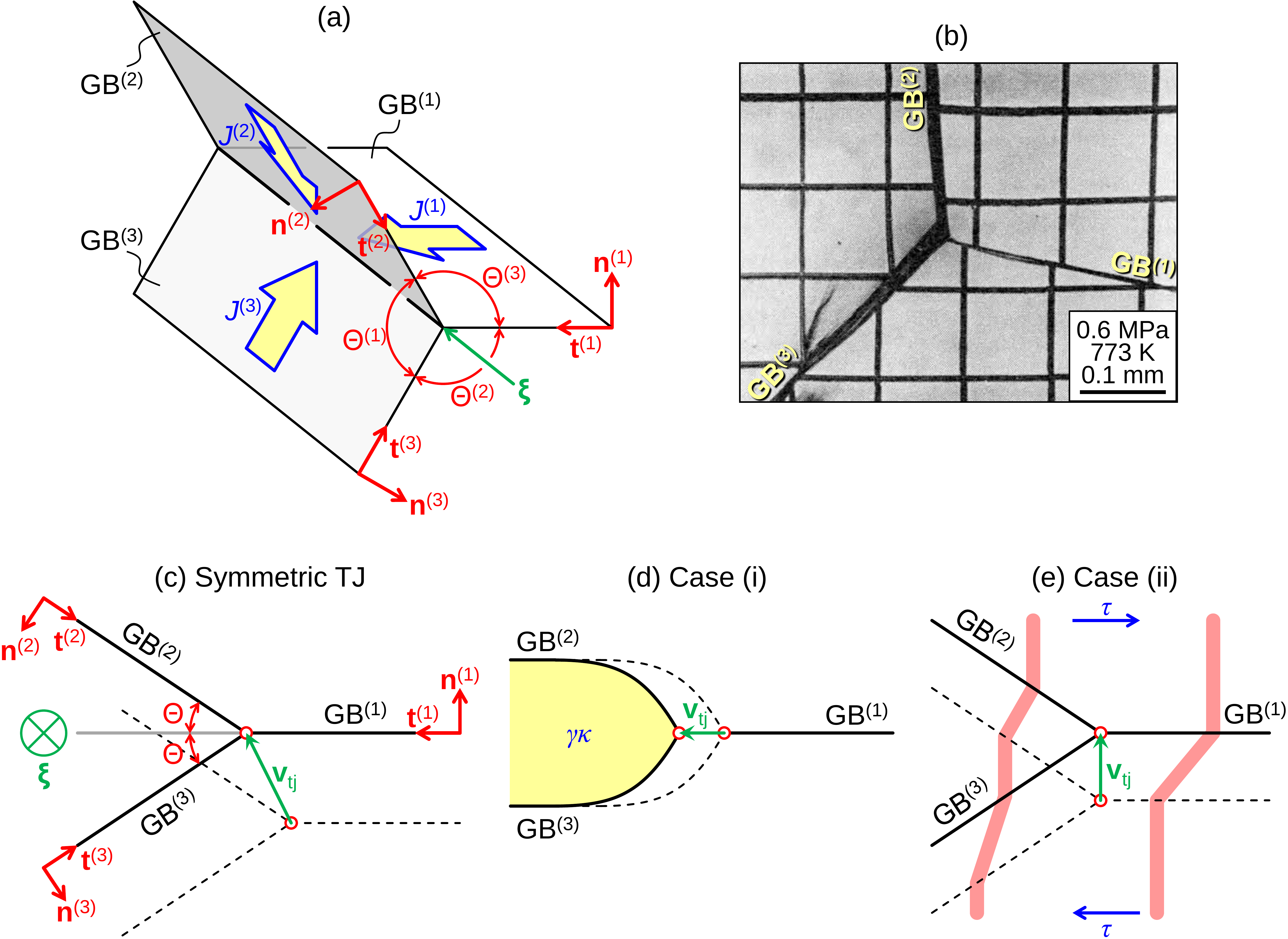}}
\caption{(a) The geometry of a TJ (line direction $\boldsymbol{\xi}$), where GB$^{(1)}$, GB$^{(2)}$ and GB$^{(3)}$ meet. 
$J^{(1)}$, $J^{(2)}$ and $J^{(3)}$ are the disconnection fluxes from the three GBs into the TJ. 
(b) GB sliding and migration around a TJ under constant stress of $0.6$ MPa at $773$ K, observed by Miura, Hashimoto and Fujii~\cite{miura1988effect}. %; the figure is reproduced from Ref.~\cite{miura1988effect}. 
(c) The geometry of a TJ configuration which is symmetric about GB$^{(1)}$. 
The dashed line indicates the initial TJ configuration. 
(d) Case (i), where the Burgers vectors of the disconnections along the GBs are tangent to the TJ line and the TJ/GB motion is driven by the capillary force. 
(e) Case (ii), where the Burgers vectors of the disconnections along the GBs are perpendicular to the TJ line and the TJ/GB motion is driven by the shear stress; the light red lines are fiducial lines.    } 
\label{TJ}
\end{center}
\end{figure}

As discussed in Sections~\ref{prediction_beta_stress} and~\ref{prediction_beta_synthetic_force}, GB migration occurs by the nucleation and glide of disconnections along the GB. 
TJs can be thought of as  sources/sinks for  the disconnections moving along the GBs~\cite{mompiou2013inter,rajabzadeh2013evidence}.  
When disconnections flow into/out of TJs, they must either accumulate there or react with other disconnections there, since disconnection step heights and Burgers vectors must be conserved.
TJ migration can be described in terms of the accumulation of disconnection steps at the TJ -- this is a purely geometrical effect.
Geometric considerations require that the following  \textit{zero displacement incompletion} condition~\cite{sisanbaev1992effect} must be met at the TJ:
\begin{equation}
\sum_{i=1}^3 J^{(i)}h^{(i)}\sin\Theta^{(i)} = 0, 
\label{TJH}
\end{equation}
where $J^{(i)}$ is the disconnection flux from GB$^{(i)}$ into the TJ, $h^{(i)}$ is the step height of the disconnection on GB$^{(i)}$, and $\Theta^{(i)}$ is the dihedral angle opposite GB$^{(i)}$ -- see Fig.~\ref{TJ}a. 
The zero displacement incompletion condition Eq.~\eqref{TJH} implies that $J^{(1)}$, $J^{(2)}$ and $J^{(3)}$ are not independent; there are two, rather than three, free variables. 
Hence, Eq.~\eqref{TJH} represents a TJ constraint on the migration of the three GBs meeting there (via $J^{(i)}$). 
Correspondingly, the velocity (vector) of the TJ is 
\begin{equation}
\mathbf{v}_\text{tj} = \frac{1}{3} \sum_{i,j,k} \varepsilon_{ijk} \left[\frac{J^{(k)}h^{(k)}}{\sin\Theta^{(j)}} - \frac{J^{(j)}h^{(j)}}{\sin\Theta^{(k)}}\right] \mathbf{t}^{(i)}, 
\label{vtj}
\end{equation}
where $\varepsilon_{ikj}$ is the permutation symbol and $\mathbf{t}^{(i)}$ is the unit tangent vector to GB$^{(i)}$ (perpendicular to the TJ direction $\boldsymbol{\xi}$) adjacent to the TJ. 

In order for the TJ to continue to move, another condition should be satisfied: there should be no net disconnection Burgers vector accumulation at the TJ.
If this condition is not met, Burgers vectors will accumulate at the TJ, the TJ will develop dislocation character, and the disconnections moving along the GBs will tend to be repelled from the TJ (via elastic Peach-Koehler forces).
If this occurs, TJ motion will stop and, eventually, GB migration will also stagnate because the disconnections will be unable to flow. 
This gives rise to an \textit{equilibrium of shear} condition~\cite{miura1988effect} which can be expressed as 
\begin{equation}
\sum_{i=1}^3 J^{(i)} \mathbf{b}^{(i)} +\sum_{l=1}^N J^{\text{L}_l}\mathbf{b}^{\text{L}_l} = \mathbf{0}, 
\label{TJb}
\end{equation}
where $\mathbf{b}^{(i)}$ is the Burgers vector of the disconnection on GB$^{(i)}$. 
The second term in this equation is associated with plasticity in the grains; $\mathbf{b}^{\text{L}_l}$ and $J^{\text{L}_l}$ are the Burgers vector and the flux, respectively, of lattice dislocations belonging to the $l$-th slip system from the grain interior flowing into the TJ. 
Note, of course, that lattice dislocations can also be absorbed into/emitted from the GBs themselves, reacting with the disconnections there and modifying the disconnection flux from the GBs into the TJs (as already captured by the $\mathbf{b}^{(i)}$ terms). 

Combining Eqs.~\eqref{TJH} and~\eqref{TJb}, we obtain a system of linear equations for the disconnection fluxes on the  GBs meeting at the TJ: 
\begin{align}
\mathbf{0}&=\left(\begin{array}{ccc|ccc}
b_x^{(1)}	& b_x^{(2)}		& b_x^{(3)}		& b_x^{\text{L}_1}	& \cdots 	& b_x^{\text{L}_N} \\
b_y^{(1)}	& b_y^{(2)}		& b_y^{(3)}		& b_y^{\text{L}_1}	& \cdots 	& b_y^{\text{L}_N} \\
b_z^{(1)}	& b_z^{(2)}		& b_z^{(3)}		& b_z^{\text{L}_1}	& \cdots 	& b_z^{\text{L}_N} \\
\hline
h^{(1)}\sin\Theta^{(1)}	& h^{(2)}\sin\Theta^{(2)}	& h^{(3)}\sin\Theta^{(3)}	& 0	& \cdots 	& 0 \\
\end{array}\right)
\left(\begin{array}{c}
J^{(1)} \\ J^{(2)} \\ J^{(3)} \\ \hline J^{\text{L}_1} \\ \vdots \\ J^{\text{L}_N} \\
\end{array}\right)
\nonumber \\
&= \left(\begin{array}{cc}
\mathbb{B}	& \mathbb{B}^\text{L} \\
\mathbb{H}	& \mathbf{0}
\end{array}\right)
\left(\begin{array}{c}
\mathbb{J}	\\ \mathbb{J}^\text{L}
\end{array}\right), 
\label{TJeqF}
\end{align}
where the $\mathbb{B}$ and $\mathbb{H}$ matrices are formed by the coefficients related to the disconnection Burgers vectors and step heights on the GBs, respectively, $\mathbb{J}$ is the disconnection flux vector, and the superscript ``L'' labels the terms related to the lattice dislocations. 
A TJ is mobile only when $\mathbb{J}$ has nontrivial solutions (i.e., solutions for which the disconnection fluxes into the TJ are nonzero); of course, if $h^{(1)}$, $h^{(2)}$ and $h^{(3)}$ are all zero, the TJ also does not move. 
If there is no lattice dislocation flowing into/out of the TJ, Eq.~\eqref{TJeqF} simplifies to
\begin{equation}
\mathbf{0}=\left(\begin{array}{c}
\mathbb{B}	\\ \mathbb{H}
\end{array}\right) \mathbb{J}, 
\label{TJeq}
\end{equation}
which is a homogeneous system of linear equations. 
In general, the coefficient matrix of Eq.~\eqref{TJeq} is of rank 4 and, thus, Eq.~\eqref{TJeq} only has the trivial solution $\mathbb{J}=\mathbf{0}$. In other words, the general (preliminary) conclusion is that TJs cannot move. 

In special cases (associated with special $\mathbf{b}^{(i)}$ and $h^{(i)}$), however, the rank of the tensor in Eq.~\eqref{TJeq} may be lower and solutions may exist. 
Consider, for example, the symmetric TJ geometry illustrated in Fig.~\ref{TJ}c, where GB$^{(2)}$ and GB$^{(3)}$ are identical and mirrored about the GB$^{(1)}$ plane. 
We examine two special cases: 
\begin{itemize}
\item[(i)] \textit{Burgers vectors tangent to the TJ line}: If $\mathbf{b}^{(i)} = b^{(i)}\boldsymbol{\xi}$ ($i=1,2,3$), there will be infinitely many nontrivial solutions to Eq.~\eqref{TJeq}, which means that in this case the TJ is always mobile ($\boldsymbol{\xi}$ is the TJ line direction). 
If it happens that $\beta^{(2)} = \beta^{(3)} = \frac{1}{2}\beta^{(1)} \sec\Theta$, the only constraint on the choice of $J^{(i)}$ is Eq.~\eqref{TJH} (two free variables). 
If this equality fails, the only nontrivial solution is $J^{(1)} = 0$ and $J^{(2)} = -J^{(3)}$ (one free variable); this means that the TJ can only move in one direction:  $\mathbf{v}_\text{tj} = J^{(2)}h^{(2)} \csc\Theta \mathbf{t}^{(1)}$. 

\item[(ii)] \textit{Burgers vectors perpendicular to the TJ line}: If $\mathbf{b}^{(i)} = b^{(i)}\mathbf{t}^{(i)} \ne \mathbf{0}$ ($i=1,2,3$), the only general solution is the trivial one -- the TJ is immobile. 
The TJ can move in the very special case where $\beta^{(2)} = \beta^{(3)} = -\frac{1}{2}\beta^{(1)} \sec^2\Theta$, for which $J^{(2)} = J^{(3)} = \frac{1}{2} J^{(1)} (b^{(1)}/b^{(2)}) \sec\Theta$ and the TJ velocity is $\mathbf{v}_\text{tj} = J^{(1)}h^{(1)} \mathbf{n}^{(1)}$. 
In this special case the boundary condition Eq.~\eqref{ztBC} is exactly correct for solving the evolution of GB$^{(1)}$ terminated at the TJ. 
\end{itemize}

As discussed above, Eq.~\eqref{TJeq} implies that it is difficult for TJs to move appreciable distances, except in a restrictive set of special cases. 
This can be thought of as the result of incompatible GB displacements and/or the accumulation of Burgers vectors at the TJ, which effectively shuts off the motion of disconnections along the GBs. 
However, we know that in real polycrystals, most TJs can, in fact, move (otherwise, grain growth would never be observed). 
There are several possible mechanisms by which TJs may overcome the restrictions discussed above.
One possible mechanism is the emission of lattice dislocations into the grains during TJ migration, as explicitly allowed for in the more general form of Eq.~\eqref{TJeq} [i.e., Eq.~\eqref{TJeqF}]. 
If one or more of the grains meeting at the TJ  have  viable slip systems, the TJ may have sufficient flexibility to emit the right combination of lattice dislocations to effectively keep the TJs Burgers vector-free.
This would imply that grain growth is necessarily tied to crystal plasticity.

Another possible mechanism for continued TJ migration is associated with the fact that each GB has multiple disconnection modes (see Section~\ref{multiple_modes}). 
Multiple disconnection modes can be activated on the three GBs meeting at the TJ such that the Burgers vectors and step heights in Eq.~\eqref{TJeq} are effectively replaced with  the average Burgers vectors $\bar{\mathbf{b}}^{(i)}$ and the average step heights $\bar{h}^{(i)}$ for all of the disconnections flowing on GB$^{(i)}$ (the bar denotes average over the modes on each GB). 
Given multiple disconnection modes on each GB (which is probably the scenario at high temperature), the system can choose the right combination to ensure that the average Burgers vectors and step heights are such that the rank of the coefficient matrix of Eq.~\eqref{TJeq} is $<3$. 
If this is indeed the case, there will be an infinite set of nontrivial solutions to Eq.~\eqref{TJeq}, suggesting that the TJ can move without emission of lattice dislocations into the grains. 
For example, in special case (i) above, if the disconnection modes activated on the three GBs are adjusted such that the \textit{average} Burgers vectors are parallel to the TJ line direction $\boldsymbol{\xi}$, the TJ will be mobile. 
This situation is applicable to the configuration in Fig.~\ref{TJ}d, which has been widely used in experiments [e.g., Ref.~\cite{czubayko1998influence,gottstein1999effect}] and MD simulations [e.g., Ref.~\cite{upmanyu2002molecular}] to deduce the TJ mobility for the case of capillarity-driven GB migration. 
Similarly, the case in which the TJ moves as illustrated in Fig.~\ref{TJ}e [corresponding to special case (ii)] implies that the \textit{average} of the activated disconnection modes satisfy $\bar{\beta}^{(2)} = \bar{\beta}^{(3)} = -\frac{1}{2}\bar{\beta}^{(1)} \sec^2\Theta$.  
Such cases were also observed in MD simulations~\cite{aramfard2014influences}.

The disconnection model of TJ motion finds support in both existing experimental and MD simulation observations~\cite{miura1988effect,hashimoto1987grain,pond1978role,sisanbaev1992effect}. 
For example, as shown in Fig.~\ref{TJ}b, by tracing the deformation of a tri-crystal via a set of fiducial lines in the creep experiments (i.e., constant stress and temperature) of Miura, Hashimoto and Fujii~\cite{miura1988effect}. They observed both shear along the three GBs meeting at a TJ and simultaneous migration of two of these (GB$^{(2)}$ and GB$^{(3)}$ in Fig.~\ref{TJ}b). 
Based on this observation, they suggested a relationship among the shear (or sliding) distances along the three GBs. 
This relationship accounts for the equilibrium of shear condition [i.e., Eq.~\eqref{TJb}]. 
Other studies~\cite{miura1988effect,hashimoto1987grain,yamakov2001length} report the emission of lattice dislocations from migrating TJs, leading to a localized deformation configuration which they call a ``TJ fold''. 
MD simulations of grain growth in polycrystals also revealed the emission of lattice (or twinning) dislocations from the TJ in the absence of an applied stress~\cite{lazar2015topological,thomas2016twins}. 
As suggested by Eq.~\eqref{TJeq}, such dislocation emission from a TJ (corresponding to the inclusion of $\{J^{\text{L}_l}\}$) helps alleviate the rigid constraints on TJ motion (where only one set of disconnections is activated on the corresponding GBs).

The kinetic process of TJ motion along with the migration of the three GBs meeting at the TJ can be simulated by solving Eq.~\eqref{continuum_eq} for each of the three GBs. 
All of the quantities in Eq.~\eqref{continuum_eq} are labeled by superscript ``$(i)$'' for GB$^{(i)}$, and the coordinate system is attached to each GB such that the $y^{(i)}$-axis is parallel to $-\mathbf{t}^{(i)}$, the $z^{(i)}$-axis is parallel to $\mathbf{n}^{(i)}$ and the TJ is located at $y^{(i)} = 0$ (see Fig.~\ref{TJ}a). 
Such a simulation algorithm is proposed in \ref{APPtj}. 
This algorithm guarantees that (i) the GBs always meet at the TJ and (ii) the stress of the Burgers vector accumulated at the TJ is included in the equations for GB migration.

\subsubsection{Grain growth in polycrystals}
\begin{figure}[!t]
\begin{center}
\scalebox{0.278}{\includegraphics{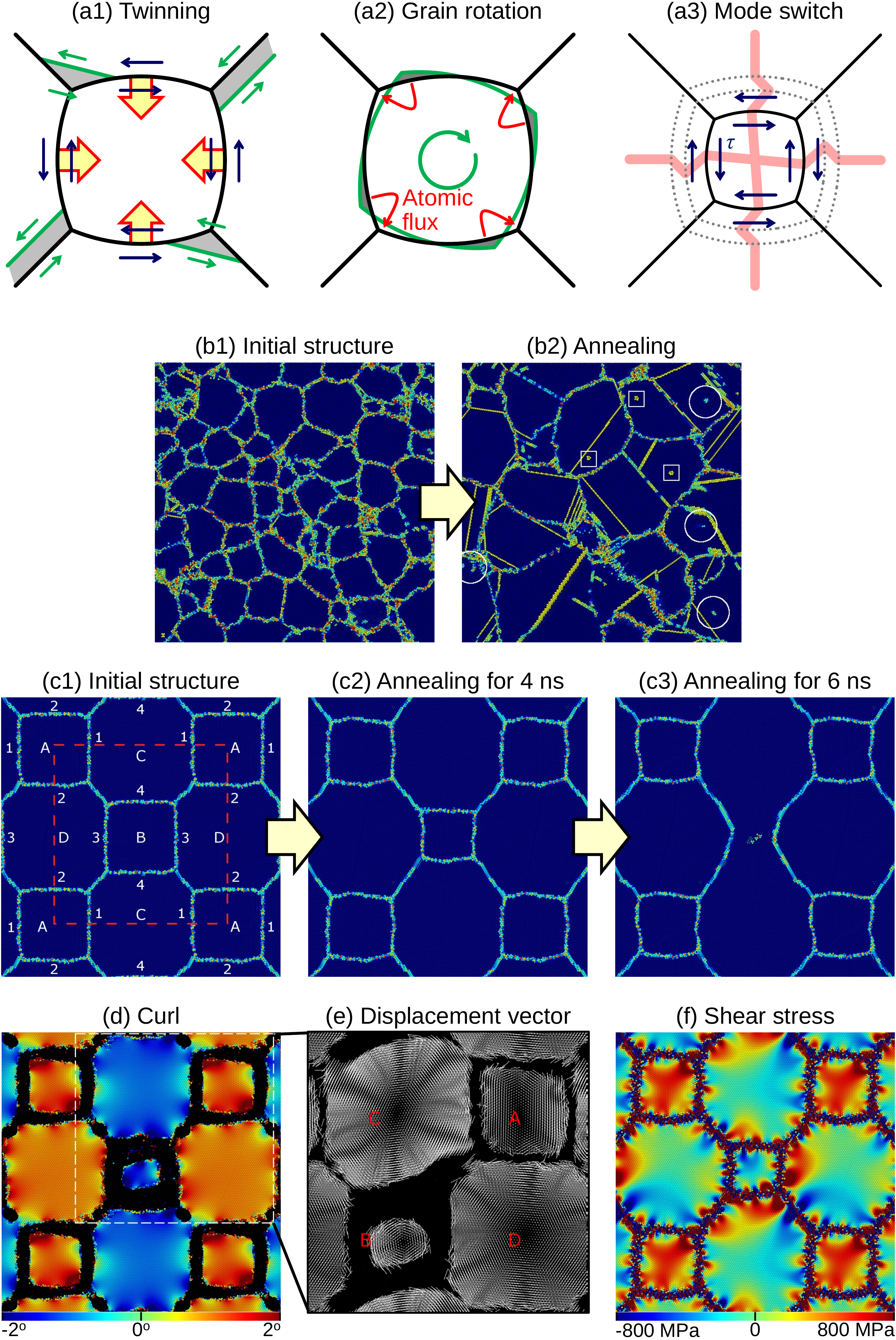}}
\caption{(a1-a3) The possible mechanisms to relieve stress induced during the shrinking of a square grain in a polycrystal. 
(a1) Induced stress is released via plastic deformation, such as twinning. 
(a2) Induced stress is released via grain rotation, which requires atomic diffusion. 
(a3) induced stress is released via repeated switch between two disconnection modes of opposite sign of shear-coupling factor; the red lines are fiducial marks.
(b1-b2) Cross-sections of polycrystalline microstructure from an MD simulation of grain growth in a 3D polycrystal of EAM Ni at $0.85 T_\text{m}$. 
Atoms are colored based on centrosymmetry, which helps visualize the defects.  
(b1) The initial (as-prepared) polycrystal configuration. 
(b2) The configuration after annealing for $2.5$ ns; white circles indicate lattice dislocations, white squares indicates vacancies, and thin yellow lines are TBs. 
(c1-c3) Time evolution of microstructure from the MD simulation of grain growth %/shrinkage 
in an idealized 2D polycrystal.  
The dashed red square in (c1) indicates the periodic simulation cell. 
(d) Atoms are colored based on the curl of the displacement vector, projected in the direction normal to the paper; blue color indicates clockwise rotation, while red indicates counterclockwise rotation. 
(e) Displacement field in the region denoted by the dashed white square in (d). 
(f) Atoms are colored by shear stress. 
(b1-f) are reprinted from Ref.~\cite{thomas2017reconciling} (D. J. Srolovitz, 2017). } 
\label{grainshrink}
\end{center}
\end{figure}

Since GB migration is, in general, coupled with the shear across the GB, shear coupling may influence the process of (capillarity driven) grain growth in polycrystals. 
For the purpose of illustration, we consider a columnar polycrystal which contains a square grain as shown in Fig.~\ref{grainshrink}a. 
According to the von Neumann~\cite{von1952metal} and Mullins~\cite{mullins1956two} theory for isotropic capillarity-driven grain growth, the rate of change of the size of a grain is $\dot{\mathcal{A}}_\text{g} = -2\pi M_\perp\gamma(1-\mathcal{N}_\text{g}/6)$, where $\mathcal{A}_\text{g}$ is the area of the grain and $\mathcal{N}_\text{g}$ is the number of edges of the grain. 
For the square grain in Fig.~\ref{grainshrink}a, $\mathcal{N}_\text{g}=4$ and, thus, $\dot{\mathcal{A}}_\text{g}<0$, implying that this grain will shrink under a capillary force. 
However, we have already demonstrated that GB migration is usually accompanied by shear  across the GB. 
Therefore, with the shrinkage of the square grain,  stresses will inevitably be established within this grain. 
Such a situation is similar to the mixed-force-driven GB migration case we discussed in Section~\ref{mixed_force_driven_gb_migration}. 
The accumulation of stresses within the grain, leads us to expect that the rate at which the square grain shrinks will deviate from that expected based upon the simple von Neumann-Mullins relation. 
If the stresses within the grain are not relaxed by some process, the growing stresses associated with GB migration will eventually  cancel the capillary force and prevent further GB migration/square grain shrinkage. 
Three possible mechanisms for relaxing the stress associated with GB migration are discussed below. 

The first possible mechanism to relax the stress stored in the grain is plastic deformation (dislocation motion or twinning within the grain). 
As illustrated in Fig.~\ref{grainshrink}a1, when the square grain  shrinks, the resultant stress can be realxed by twinning (or, equivalently, emission of twinning dislocations) at the TJs. 
Such a phenomenon was observed in the MD simulations of nanocrystalline Ni of Thomas, et al.~\cite{thomas2016twins,thomas2017reconciling};  \textit{cf.} the initial and final microstructures after $2.5$~ns at $0.85 T_\text{m}$ in Fig.~\ref{grainshrink}b, where $T_\text{m}$ is the melting temperature. 
This microstructure revealed the formation  of multiple twins  near the TJs. 
The grain growth experiments of Jin et al.~\cite{jin2014annealing} showed that the evolution of grain size and the evolution of twin density are proportional with one another; this is consistent with twin generation being a natural consequence of GB motion in a polycrystal.

A second  mechanism for relaxing stress during grain shrinkage is grain rotation; this is illustrated in Fig.~\ref{grainshrink}a2.
Grain rotation was also observed in the MD simulations of Thomas~\cite{thomas2017reconciling}. 
For the microstructure constructed in this MD simulation (Fig.~\ref{grainshrink}c1), the square grains (labelled A and  B) should shrink under  capillary forces. 
Instead, they found that while Grain B did indeed shrink and disappear, Grain A shrank by only a small amount and stagnated at finite grain size after $6$~ns (see Fig.~\ref{grainshrink}c). 
In Fig.~\ref{grainshrink}d the atoms are colored by the curl of the displacement vector (projected into the direction normal to the image); the blue color indicates clockwise rotation while red  indicates counterclockwise rotation, which can be clearly verified by the displacement field in Fig.~\ref{grainshrink}e. 
By examining Fig.~\ref{grainshrink}f, where the atoms are colored by shear stress, we see that in the shrinking grain, Grain B, clockwise rotation occurs and the shear stress remains small. On the other hand, Grain A does not significantly shrink or rotate but rather develops a large shear stress. 
This demonstrates that grain rotation provides a mechanism for relaxing the stress produced by GB migration and hence facilitates grain shrinkage. 
Grain rotation during capillarity-driven grain growth was also observed experimentally~\cite{yamasaki1996grain,harris1998grain,shan2004grain} and in other MD simulations~\cite{farkas2007linear,thomas2017reconciling}. 
We note that, while some grain rotation may be associated with the misorientation dependence of the GB energy, this does not suffice to explain the grain rotation observed in Fig.~\ref{grainshrink}, nor does it describe all of the observed rotation in other MD simulations~\cite{trautt2012grain}. In short, grain rotation is usually controlled by shear coupling rather than GB energy anisotropy.

Grain rotation requires atomic diffusion; Fig.~\ref{grainshrink}a2 demonstrates a case for which atoms must be transported from the overlapped wedge region (gray) to the hollow wedge region (white). 
In analogy to the analysis of Harris, Singh and King~\cite{harris1998grain}, we find that GB diffusional transport-limited grain growth should follow: 
\begin{equation}
r^2 = r_0^2 - 2 \mathcal{M} \gamma t, 
\text{ and }
\mathcal{M}
\equiv \frac{\mathsf{g} \lambda D_\text{gb}\Omega}{(r_0\theta_0)^2 k_\text{B}T}, 
\label{r2_2}
\end{equation}
where $\mathsf{g}$ is a dimensionless geometric factor, $\lambda$ is the effective GB width~\cite{SuttonBalluffi}, $\Omega$ is the atomic volume, $D_\text{gb}$ is the GB self-diffusivity, $r_0$ and $\theta_0$ are the initial grain size and orientation of the original polycrystal, respectively, and $\mathcal{M}$ is the apparent GB mobility. 
The derivation of Eq.~\eqref{r2_2} and its assumptions are given in \ref{APPgrainrotation}. 
The main assumptions are (i) GB migration is coupled with shear across the GB with a shear-coupling factor $\beta \approx \theta$ [referring to Eq.~\eqref{branch12} with $\theta \ll 1$] and (ii) grain rotation is driven by shear-coupled GB migration (rather than the reduction of the average GB energy).
Equation~\eqref{r2_2} has a similar form to the classical parabolic law Eq.~\eqref{r2_1} with the main difference that the apparent GB mobility $\mathcal{M}$ is directly related to the GB self-diffusivity $D_\text{gb}$. 
GB self-diffusion and GB migration involve distinct mechanisms -- the former is associated with the atomic flux along the GB plane while the latter is associated with atomic shuffling at the disconnection core. 
In general, therefore, the (intrinsic) GB mobility and the GB self-diffusivity should be unrelated~\cite{gottstein2009grain}.  
Many MD simulations and experiments on bicrystals indeed demonstrate that the activation energy for GB migration and for GB self-diffusion are different~\cite{schonfelder2005comparative,schonfelder1997molecular,molodov1994effect}. 
However, Eq.~\eqref{r2_2} suggests that in polycrystals the apparent GB mobility may be coupled with the GB self-diffusivity through shear-coupled GB migration.  
On the other hand, some grain growth experiments in nanocrystalline materials showed that the activation energy for GB migration is similar to that for GB self-diffusion~\cite{iordache1999grain,wang1997isokinetic}. 
An MD study of polycrystalline grain growth  by Yamakov et al.~\cite{yamakov2006relation} also reported identical activation energies for grain growth and GB self-diffusion. 
These results suggest that grain rotation may dominate grain growth in nanocrystalline materials, while in bulk polycrystalline materials other stress-relaxation mechanisms dominate.

A third possible mechanism for the relaxation of the stress generated during GB migration is repeatedly switching between distinct disconnection modes that exhibit  $\beta$ values of opposite sign. 
This mechanism was discussed above on the basis of a bicrystal model with fixed surfaces (see Fig.~\ref{coupling_schematic}c3 and Section~\ref{mixed_force_driven_gb_migration}). 
A similar situation is illustrated in Fig.~\ref{grainshrink}a3 for a polycrystal. 
The illustration shows that the square grain may shrink continuously via repeated switch between two disconnection modes. 
The process of grain shrinkage in this two-mode scenario corresponds to a trajectory in the $\psi$-$\tau$ phase space, which is schematically shown as the curved arrow in Fig.~\ref{mixed_map}b. 
Initially, the grain shrinks with a driving force $\psi \sim\gamma/L$ via a single mode, where $L$ is the edge length of the square grain. 
As the grain continues to shrink, shear stress will build up via shear-coupled GB migration leading to a constant ratio of motion by mode one and mode two, at which point the trajectory will reach a point in phase space  where $\bar{\beta} = 0$. 
In the steady-state two-mode case, the driving force has two contributions, capillarity and stress, such that the driving force for grain shrinkage is proportional to $\mathcal{C}\gamma/L$, where $\mathcal{C}$ is a function of the shear-coupling factors of the two modes $(\beta_2/\beta_1)$. 
As shown in the derivation in \ref{APPmodeswitch}, $\mathcal{C}$ is strictly less than one.  
Hence, the rate of GB migration will be lower when two disconnection modes are necessary than when it occurs via a single disconnection mode. 
The single disconnection mode can only occur if that disconnection mode corresponds to a pure step (i.e., disconnections with no dislocation content, $\mathbf{b}=0$). 

Based on the discussion of the shrinkage of a square grain shown in Fig.~\eqref{grainshrink}a, we should not expect grain growth to necessarily follow the classical von Neumann-Mullins relation (or its anisotropic extensions~\cite{abbruzzese1986theory,holm2001misorientation}). 
Since stress is a natural consequence of shear coupling to GB migration, a more realistic grain growth law should include the effect of the stress on the driving force for GB migration as well as the three stress dissipation mechanisms described above. 
This has not been achieved to-date; however, there has been recent progress in building an equation of GB motion that includes many of the features of the disconnection model~\cite{zhang2017equation,thomas2017reconciling}. 
Nonetheless, the fact that the von Neumann-Mullins is well established in coarse-grained polycrystalline materials suggests that in such systems nature is able to easily take advantage of the plethora of mechanisms for relaxing stresses generated by shear coupling.

%%%%%%%%%%%%%%%%%%%%%%%%%%%%%%%%%%%%%%%%%%%%%%%%%%%%%%%
\section{Non-conservative grain-boundary kinetics}
\label{sec6}

The non-conservative kinetic behavior of a GB involves the interaction between the GB and defects in the bounding grains. Here, we focus on the interaction of GBs specifically with point defects and with lattice dislocations in the framework of the disconnection model.
We note that while there has been a thorough review of this topic recently~\cite{beyerlein2015defect}, we specifically  focus on aspects of these topics from the point of view of the disconnection model. 

\subsection{Absorption/emission of point defects}
\label{interaction_pd}

We note that disconnections with nonzero Burgers vector component perpendicular to the GB plane $\mathbf{b}_\perp$ (see Fig.~\ref{disconnection}c) can move by the addition of atoms or vacancies to the discontinuous DSC lattice plane parallel to the GB  (see the gray lattice in Fig.~\ref{disconnection}c2). 
Such disconnections move along the GB by a disconnection climb mechanism. 

\subsubsection{Disconnection description}
\label{disconnection_description}
By employing concepts from bicrystallograpy, we will show that (i) the absorption/emission of point defects into a GB necessarily introduces disconnections with Burgers vectors perpendicular to the GB plane $\mathbf{b}_\perp$, (ii) the Burgers vector $\mathbf{b}_\perp$ will (in most cases) be associated with a step height $h$, implying that the absorption/emission of point defects into the GB will lead to GB migration, and (iii) a GB can continuously absorb/emit point defects (by disconnection climb) along the GB. 

We illustrate the disconnection model based on the example of a bicrystal containing a $\Sigma 5$ $[100]$ STGB in a simple cubic crystal as shown in Fig.~\ref{disconnection}a. 
Figure~\ref{disconnection}c1 indicates that, if we displace the black lattice in the yellow region with respect to the white lattice according to the DSC vector $\mathbf{b}_\perp = b_\perp\mathbf{n}$, the GB plane in the yellow region will shift downwards by $h$ in order to make the GB structure in the yellow region exactly same as that in the right region.  
If we remove the black lattice below the GB plane (the blue line in Fig.~\ref{disconnection}c) and the white lattice above the GB plane, we will introduce the disconnection $(\mathbf{b}_\perp, h)$ into the GB (see Fig.~\ref{disconnection}c2). 
We note that this disconnection is associated with an extra DSC lattice plane lying to the left of the disconnection in the GB or, equivalently, a missing DSC lattice plane to its right. 
From this example, we see that the absorption/emission of point defects into/out of the GB leads to the formation or motion of the extra/missing DSC lattice plane segment parallel to the GB plane. The disconnection located at the extra/missing DSC lattice plane is associated with a Burgers vector perpendicular to the GB plane. 

\begin{figure}[!t]
\begin{center}
\scalebox{0.29}{\includegraphics{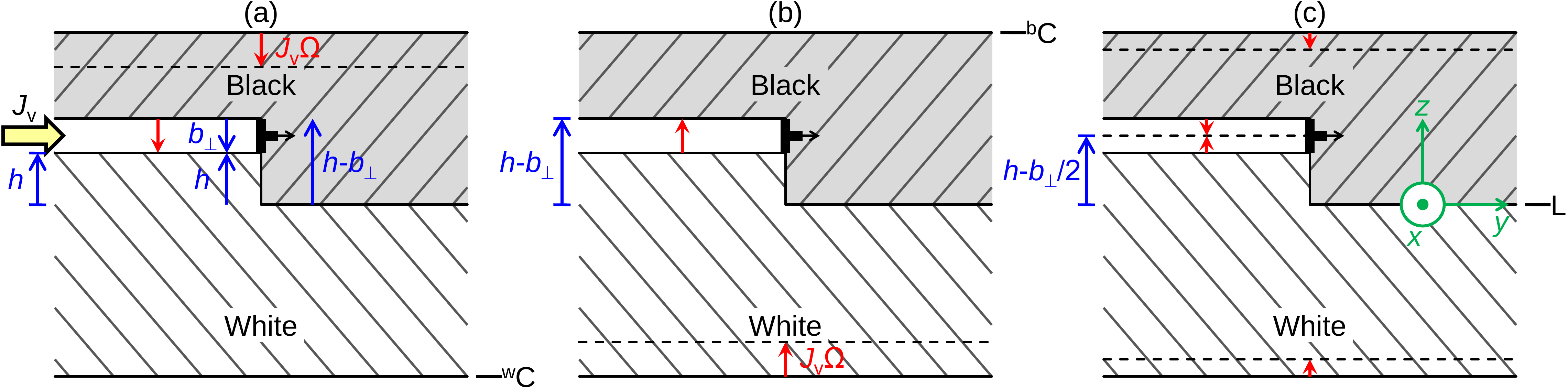}}
\caption{Schematics of a disconnection associated with Burgers vector normal to the GB plane ($b_\perp$) and nonzero step height ($h$) in a bicrystal (the upper and lower grains are colored black and white, respectively). 
The disconnection is formed by closing the gap between the black and the white grains by the red arrow. 
The GB migration by the lateral disconnection motion is measured with respect to (a) the white crystal ($^\text{w}$C), (b) the black crystal ($^\text{b}$C) and (c) the middle plane of the bicrystal (L).  }
\label{disconnection_diffusive_flux}
\end{center}
\end{figure}

The Burgers vector $\mathbf{b}_\perp$ is associated with a step height $h$, suggesting that the propagation of the disconnection with Burgers vector $\mathbf{b}_\perp$ along the GB is accompanied by GB migration. 
In the case of conservative GB migration (see above), one grain grows at the expense of the other grain; hence, the GB migration distance is the same whether it is measured with reference to the laboratory frame (L-frame), the white crystal ($^\text{w}$C-frame) or the black crystal ($^\text{b}$C-frame); the GB migration distance is simply $h$. 
However, in the case of non-conservative GB migration, the GB migration distance depends on the frame with respect to which it is measured. 
In the procedure for determining $h$ for a particular $\mathbf{b}_\perp$ (see above), we displaced the black lattice with respect to the white lattice (i.e., the scenario shown in Fig.~\ref{disconnection}c), so $h$ corresponds to the GB migration distance measured in the $^\text{w}$C-frame, denoted $h^\text{w}\equiv h$ (see the geometry shown in Fig.~\ref{disconnection_diffusive_flux}a). 
Similarly, the GB migration distance measured with respect to the $^\text{b}$C-frame is $h^\text{b} \equiv h-b_\perp$ (see Fig.~\ref{disconnection_diffusive_flux}b). 
However, the net effect of GB migration cannot be well described in either the $^\text{w}$C- or $^\text{b}$C-frame in the non-conservative case. 
For example, if the absorbed vacancies/interstitials are added (in equal concentration) to the black and the white lattice planes immediately adjacent to the GB, the GB plane will  always remain at the center of the  bicrystal -- a case that implies no (net) GB migration; in this case, however, the GB migration distance measured in either the $^\text{w}$C- or $^\text{b}$C-frame is nonzero since either the white or black grain grows. 
The net GB migration distance (unbiased by the influence of vacancy/interstitial addition) can be obtained by always measuring the GB position with reference to the center plane of the bicrystal (i.e., L-frame). 
In the L-frame, the GB migration distance is $h^\text{L} \equiv h - b_\perp/2 = (h^\text{b}+h^\text{w})/2$ (see Fig.~\ref{disconnection_diffusive_flux}c). 

\begin{figure}[!t]
\begin{center}
\scalebox{0.215}{\includegraphics{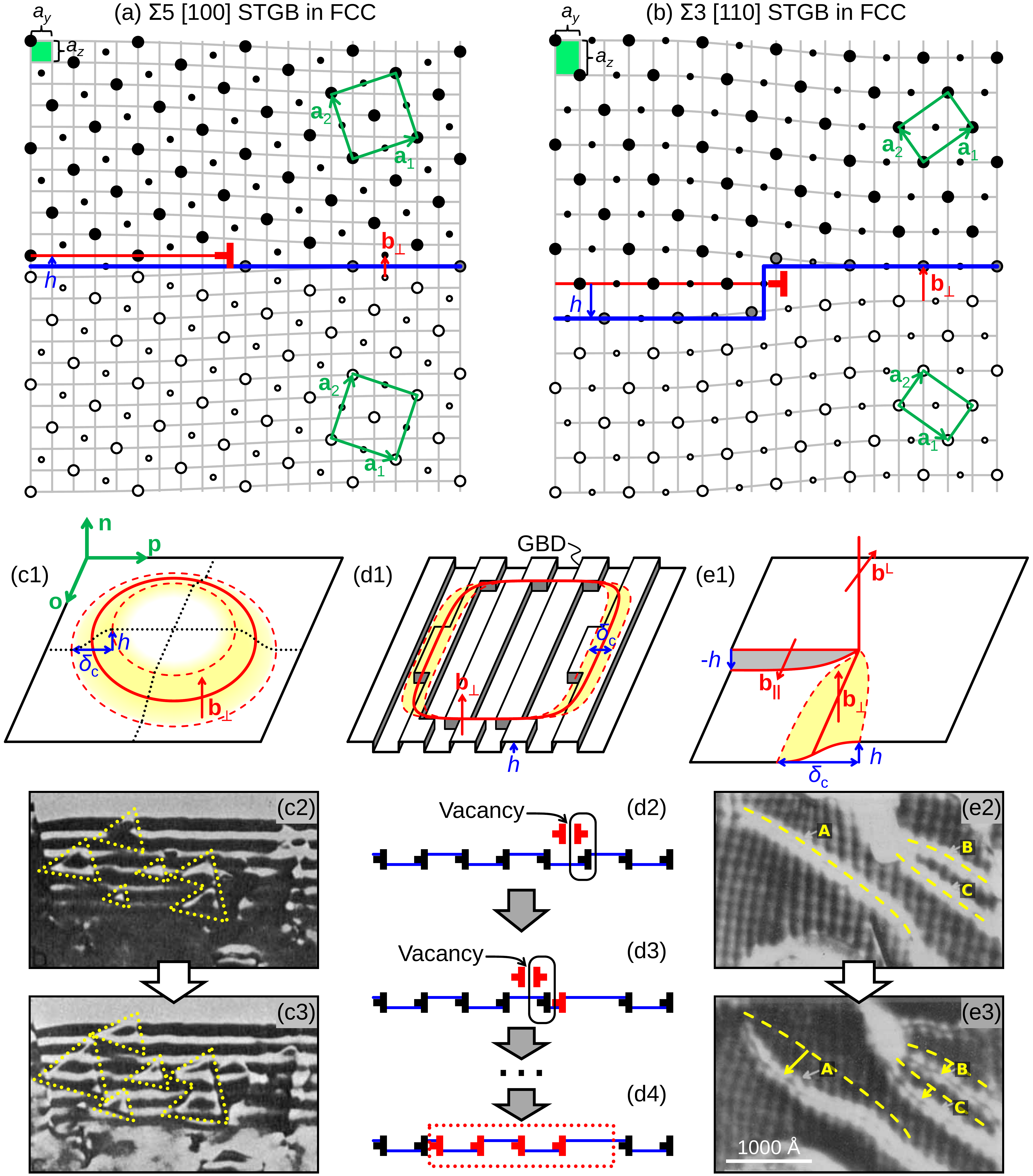}}
\caption{(a) A disconnection characterized by $\mathbf{b} = (0,0,1)a_z$ and $h=(0,0,1/2)a_z$ in a $\Sigma 5$ $[100]$ $(013)$ STGB in an FCC crystal. 
(b) A disconnection characterized by $\mathbf{b} = (0,0,1)a_z$ and $h=(0,0,-1/2)a_z$ in a $\Sigma 3$ $[110]$ $(\bar{1}11)$ STGB in an FCC crystal. 
In (a) and (b), the gray lines indicate the DSC lattice, the blue lines denote the GB position, and the red lines indicate the extra DSC lattice plane in the GB. 
(c1) Schematic of homogeneous nucleation of a disconnection loop in a high-angle GB, where SGBDs cannot be well defined. 
The solid red line indicates the disconnection line and the dashed red lines denote the disconnection core size. 
(c2-c3) Growth of disconnection loops in a coherent TB in Al under electron-beam irradiation, observed in the experiment performed by King and Smith~\cite{king1980mechanisms}. 
The triangular islands, labeled by the yellow dotted lines, correspond to the disconnection loops. 
(d1) Schematic of homogeneous nucleation of a disconnection loop in a GB where primary/secondary GBDs can be well defined. 
In general, the GBDs are associated with nonzero step height, so the steps on the GB denote the position of GBDs. 
(d2-d4) Schematic showing that the misalignment of the GBDs is equivalent to the introduction of disconnection pairs by absorption of vacancies. 
(e1) Schematic of dissociation of a lattice dislocation ($\mathbf{b}^\text{L}$) into a disconnection with Burgers vector parallel to the GB ($\mathbf{b}_\parallel$) and a disconnection with Burgers vector perpendicular to the GB ($\mathbf{b}_\perp$). 
It is assumed that the disconnection $\mathbf{b}_\perp$ is dissociated with the disconnection core $\delta_\text{c}$. 
(e2-e3) Irradiation-induced climb of extrinsic GBDs, labeled by ``A'', ``B'' and ``C'', observed by Komem et al. for a $[001]$ TwGB in Au~\cite{baluffi1972electron,komem1972direct}. }
%(c2-c3) are reproduced from Ref.~\cite{king1980mechanisms}; (e2-e3) are reproduced from Ref.~\cite{baluffi1972electron,komem1972direct}.  } 
\label{pdabsorption}
\end{center}
\end{figure}

As a first example, consider a $\Sigma 5$ $[100]$ $(013)$ STGB in an FCC crystal, as shown in Fig.~\ref{pdabsorption}a. 
For this bicrystallography, the disconnection has a Burgers vector perpendicular to the GB plane and the disconnection can be described by Burgers vector $\mathbf{b}_\perp = (0,0,a_z)$ and step height $h=a_z/2$ ($a_z$ is the size of a DSC unit cell in the $z$-direction). 
We see that the motion of this disconnection makes the GB migrate by a distance $h^\text{w} = a_z/2$ or $h^\text{b} = -a_z/2$ with respect to the $^\text{w}$C- or $^\text{b}$C-frame, respectively, or by $h^\text{L} = h-b_\perp/2 = 0$ in the L-frame. 
Therefore, the introduction of this disconnection will locally change the misorientation but  induce no net GB migration (as seen in the geometry of Fig.~\ref{pdabsorption}a). 
The smallest possible step height (in L-frame) is zero for any disconnection with Burgers vector $\mathbf{b}\parallel \mathbf{n}$ in any $[100]$ STGB in an FCC crystal. 
Of course, this is not necessarily true for disconnections with $\mathbf{b} \nparallel \mathbf{n}$; even for those with $\mathbf{b}\parallel \mathbf{n}$ the step height could be finite, i.e., integer multiple of the pure step height.  

As a second example, consider the special case of a coherent twin boundary (TB) in an FCC crystal (i.e., a $\Sigma 3$ $[110]$ $(\bar{1}11)$ STGB), as shown in Fig.~\ref{pdabsorption}b. 
The disconnection shown in Fig.~\ref{pdabsorption}b is characterized by Burgers vector $\mathbf{b}_\perp = (0,0,a_z)$ and step height $h=-a_z$. 
Therefore, the GB migration distance is $h^\text{w}=-a_z$ or $h^\text{b}=-2a_z$ in the $^\text{w}$C- or $^\text{b}$C-frame, respectively, such that the net GB migration distance is nonzero, $h^\text{L} = -3a_z/2$. 
In both experiments and atomistic simulations the disconnection structure (Fig.~\ref{pdabsorption}b) was indeed observed in $\Sigma 3$ TBs of Au~\cite{merkle1994atomic,marquis2005structural}. 
The formation of this disconnection can be easily understood from the stacking sequence of the closed packing planes. 
The stacking sequence for a coherent TB in an  FCC crystal is often represented as~\cite{HirthLothe} 
\[
\begin{array}{*{17}{@{}c@{}}}
\cdots &\phantom{\text{(A)}}& \text{C} &\phantom{\text{(A)}}& \text{A} &\phantom{\text{(A)}}& \text{B} &\phantom{\text{(A)}}& \underline{\text{C}} &\phantom{\text{(A)}}& \text{B} &\phantom{\text{(A)}}& \text{A} &\phantom{\text{(A)}}& \text{C} &\phantom{\text{(A)}}& \cdots \\
\end{array}
\]
where the underlined letter denotes the TB plane (about which the stacking sequence is symmetric). 
If we insert an extra plane to the left or right side of the TB plane, the stacking sequence  becomes
\[
\begin{array}{*{17}{@{}c@{}}}
\cdots &\phantom{\text{(A)}}& \text{C} &\phantom{\text{(A)}}& \text{A} &\phantom{\text{(A)}}& \underline{\text{B}} & \text{(A)} & \text{C} &\phantom{\text{(A)}}& \text{B} &\phantom{\text{(A)}}& \text{A} &\phantom{\text{(A)}}& \text{C} &\phantom{\text{(A)}}& \cdots \\
\end{array}
\]
or 
\[
\begin{array}{*{17}{@{}c@{}}}
\cdots &\phantom{\text{(A)}}& \text{C} &\phantom{\text{(A)}}& \text{A} &\phantom{\text{(A)}}& \text{B} &\phantom{\text{(A)}}& \text{C} & \text{(A)} & \underline{\text{B}} &\phantom{\text{(A)}}& \text{A} &\phantom{\text{(A)}}& \text{C} &\phantom{\text{(A)}}& \cdots \\
\end{array}
\]
where the letter enclosed in parentheses denotes the inserted plane. 
We see that the introduction of an extra plane must shift the TB plane position; this is (non-conservative) TB migration. 
On the other hand, an edge dislocation with Burgers vector $\mathbf{b}_\perp$ must exist at the termination of this extra plane (akin to Frank loop~\cite{HirthLothe}). 
Hence, the termination of this extra plane is a line defect featuring both a Burgers vector $\mathbf{b}_\perp$ and step height $h$ (change of TB plane); i.e., the disconnection illustrated in Fig.~\ref{pdabsorption}b. 
Note that, in the absence of any applied driving force, there is no difference whether the extra plane is inserted on either side of the original TB plane (see the stacking sequences shown above), such that the TB can equally migrate in either direction. 

The most important difference between the first (Fig.~\ref{pdabsorption}a) and the second examples (Fig.~\ref{pdabsorption}b) is that, in the former case, the insertion/removal of an extra DSC lattice plane (parallel to the GB plane)  induces no GB migration (in the L-frame) while, in the latter case, it  necessarily induces GB migration. 
Such a disconnection [i.e., $(\mathbf{b}_\perp, h)$ shown in Fig.~\ref{pdabsorption}a or b] can be an intrinsic secondary GB dislocation (SGBD)~\cite{SuttonBalluffi,Sutton83a,han2017grain} in a vicinal STGB~\cite{SuttonBalluffi} of near-$\Sigma 5$ or near-$\Sigma 3$ misorientation. 
When $h = 0$ (such as in Fig.~\ref{pdabsorption}a), the vicinal STGB will be atomically flat upon the  introduction of SGBDs. 
However, when $h \ne 0$ (such as in Fig.~\ref{pdabsorption}b), introduction of an arrays of SGBDs will make the vicinal GB non-flat (on the atomic scale). 
Note that, in the second example (Fig.~\ref{pdabsorption}b), for the same $\mathbf{b}_\perp$ the step height can be equally $h$ or $-h$. 
Hence, the vicinal STGB may look like the stepped structure as illustrated in Fig.~\ref{pdabsorption}c2 (note the black dislocations and the blue line); the existence of such stepped structure was indeed demonstrated by atomistic simulation of the near-$\Sigma 3$ STGBs~\cite{o2016misoriented}. 

Since a disconnection with Burgers vector perpendicular to the GB plane involves addition/removal of a DSC lattice plane lying in the GB, continuous absorption/emission of vacancies into/out of a GB involves climb of such disconnections along the GB~\cite{ashby1969interface}. 
Balluffi and Granato pointed out that this process can be viewed as internal crystal growth/dissolution, which is similar to the growth/evaporation of crystals at free surfaces by the movement of steps; the difference is that the step in the internal surface (i.e., GB) is, in general, associated with a Burgers vector~\cite{balluffi1980dislocations}. 

\begin{figure}[!t]
\begin{center}
\scalebox{0.3}{\includegraphics{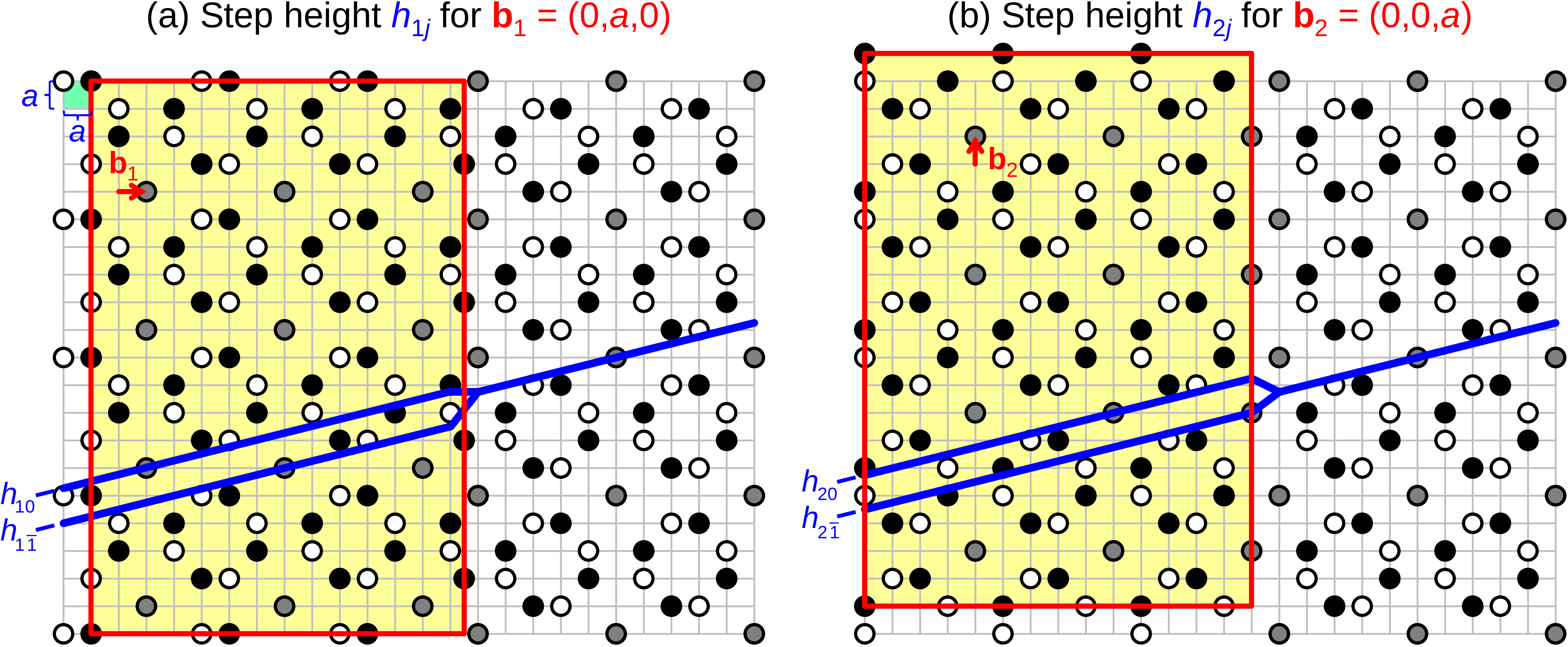}}
\caption{Dichromatic pattern with a $\Sigma 5$ misorientation in an SC crystal; the blue line indicates the GB position such that the GB is an ATGB.  
(a) The disconnection is formed by displacing the black lattice in the yellow region with respect to the white lattice by the Burgers vector $\mathbf{b}_1$. 
Associated with $\mathbf{b}_1$, two possible step heights, $h_{10}$ and $h_{1\bar{1}}$, are shown. 
(b) The disconnection is formed by displacing the black lattice in the yellow region with respect to the white lattice by the Burgers vector $\mathbf{b}_2$. 
Associated with $\mathbf{b}_2$, two possible step heights, $h_{20}$ and $h_{2\bar{1}}$, are shown.    } 
\label{nonconservative_ATGB}
\end{center}
\end{figure}

We note that the examples of the disconnection structure illustrated in Fig.~\ref{disconnection}c, Fig.~\ref{pdabsorption}a and Fig.~\ref{pdabsorption}b are all for low-$\Sigma$ STGBs. 
Below we will discuss the cases for high-$\Sigma$ STGBs and non-symmetric-tilt GBs. 
\begin{itemize}
\item[(i)] \textit{High-$\Sigma$ STGBs.} 
We recall that the higher $\Sigma$ value a GB has, the smaller the atomic density (or the larger the atomic spacing) is in a DSC lattice plane parallel to the GB. 
The atomic density determines the disconnection core size. 
For example, consider a high-$\Sigma$ $[100]$ STGB in a SC crystal; the atomic spacing in the $\mathbf{p}$- and $\mathbf{o}$-directions are $\sqrt{\Sigma}a_0$ and $a_0$, respectively (referring to the geometry shown in Figs.~\ref{disconnection}a and c, albeit this GB is of low $\Sigma$). 
The absorption/emission of atoms introduces a pair of disconnections (with Burgers vectors $\mathbf{b}_\perp$). 
If the disconnection line is parallel to the $\mathbf{o}$-axis, the disconnection core will be highly dissociated in the $\mathbf{p}$-direction (scaled by $\sqrt{\Sigma}a_0$); if the disconnection line is parallel to the $\mathbf{p}$-axis, the disconnection core will be localized in the $\mathbf{o}$-direction (scaled by $a_0$). 
This implies that the introduction of an array of atoms along the $\mathbf{o}$-direction into a high-$\Sigma$ GB will generate a pair of disconnections (with line direction parallel to the $\mathbf{o}$-axis) separated by a large distance $\sqrt{\Sigma}a_0$. 
On the other hand, a pair of disconnections with line direction parallel to the $\mathbf{o}$-axis separated by a distance smaller than $\sqrt{\Sigma}a_0$ will annihilate with each other without the addition of any point defects. 

\item[(ii)] \textit{Non-symmetric-tilt GBs.} 
We have already determined that for some STGBs the addition/removal of a DSC lattice plane will generate a disconnection with zero step height (in the L-frame) and imply zero net GB migration. Such is the case in Fig.~\ref{pdabsorption}a. 
For other STGBs, although the step height is not zero, the step height could be either $h$ or $-h$ with equal probability (if no driving force is applied). Such is the case in Fig.~\ref{pdabsorption}b. 
We also find that the same situation applies to pure TwGBs. 
However, the situation for ATGBs is different. 
We will only focus on non-faceted ATGBs, since faceted ones can be viewed as composed by STGB segments (see Fig.~\ref{beta_mixed_geometry}e).
Figure~\ref{nonconservative_ATGB} shows the example for a $\Sigma 5$ $[100]$ $(029)/(0\bar{7}6)$ ATGB in a SC crystal. 
Here, the Burgers vector with a small magnitude cannot be purely perpendicular or parallel to the GB plane -- e.g. $\mathbf{b}_1$ and $\mathbf{b}_2$ are the two with the smallest magnitude. 
For $\mathbf{b}_1$ (see Fig.~\ref{nonconservative_ATGB}a), the positive and negative step heights are $h_{10} = a/\sqrt{17}$ and $h_{1\bar{1}} = -4a/\sqrt{17}$, respectively, and the corresponding net GB migration distances are $3a/(2\sqrt{17})$ and $-7a/(2\sqrt{17})$, respectively. 
The upward and downward GB migrations are not equivalent for this particular Burgers vector $\mathbf{b}_1$. 
Similarly, for $\mathbf{b}_2$ (see Fig.~\ref{nonconservative_ATGB}b), $h_{20} = 3a/\sqrt{17}$ and $h_{2\bar{1}} = -2a/\sqrt{17}$, and the corresponding net GB migration distances are $a/\sqrt{17}$ and $-4a/\sqrt{17}$, respectively. 
Again, the GB migration in two directions is unbalanced. 
Additionally, in ATGBs the disconnection core is highly dissociated in the $\mathbf{p}$-direction even when $\Sigma$ is large because the atomic density in a DSC plane parallel to the inclined GB plane can be small. 
\end{itemize}

In the following two sections, we apply this disconnection description to several non-conservative GB phenomena. 

\subsubsection{A grain boundary as a vacancy/self-interstitial sink}
\label{GB_as_sink}
Grain boundaries can act as sinks for point defects, such as vacancies and self-interstitial atoms (SIAs). 
This is an important phenomenon  under (electron, neutron or ion) irradiation conditions where a supersaturation of vacancies/SIAs readily accumulate in the material. 
However, many experimental and atomistic simulation observations indicate that the concentration of point defects or point defect clusters (e.g., Frank vacancy/SIA loops, voids and stacking-fault tetrahedra) is lower near GBs (e.g. Refs.~\cite{sakaguchi2001heterogeneous,el2017direct,han2012effect,khan2012irradiation,siegel1980vacancy,zhang2012atomistic}); the region near each GB where the defect concentration remains low during irradiation is often called the defect denuded zone (DZ). 
The existence of a DZ is a manifestation of  GB sink behavior for vacancies/SIAs. 
Such a sink effect plays an important role in the reduction of radiation damage in polycrystalline materials, as verified by the observation that defect densities tend to decrease with decreasing grain size  (e.g. Ref.~\cite{rose1997instability,sun2013situ,el2017grain}). 

Based on the disconnection description discussed in the last section, we propose that the microscopic mechanism behind the GB sink effect is associated with the formation and climb of disconnections (with nonzero Burgers vector components in the direction normal to the GB plane). 
First, we consider a GB that is free of extrinsic GB dislocations (GBDs, i.e., those which are not part of the Burgers vector content required by the GB misorientation~\cite{SuttonBalluffi}). 
In this case, the absorption of point defects will induce homogeneous nucleation and growth of a disconnection loop; this loop has a Burgers vector perpendicular to the GB plane (akin to a Frank loop in a crystal). 
This scenario is illustrated in Fig.~\ref{pdabsorption}c1. 
In general, there is step character associated with the disconnection loop although the step height may be zero in special cases (e.g., see Fig.~\ref{pdabsorption}a). 
The irradiation-induced homogeneous nucleation and growth of disconnection loops were observed in experiment by King and Smith for a coherent TB in Al~\cite{king1980mechanisms} (see Figs.~\ref{pdabsorption}c2-c3). 
In Fig.~\ref{pdabsorption}c2, the triangular islands correspond to disconnection loops. 
According to the geometry shown in Fig.~\ref{pdabsorption}b, these disconnection loops have Burgers vector $\mathbf{b}_\perp = [\bar{1}11]a_0/3$ and step height $h = -a_0/\sqrt{3}$ (in the $^\text{b}$C-frame). 
Under electron-beam irradiation, the triangular islands (i.e., the disconnection loops) grow, as seen in Fig.~\ref{pdabsorption}c3. 

There are well-established models for predicting GB sink efficiency that are based upon point defects being absorbed by the climb of intrinsic GBDs, such as primary GB dislocations (PGBDs) in low-angle GBs and SGBDs in high-angle GBs~\cite{gleiter1979grain,siegel1980vacancy,jiang2014effect,gu2017point}. 
Evidence for irradiation-induced SGBD climb was provided by King and Smith for a near-$\Sigma 15$ high-angle GB in Al, although the climb distance observed during the course of the experiment was very small~\cite{king1980mechanisms}. 
The GBD climb mechanism is different from the nucleation and growth of disconnection loops (as mentioned above and illustrated in Fig.~\ref{pdabsorption}c1). 
However, below we will see that the GBD climb mechanism is actually consistent with the disconnection loop nucleation and growth mechanism.
Since GBDs are identical to the disconnections with Burgers vector $\mathbf{b}_\perp$ in the reference GB structure for which the SGBDs are defined (e.g. see Fig.~\ref{pdabsorption}a), the climb of a GBD can be viewed as being the result of the reaction between the initial GB structure (composed by evenly spaced GBDs) and a pair of disconnections. 
This process is illustrated in Figs.~\ref{pdabsorption}d2-d4, which indicates that the GBD climb along the GB is equivalent to nucleation and growth of a disconnection pair. 
The scenario of a 2D GB is illustrated in Fig.~\ref{pdabsorption}d1, where GBD climbing occurs in an enclosed region of the GB, the periphery of which corresponds to a disconnection loop characterized by $(\mathbf{b}_\perp, h)$. 

If extrinsic GBDs exist within a GB, point defects can be absorbed by their climb. 
The extrinsic GBDs are usually the products of the reaction between lattice dislocations and GBs; the details of which will be discussed in Section~\ref{GBLattDisl}. 
As seen in Fig.~\ref{pdabsorption}e1, when a lattice dislocation with Burgers vector $\mathbf{b}^\text{L}$ is absorbed by a GB, it may dissociate into two extrinsic GBDs: $(\mathbf{b}_\parallel, -h)$ and $(\mathbf{b}_\perp, h)$; note  the  Burgers vector and step height conservation: $\mathbf{b}^\text{L} = \mathbf{b}_\parallel + \mathbf{b}_\perp$ and  $0=h+(-h)$. 
The former GBD can  glide conservatively along the GB, whereas the latter GBD can only climb by absorption/emission of point defects (i.e., a non-conservative process). 
In short, point defects can be absorbed into the GB via the climb of the latter GBD. 
The irradiation-induced climb of extrinsic GBDs was observed by Komem et al.~\cite{baluffi1972electron,komem1972direct} for a $[001]$ TwGB in Au (see Figs.~\ref{pdabsorption}e2-e3). 
We can clearly see the motion of the extrinsic edge GBDs (labeled  ``A'', ``B'' and ``C'') under ion irradiation. 

\subsubsection{Diffusive interface migration}

The second class of non-conservative GB phenomenon is GB migration that requires diffusive fluxes of solute atoms. 
Following the analysis of Hirth, Pond and Sarrazit~\cite{hirth1996steps,pond1997diffusive}, we considered the migration of a GB in a bicrystal as illustrated in Fig.~\ref{disconnection_diffusive_flux} but with different alloy compositions in the black and white grains. 
Due to this composition difference, GB migration may be accompanied by diffusion of solute atoms into/out of the GB. 
We focus on the case of a substitutional binary alloy AB and denote the solute atom B content (atom fraction) in the black (white) grain as $X_\text{B}^\text{b}$ ($X_\text{B}^\text{w}$). 
If we assume that the GB migrates via the climb of disconnections with Burgers vector perpendicular to the GB plane $\mathbf{b}_\perp = b_\perp\mathbf{n}$,  disconnection climb may require the flux of solute atoms and vacancies into/out of the GB. 
Based on the geometry shown in Fig.~\ref{disconnection_diffusive_flux}a, when the disconnection climbs in the positive $y$-direction (i.e., from left to right) by distance $\Delta y$, 
the black crystal is replaced by the white crystal within the volume element $h \Delta y L_x$ ($L_x$ is the length of the disconnection line) and the black crystal is removed in the volume $-b_\perp \Delta y L_x$. 
Therefore, the total number of solute atoms added to the GB during this disconnection climb is 
\begin{equation}
\Delta N_\text{B}
= \left[(X_\text{B}^\text{w}-X_\text{B}^\text{b})h + X_\text{B}^\text{b}b_\perp\right] \Delta y L_x/\Omega, 
\end{equation}
where $\Omega$ is an atomic volume. 
Recalling that $h^\text{w}\equiv h$ is the step height defined by displacing the black lattice with respect to the white lattice and, thus, represents the step height measured in the $^\text{w}$C-frame. 
The step heights measured in the $^\text{w}$C- and L-frames are $h^\text{b}\equiv h-b_\perp$ and $h^\text{L}\equiv h-b_\perp/2$, respectively. 
The current of solute atoms per unit length of the disconnection is 
\begin{equation}
I_\text{B} 
= \dot{N}_\text{B}/L_x
= \left(X_\text{B}^\text{b/w}b_\perp - \Delta X_\text{B}h^\text{w/b}\right) v_\text{cl} /\Omega
= \left(\bar{X}_\text{B} b_\perp - \Delta X_\text{B} h^\text{L}\right) v_\text{cl} /\Omega, 
\label{IB}
\end{equation}
where $\Delta X_\text{B} \equiv X_\text{B}^\text{b}-X_\text{B}^\text{w}$ and $\bar{X}_\text{B}\equiv (X_\text{B}^\text{b}+X_\text{B}^\text{w})/2$. 
For this substitutional alloy, the currents of A  and B atoms and vacancies should satisfy network constraints~\cite{larche1973linear}, from which we  obtain the vacancy current into the disconnection line: 
\begin{equation}
I_\text{v} = - (I_\text{A}+I_\text{B})
= - b_\perp v_\text{cl} /\Omega. 
\label{Iv}
\end{equation}
Note that Eqs.~\eqref{IB} and \eqref{Iv} are valid for both positive and negative values of   $b_\perp$ and $h^\text{L}$. 

We can construct appropriate boundary conditions for describing the interface sink effect based on Eqs.~\eqref{IB} and \eqref{Iv}. 
We assume that the origin of the coordinate system is at the center plane of the bicrystal (i.e.,  all quantities refer to L-frame). 
Mass conservation at the interface ($z = z_\text{i}$) gives 
\begin{equation}
J_\text{B}^\text{w}(z_\text{i}) + J_\text{B}^\text{b}(z_\text{i}) + J_\text{B}^\text{i} = \varrho I_\text{B} 
\text{ and }
J_\text{v}^\text{w}(z_\text{i}) + J_\text{v}^\text{b}(z_\text{i}) + J_\text{v}^\text{i} = \varrho I_\text{v}, 
\label{JBC}
\end{equation}
where $J_\text{B}^\text{w/b}$ is the flux of B atoms from the white/black grain into the interface, $J_\text{v}^\text{w/b}$ is the vacancy flux from the white/black grain into the interface, $J_\text{B}^\text{i}$ ($J_\text{v}^\text{i}$) is the flux of B atoms (vacancies) from free surfaces (or  adjoining interfaces), and $\varrho$ is the density of the disconnections in the interface which absorbs/emits point defects. 
These equations  are coupled via $v_\text{cl}$ in $I_\text{B}$ and $I_\text{v}$. 
We assume that disconnection motion is overdamped such that their climb velocity may be written as 
\begin{align}
v_\text{cl}
=~& M_\text{d}F
= M_\text{d}
\big\{ 
f_\text{v}^\text{w} [X_\text{v}(z_\text{i}^-)-X_\text{v}^\text{w,eq}]
+ f_\text{B}^\text{w} [X_\text{B}(z_\text{i}^-)-X_\text{B}^\text{w,eq}]
\nonumber \\
&+ f_\text{v}^\text{b} [X_\text{v}(z_\text{i}^+)-X_\text{v}^\text{b,eq}]
+ f_\text{B}^\text{b} [X_\text{B}(z_\text{i}^+)-X_\text{B}^\text{b,eq}]
+ \sigma_{nn} b_\perp
\big\}, 
\label{vcl_general}
\end{align}
where $M_\text{d}$ is the disconnection mobility, $F$ is the driving force, $\sigma_{nn} = \mathbf{n}\cdot\boldsymbol{\sigma}\mathbf{n}$ is the component of the stress normal to the interface, $X_\text{v/B}^\text{w/b,eq}$ is the equilibrium concentration (site fraction) of vacancies/B atoms at the white-/black-grain side of the interface (rather than in grain) and  $f_\text{v/B}^\text{w/b}$ is the corresponding coefficient. 
Equation~\eqref{vcl_general} can be derived based on a ternary regular solution model with the assumption that $|X_\text{v/B} - X_\text{v/B}^\text{w,eq}|\ll 1$ at the interface (see the derivation in \ref{APPdriving}). 
From this derivation, we note that the coefficient $f_\text{v/B}^\text{w/b}$ is, in general, a function of the regular solution interaction parameters, the disconnection mode $(b_\perp, h^\text{L})$ and temperature. 
We  discuss the detailed form of the boundary condition Eq.~\eqref{JBC} for several problems of practical interest below.  

(i) \textit{GB sink effect under irradiation or creep}.
For a GB (i.e., a homophase interface), $\Delta X_\text{B} = 0$ and the bicrystal configuration is symmetric at continuum level about the interface. 
According to Eqs.~\eqref{JBC} and \eqref{vcl_general} (see \ref{APPdriving} for detailed derivation), the boundary condition (BC) for vacancy diffusion at the GB can be written as 
\begin{equation}
\frac{D_\text{v}}{\Omega} \frac{\mathrm{d}X_\text{v}}{\mathrm{d}z}\bigg|_{z_\text{i}^\pm}
= \pm M_\text{d}\varrho \left(\frac{b_\perp}{\Omega}\right)^2
\left\{\mathcal{E}_\text{v}[X_\text{v}(z_\text{i}) - X_\text{v}^\text{eq}] + \mathcal{E}_\text{B}[X_\text{B}(z_\text{i}) - X_\text{B}^\text{eq}] - \sigma_{nn}\Omega \right\},  
\label{BCv0}
\end{equation}
where the coefficient $\mathcal{E}_\text{v/B}$ (dimensions of energy) is a function of the interaction parameters (in regular solution model) and temperature. 
If the interface moves slowly such that the equilibrium segregation is almost maintained [i.e., $X_\text{B}(z_\text{i}) = X_\text{B}^\text{eq}$], Eq.~\eqref{BCv0} reduces to the Robin BCs~\cite{erban2007reactive,gu2017point}: 
\begin{equation}
\mathfrak{L}(\mathrm{d}X_\text{v}/\mathrm{d}z)_{z_\text{i}} 
= X_\text{v}(z_\text{i}) - (X_\text{v}^\text{eq} + \tilde{\sigma}), 
\label{RobinBC1}
\end{equation}
where $\mathfrak{L}\equiv D_\text{v}\Omega/(M_\text{d}\varrho b_\perp^2\mathcal{E}_\text{v})$ (dimensions of length) and $\tilde{\sigma}\equiv \sigma_{nn}\Omega/\mathcal{E}_\text{v}$ (dimensionless). 
The parameter $\mathfrak{L}$ includes all the information about GB structure and chemical effects. 

\begin{figure}[!t]
\begin{center}
\scalebox{0.39}{\includegraphics{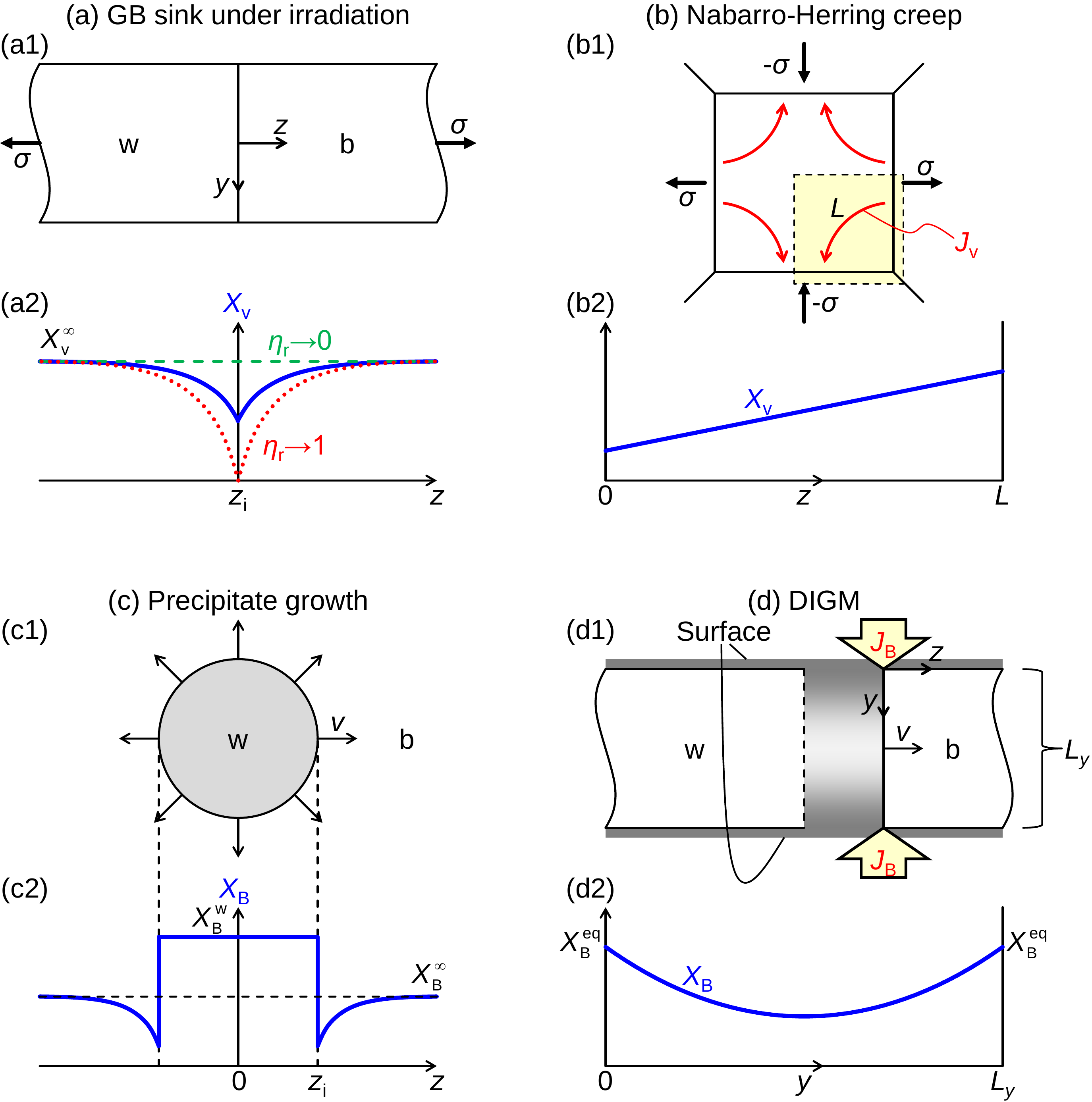}}
\caption{(a1) Model for studying the GB sink effect under continuous irradiation. 
(a2) Steady-state vacancy concentration. The green dashed line and the dotted red line correspond to the case of zero sink efficiency and perfect sink, respectively.
(b1) Model for studying Nabarro-Herring creep under applied stress. 
(b2) Steady-state vacancy composition along the diffusion path. 
(c1) Model for studying precipitate growth, where we denote the precipitate as a white grain while the matrix is the black grain. 
(c2) Steady-state composition of solute atoms (i.e., B atoms) from the solution of the Stefan problem. 
(d1) Model for studying DIGM in a thin film containing a vertical GB. 
The solute atoms are deposited on the two surfaces and diffuse into the GB. 
The region where the composition of solute atoms is large is shadowed gray. 
The dashed and solid vertical lines denote the GB positions before and after the solute atom diffusion along the GB. 
Induced by the diffusion, the GB moves to the right and the alloying region follows the GB migration.
(d2) Steady-state composition of solute atoms (i.e. B atoms) along the GB (i.e., the $y$-axis).  } 
\label{GBdiffusion}
\end{center}
\end{figure}

Gu et al. have used this Robin BC to analytically study the point-defect sink efficiency of low-angle tilt GBs under irradiation~\cite{gu2017point}. 
The model for this problem and the coordinate system are shown in Fig.~\ref{GBdiffusion}a1. 
The steady-state vacancy diffusion equation in the presence of irradiation can be written as~\cite{beyerlein2015defect,bullough1980sink,demkowicz2011influence} 
\begin{equation}
D_\text{v} \mathrm{d}^2 X_\text{v}/\mathrm{d} z^2 = D_\text{v}\mathfrak{K}^2 X_\text{v} - \mathfrak{S}, 
\label{irr}
\end{equation}
where $\mathfrak{S}$ is the production rate of vacancies due to irradiation and $\mathfrak{K}^2$ is the total sink strength within the grain ($\mathfrak{K}^{-1}$ represents the mean free path of a vacancy before being trapped by sinks in the grain)~\cite{was2016fundamentals}. 
The solution to Eq.~\eqref{irr} based on the BC Eq.~\eqref{RobinBC1} with $\tilde{\sigma}=0$ gives the steady-state vacancy concentration distribution: 
\begin{equation}
X_\text{v}(z-z_\text{i})
= \frac{X_\text{v}^\text{eq} - X_\text{v}^\infty}{1+\mathfrak{K}\mathfrak{L}} e^{-\mathfrak{K}|z-z_\text{i}|} + X_\text{v}^\infty, 
\label{irrsol}
\end{equation}
where $X_\text{v}^\infty\equiv \mathfrak{S}/(D_\text{v}\mathfrak{K}^2)$ is the equilibrium vacancy concentration in bulk. 
This solution is schematically plotted in Fig.~\ref{GBdiffusion}a2. 
The GB sink efficiency is defined as the ratio of the steady-state vacancy currents into the GB obtained from the realistic model (explicitly considering GB structure) and the ideal model (taking the GB as a perfect sink)~\cite{SuttonBalluffi}; from the solution Eq.~\eqref{irrsol}, we can find that the GB sink efficiency is 
\begin{equation}
\eta_\text{r} = \left(1+\mathfrak{K}\mathfrak{L}\right)^{-1}.  
\label{eta_irrsol}
\end{equation}
We note that, when $M_\text{d} \to \infty$ (thus, $\mathfrak{L}\to 0$), Eq.~\eqref{RobinBC1} will become the Dirichlet BC, $X_\text{v}(z_\text{i}) = X_\text{v}^\text{eq}$ and $\eta_\text{r} \to 1$, which simply indicates that the disconnection can climb without resistance, the GB is a perfect sink to vacancies, and the whole kinetic process is diffusion-controlled; on the contrary, when $M_\text{d} \to 0$ (thus, $\mathfrak{L}\to \infty$), Eq.~\eqref{RobinBC1} will become the Neumann BC, $X_\text{v} = X_\text{v}^\infty$ everywhere and $\eta_\text{r} \to 0$, suggesting the reaction-controlled (or ``interface-controlled'') kinetics. 
Therefore, the Robin boundary value problem is actually a simple, apt model for describing mixed kinetics; this is the central conclusion reached from the work by Gu et al.~\cite{gu2017point}

The Robin boundary value problem Eq.~\eqref{RobinBC1} with finite $\tilde{\sigma}$ can also be used to model Nabarro-Herring creep~\cite{nabarro1948deformation,herring1950diffusional}. 
As shown in Fig.~\ref{GBdiffusion}b1, when a square grain (embedded in a polycrystal) undergoes constant tensile stress $\sigma$ and $-\sigma$ in the horizontal and vertical directions, respectively, a vacancy flux will form. This flux flows from the vertical GBs to the horizontal GBs, since the vacancy chemical potential at the GB under tension is higher than that at the GB under compression. 
If the vacancy flux is primarily through the grain (in the case where the difference between the diffusivity of lattice diffusion and that of GB diffusion is small), such a phenomenon (plastic deformation via lattice diffusion under constant stress) corresponds to Nabarro-Herring creep. 
There has been some effort to build the constitutive relationship for creep explicitly based on the GBD climb model~\cite{ashby1969interface,burton1972interface,arzt1983interface}. 
Here, we assume that the effective distance between two neighboring GBs is $L$, which is scaled by the grain size (we ignore the complicated dependence on grain shape, which is not important) and simplify the model to be a 1D diffusion problem as illustrated in Fig.~\ref{GBdiffusion}b. 
One GB is located at $z=0$ and modeled by the BC Eq.~\eqref{RobinBC1} with stress $-\tilde{\sigma}$, while the other GB is located at $z=L$ and modeled by the BC Eq.~\eqref{RobinBC1} with stress $\tilde{\sigma}$. 
The steady-state diffusion equation is $\mathrm{d}^2X_\text{v}/\mathrm{d}z^2 = 0$ and, along with the applied Robin BCs, the solution for the vacancy concentration is 
\begin{equation}
X_\text{v}(z) = X_\text{v}^\text{eq} + \frac{\tilde{\sigma}(z-L/2)}{\mathfrak{L}+L/2},  
\end{equation}
which is schematically plotted in Fig.~\ref{GBdiffusion}b2. 
From this solution, we can obtain the strain rate: 
\begin{align}
\dot{\epsilon} 
&= \frac{\Omega}{L}J_\text{v} 
= \frac{\Omega}{L} \frac{D_\text{v}}{\Omega} \frac{\mathrm{d}X_\text{v}}{\mathrm{d}z}
= \frac{D_\text{v}}{\mathfrak{L}+L/2} \frac{\tilde{\sigma}}{L}
=  \left( \frac{k_\text{B}TL}{2D_\text{L}} + \frac{\Omega}{M_\text{d}\varrho b_\perp^2} \right)^{-1} \frac{\sigma \Omega}{L}
\nonumber \\
&= \left\{\begin{array}{ll}
\dfrac{2D_\text{L}\Omega\sigma}{k_\text{B}TL^2}, 		& \text{when } \dfrac{D_\text{L}}{k_\text{B}TL} \ll \dfrac{M_\text{d}\varrho b_\perp^2}{\Omega} \text{ (diffusion-controlled)}\\
\dfrac{M_\text{d}\varrho b_\perp^2\sigma}{L}, 				& \text{when } \dfrac{D_\text{L}}{k_\text{B}TL} \gg \dfrac{M_\text{d}\varrho b_\perp^2}{\Omega} \text{ (reaction-controlled)}
\end{array}\right., 
\label{depsilon}
\end{align}
where we have used the relation: $\mathcal{E}_\text{v} \approx k_\text{B}T/X_\text{v}^\text{eq}$ as $X_\text{v}^\text{eq}\ll 1$ (see \ref{APPdriving} for derivation) and the self-diffusivity in lattice: $D_\text{L} = D_\text{v}X_\text{v}^\text{eq}$. 
Equation~\eqref{depsilon} is exactly the same as that derived by Herring~\cite{herring1950diffusional} and Ashby~\cite{ashby1969interface}. 
Unlike Nabarro-Herring creep, Coble creep refers to plastic deformation predominantly via GB diffusion under constant stress~\cite{coble1963model}. 
Simply by replacing $D_\text{L}$ with $\lambda D_\text{gb}/L$ in Eq.~\eqref{depsilon} ($\lambda$ is the effective GB width and $D_\text{gb}$ is the self-diffusivity in GB), we may obtain the constitutive relationship for Coble creep~\cite{coble1963model}. 

Recall that we establish the BC Eq.~\eqref{BCv0} [and thus Eq.~\eqref{RobinBC1}] based on the disconnection climb mechanism and the disconnection character $(b_\perp, h^\text{L})$ rather than the lattice dislocation array which constitutes a low-angle tilt GB. Hence, the analytical solutions for the irradiation problem and the creep problem obtained above are, in principle, not limited to low-angle tilt GBs; they can be applied to general GBs if we choose $\varrho$ and $b_\perp$ properly [which actually only influences the value of $\mathfrak{L}$ in Eq.~\eqref{RobinBC1}]. 
In the case of a high-angle GB the structure of which can be well described as an array of SGBDs (as in Fig.~\ref{pdabsorption}d), $b_\perp$ is the Burgers vector of the SGBDs and $\varrho$ is the density (inverse of the spacing) of the SGBDs. 
In the case of a low-$\Sigma$ high-angle GB for which SGBDs cannot be well-defined (as in Fig.~\ref{pdabsorption}c), $b_\perp$ is the normal component of the disconnection Burgers vector (consistent with the DSC lattice vector) and $\varrho$ is the inverse of the correlation length between nuclei (i.e., the homogeneously nucleated disconnection loops). 
In the case where extrinsic GBDs exist on a GB, e.g. in a highly plastically deformed polycrystal (scenario in Fig.~\ref{pdabsorption}e), $b_\perp$ is the normal component of the extrinsic GBD Burgers vector and $\varrho$ is the density of the extrinsic GBDs. 

(ii) \textit{Precipitate growth}. 
We assume that a spherical precipitate (corresponding to a stoichiometric compound) nucleates and grows from a matrix (supersaturated solid solution) and consider the concentration of solute atoms (i.e., B atoms) in the matrix during this process. 
We denote the precipitate and the matrix as white and black grains, respectively, as illustrated in Fig.~\ref{GBdiffusion}c1. 
Local mass balance requires that the diffusion of B atoms at the interface should satisfy the first equation in Eq.~\eqref{JBC} with $J_\text{B}^\text{w/i} = 0$. This is called the Stefan condition~\cite{rubinvsteuin2000stefan}. 
When $b_\perp = 0$ (i.e., assume that the disconnection is a pure step), Eq.~\eqref{JBC}, associated with Eq.~\eqref{IB}, will reduce to $J_\text{B}(z_\text{i}) = [X_\text{B}^\text{w} - X_\text{B}(z_\text{i})]v/\Omega$, where $v = \varrho v_\text{cl} h^\text{L}$ is the interface migration velocity; this is the classical form of Stefan condition which is commonly used in solving the problem of precipitate growth (e.g. Refs.~\cite{balluffi2005kinetics,vermolen2007three,lahiri2017theoretical,barba2017microtwinning}). 
However, it is not reasonable to assume that $b_\perp = 0$; hence, Eq.~\eqref{JBC} along with Eq.~\eqref{IB} describes a more general Stefan condition. 
By using the expression of Eq.~\eqref{vcl_general}, we can write Eq.~\eqref{JBC} as
\begin{align}
\frac{D_\text{B}}{\Omega} \frac{\mathrm{d}X_\text{B}}{\mathrm{d}z}\bigg|_{z_\text{i}}
=~& M_\text{d}\varrho \left(\frac{h^\text{L}}{\Omega}\right)^2
\nonumber \\
&\times \left\{ 
\mathcal{E}'_\text{v}\left[X_\text{v}(z_\text{i})-X_\text{v}^\text{eq}\right] 
+ \mathcal{E}'_\text{B}\left[X_\text{B}(z_\text{i})-X_\text{B}^\text{eq}\right]  
- \mathcal{E}'_\text{s} \sigma_{nn}\beta_\perp\Omega 
\right\}, 
\label{BCvii}
\end{align}
where the coefficient $\mathcal{E}'_\text{v/B/s}$ is determined by the interaction parameters in regular solution model, the concentration of the precipitate $X_\text{B}^\text{w}$ (which is a constant), the coupling factor $\beta_\perp\equiv b_\perp/h^\text{L}$, the stress $\sigma_{nn}$ and temperature (see \ref{APPdriving} for derivation). 
Note that $X_\text{B}^\text{eq}$ is the equilibrium concentration which can be obtained from the equilibrium phase diagram. 
If we assume that $X_\text{v}(z_\text{i})\approx X_\text{v}^\text{eq}$ and no stress is applied (i.e., $\sigma_{nn}=0$), then Eq.~\eqref{BCvii} reduces to the Robin BC: 
\begin{equation}
\mathfrak{L}_\text{B} (\mathrm{d}X_\text{B}/\mathrm{d}z)_{z_\text{i}}
= X_\text{B}(z_\text{i}) - X_\text{B}^\text{eq}, 
\label{RobinBC2}
\end{equation}
where $\mathfrak{L}_\text{B}\equiv D_\text{B}\Omega/(M_\text{d}\varrho {h^\text{L}}^2 \mathcal{E}'_\text{B})$. 
The parameter $\mathcal{E}'_\text{B}$ is a function of the interaction parameters in regular solution model, $X_\text{B}^\text{w}$, $\beta_\perp$ and temperature (the detailed formula of $\mathcal{E}'_\text{B}$ is given in \ref{APPdriving}). 
When the disconnection is a pure step $b_\perp = 0$, $\mathcal{E}'_\text{B} = 
\mathcal{E}_\text{BB}(X_\text{B}^\text{eq}-X_\text{B}^\text{w})^2$, where $\mathcal{E}_\text{BB}$ only depends on the interaction parameters and temperature. 

The problem of precipitate growth is usually solved based on the assumption that the precipitate is spherical, it grows in an infinitely large matrix and the interface moves sufficiently slowly such that the steady-state solute concentration profile exists~\cite{jackson2006kinetic}. 
Of course, there exist more sophisticated treatments with less assumption. 
But we won't study the problem itself in depth; we will adopt the simplest model of precipitate growth and just focus on how the Robin BC Eq.~\eqref{RobinBC2} works in this problem. 
Due to the spherical symmetry of the problem, we put the origin at the center of the precipitate and use the steady-state diffusion equation in the matrix: $\mathrm{d}^2X_\text{B}/\mathrm{d}z^2 + (2/z)(\mathrm{d}X_\text{B}/\mathrm{d}z) = 0$ where $z\ge 0$ is the radial distance. 
One BC is $X_\text{B}(z\to\infty) = X_\text{B}^\infty$, where $X_\text{B}^\infty$ is the initially prepared concentration before the precipitate is nucleated, and the other BC is Eq.~\eqref{RobinBC2}, where $z_\text{i}$ corresponds to the size of the precipitate (see Fig.~\ref{GBdiffusion}c2). 
The steady-state solution is 
\begin{equation}
X_\text{B}(z)
= (1+\mathfrak{L}_\text{B}/z_\text{i})^{-1} (X_\text{B}^\text{eq}-X_\text{B}^\infty)\frac{z_\text{i}}{z} + X_\text{B}^\infty,  
\label{XBzii}
\end{equation}
which is schematically plotted in Fig.~\ref{GBdiffusion}c2. 
From Eqs.~\eqref{JBC} and \eqref{IB} and the replacement: $\varrho h^\text{L} v_\text{cl} = \mathrm{d}z_\text{i}/\mathrm{d}t$ (i.e., the interface migration velocity), we can obtain the evolution of the GB position over time: 
\begin{equation}
z_\text{i}
= \sqrt{2\mathcal{C}_1 D_\text{B}t +(\mathcal{C}_2\mathfrak{L}_\text{B})^2} -  \mathcal{C}_2\mathfrak{L}_\text{B}, 
\end{equation}
where the coefficients are 
\begin{equation}
\mathcal{C}_1 
\equiv \frac{X_\text{B}^\infty-X_\text{B}^\text{eq}}{(1+\frac{\beta_\perp}{2})X_\text{B}^\text{w}-(1-\frac{\beta_\perp}{2})X_\text{B}^\text{eq}}
\text{ and }
\mathcal{C}_2 
\equiv \frac{(1+\frac{\beta_\perp}{2})X_\text{B}^\text{w}-(1-\frac{\beta_\perp}{2})X_\text{B}^\infty}
{(1+\frac{\beta_\perp}{2})X_\text{B}^\text{w}-(1-\frac{\beta_\perp}{2})X_\text{B}^\text{eq}}. 
\nonumber
\end{equation}
From Eq.~\eqref{XBzii}, we also know that the interface sink efficiency is 
\begin{equation}
\eta_\text{p} = J_\text{B}(z_\text{i};\mathfrak{L}_\text{B})/J_\text{B}(z_\text{i};\mathfrak{L}_\text{B}=0)
= (1+\mathfrak{L}_\text{B}/z_\text{i})^{-1}. 
\end{equation}
The sink efficiency is consistent in form with that derived by Balluffi et al. in form~\cite{SuttonBalluffi,balluffi1983sinks}. 
When $M_\text{d}\to \infty$ (or $\mathfrak{L}_\text{B}\to 0$), then $\eta_\text{p}=1$, which indicates perfect sink behavior. 
Close to this limit, the precipitate growth is diffusion-controlled; this is what was usually assumed in theory~\cite{jackson2006kinetic,balluffi2005kinetics,vermolen2007three}. 
Based on this assumption, the classical growth law follows: $z_\text{i} \sim \sqrt{D_\text{B}t}$. 
When $M_\text{d}$ is small (or $\mathfrak{L}_\text{B}\to \infty$), then $\eta_\text{p}$ is small, corresponding to reaction-controlled precipitate growth. 
At the limit of $\mathfrak{L}_\text{B}\to \infty$, the precipitate growth follows a linear law: $z_\text{i} \sim D_\text{B}t/\mathfrak{L}_\text{B}$. 

From the analysis above, we find that the use of the Robin BC Eq.~\eqref{RobinBC2} in solving the problem of precipitate growth generalizes the classical solution in two aspects: (1) the growth process is controlled by a combination of reaction and diffusion (rather than purely diffusion-controlled, as usually assumed); (2) the interface migrates via the climb of disconnections associated with a Burgers vector (rather than pure steps, as usually assumed). 
If the Burgers vector is not zero, the precipitate growth will be necessarily accompanied by vacancy diffusion. 
This suggests that the precipitate growth (or migration of the heterophase interface) may influence the preference of void formation near the precipitate. Conversely, the presence of vacancy supersaturation may facilitate precipitate formation and growth. 
This implication was indeed supported by some experimental observations (e.g. Refs.~\cite{takeyama1980effect,xu2006dose}). 

(iii) \textit{Diffusion-induced GB migration}.
At low temperature, the diffusivity of solute atoms along a GB is overwhelmingly larger than that in the lattice. 
When a polycrystal is exposed to an environment containing a high concentration of solute atoms, the solute atoms will diffuse from the free surface into the polycrystal through the GBs. 
It was found in many experiments that, with the diffusion of the solute atoms along the GBs (in the direction parallel to each GB), the GBs would migrate (in the direction normal to each GB)~\cite{rhines1938new,den1972interface,hillert1978chemically,cahn1979diffusion,langelier2017effects}; such a phenomenon is called diffusion-induced GB migration (DIGM). 
Since the fundamental driving force for DIGM is the reduction of free energy of mixing/alloying, DIGM is sometimes termed chemically induced GB migration~\cite{hillert1978chemically,tashiro1987observations}. 
The direct, mechanistic reason for DIGM is still under debate; among the proposals are: (1) GB migration is driven by a coupling effect of GB diffusion due to disconnection climb~\cite{balluffi1981mechanism,smith1981mechanism} and (2) GB migration is driven by the coherency strain energy density difference arising from the introduction of misfitting solute atoms from the GB~\cite{hillert1983driving}.
According to Sutton and Balluffi, a fair conclusion at present is that ``no single one of these models, at least in a simple form, has been able to explain all of the complex features of DIGM which have been observed in different systems''~\cite{SuttonBalluffi}. 
Here, we will not review these mechanistic reasons/models in detail (such review can be found in the literatures~\cite{SuttonBalluffi,king1987diffusion}). 
No matter the reason for DIGM, we believe that it involves disconnection climb. 

For simplicity, we consider the DIGM problem based on a thin film configuration as illustrated in Fig.~\ref{GBdiffusion}d1.  
Starting from a single-component thin film (purely composed by A atoms) with two parallel horizontal free surfaces and a vertical GB, when the solute atoms (B atoms) diffuse from the surfaces into the GB, the GB will migrate from the white grain to the black grain. 
In the region swept by the GB, an AB alloy (assumed substitutional) is formed. 
In this alloying region, the concentration of B atoms is not uniform [which can be analytically determined by Eq.~\eqref{JBC}] according to the model and analysis originally made by Cahn~\cite{cahn1959kinetics}. 
The steady-state equation of the diffusion of B atoms along the GB can be written as 
\begin{equation}
\frac{\lambda D_\text{B}^\text{gb}}{\Omega}\frac{\mathrm{d}^2 X_\text{B}^\text{gb}}{\mathrm{d} y^2} - \varrho I_\text{B} = 0, 
\label{DIGM_eqn1}
\end{equation}
where the coordinate system is attached to the GB and shown in Fig.~\ref{GBdiffusion}d1, $X_\text{B}^\text{gb}(y)$ and $D_\text{B}^\text{gb}$ are the concentration and diffusivity of B atoms along the GB (i.e. the $y$-axis). 
In this equation, the GB sink effect on the solute atoms is not treated as a BC but as a uniform sink along the $y$-axis.   
We assume that the segregation coefficient $s \equiv X_\text{B}^\text{gb}(y)/X_\text{B}(y)$ is a constant. 
If there is steady-state GB migration, the migration velocity $v = \varrho(y)h^\text{L}v_\text{cl}(y)$ is constant with respect to the coordinate $y$. 
$I_\text{B}$ can then be obtained by Eq.~\eqref{IB} and set constant $v$ and $X_\text{B}^\text{b}=0$. 
Based on the above assumptions, Eq.~\eqref{DIGM_eqn1} becomes
\begin{equation}
\frac{\mathrm{d}^2 X_\text{B}}{\mathrm{d} y^2} - \frac{X_\text{B}}{y_\text{c}} = 0, 
\label{DIGM_eqn2}
\end{equation}
where 
\begin{equation}
y_\text{c} \equiv \sqrt{\frac{s\lambda D_\text{B}^\text{gb}}{(\beta_\perp/2+1)v}}. 
\end{equation}
For the thin film configuration shown in Fig.~\ref{GBdiffusion}d1, the BC for this equation can be represented as $X_\text{B}(0)=X_\text{B}(L_y) =X_\text{B}^\text{eq}$, where $L_y$ is the thickness of the film. 
The solution to Eq.~\eqref{DIGM_eqn2} is then
\begin{equation}
X_\text{B}(y)
= X_\text{B}^\text{eq} \frac{\cosh(y/y_\text{c}-L_y/2y_\text{c})}{\cosh(L_y/2y_\text{c})}, 
\label{DIGM_sol}
\end{equation}
which is schematically shown in Fig.~\ref{GBdiffusion}d2. 
This solution has been demonstrated to be consistent with the concentration profile observed in many experiments for different materials~\cite{chongmo1982diffusion,hillert1978chemically}. 
If the film shown in Fig.~\ref{GBdiffusion}d1 is infinitely thick (i.e., at the limit of $L_y\to \infty$), the solution Eq.~\eqref{DIGM_sol} will become
\begin{equation}
X_\text{B}(y)
= X_\text{B}^\text{eq} e^{-y/y_\text{c}}. 
\end{equation}
This solution is also consistent with some experimental observations~\cite{chongmo1982diffusion,giakupian1990misorientation,schmelzle1992diffusion}. 
We note that the characteristic length $y_\text{c}$ is determined by the competition between the GB diffusivity and the GB migration velocity. 
When $v\gg \lambda D_\text{B}^\text{gb}$, the penetration depth of the solute atoms into the GB (characterized by $y_\text{c}$) will be small and vice versa. 
The concentration of B atoms may not be uniform in the $z$-direction either, but this cannot be analytically solved and the situation varies largely for different materials~\cite{cahn1979diffusion,ma1995kinetic,lee2000diffusion,lee1990chemically,liang1996diffusion,guan1993characterization}. 

Aside from the diffusion equation for the solute atoms, i.e., Eq.~\eqref{DIGM_eqn1}, there should actually be another diffusion equation for vacancies if the Burgers vector $b_\perp$ of the disconnections involved in the GB migration is not zero. 
The vacancies may come from the unequal diffusion of B atoms (solute) and A atoms (solvent) along the GB (the ``GB Kirkendall effect''~\cite{balluffi1981mechanism,klinger2011theory}).

\subsubsection{Limitation of the disconnection description}
In the above several sections, we understood non-conservative GB kinetics based on disconnection climb along GBs, and based on the disconnection model we have constructed boundary conditions for describing the GB/interface sink effect accompanied by irradiation or GB/interface migration. 
Finally, we should point out the limitation of the application of the disconnection model to non-conservative GB kinetics. 
The fundamental assumption behind the disconnection model is that the GB structure would not change before or after the absorption of point defects; a disconnection is a line defect separating two GB regions which have identical GB structure but distinct relative displacement of two grains by a DSC vector. 
However, it has been demonstrated that the absorption/emission of certain amount of point defects into/out of a GB may change the GB structure. 
For example, via MD simulations, Frolov et al. found new GB structures by varying the atomic fraction in the GBs (with fixed macroscopic DOFs) for $\Sigma 5$ $[100]$ STGBs in FCC metals. 
This structural change due to the variation of microscopic DOFs is called a ``GB structural phase transformation'' by Frolov et al.~\cite{Frolov13} and ``GB structural multiplicity'' by Vitek et al.~\cite{vitek1983multiplicity,wang1984computer,Vitek85}
There are many observations of GB structural change for other types of GBs and materials in MD simulations~\cite{Oh86,majid1987dynamical,Alfthan06,von2007molecular,han2016grain,hickman2017extra} and experiments~\cite{merkle1987atomic,Krakow91,hu2002hrtem,ayache2005determination}. 
We do not expect that the disconnection model can describe the change of GB atomic structure. 

We will briefly discuss the possible influence of GB structural change on non-conservative GB kinetics. 
First, the change of GB atomic structure does not influence the character of disconnections [i.e., $(\mathbf{b},h)$] because a disconnection is a topological defect which is only determined by the translational symmetry of the DSC lattice (fixed by macroscopic DOFs). 
There are two possible mechanisms for a GB to absorb point defects (e.g. vacancies) when they are supersaturated due to irradiation: (1) disconnection climb and (2) transformation from one GB structure to another (when the two GB structures have different atomic fractions). 
The first mechanism has been discussed in Section~\ref{GB_as_sink}, but the second one cannot be considered based on the disconnection model. 
One simple idea is that when the disconnection climb velocity wins over the structural transformation rate, the first mechanism will dominate over the second one; and vice versa. 
Based on this simple idea, we can derive a qualitative criterion that determines which mechanism dominates. 
Consider a part of GB with area $\mathcal{A}_\text{c}$, which will be completely transformed into another structure via cooperative atomic relaxation if this GB segment absorbs $n_\text{v}$ vacancies. Note that the area $\mathcal{A}_\text{c}$ may correspond to the critical size for nucleation of the new structure. 
If the second mechanism operates, we can assume that the atomic relaxation time (for nucleation of the new structure) is $t_\text{R}$, which is only a function of temperature. 
Now, we consider how much time is required if the same amount of vacancies are digested in the same area of the GB via the climb of disconnections (PGBDs in low-angle tilt GBs, SGBDs in high-angle tilt GBs, or extrinsic GBDs -- see discussion in Section~\ref{GB_as_sink}). 
After absorption of $n_\text{v}$ vacancies in the area $\mathcal{A}_\text{c}$, the disconnection climb distance is $\Delta y = n_\text{v}\Omega/(\mathcal{A}_\text{c}\varrho b_\perp)$ and the disconnection climb velocity is $v_\text{cl} = |J_\text{v}|\Omega/(\varrho b_\perp)$. Thus, the time for this process is $t_\text{C}=\Delta y/v_\text{cl}$. 
We can then define the measure for the tendency of GB structural transformation as 
\begin{equation}
\Lambda 
\equiv \frac{t_\text{C}}{t_\text{R}}
= \frac{n_\text{v}}{t_\text{R} \mathcal{A}_\text{c} |J_\text{v}(z_\text{i})|}
\approx \frac{n_\text{v} \Omega \mathfrak{K}}{t_\text{R} \mathcal{A}_\text{c} \eta_\text{r} \mathfrak{S}}, 
\end{equation}
where we have used Eqs.~\eqref{irrsol} and \eqref{eta_irrsol} and the approximation that $X_\text{v}^\text{eq}\approx 0$. 
When $\Lambda > 1$, the structural transformation mechanism (i.e., the second mechanism) dominates over the disconnection climb mechanism (i.e., the first mechanism); and vice versa. 
From the observation that $\Lambda \propto \eta_\text{r}^{-1}$ and combined with Eq.~\eqref{eta_irrsol}, we know that the structural transformation mechanism will be important in the case where the disconnection climb mobility $M_\text{d}$ is small and/or the disconnection density $\varrho$ is small. 
If vacancies are absorbed by a GB via the structural transformation mechanism, then the BC for describing the GB sink effect can be written as $J_\text{v}(z_\text{i}) = n_\text{v}/(t_\text{R}\mathcal{A}_\text{c})$, which is a Neumann BC (rather than a Robin BC). 
The application of Neumann BCs will simply lead to the steady-state solution: $X_\text{v}(z) = X_\text{v}^\infty$. 
This solution implies that the kinetics is purely reaction-controlled and there is no vacancy denuded zone (DZ) around the GB. 
Negligible DZ width was indeed observed for some GBs in polycrystals in experiments~\cite{han2012effect,el2017direct};  the extreme case is the coherent TB in FCC metals.

\subsection{Interaction between grain boundaries and lattice dislocations}
\label{GBLattDisl}

The absorption/emission of a lattice dislocation into/out of a GB can be decomposed into a series of disconnections such that the total Burgers vector is conserved.
Since Burgers vectors of lattice dislocations are not parallel to GBs, this absorption/emission will, in general, lead to line defects with nonzero components of the Burgers vector normal to the GB (i.e., $b_\perp \ne 0$) in the GB. 
As discussed in Section~\ref{disconnection_description} and depicted in Figs.~\ref{disconnection}c1 and \ref{disconnection}c2, the existence of a disconnection with $b_\perp \ne 0$ implies the introduction of an additional/missing DSC lattice plane along the GB. 
Therefore, the absorption/emission of lattice dislocations into/out of a GB can be thought of as a type of non-conservative  process. 

The interaction between GBs and lattice dislocations is an important problem in materials science, since it is the mechanism behind many types of mechanical behavior. 
For example, the well-known Hall-Petch relation~\cite{hall1951deformation,petch1953cleavage} (relationship between strength and grain size) rests on the assumption that lattice dislocations pile up against GBs (smaller grain sizes imply more closely spaced dislocation barriers and hence larger strength)~\cite{kumar2003mechanical}. 
This requires GBs to effectively block lattice dislocations rather than absorb them.
Some special GBs (e.g., coherent twin boundaries) not only block lattice dislocations but also slow dislocation propagation along the GBs as well; introduction of such special GBs can enhance material strength without significant  loss of ductility~\cite{lu2009strengthening}. 
Plastic deformation of a nanocrystalline metals (i.e., polycrystals with nanometer-scale grain sizes) may also involve nucleation of lattice dislocations at the GBs and dislocation motion threaded into GBs~\cite{van2004stacking,van2006nucleation,you2013plastic}. 
Another example of the phenomenon where the interaction of lattice dislocations with GBs plays the role is the primary recrystallization of heavily deformed polycrystals~\cite{rollett2004recrystallization}. 
Primary recrystallization requires the absorption of lattice dislocations into GBs. 
Hence,  GB/lattice dislocation interactions are central to understanding, predicting and designing the mechanical response of materials and the processing that affects it. 

\begin{figure}[!t]
\begin{center}
\scalebox{0.28}{\includegraphics{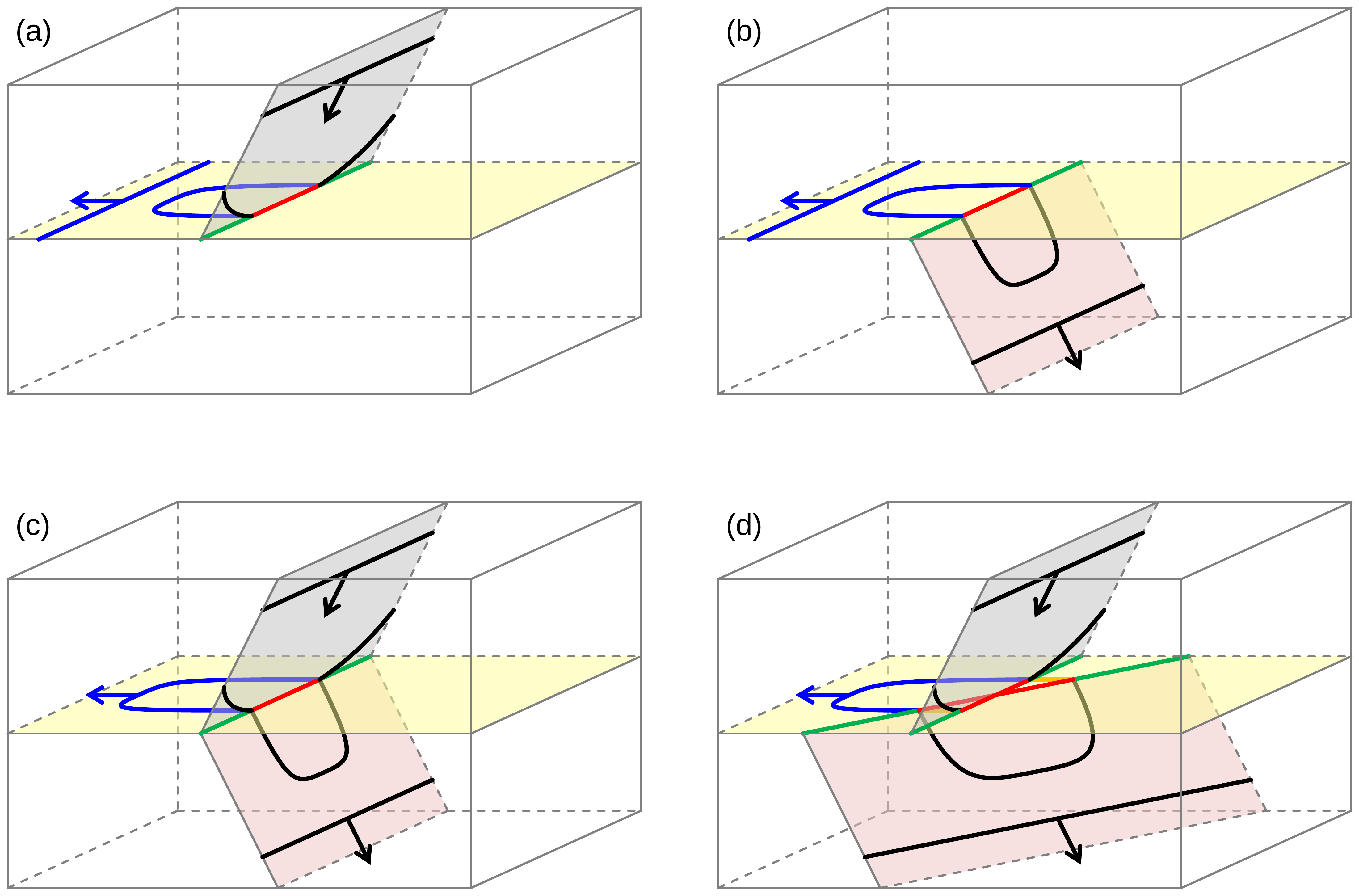}}
\caption{Scenarios of interaction between lattice dislocations and a GB in a bicrystal: (a) absorption of the lattice dislocations from the upper grain into the GB, (b) emission of the lattice dislocations from the GB into the lower grain, (c) Direct transmission of the lattice dislocations across the GB (the slip planes and the GB plane intersect in a line), and (d) indirect transmission of the lattice dislocations across the GB (the slip planes and the GB plane do not intersect in a line). 
The GB plane is colored yellow. 
The slip planes in the upper and lower grains are colored gray and pink, respectively. 
The lines of lattice dislocations, glissile disconnections and sessile disconnections are colored black, blue and other colors. } 
\label{DislGB_interaction}
\end{center}
\end{figure}

GB/lattice dislocation interactions may be categorized into  four types, shown schematically in Fig.~\ref{DislGB_interaction}. 
In the bicrystal shown in Fig.~\ref{DislGB_interaction} we consider a lattice dislocation being absorbed into a GB from the upper grain and emitted into the GB and/or the lower grain. 
\begin{itemize}
\item[(i)] Lattice dislocations may be absorbed into the GB (illustrated in Fig.~\ref{DislGB_interaction}a). 
In general, such absorption occurs via dissociation of the lattice dislocations into extrinsic GB dislocations within the GB (see below). 
The resultant GB dislocations may be glissile or sessile, depending on whether the Burgers vector is parallel to the GB or otherwise. 
All the GB dislocations (resulting from lattice dislocation dissociation) will be glissile only when the Burgers vector of the lattice dislocation was of screw type (i.e., $\mathbf b$ is parallel to the intersection line between the slip plane and the GB plane). 
The absorption of lattice dislocations into GBs was observed in MD simulations~\cite{tsuru2009fundamental} and experiments~\cite{shen1988dislocation,minor2004direct,kondo2016direct,wang2011transmission}. 

\item[(ii)] Lattice dislocations may be emitted from the GB (illustrated in Fig.~\ref{DislGB_interaction}b). 
The emission of lattice dislocations will inevitably leave residual GB dislocations at the GB. 
Such a scenario was observed both in MD simulations~\cite{tschopp2008grain,spearot2007tensile,tucker2010evolution,tucker2011non,sangid2011energy,zhang2014atomistic,wyman2017variability} and experiments~\cite{murr1981strain,liao2003deformation,liao2004deformation,wang2011transmission}. 

\item[(iii)] Lattice dislocations may be directly transmitted from one grain into the other through the GB (illustrated in Fig.~\ref{DislGB_interaction}c). 
Direct transmission can occur only when the two slip planes and the GB plane intersect along a line. 
This scenario was observed in both MD simulations~\cite{cheng2008atomistic} and experiments~\cite{lim1984slip,shen1988dislocation,lee1990situ}. 
Note that direct, complete transmission (without residual GB dislocations) will be possible only when the Burgers vector of the lattice dislocation is parallel to the intersection line between the slip plane and the GB plane (i.e., cross slip). 

\item[(iv)] In general, the slip planes from the two grains and the GB  do not intersect along a line. 
In this case,  incoming lattice dislocations can pile up at the GB. 
With the accumulation of dislocations in one grain at the GB, the induced stress may trigger  lattice dislocation nucleation in the other grain. 
This  is the indirect transmission depicted in Fig.~\ref{DislGB_interaction}d. 
For example,  indirect transmission is expected at a twist GB. 
This scenario was frequently observed in both MD simulations~\cite{sangid2011energy} and experiments~\cite{kacher2011visualization,kacher2012quasi,malyar2017size}. 
\end{itemize}
There is another scenario discussed in the literature: absorption of dislocations at some location along a GB followed by emission of dislocations at another location along the GB~\cite{shen1988dislocation,lee1990situ}. 
This case may be simply decomposed into Cases (i) and (ii) mentioned above. 
It should be emphasized that the residual GB dislocations are disconnections which are characterized by both Burgers vector and step height (the step character is not drawn in Fig.~\ref{DislGB_interaction}). 
We note that the term ``lattice dislocation transmission''  at a GB should be thought of as dislocation absorption at the GB followed by dislocation emission, since  two grains will generally have differently oriented slip systems and hence different Burgers vectors. 

\begin{figure}[!t]
\begin{center}
\scalebox{0.4}{\includegraphics{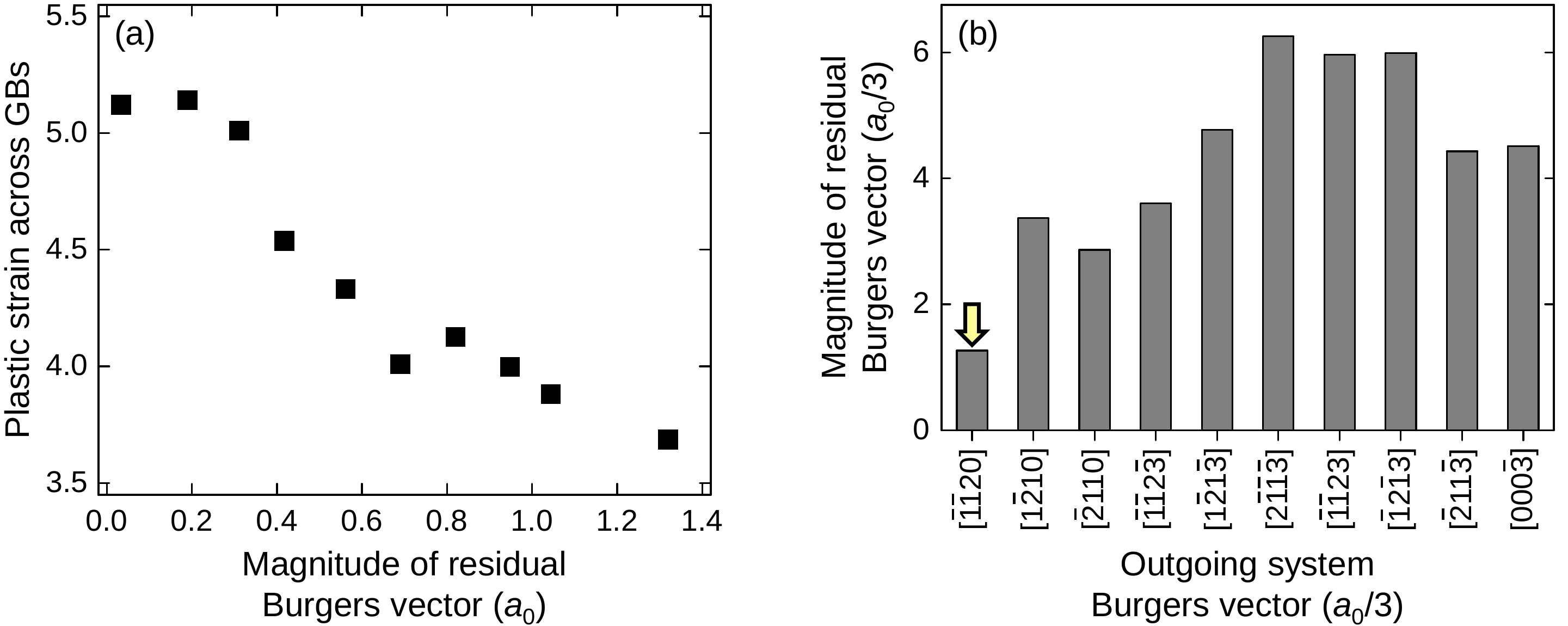}}
\caption{(a) Plastic strain across GBs vs. magnitude residual Burgers vector $|\mathbf{b}_\text{r}|$ measured in a polycrystalline Ni-based superalloy. 
(a) is reproduced from Ref.~\cite{abuzaid2012slip} (H. Sehitoglu, Elsevier 2012). 
(b) Magnitude of residual Burgers vector associated with all the potential slip systems of the outgoing lattice dislocations in $\alpha$-Ti. 
The actually operative slip system of the outgoing lattice dislocations is indicated by the arrow.
The incoming lattice dislocation is pure screw and has Burgers vector $[\bar{1}\bar{1}20]a_0/3$. 
The two grains adjacent to the GB are related by a $46^\circ$ rotation about the $[8,\bar{20},12,5]$ axis. 
(b) is reproduced from Ref.~\cite{kacher2014dislocation} (I. M. Robertson, Elsevier 2014).  } 
\label{residual}
\end{center}
\end{figure}

There are several criteria established for measuring the difficulty of  lattice dislocation transmission through a GB in a bicrystal. 
These criteria are based on the misorientation of two grains~\cite{chalmers1937influence,aust1954effect,clark1954mechanical}, the slip geometry (as shown in Fig.~\ref{DislGB_interaction}d)~\cite{livingston1957multiple,shen1986dislocation,shen1988dislocation,beyerlein2012structure,luster1995compatibility,werner1990slip} and the residual Burgers vector at the GB~\cite{lim1985continuity,lee1989prediction,abuzaid2012slip} (see~\ref{APPcriteria} for a brief review).  
We focus on the role of the residual Burgers vector in the GB, since this is directly related to the disconnection model (the focus of this review). 
The residual Burgers vector is $\mathbf{b}_\text{r} = \mathbf{b}^\text{i}-\mathbf{b}^\text{o}$, where $\mathbf{b}^\text{i}$ and $\mathbf{b}^\text{o}$ are the Burgers vectors of the incoming and outgoing lattice dislocations, respectively, and all the Burgers vectors are represented with reference to the DSC lattice. 
Experimental observations suggest that, for a particular incoming slip system, the outgoing slip system tends to minimize the magnitude of the residual Burgers vector at the GB. 
For example, mapping  the local plastic strains in a plastically deformed polycrystalline Ni-based superalloy (Hastelloy X), Abuzaid et al.~\cite{abuzaid2012slip} found that the strain is continuous across some GBs and discontinuous across others. 
They measured the strain across some GBs and calculated the residual Burgers vector for these GBs. 
They found that the strain across the GBs decreases with  increasing the residual Burgers vector, as shown in Fig.~\ref{residual}a. 
This is evidence that the slip systems associated with dislocation transmission correspond to those with the minimum residual Burgers vector magnitude. 
Another example is from the in situ TEM straining experiment and tomographic analysis performed by Kacher and Robertson~\cite{kacher2014situ}. 
They observed transmission of lattice dislocations across several GBs in $\alpha$-Ti. 
They found that, for a particular GB and a particular incoming slip system, the operative outgoing slip system is that which led to the minimum residual Burgers vector magnitude. 
As shown in Fig.~\ref{residual}b, when the screw lattice dislocations with $\mathbf b = [\bar{1}\bar{1}20]a_0/3$ come into a GB characterized by a $46^\circ$ rotation of two grains about the $[8,\bar{20},12,5]$ axis, the outgoing lattice dislocations will be associated with $\mathbf b = [\bar{1}\bar{1}20]a_0/3$ which corresponds to the minimum residual Burgers vector magnitude. 

\begin{figure}[!t]
\begin{center}
\scalebox{0.18}{\includegraphics{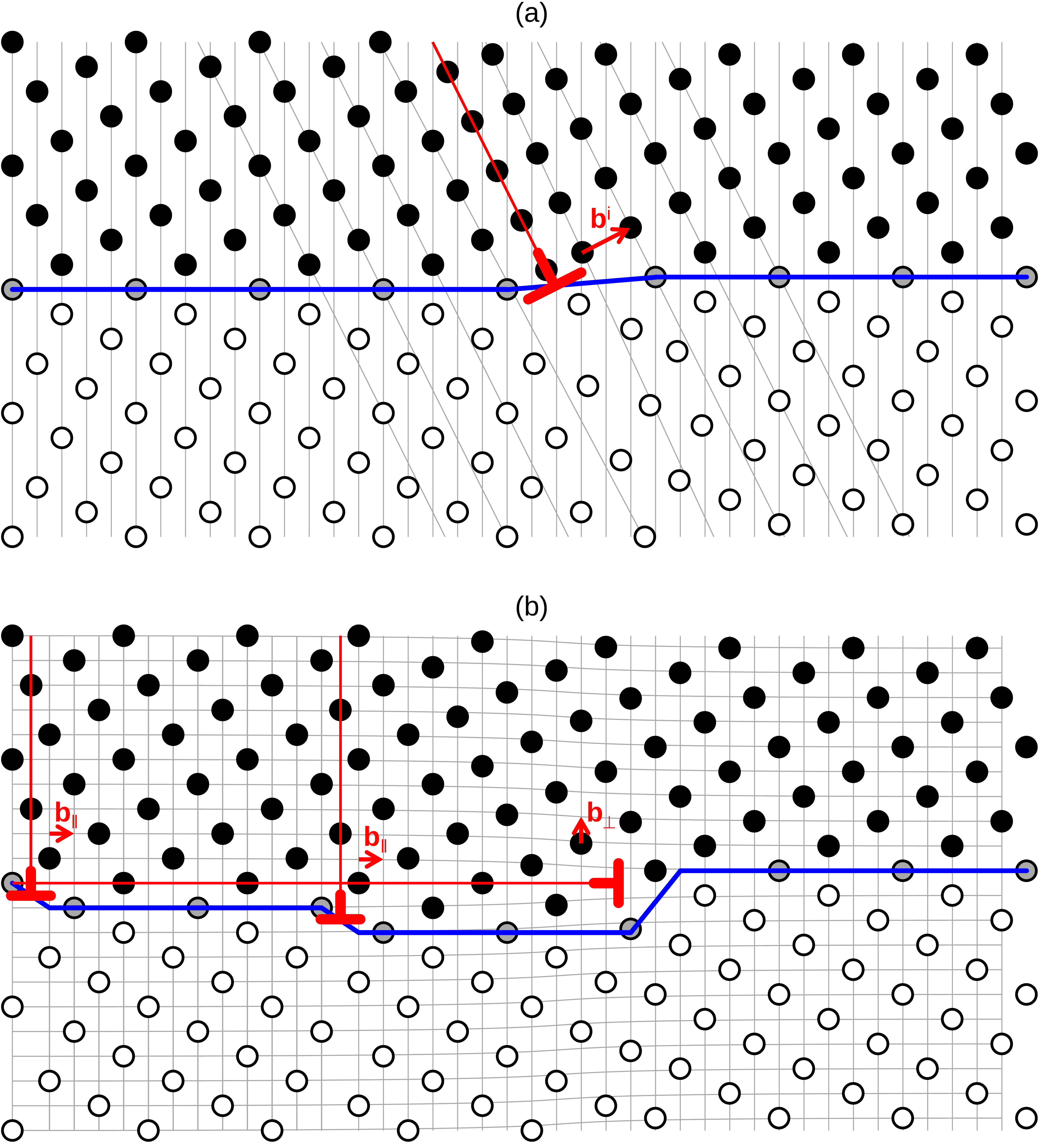}}
\caption{Schematic of absorption of a lattice dislocation into a $\Sigma 5$ $[100]$ $(012)$ STGB in a SC crystal. 
(a) A lattice dislocation with Burgers vector $\mathbf{b}^\text{i}$ from the black grain trapped at the GB. 
(b) Dissociation of the lattice dislocation in (a) into two glissile disconnections $\mathbf{b}_\parallel$ and one sessile disconnection $\mathbf{b}_\perp$. 
The GB plane is denoted by the blue line and the extra DSC lattice planes are denoted by the red lines.  } 
\label{ExDisl}
\end{center}
\end{figure}

We now discuss the interaction between lattice dislocations and GBs based on the disconnection model; we focus on the simple example of a simple cubic crystal. 
Figure~\ref{ExDisl}a shows the case of a lattice dislocation with Burgers vector $\mathbf{b}^\text{i}$ trapped and absorbed into a GB [Scenario (i)] in which the bicrystal contains a $\Sigma 5$ $[100]$ $(012)$ STGB (i.e.,  the example used in Sections~\ref{CSL_DSC} and \ref{GB_disconnection} to introduce  bicrystallography and disconnections). 
The trapped lattice dislocation may  dissociate into several disconnections along the GB. 
Figure~\ref{ExDisl}b shows that such dissociation could be: $\mathbf{b}^\text{i} \rightarrow 2\mathbf{b}_\parallel + \mathbf{b}_\perp$, where $\mathbf{b}_\parallel$ and $\mathbf{b}_\perp$ are the Burgers vectors of disconnections as shown in Figs.~\ref{disconnection}b and c. 
This proposed dissociation is consistent with the conservation of total Burgers vector. 
The two disconnections with Burgers vector $\mathbf{b}_\parallel$ are glissile; they may glide away from the position where the the lattice dislocation is absorbed. 
The disconnection with Burgers vector $\mathbf{b}_\perp$ is sessile; it cannot glide but can climb if vacancies or self-interstitial atoms (SIAs) diffuse into this disconnection. 
We emphasize that, since each disconnection has an associated step height, the dissociation of the lattice dislocation into the disconnections and their motion inevitably leads to local GB migration (or the change of GB profile). 
The step heights also follow the conservation law: $0 \rightarrow 2h_\parallel + h_\perp$~\cite{king1980effects}. 
Similar  lattice dislocation dissociation at GBs accompanied by local change of GB profile has been widely observed in MD simulations (e.g. STGBs in Mg~\cite{wang2014reactions}; $\Sigma 3$, $\Sigma 11$ and $\Sigma 9$ STGBs in Al~\cite{dewald2006multiscale,dewald2007multiscale}; coherent TBs in Al~\cite{jin2006interaction,jin2008interactions,tsuru2009fundamental}; $\Sigma 3$ and $\Sigma 9$ STGBs in W~\cite{cheng2008atomistic}). 
While the atomic-level details of this lattice dislocation absorption has been difficult to observe experimentally (however, see Ref.~\cite{elkajbaji1988interactions} for an example where it was seen), experimental observations have been made without such atomic-scale detail~\cite{minor2004direct,de2006situ}.
Minor~\cite{minor2004direct} performed an in situ nanoindentation experiment on a polycrystalline Al thin film and observed that the continuous annihilation of the lattice dislocations created by indentation at a GB led to the migration of the GB (consistent with Fig.~\ref{ExDisl}). 
We note, however, that this observation does not prove the proposed  scenario (Fig.~\ref{ExDisl}) since it is also possible that GB migration was caused by stress rather than the absorption of  lattice dislocations. 

\begin{figure}[!t]
\begin{center}
\scalebox{0.4}{\includegraphics{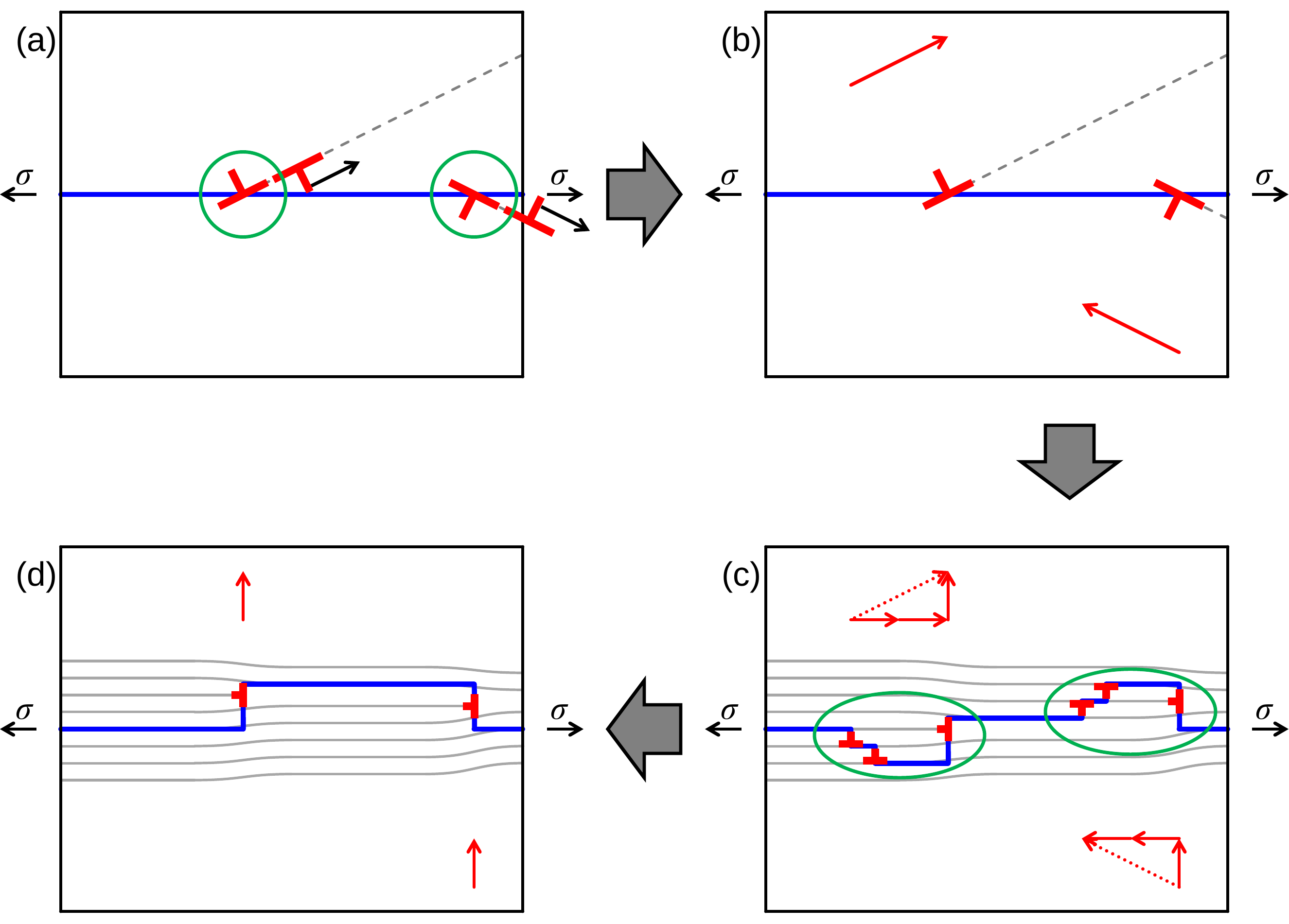}}
\caption{Schematic of emission of lattice dislocations into the upper and lower grains under tensile stress $\sigma$. 
(a) Emission of lattice dislocations. 
(b) Residual lattice dislocations left at the GB after emission. 
(c) Dissociation of the residual lattice dislocations into GB disconnections following the reaction shown in Fig.~\ref{ExDisl}. 
(d) Annihilation of the glissile disconnections. 
The blue line denotes the GB plane; the gray lines denote the DSC lattice planes parallel to the GB; the red arrows denote the Burgers vectors.  } 
\label{DislEmission}
\end{center}
\end{figure}

Based on the disconnection mechanism associated with lattice dislocation absorption depicted in Fig.~\ref{ExDisl}, we can further understand the mechanism of Scenario (ii), i.e., lattice dislocation emission from a GB. 
As illustrated in Figs.~\ref{DislEmission}a and b, when  lattice dislocations are emitted from the GB under the applied tensile stress $\sigma$, according to the conservation of Burgers vector, there must be lattice dislocations (with Burgers vectors of opposite sign) left at the GB. 
Of course, a dislocation may be emitted into the upper or lower grain, each leading to different residual lattice dislocations at the GB. 
As shown in Fig.~\ref{DislEmission}c, the residual lattice dislocations may dissociate into three disconnections in the GB, following the same reaction as that depicted in Fig.~\ref{ExDisl}. 
The glissile disconnections can glide along the GB and annihilate with each other. 
The net result of dislocation emission, dissociation and disconnection annihilation is depicted in Fig.~\ref{DislEmission}d: (1) the GB profile is changed due to introduction of the steps associated with the sessile disconnections and (2) the misorientation of the GB is changed since the two sessile disconnections possess the same Burgers vector perpendicular the GB. 
This emission mechanism has been demonstrated in the MD simulations of Spearot et al. on a $\Sigma 9$ $[110]$ $(221)$ STGB in Cu bicrystal~\cite{spearot2007tensile}. 
They found that, under an applied tensile stress, lattice dislocations (in this case, partial dislocations leading extended stacking faults) were emitted from the GB into both grains leading to the formation of a serrated GB profile. 

\begin{figure}[t]
\begin{center}
\scalebox{0.4}{\includegraphics{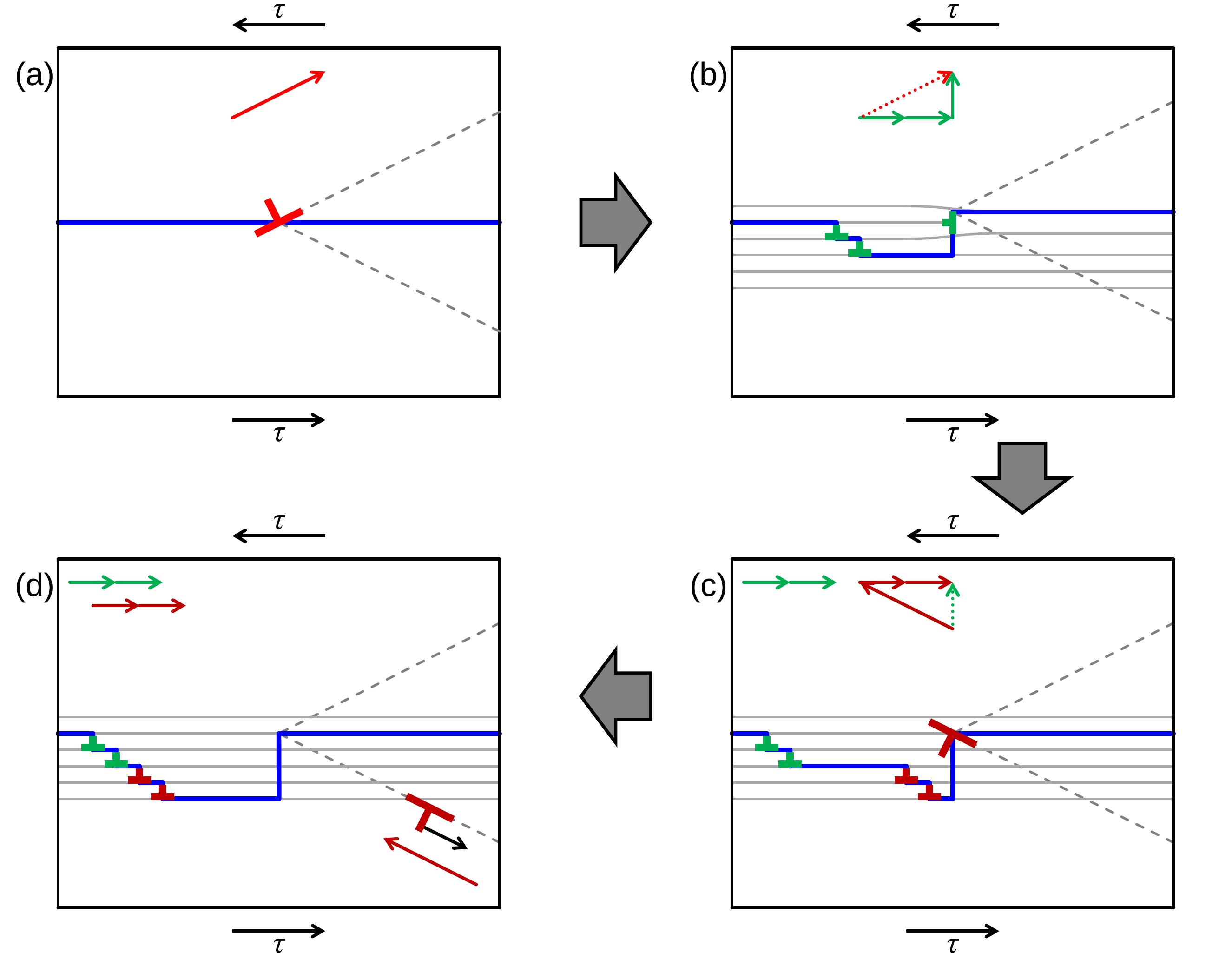}}
\caption{Schematic of direct transmission of a lattice dislocation from the upper grain to the lower one under shear stress $\tau$. 
(a) A lattice dislocation coming from the upper grain and trapped at the GB. 
(b) Dissociation of the lattice dislocation into GB disconnections following the reaction shown in Fig.~\ref{ExDisl}. 
(c) Dissociation of the sessile disconnection into two glissile disconnections and one lattice dislocation in the lower grain. 
(d) Emission of the lattice dislocation in (c) into the lower grain. 
The blue line denotes the GB plane; the gray lines denote the DSC lattice planes parallel to the GB; the red, dark red and green arrows denote the Burgers vectors.  } 
\label{DislTransmission}
\end{center}
\end{figure}

We can also understand the mechanisms of Scenarios (iii) and (iv) (i.e., lattice dislocation transmission across a GB) based on the disconnection model. 
As illustrated in Fig.~\ref{DislTransmission}a, a lattice dislocation can glide from the upper grain under an applied shear stress and be driven into/trapped at the GB. 
Figure~\ref{DislTransmission}b shows that this lattice dislocation may undergo dissociation into disconnections following the reaction shown in Fig.~\ref{ExDisl}. 
If the shear stress is sufficiently large, the sessile disconnection may further dissociate into two glissile disconnections and one lattice dislocation of the lower grain, as shown in Fig.~\ref{DislTransmission}c. 
Then, under a shear stress, the lattice dislocation formed from the dissociation can glide from the GB into the lower grain and all the glissile disconnections can move to the left. 
The net result of such dislocation transmission is the creation of a pure step in the GB, as shown in Fig.~\ref{DislTransmission}d. 
Such a transmission mechanism has been proposed by Hirth et al.~\cite{hirth1974defect,hirth2006disconnections} and by Zhu et al.~\cite{zhu2011dislocation} and Abuzaid et al.~\cite{abuzaid2012slip} particularly for coherent TBs. 
The creation of GB steps due to dislocation transmission has been observed in both experiments~\cite{li2011twinning,abuzaid2012slip,malyar2017size} and MD simulations~\cite{bachurin2010dislocation,sangid2011energy,dewald2011multiscale,chowdhury2013modeling}. 
Lim and Raj also observed that the slip of many dislocations across GBs in bicrystals could cause GB migration, which, on the scale of macroscopic plasticity, implies GB step creation due to dislocation transmission~\cite{lim1985role}. 

Based on the disconnection description of interaction between lattice dislocations and GBs above, we can establish a criterion for the occurrence of dislocation absorption, emission and transmission. 
Since the energy of a dislocation is proportional to $b^2$ ($b$ is the magnitude of Burgers vector),
a dislocation reaction is energetically favorable if the sum of  $b^2$ over all of the product dislocations/disconnections is smaller than that of the reactant dislocations~\cite{frank1949discussion,HirthLothe}.
This idea can be simply extended to the case of disconnections. 
The absorption of a lattice dislocation into a GB [Scenario (i), Fig.~\ref{DislGB_interaction}a] will occur if the energy change due to this absorption event is negative, i.e.,  
\begin{equation}
\Delta E_\text{a}
= \sum_k \left( K b_k^2 + \mathit{\Gamma}_\text{s} |h_k| \right)
- \tau \ell_0 \sum_{\mathbf{b}_k\cdot\mathbf{n}=0} b_k
- K {b^\text{i}}^2 
< 0, 
\label{reaction_absorption}
\end{equation}
where $b^\text{i}$ is the magnitude of Burgers vector of the incoming lattice dislocation, $b_k$ and $h_k$ are, respectively, the magnitude of the Burgers vector and the step height of the $k$-th GB disconnection (i.e., residual dislocation), $\tau$ is the shear stress resolved onto the GB plane, and $\ell_0$ corresponds to the nucleation size of the disconnections; the first summation is over all of the disconnections while the second is over those whose Burgers vectors are parallel to the GB (see Fig.~\ref{ExDisl}). 
We note that, unlike for lattice dislocations, disconnections also have a step energy which must be included.  
Dewald and Curtin~\cite{dewald2006multiscale,dewald2007multiscale} performed multiscale simulations of the interaction between lattice dislocations and $\Sigma 3$, $\Sigma 11$ and $\Sigma 9$ STGBs in Al and found small steps associated with residual GB dislocations (i.e., disconnections).
Equation~\eqref{reaction_absorption}  also includes a contribution from the work done by the stress (there are stress contributions from both the applied stress and the stress from other dislocations). 

Emission of a lattice dislocation from a GB [Scenario (ii); see Fig.~\ref{DislGB_interaction}b] will occur if the energy change due to this emission event is negative, i.e.,  
\begin{equation}
\Delta E_\text{e}
= \sum_k \left( K b_k^2 + \mathit{\Gamma}_\text{s} |h_k| \right)
- \tau \ell_0 \sum_{\mathbf{b}_k\cdot\mathbf{n}=0} b_k
+ K{b^\text{o}}^2
- \tau^\text{o} \ell_0 b^\text{o}
< 0, 
\end{equation}
where $\tau^\text{o}$ is the shear stress resolved to the slip plane of the outgoing lattice dislocation and $b^\text{o}$ is the magnitude of Burgers vector of the outgoing lattice dislocation. 
Obviously,  emission cannot occur (always $\Delta E_\text{e}>0$) unless a stress  (of sufficient magnitude) is applied. 
Similarly, direct transmission of a lattice dislocation across a GB [Scenario (iii); see Fig.~\ref{DislGB_interaction}c] will occur if the energy change due to this transmission event is negative, i.e.,   
\begin{equation}
\Delta E_\text{t}
= \sum_k \left( K b_k^2 + \mathit{\Gamma}_\text{s} |h_k| \right)
- \tau \ell_0 \sum_{\mathbf{b}_k\cdot\mathbf{n}=0} b_k
+ K{b^\text{o}}^2
- \tau^\text{o} \ell_0 b^\text{o}
- K{b^\text{i}}^2
< 0. 
\end{equation}
Indirect transmission of a lattice dislocation across a GB [Scenario (iv); see Fig.~\ref{DislGB_interaction}d] can be decomposed into two steps: absorption and emission. 
Indirect transmission will occur if $\Delta E_\text{a}<0$ and $\Delta E_\text{e}<0$. 

Since the criteria above are based on  energy changes, they only predict whether a reaction is thermodynamically possible. 
On the other hand, whether such a reaction will occur within a fixed time period is determined by kinetics.
The kinetics are determined by the relevant energy landscape/reaction barriers.
Unfortunately, these energy barrier cannot be determined analytically but must be determined from atomic scale calculations that include an adequate description of the bonding.
Such barriers may be determined, for example, using NEB method as done by Tsuru et al.~\cite{tsuru2009fundamental} or via MD simulations as done by Sangid et al.~\cite{sangid2011energy}.

\section{Final remarks}
\label{sec7}

\subsection{Summary of disconnection-based approach for grain-boundary kinetics}
\label{summary}

Grain-boundary (GB) kinetics describes the evolution of the (broadly defined) state of a GB (structure, composition, position and shape). 
Such change can be induced by stress, chemical potential gradient, interaction with defects from the grain interior, etc. 
Change of GB state (driven by these generalized forces) occurs through  the addition, emission, and/or motion of defects  within GBs. 
In this review, we focused on line defects within the GBs (i.e., disconnections).
We investigated many classes of GB kinetic phenomena (related to change of GB structure, position and shape) based on the nucleation, annihilation and/or motion of disconnections. 

A disconnection is a type of line defect that only exists in interfaces (including GBs) and is characterized by both a Burgers vector and a step height. 
Since GBs are crystalline interfaces, the translational symmetry of the DSC lattice (associated with the dichromatic pattern formed by two crystalline grains) implies the possibile existence of line defects which have dislocation character. 
The translational symmetry of the DSC lattice only conserves the dichromatic pattern but allows for the shift the origin of the pattern. Such shifts imply that disconnections may also have GB step character. 
The Burgers vector/step height pair represents a disconnection mode. 
The disconnection mode is determined by the macroscopic degrees of freedom of the GB. 
For a particular bicrystallography, the disconnection mode is not unique. 
We have proposed an algorithm that enumerates all possible disconnection modes for any macroscopic GB geometry. 
We also discussed the disconnection energy that depends on the disconnection mode. 

Based on the disconnection model, we  examined  the thermal equilibration of the GB profile (roughness) and the intrinsic GB mobility. 
Reports in the literature suggest that the GB roughening transition temperature, stiffness and intrinsic GB mobility are influenced by disconnection character; since disconnection modes are determined by bicrystallography, evaluation of these properties based on the disconnection model implies a dependence on bicrystallography. 
Some GBs show  thermal fluctuation behavior associated with the formation and motion of pure steps (disconnections with zero Burgers vector); while others show  behavior dominated by disconnections with nonzero Burgers vectors. 
The existence of disconnections with Burgers vector character implies that equilibrium roughening of  GBs is intrinsically different from the roughening of free surfaces and liquid-solid interfaces (for which the pertinent line defects are pure steps), as demonstrated through Monte Carlo and molecular dynamics~\cite{karma2012relationship} simulations of thermally equilibrated GB. 

Disconnection models can also be used to study conservative GB kinetics, i.e., driven GB migration and sliding without absorption/emission of defects from within the grains. 
Since the mechanism of (diffusionless) GB motion is the nucleation and glide of disconnections characterized by both Burgers vector and step height, GB migration and sliding are coupled (the shear-coupling factor is the ratio of GB sliding velocity to migration velocity)~\cite{cahn2004unified,cahn2006coupling}, or equivalently, the ratio of the GB sliding  to migration distances during steady-state GB motion~\cite{ashby1972boundary}. 
GB sliding can be directly driven by shear stress (resolved to the GB plane). 
Shear stress can induce the nucleation and glide of disconnections with nonzero Burgers vectors; the glide of such disconnections necessarily leads to GB sliding (plastic shear deformation localized at the GB). 
Since these disconnections may be associated with finite step heights,  disconnection glide may  also be accompanied by the propagation of steps along the GB and, thus, GB migration. 
Therefore, the step character of disconnections that are driven by the shear stress can indirectly drive GB migration. 
GB migration can be directly driven by the bulk energy density difference in the two grains (e.g., associated with elastic anisotropy, defect density difference in two grains, magnetic fields, capillarity, etc.). 
The energy density difference can induce the nucleation and glide of disconnections with finite step height; the motion of such disconnections necessarily leads to GB migration. 
Since these disconnections may also have an associated (nonzero) Burgers vector,  disconnection glide may  be accompanied by the propagation of Burgers vector along the GB and, thus, GB sliding. 
Hence, due to the dislocation character of disconnections, the bulk energy density difference can indirectly drive GB sliding. 

Reports in the literature demonstrate that shear-coupled GB migration transforms to pure GB sliding with increasing temperature in some GBs~\cite{cahn2006coupling}; while shear coupling under an applied bulk energy density difference switches to pure GB migration (no sliding) with increasing temperature in some GBs~\cite{homer2013phenomenology}. 
We can understand the temperature dependence of GB shear-coupling behavior based on the statistics of multiple disconnection modes. 
The idea is that, at low temperature, only a single disconnection mode associated with the lowest nucleation energy barrier occurs, while, at high temperature, multiple disconnection modes can be activated and the average step height goes to zero under shear stress or the average Burgers vector goes to zero under the bulk energy density difference. 
In this way, we can understand pure GB sliding and pure GB migration at high temperature based on the disconnection model. 
Since the energy spectrum of disconnection modes is determined by GB geometry, the temperature dependence of the shear-coupling factor results from the statistics of multiple disconnection modes, varies with bicrystallography. 
The predicted temperature dependence from the disconnection model is consistent with MD simulations~\cite{cahn2006coupling,thomas2017reconciling}. 

Based on the statistics of multiple disconnection modes, we  derive a generalized kinetic equation for the motion of GBs in bicrystals~\cite{thomas2017reconciling}: $\mathbf{V}=\mathbf{M}\mathbf{F}$, where $\mathbf{V}=(v_\parallel,v_\perp)^T$ ($v_\parallel$ and $v_\perp$ are the GB sliding and migration velocities), $\mathbf{F}=(\tau,\psi)^T$ ($\tau$ and $\psi$ are the shear stress and the bulk energy density difference) and $\mathbf{M}$ is the temperature-dependent, symmetric GB mobility tensor.
Unlike the classical kinetic equation for capillarity-driven GB migration ($v_\perp=M\psi = M\gamma\kappa$), the generalized kinetic equation considers coupled GB migration and sliding driven by shear stress, bulk energy density difference in any combination. 

While the discrete disconnection model is able to describe many features of GBs, a continuum model is more amenable  to applications in microstructure evolution. We developed such a continuum kinetic equation for GB migration based upon disconnection glide~\cite{zhang2016review}, i.e., $v_\perp = v_\text{d}\varrho^h=M_\text{d}F_\text{d}\varrho^h$, where $v_\text{d}$ is the disconnection glide velocity, $\varrho^h$ is the step density, $M_\text{d}$ is the disconnection mobility and $F_\text{d}$ is the driving force for disconnection motion.
The driving force $F_\text{d}$ may include shear stresses, bulk energy density differences, and GB curvature; the  shear stress term is nonlocal (because the stress is determined by the distribution of Burgers vector along the entire GB) while the other terms are local. 
This continuum equation enables us to numerically study  changes in GB position and shape induced by various types of driving forces~\cite{zhang2016review}. 

While much can be deduced by studying GB motion in the relatively ``tame'' bicrystal geometry, GBs in the ``wild'' are more commonly part of polycrystalline microstructures. 
The kinetics of GB motion in a polycrystal differs from that in a bicrystal since GBs in a polycrystal are joined by triple junctions (TJs) and the grains in a polycrystal cannot undergo arbitrary shear because of the constraint of the surrounding grains. 
The motion of a TJ can be understood as the result of reactions amongst  disconnections flowing into or out of the TJ from the three GBs that meet there. 
In the case where the  disconnections that react are pure steps,  TJ motion is driven by the balance of surface tension (equivalently, reduction of total GB energy); this case is classical. 
However, in the more common case, where the reacting disconnections have nonzero Burgers vector,  TJ motion can be driven by stress, as observed in both experiments~\cite{miura1988effect} and MD simulations~\cite{aramfard2014influences}. 
Whether  microstructure evolution is dominated by GB or TJ motion depends on the competition between the disconnection velocity (scaled by grain size) along the GBs and the rate of their reactions at the TJ. 
In most cases, TJ motion will only be possible if multiple disconnection modes operate along the GBs or the system is able to expel or emit lattice dislocations within the vicinity of the TJs.

The disconnection mechanism, by which shear-coupled GB migration occurs, implies some specific features for polycrystal microstructure evolution, such as capillarity-driven grain growth.
Since GB migration is shear coupled, grain growth leads to stress accumulation. 
For grain growth to occur continuously, one or more mechanisms must operate to relieve the stresses generated by shear coupling. 
Three possible mechanisms are proposed: (i) plastic deformation (e.g.,  twinning), (ii) grain rotation, and (iii) adjustment of the GB shear-coupling factor on each GB such that the effective shear-coupling factor is infinity. 
Mechanisms (i) and (ii) have been readily observed in experiments~\cite{jin2014annealing,yamasaki1996grain,harris1998grain,shan2004grain} and MD simulations~\cite{farkas2007linear,thomas2016twins,thomas2017reconciling}. 
There is no evidence for mechanism (iii); this mechanism implies that the rate of GB migration is low in comparison with that in a bicrystal.  

Non-conservative GB kinetics involves the interaction between GBs and point defects or lattice dislocations. 
The absorption/emission of point defects or lattice dislocations into/out of a GB will introduce disconnections with nonzero Burgers vector component in the direction normal to the GB plane or the motion of such disconnections. These disconnections imply extra/missing DSC lattice plane parallel to the GB. 

A GB can be a sink/source to vacancies/self-interstitial atoms (SIAs); the mechanism is associated with the climb of disconnections with nonzero normal component of Burgers vector along the GB. 
A GB can also be a sink/source of solute atoms; the mechanism is associated with the motion of disconnections with nonzero step height along the GB (unlike the case of GB segregation, where GBs can only absorb a finite quantity of solute atoms). 
Based on the disconnection mechanism, a boundary condition (BC) can be proposed to model the sink effect of a GB~\cite{hirth1996steps,pond1997diffusive,hirth2013interface}. 
This BC is of the type of Robin BC and includes both disconnection Burgers vector and step height as parameters. 
We have shown that this BC can be used to study the problems such as: (i) GB sink effect to vacancies and SIAs under irradiation and creep, (ii) interface sink effect to solute atoms during precipitate growth, and (iii) diffusion-induced GB migration. 

A GB can also be a sink/source of lattice dislocations. 
The mechanism is associated with reactions of disconnections occurring at the GB. 
This reaction may result in disconnections with Burgers vectors parallel to the GB; the introduction or annihilation of these disconnections can lead to local GB migration and changes in the GB shape. 
The disconnection reaction may also result in the formation or annihilation of disconnections with Burgers vectors perpendicular to GB; these disconnections are sessile and lead to local changes in GB misorientation. 
For any particular bicrystallography, we can understand the mechanism of absorption/emission/transmission of lattice dislocations into/out of/across the GB based on the disconnection model. 
The reactions associated with dislocation transmission across a GB commonly leads to a decrease (minimization) of the residual disconnections in the GB~\cite{abuzaid2012slip,kacher2014situ}.

\subsection{Advantages of the disconnection-based approach to GB kinetics}
\label{advantage}

We summarize several of the advantages of disconnection-based approaches for the study of GB kinetics. 
\begin{itemize}
\item[(i)] The GB disconnection approach is a mesoscale model; residing between atomic-scale and continuum models. 
In continuum models, GBs are described as continuum surfaces, the properties associated with such continuum surfaces are GB energy and mobility, and GB motion is modeled as the evolution of such continuum surfaces. 
Typical continuum models provide no link between GB bicrystallography and GB properties, including stress.  
On the other hand,  atomic-scale models explicitly represent GBs in terms of the geometric arrangement of atoms at the GB (such as through the structural unit model) and GB motion is naturally modeled as the cooperative motion of atoms at the GBs or rearrangements of their atomic-scale structural units. 
However, the atomic structure of most GBs are difficult to determine experimentally and atomistic simulation are challenged to describe GB kinetics on appropriate time scales or microstructures on appropriate length scales.
Hence, the atomic-scale models, while informative, rarely suffice to describe GB kinetics.
The disconnection model requires no detail of the GB structure on the  atomic-scale  but does incorporate the effect of bicrstyallography; disconnection character is fixed by the macroscopic degrees of freedom of a GB. 
In the disconnection model, GB motion is described as the activity of disconnections along the GB; hence, it incorporates many more features than continuum models of GB motion but without the necessity of tracking the motion of each atom. 

\item[(ii)] Based on the disconnection model, we can understand several GB-related kinetic phenomena not  previously understood.  
Perhaps the most important of these is shear-coupled GB migration. 
In experiments and MD simulations, many researchers found that grain growth could be driven by applied stresses~\cite{rupert2009experimental,mompiou2009grain,rajabzadeh2013evidence}. 
At the same time, others reported that capillarity-driven grain growth was often accompanied by accumulation of shear stress or grain rotation~\cite{harris1998grain,shan2004grain,thomas2017reconciling}. 
These observations can be explained based on the coupling between GB migration and GB sliding. 
The microscopic mechanism of this coupling effect is disconnection motion along the GB. 
The fact that each disconnection is characterized by both a Burgers vector and a step height implies that the glide of disconnections along the GB corresponds to simultaneous propagation of both Burgers vector and step height, leading to coupled GB sliding and GB migration. We note that non-disconnection-based approaches have also shown how coupling may occur~\cite{read1950dislocation,cahn2006coupling,cahn2006duality}, albeit in a less general manner.

\item[(iii)]  We have also shown how the disconnection model can be applied to understand some GB-related phenomena which have been previously been understood in terms of specialized models. The disconnection model provides a means to understand many aspects of GB kinetic behavior within a single, unified approach. 
These phenomena include thermal fluctuation of the GB plane (Section~\ref{sec4}), GB roughening transition (Section~\ref{GB_roughening}), intrinsic GB mobility (Section~\ref{intrinsicGBmobility}), pure GB sliding (Section~\ref{temperature_stress}), pure GB migration (Section~\ref{temperature_synthetic}), GB sink/source strength for point defects (Section~\ref{interaction_pd}) and the interaction between GBs and lattice dislocations (Section~\ref{GBLattDisl}). 

\item[(iv)]  The disconnection model can be readily extended to predict the kinetics of polycrystalline microstructure. 
Besides GBs,  important elements of such polycrystalline microstructure are triple junctions. 
Triple junction motion can be viewed as resulting from the reactions of  disconnections from the three GBs which meet at the TJs (Section~\ref{TJconstraints}). 
The GB disconnection model may also be applied to relate plasticity within grains to GB dynamics, since lattice dislocation may be viewed as a special class of disconnections (zero step height and crystal translation vector as Burgers vector). 
In this view,  interactions between GBs and lattice dislocations are fundamentally the dissociation and/or recombination of disconnections (Section~\ref{GBLattDisl}). 
Based on the disconnection climb mechanism,  GB sink behavior for point defects can be modeled as Robin boundary condition, parameterized by the disconnection character (Section~\ref{interaction_pd}).  
Then, diffusion in polycrystals (such as in radiation damage, creep, etc.) can be modeled on the continuum scale by treating the GBs as Robin boundary conditions. 
\end{itemize}

\subsection{Limitations of the disconnection-based approach to GB kinetics}
\label{limitation}

\begin{figure}[!t]
\begin{center}
\scalebox{0.38}{\includegraphics{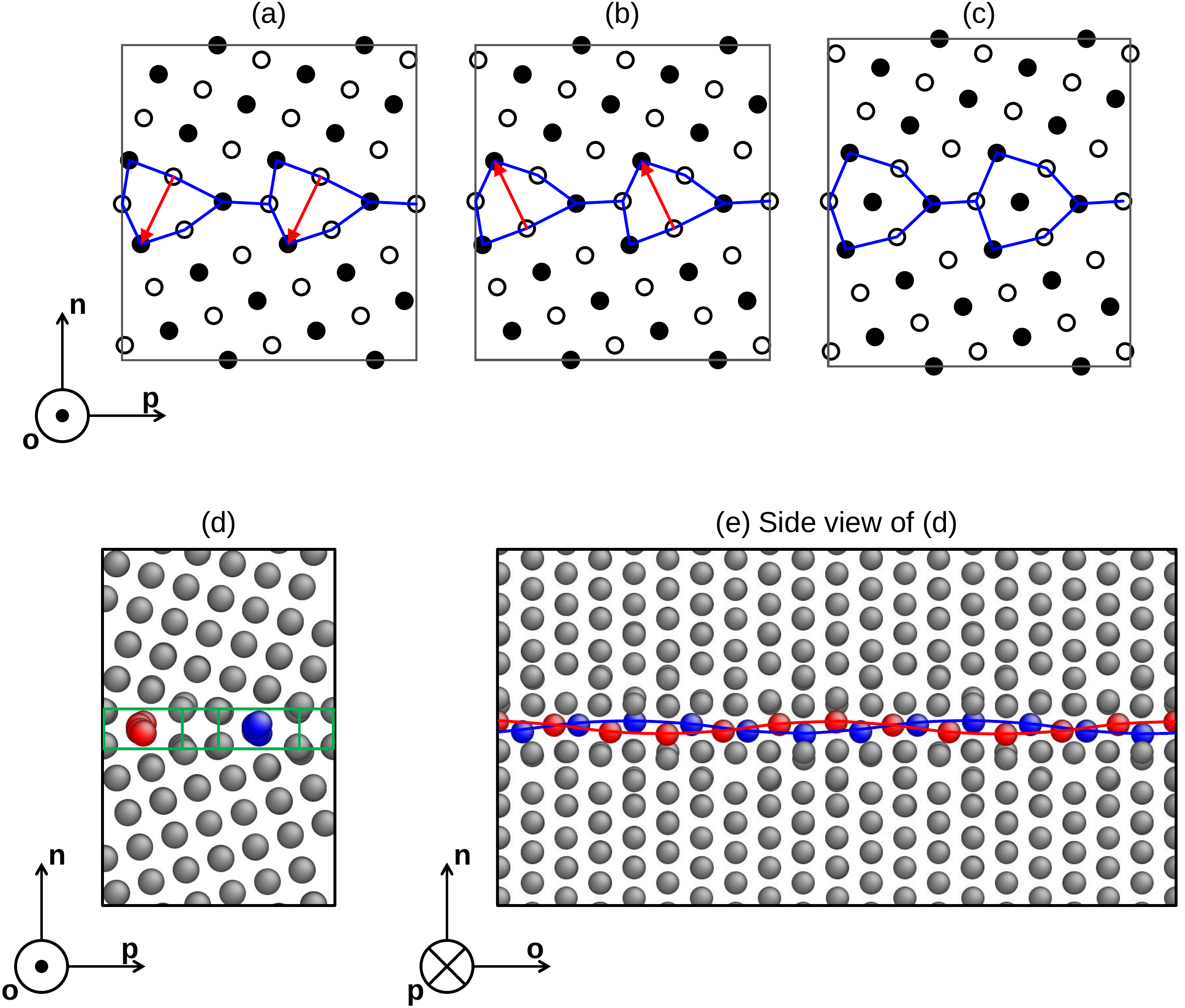}}
\caption{Stable/metastable structures of $\Sigma 5$ $[100]$ $(012)$ STGBs obtained by atomistic simulations. 
The structures in (a), (b) and (c) are obtained by varying the relative displacement of two grains. 
The blue solid lines outline the structural units; the red arrows indicate that the structures in (a) and (b) are related by mirror symmetry. 
The structure in (d) is obtained by varying the atomic fraction in the GB based on the structure in (a). 
(e) shows the side view (i.e., the view along the tilt axis) of (d). 
The green solid lines outline the structural units. 
An array of atoms centered in a structural unit are colored red and another array of atoms in the neighboring structural unit are colored blue. 
(e) shows two arrays of the colored atoms that exhibit a wavy arrangement along the tilt axis.   } 
\label{metastability}
\end{center}
\end{figure}

We summarize several of the limitations of  disconnection-based approaches for the study of GB kinetics. 
\begin{itemize}
\item[(i)] Since the disconnection construction is based upon  the translational symmetry of the DSC lattice, disconnection are only well-defined for CSL GBs but not for non-CSL (irrational) GBs -- in principle. 
Hence, the application of disconnection model might be limited to a small group of special GBs. 
However, as discussed in Section~\ref{disconnection_for_CSL},  any non-CSL set of GB misorientations can modeled as a very close CSL misorientation, with the non-CSL features described by the addition of an aperiodic set of disconnections with Burgers vector perpendicular to the GB. This limitation should be thought of as more conceptual than real.

\item[(ii)] Disconnections are only one type of line defect in a GB which is allowed by the macroscopic DOFs. 
There are other types of line defects associated with the microscopic DOFs of a GB; i.e., the relative displacement of two grains with respect to one another and atomic fraction. 
By changing the relative displacement (for fixed macroscopic DOFs), we may obtain multiple metastable GB structures \cite{han2016grain}. 
For example, Figs.~\ref{metastability}a, b and c show the stable/metastable GB structures obtained by varying the relative displacement for $\Sigma 5$ $[100]$ $(012)$ STGBs in W via atomistic simulations~\cite{han2016grain}. 
The structures in Figs.~\ref{metastability}a and b are related by mirror symmetry and have identical energy. 
The line defect formed between these two structures is a partial GB dislocation with zero stacking-fault energy. 
The Burgers vector of the partial GB dislocation does not belong to any DSC lattice vector. 
The structures in Figs.~\ref{metastability}a and c are two stable/metastable structures and have different energy.  
The line defect formed between these two structures is a partial GB dislocation with finite stacking-fault energy. 
By changing the atomic fraction within a GB (for fixed macroscopic DOFs), we may also obtain multiple metastable GB structures. 
For example, Fig.~\ref{metastability}d shows the metastable GB structure obtained by varying the atomic fraction~\cite{han2017grain}. 
Figure~\ref{metastability}e shows a view of this structure along the $\mathbf{o}$-axis (tilt axis). 
We see that this structure has a larger period along both the $\mathbf{o}$- and $\mathbf{p}$-axes. 
The line defect formed between two such structures differ by a shift along the $\mathbf{o}$- or $\mathbf{p}$-axis and can be thought of as an anti-phase boundary (a boundary in the GB plane is a line defect). 
In summary, the types of line defects in GBs include disconnections, partial GB dislocations with zero stacking-fault energy, partial GB dislocations with finite stacking-fault energy, anti-phase boundaries, and their mixtures~\cite{han2016grain}. 
GB kinetics will, in general, involve all types of line defects. 
However, disconnections are expected to play the most important role in GB kinetics in most situations. 
All the line defects, except for disconnections, have no associated step height and, thus, do not contribute to GB migration. 
Anti-phase boundary are not associated with a Burgers vector and, thus, do not contribute to shear deformation. 
Although partial GB dislocations do have Burgers vectors, these Burgers vector do not correspond to translational symmetries and, so, a partial GB dislocation may be viewed as a part of a complete disconnection~\cite{khater2012disconnection} (in the same way that a partial dislocation is related to a full dislocation in crystal plasticity). 

\item[(iii)] Prediction of many types of GB kinetic behavior based on disconnection model requires knowledge of the active disconnection modes and their associated activation energies for motion as input. 
The list of possible disconnection modes can be determined from bicrystallography alone, but the relative ease with which different disconnection modes may be formed or move depending on details of bonding and core information that is only available from atomic-scale approaches (such as first-principles or empirical potential-based atomistic simulations).
Hence, the application of the disconnection model for concrete predictions of the kinetic behavior of specific GBs in specific materials depends on such atomic-scale information. 
Nonetheless, just as mesoscale dislocation models have been integral in the formation of our understanding of plasticity in crystals in the absence of such information, we expect that mesoscale disconnection model should form the basis of how we understand microstructure evolution in polycrystalline materials. 
\end{itemize}

\subsection{Outlook}
\label{outlook}

Finally, we  discuss several outstanding challenges and directions for the application of disconnection-based approach for GB kinetics and avenues for future research. 
\begin{itemize}
\item[(i)] Most of the existing disconnection-based approaches discussed above have focused on 1D or quasi-1D GBs in a 2D bicrystal. 
It is of practical interest to extend these approaches more completely to the study of 2D GBs in 3D bicrystals. 
Roughening of 2D free surfaces has been studied based on pure step model (solid-on-solid model) (e.g. Ref.~\cite{swendsen1977monte,mon1990interface}). 
The challenge of extending this approach to 2D GBs lies in the fact that disconnections are line defects in GBs characterized by both  step height and Burgers vector  and that the interaction between disconnections are long-ranged.  
Calculation of the long-range interaction between generally curved disconnections in  non-flat 2D GBs may  be implemented using existing techniques. 
For a lattice model (such as the model in Section~\ref{roughening_MC_simulation}), fast-multipole expansion~\cite{lesar2002multipole,wang2004multipole} and discrete dislocation dynamics approaches can be used (e.g. Refs.~\cite{kubin1992dislocation,zbib1998plastic,arsenlis2007enabling,xiang2003level,quek2014polycrystal}), while for continuum models (such as that in Section~\ref{continuum_eqn_for_GB_migration}), continuum dislocation dynamics may be used (e.g. Refs.~\cite{zhu2010continuum,zhu2014continuum}). 

\item[(ii)] Most existing disconnection-based approaches have focused on the ``tame'' bicrystal configuration, rather than GBs in the ``wild'' (i.e., polycrystalline materials). 
Methods for dealing with the complexity of disconnection motion within a polycrystalline context are in their infancy.
We propose three  approaches towards this goal. 
\begin{itemize}
\item[(1)] \textit{Sharp-interface model.} 
GBs can be described as continuum surfaces. 
For example, several researchers (e.g. Refs.~\cite{lazar2011more,mason2015geometric}) studied grain growth in three dimensions based on this approach. 
Such simulations describe the evolution of the GB network following the classical equations for capillarity-driven GB motion. 
However, the classical equations for GB motion do not incorporate the important effects of shear coupling. 
Zhang et al.~\cite{zhang2017equation} recently proposed a continuum equation for GB motion based on the disconnection mechanism (Section~\ref{continuum_eqn_for_GB_migration}) that includes the combined effects of  capillarity, stress and other driving forces. 
This equation can be applied to each GB in a polycrystal and the interaction amongst the three GBs meeting at each TJ can be properly considered (Section~\ref{TJconstraints}); however, this has yet to be implemented.
 
\item[(2)] \textit{Diffuse-interface model.}
A GB can be described by the switching of a continuum field, such as the gradient of crystal orientation~\cite{kobayashi2000continuum}. 
Based on the formal GB dislocation model, the change of orientation can be viewed as resulting from the distribution of dislocations localized around the GB. 
Admal et al. described a GB by the geometrically necessary dislocation density tensor field; such an approach  naturally accounts for  GB  migration under stress~\cite{admal2017unified}. 
In addition, they included GB energy in the free energy by the dependence of the norm of geometrically necessary dislocation density tensor in the phase field model such that they could simulate GB migration driven by capillarity within the same framework~\cite{kobayashi2000continuum}. 
Although  their simulations  only focused on the evolution of GBs in bicrystals, this approach may  be extended to  to investigate the evolution of polycrystal microstructures in a disconnection context. 

\item[(3)] \textit{Potts model.} 
Grain growth can also be modeled using the long-established Potts model combined with Monte Carlo methods (e.g. Refs.~\cite{anderson1984computer,frazier2015abnormal}). 
In the conventional Potts model each lattice point is assigned a value to represent local crystal orientation. 
The energy is described by  interactions between  neighboring lattice points with different orientation values. 
Hence, in this model only GB energy is accounted for; GB migration is driven only by capillarity. 
However, the disconnection mechanism can be incorporated by incorporating shear-coupled GB migration. 
A possible approach is to assign an eigenstrain to the lattice points and include the induced elastic energy in the total  energy (perhaps by microelasticity approach~\cite{khachaturyan2013theory} with discretization according to the lattice in Potts model) that drives the Monte Carlo simulation. 
\end{itemize}

\item[(iii)] More experiments and atomistic simulations are required to describe GBs in more general situations. 
Most current studies of shear-coupled GB migration focus on symmetric tilt boundaries in single-component metallic bicrystals. 
The disconnection approach should be systematically generalized be examination of the following cases. 
\begin{itemize}
\item[(1)] There has been relatively few studies of the motion of general GBs, including asymmetric tilt and mixed tilt-twist boundaries. 
However, several experimental and atomistic simulation studies focused on shear-coupled GB migration in polycrystals and embedded cylindrical grains~\cite{thomas2017reconciling,barrales2016capillarity}. 
Nonetheless, it would be of  interest to study shear coupling (shear-coupling factor and atomic-scale mechanisms) for isolated, flat GBs where inclination is systematically varied. 
One recent exception is the study of Trautt et al.~\cite{trautt2012coupled}, where  shear-coupling factors were determined for a series of $\Sigma 5$ and $\Sigma 25$ asymmetric tilt boundaries using  MD and phase-field cyrstal simulations. 
While the disconnection model was able to accurately reproduce their results (see Section~\ref{ATGB}), this is a limited set. Additionally, little is known about the local mechanism of GB migration in such asymmetric tilt GB cases or anything at all about the motion of mixed tilt-twist boundaries. 

\item[(2)] Most theoretical and simulation studies of GB kinetics focussed on pure materials with primitive crystal lattices. 
An opportunity exists for experiments, simulation and theory on systems with more complex structure; e.g. alloys (disordered or ordered) and the more complex crystal structures associated with intermetallics, ionic crystals, covalent crystals, van der Waals materials, etc. 
In principle, the disconnection model should be equally applicable to such more complex materials, as discussed briefly in Section~\ref{microscopicDOFs}. 
In experiments, there have been observations of disconnections in the primarily ionic  materials Al$_2$O$_3$~\cite{heuer2015disconnection,heuer2016band} and SrTiO$_3$~\cite{sternlicht2015mechanism}.
However, we are unaware of systematic investigations of the role of  disconnections in the GB  migration and shear deformation in such systems or in understanding the role of charged disconnections.
Other opportunities exist in understanding  shear-coupled GB migration in solid solution systems ~\cite{schafer2012influence,schafer2012competing,wang2014shear}.

\end{itemize}
\end{itemize}

%%%%%%%%%%%%%%%%%%%%%%%%%%%%%%%%%%%%%%%%%%%%%%%%%%%%%%%
\section*{Acknowledgments}
\addcontentsline{toc}{section}{\protect\numberline{}Acknowledgments}%
The authors gratefully acknowledge useful discussions with Vaclav Vitek, Gregory S. Rohrer, Yang Xiang, Marc Legros, Yongwei Zhang, Prashant K. Purohit, Kongtao Chen, Phillip Duxbury, Jeffrey M. Rickman, Chaozhen Wei, Luchan Zhang, Chuang Deng, Jaime Marian, Sherri Hadian, and Elizabeth A. Holm.
D.J.S. acknowledges partial support of the National Science Foundation Division of Materials Research through Award 1507013 and Penn's NSF MRSEC through Award DMR-112090. S.L.T. acknowledges support of the Department of Education GAANN program, grant number P200A160282.

%%%%%%%%%%%%%%%%%%%%%%%%%%%%%%%%%%%%%%%%%%%%%%%%%%%%%%%
\section*{References}
\addcontentsline{toc}{section}{\protect\numberline{}References}%
\bibliography{manuscript}

%%%%%%%%%%%%%%%%%%%%%%%%%%%%%%%%%%%%%%%%%%%%%%%%%%%%%%%
\clearpage
\appendix

\section{Practical algorithm for determining disconnection modes for symmetric tilt boundaries in FCC crystals}
\label{APPalgorithm}

We give an example of $[100]$ STGBs in a FCC crystal. 
A $[100]$ STGB in a FCC crystal can be represented by $(0,-m,n)_\text{w}/(0,m,n)_\text{b}$ $[0,n,m]_\text{w}/[0,n,-m]_\text{b}$, where $m$ and $n$ are integers and $n>m>0$; ``w'' and ``b'' denote white and black grains, respectively. 
Below, all indices refer to the black grain. 
The DSC lattice is cubic (with reference to Fig.~\ref{Diophantine}). 
The lattice parameters of the DSC lattice along $x$-, $y$- and $z$-axes are 
\begin{equation} 
a_x = \frac{a_0}{2}, ~
a_y = d_{(0,n,-m)} = \frac{a_0}{2\sqrt{n^2+m^2}}, ~
a_z = d_{(0,m,n)} = \frac{a_0}{2\sqrt{n^2+m^2}}, 
\end{equation}
where $a_0$ is the lattice parameter of conventional FCC unit cell, $d_{(0,n,-m)}$ and $d_{(0,m,n)}$ are interplanar spacings of the $(0,n,-m)$ planes and the $(0,m,n)$ planes, respectively. 
Then, we will determine $\tilde{d}_\text{u}$, $\tilde{d}_\text{1L}$ and $\tilde{d}_\text{gb}$ in the Diophantine equation Eq.~\eqref{tilded}. 
The projection of $[010]a_0$ on the GB plane is 
\begin{equation}
d_y
= [010]a_0 \cdot \frac{1}{\sqrt{n^2+m^2}} [0,n,-m]
= \frac{n a_0}{\sqrt{n^2+m^2}}
\Rightarrow 
\tilde{d}_y 
= \frac{d_y}{a_y}
= 2n. 
\end{equation}
The projection of $[010]a_0$ on the GB normal is 
\begin{equation}
d_z
= [010]a_0 \cdot \frac{1}{\sqrt{n^2+m^2}} [0,m,n]
= \frac{m a_0}{\sqrt{n^2+m^2}}
\Rightarrow 
\tilde{d}_z 
= \frac{d_z}{a_z}
= 2m. 
\end{equation}
The minimum displacement (reduced by $a_y$) which is parallel to the GB and preserves CSL is
\begin{equation}
\tilde{d}_\text{gb}
= \left\{\begin{array}{ll}
n^2+m^2, & n+m=2k \\
2(n^2+m^2), & n+m=2k+1 
\end{array}\right., 
\end{equation} 
where $k$ is any integer ($k$ always denotes integer below). 
The displacement (reduced by $a_y$) which is parallel to the GB and leads to a one-layer upward shift of the GB plane is 
\begin{equation}
\tilde{d}_{\text{1L}}
= 2\left[
\frac{k\tilde{d}_\text{gb}-2\tilde{d}_y}{2\tilde{d}_z}
\right]_k, 
\label{app_d1L}
\end{equation}
where the operation $[A]_k$ is to raise the value of $k$ from one until $A$ becomes an integer and return the value of $A$. 
The displacement (reduced by $a_y$) which is parallel to the GB and equivalent to the effect of displacing the black lattice by the Burgers vector $\mathbf{b}_{\perp} = a_z \mathbf{n}$ is 
\begin{equation}
\tilde{d}_{\text{u}}
= \left[
\frac{k\tilde{d}_\text{gb}-\tilde{d}_y}{\tilde{d}_z}
\right]_k. 
\label{app_du}
\end{equation}
Equations~\eqref{app_d1L} and \eqref{app_du} are valid for any GB geometry. 

The bases of DSC lattice are 
\begin{equation}
\mathbf{b}^{(1)} = (0,a_y,0), ~
\mathbf{b}^{(2)} =(0,0,a_z) \text{ and }
\mathbf{b}^{(3)} = \left(\frac{a_x}{2}, \frac{a_y}{2}, \frac{a_z}{2}\right) 
\end{equation}
if $n+m=2k$; or 
\begin{equation}
\mathbf{b}^{(1)} = \left(0,\frac{a_y}{2},-\frac{a_z}{2}\right), ~
\mathbf{b}^{(2)} = \left(0,\frac{a_y}{2},\frac{a_z}{2}\right) \text{ and }
\mathbf{b}^{(3)} = \left(\frac{a_x}{2},\frac{a_y}{2},0\right)
\end{equation}
if $n+m=2k+1$. 
Any DSC Burgers vector can be represented according to Eq.~\eqref{anyb}. 
The step heights $\tilde{h}^{(1)}$, $\tilde{h}^{(2)}$ and $\tilde{h}^{(3)}$ corresponding to the Burgers vectors $\mathbf{b}^{(1)}$, $\mathbf{b}^{(2)}$ and $\mathbf{b}^{(3)}$ can be obtained by solving the Diophantine equations: 
\begin{equation}
\begin{array}{l}
2 = \tilde{d}_\text{1L}\tilde{h}^{(1)} + \tilde{d}_\text{gb}N, \\
2\tilde{d}_\text{u} = \tilde{d}_\text{1L}\tilde{h}^{(2)} + \tilde{d}_\text{gb}N \text{ and } \\
2+2\tilde{d}_\text{u} = 2\tilde{d}_\text{1L}\tilde{h}^{(3)} + \tilde{d}_\text{gb}N, 
\end{array}
\end{equation}
if $n+m=2k$; or 
\begin{equation}
\begin{array}{l}
2-2\tilde{d}_\text{u} = 2\tilde{d}_\text{1L}\tilde{h}^{(1)} + \tilde{d}_\text{gb}N, \\
2+2\tilde{d}_\text{u} = 2\tilde{d}_\text{1L}\tilde{h}^{(2)} + \tilde{d}_\text{gb}N \text{ and } \\
2 = 2\tilde{d}_\text{1L}\tilde{h}^{(3)} + \tilde{d}_\text{gb}N, 
\end{array}
\end{equation}
if $n+m=2k+1$. 
The closed-form solution for each of these equations is Eq.~\eqref{h0}. 

The way to evaluate the parameters and the Diophantine equations for the other STGBs in FCC crystals are shown in Tables~\ref{Diopara}, \ref{Diobases} and \ref{Dioeqn}. 

%%%%%%%%%%%%%%%%%%%%%%%%%%%%%%%%%%%%%%
\begin{landscape}
\begin{table}[htp]
\begin{center}
\renewcommand{\arraystretch}{1.5}
\caption{\label{Diopara}
Geometric parameters used in the algorithm to calculate the disconnection modes 
for FCC crystals. $k$ is any integer. }
\begin{tabular}{|c|c|c|c|c|c|c|}
\hline
Cases & 
$a_x/a_0$ & $a_y/a_0$ & $a_z/a_0$ &
$\tilde{d}_y$ & $\tilde{d}_z$ & $\tilde{d}_{\text{gb}}$  \\
\hline
%%%%%%%%%%%%%%%%%%%%%%%%%%%%%%%%%%%
\multicolumn{7}{|c|}
{$[100]$ $(0,-m,n)_{\text{w}} / (0,m,n)_{\text{b}}$ ($0<m<n$)} \\
\hline
$n+m = 2k$		& $\frac{1}{2}$		& $\frac{1}{2\sqrt{n^2+m^2}}$		& $\frac{1}{2\sqrt{n^2+m^2}}$	&
$2n$			& $2m$				& $n^2+m^2$				\\
\hline
$n+m = 2k+1$	& $\frac{1}{2}$		& $\frac{1}{2\sqrt{n^2+m^2}}$		& $\frac{1}{2\sqrt{n^2+m^2}}$	&
$2n$			& $2m$				& $2(n^2+m^2)$	\\
\hline
%%%%%%%%%%%%%%%%%%%%%%%%%%%%%%%%%%%
\multicolumn{7}{|c|}
{$[110]$ $(-m,m,-n)_{\text{w}} / (-m,m,n)_{\text{b}}$ ($0<m$, $0<n$)} \\
\hline
$n$ is odd, $m$ is odd	&
$\frac{1}{2\sqrt{2}}$		& $\frac{1}{2\sqrt{2}\sqrt{2m^2+n^2}}$		& $\frac{1}{\sqrt{2m^2+n^2}}$		&
$4m$					& $n$										& $2(2m^2+n^2)$					\\
\hline
$n$ is odd, $m$ is even	& 
$\frac{1}{2\sqrt{2}}$		& $\frac{1}{2\sqrt{2}\sqrt{2m^2+n^2}}$		& $\frac{1}{2\sqrt{2m^2+n^2}}$	 	&
$4m$					& $2n$										& $2(2m^2+n^2)$					\\ 
\hline
$n = 4k$, $m$ is odd 	& 
$\frac{1}{2\sqrt{2}}$		& $\frac{1}{\sqrt{2}\sqrt{2m^2+n^2}}$		& $\frac{1}{2\sqrt{2m^2+n^2}}$		&
$2m$					& $2n$										& $2m^2+n^2$						\\ 
\hline
$n = 4k+2$, $m$ is odd	& 
$\frac{1}{2\sqrt{2}}$		& $\frac{\sqrt{2}}{\sqrt{2m^2+n^2}}$		& $\frac{1}{2\sqrt{2m^2+n^2}}$		&
$m$					& $2n$										& $\frac{1}{2}(2m^2+n^2)$			\\
\hline
%%%%%%%%%%%%%%%%%%%%%%%%%%%%%%%%%%%%
\multicolumn{7}{|c|}
{$[111]_{(110)}$ $(m,-n,-m+n)_{\text{w}} / (n,-m,m-n)_{\text{b}}$ ($0<n<m<2n$)} \\
\hline
$n+m = 3k$				& 
$\frac{1}{\sqrt{3}}$		& $\frac{3}{2\sqrt{6}\sqrt{m^2+n^2-mn}}$	& $\frac{3}{2\sqrt{2}\sqrt{m^2+n^2-mn}}$	&
$m+n$					& $m-n$										& $\frac{2}{3}(m^2+n^2-mn)$				\\
\hline
$n+m \ne 3k$			& 
$\frac{1}{\sqrt{3}}$		& $\frac{3}{2\sqrt{6}\sqrt{m^2+n^2-mn}}$	& $\frac{1}{2\sqrt{2}\sqrt{m^2+n^2-mn}}$	&
$m+n$					& $3(m-n)$									& $2(m^2+n^2-mn)$						\\ 
\hline
%%%%%%%%%%%%%%%%%%%%%%%%%%%%%%%%%%%%
\multicolumn{7}{|c|}
{$[111]_{(112)}$ $(m,n,-m-n)_{\text{w}} / (n,m,-m-n)_{\text{b}}$ ($0<m<n$)} \\
\hline
$n-m = 3k$				& 
$\frac{1}{\sqrt{3}}$		& $\frac{3}{2\sqrt{6}\sqrt{m^2+n^2+mn}}$	& $\frac{1}{2\sqrt{2}\sqrt{m^2+n^2+mn}}$	&
$m+n$					& $n-m$										& $\frac{2}{3}(m^2+n^2+mn)$				\\
\hline
$n-m \ne 3k$ 			& 
$\frac{1}{\sqrt{3}}$		& $\frac{1}{2\sqrt{6}\sqrt{m^2+n^2+mn}}$	& $\frac{1}{2\sqrt{2}\sqrt{m^2+n^2+mn}}$ 	&
$3(m+n)$				& $n-m$										& $2(m^2+n^2+mn)$						\\ 
\hline
\end{tabular}
\end{center}
\end{table}
\end{landscape}

%%%%%%%%%%%%%%%%%%%%%%%%%%%%%%%%%%%
\begin{landscape}
\begin{table}[htp]
\begin{center}
\renewcommand{\arraystretch}{1.5}
\caption{\label{Diobases}
Bases of DSC lattice for STGBs in FCC crystals (with geometry listed in Table~\ref{Diopara}).  }
\begin{tabular}{|c|c|c|c|}
\hline
Cases & 
$\mathbf{b}^{(1)}$ & $\mathbf{b}^{(2)}$ & $\mathbf{b}^{(3)}$ \\
\hline
%%%%%%%%%%%%%%%%%%%%%%%%%%%%%%%%%%%
\multicolumn{4}{|c|}
{$[100]$ $(0,-m,n)_{\text{w}} / (0,m,n)_{\text{b}}$ ($0<m<n$)} \\
\hline
$n+m = 2k$		& 
$\left(0,a_y,0\right)$		& 
$\left(0,0,a_z\right)$		& 
$\left(\frac{a_x}{2},\frac{a_y}{2},\frac{a_z}{2}\right)$	\\
\hline
$n+m = 2k+1$	& 
$\left(0,\frac{a_y}{2},-\frac{a_z}{2}\right)$		& 
$\left(0,\frac{a_y}{2},\frac{a_z}{2}\right)$		& 
$\left(\frac{a_x}{2},\frac{a_y}{2},0\right)$		\\
\hline
%%%%%%%%%%%%%%%%%%%%%%%%%%%%%%%%%%%
\multicolumn{4}{|c|}
{$[110]$ $(-m,m,-n)_{\text{w}} / (-m,m,n)_{\text{b}}$ ($0<m$, $0<n$)} \\
\hline
$n$ is odd, $m$ is odd	&
$\left(0,a_y,0\right)$	& 
$\left(0,0,\frac{a_z}{2}\right)$		& 
$\left(\frac{a_x}{2},\frac{a_y}{2},0\right)$		\\
\hline
$n$ is odd, $m$ is even	& 
$\left(0,a_y,0\right)$	& 
$\left(0,0,a_z\right)$		& 
$\left(\frac{a_x}{2},\frac{a_y}{2},\frac{a_z}{2}\right)$		\\ 
\hline
$n = 4k$, $m$ is odd 	& 
$\left(0,a_y,0\right)$	& 
$\left(0,0,a_z\right)$		& 
$\left(\frac{a_x}{2},\frac{a_y}{2},\frac{a_z}{2}\right)$		\\ 
\hline
$n = 4k+2$, $m$ is odd	& 
$\left(0,\frac{a_y}{2},0\right)$	& 
$\left(0,0,a_z\right)$		& 
$\left(\frac{a_x}{2},0,\frac{a_z}{2}\right)$		\\
\hline
%%%%%%%%%%%%%%%%%%%%%%%%%%%%%%%%%%%%
\multicolumn{4}{|c|}
{$[111]_{(110)}$ $(m,-n,-m+n)_{\text{w}} / (n,-m,m-n)_{\text{b}}$ ($0<n<m<2n$)} \\
\hline
$n+m = 3k$, $m-n = 3k+1$	& 
$\left(0,a_y,-\frac{a_z}{3}\right)$ &
$\left(0,a_y,\frac{a_z}{3}\right)$ &
$\left(a_x,0,\frac{2a_z}{3}\right)$		\\
\hline
$n+m = 3k$, $m-n = 3k+2$	& 
$\left(0,a_y,\frac{a_z}{3}\right)$ &
$\left(0,a_y,-\frac{a_z}{3}\right)$ &
$\left(-a_x,0,\frac{2a_z}{3}\right)$		\\
\hline
$n+m = 3k+1$ 			& 
$\left(0,a_y,-a_z\right)$ &
$\left(0,a_y,a_z\right)$ &
$\left(a_x,\frac{2a_y}{3},0\right)$	\\
\hline
$n+m = 3k+2$ 			& 
$\left(0,a_y,-a_z\right)$ &
$\left(0,a_y,a_z\right)$ &
$\left(a_x,-\frac{2a_y}{3},0\right)$	\\
\hline
%%%%%%%%%%%%%%%%%%%%%%%%%%%%%%%%%%%%
\multicolumn{4}{|c|}
{$[111]_{(112)}$ $(m,n,-m-n)_{\text{w}} / (n,m,-m-n)_{\text{b}}$ ($0<m<n$)} 	\\
\hline
$n-m = 3k$ & 
$\left(0,a_y,-a_z\right)$ &
$\left(0,a_y,a_z\right)$ &
$\left(a_x,0,\frac{2a_z}{3}\right)$	\\
\hline
$n-m \ne 3k$ & 
$\left(0,\frac{a_y}{3},-a_z\right)$ &
$\left(0,\frac{a_y}{3},a_z\right)$ &
$\left(a_x,\frac{2a_y}{3},0\right)$			\\
\hline
\end{tabular}
\end{center}
\end{table}
\end{landscape}

%%%%%%%%%%%%%%%%%%%%%%%%%%%%%%%%%%%%%%%
\begin{landscape}
\begin{table}[htp]
\begin{center}
\renewcommand{\arraystretch}{1.5}
\caption{\label{Dioeqn}
Diophantine equations for calculating the step heights corresponding to the bases of DSC lattice listed in Table~\ref{Diobases}. 
$k$ is any integer and $N$ is integer. 
The coefficient $\tilde{d}_\text{gb}$ is given in Table~\ref{Diopara}. 
The coefficients $\tilde{d}_\text{1L}$ and $\tilde{d}_\text{u}$ are obtained by $\tilde{d}_y$ and $\tilde{d}_z$ in Table~\ref{Diopara} along with Eqs.~\eqref{app_d1L} and \eqref{app_du}. }
\begin{tabular}{|c|c|c|c|}
\hline
Cases & 
Eqn. for $\tilde{h}^{(1)}$		& Eqn. for $\tilde{h}^{(2)}$ 		& Eqn. for $\tilde{h}^{(3)}$		\\
\hline
%%%%%%%%%%%%%%%%%%%%%%%%%%%%%%%%%%%%
\multicolumn{4}{|c|}
{$[100]$ $(0,-m,n)_{\text{w}} / (0,m,n)_{\text{b}}$ ($0<m<n$)} \\
\hline
$n+m = 2k$				& 
$2 = \tilde{d}_{\text{1L}}\tilde{h}_1 + \tilde{d}_{\text{gb}}N$						& 
$2\tilde{d}_{\text{u}} = \tilde{d}_{\text{1L}}\tilde{h}_2 + \tilde{d}_{\text{gb}}N$		&
$2+2\tilde{d}_{\text{u}} = 2\tilde{d}_{\text{1L}}\tilde{h}_3 + \tilde{d}_{\text{gb}}N$	\\
\hline
$n+m = 2k+1$			& 
$2-2\tilde{d}_{\text{u}} = 2\tilde{d}_{\text{1L}}\tilde{h}_1 + \tilde{d}_{\text{gb}}N$	& 
$2+2\tilde{d}_{\text{u}} = 2\tilde{d}_{\text{1L}}\tilde{h}_2 + \tilde{d}_{\text{gb}}N$	&
$2 = 2\tilde{d}_{\text{1L}}\tilde{h}_3 + \tilde{d}_{\text{gb}}N$						\\
\hline
%%%%%%%%%%%%%%%%%%%%%%%%%%%%%%%%%%%%
\multicolumn{4}{|c|}
{$[110]$ $(-m,m,-n)_{\text{w}} / (-m,m,n)_{\text{b}}$ ($0<m$, $0<n$)} \\
\hline
$n$ is odd, $m$ is odd	& 
$2 = \tilde{d}_{\text{1L}}\tilde{h}_1 + \tilde{d}_{\text{gb}}N$						& 
$\tilde{d}_{\text{u}} = \tilde{d}_{\text{1L}}\tilde{h}_2 + \tilde{d}_{\text{gb}}N$		&
$2 = 2\tilde{d}_{\text{1L}}\tilde{h}_3 + \tilde{d}_{\text{gb}}N$						\\
\hline
$n$ is odd, $m$ is even	& 
$2 = \tilde{d}_{\text{1L}}\tilde{h}_1 + \tilde{d}_{\text{gb}}N$						& 
$2\tilde{d}_{\text{u}} = \tilde{d}_{\text{1L}}\tilde{h}_2 + \tilde{d}_{\text{gb}}N$		&
$2+2\tilde{d}_{\text{u}} = 2\tilde{d}_{\text{1L}}\tilde{h}_3 + \tilde{d}_{\text{gb}}N$	\\
\hline
$n = 4k$, $m$ is odd		& 
$2 = \tilde{d}_{\text{1L}}\tilde{h}_1 + \tilde{d}_{\text{gb}}N$						& 
$2\tilde{d}_{\text{u}} = \tilde{d}_{\text{1L}}\tilde{h}_2 + \tilde{d}_{\text{gb}}N$		&
$2+2\tilde{d}_{\text{u}} = 2\tilde{d}_{\text{1L}}\tilde{h}_3 + \tilde{d}_{\text{gb}}N$	\\
\hline
$n = 4k+2$, $m$ is odd & 
$1 = \tilde{d}_{\text{1L}}\tilde{h}_1 + \tilde{d}_{\text{gb}}N$						& 
$2\tilde{d}_{\text{u}} = \tilde{d}_{\text{1L}}\tilde{h}_2 + \tilde{d}_{\text{gb}}N$		&
$2\tilde{d}_{\text{u}} = 2\tilde{d}_{\text{1L}}\tilde{h}_3 + \tilde{d}_{\text{gb}}N$	\\
\hline
%%%%%%%%%%%%%%%%%%%%%%%%%%%%%%%%%%%%
\multicolumn{4}{|c|}
{$[111]_{(110)}$ $(m,-n,-m+n)_{\text{w}} / (n,-m,m-n)_{\text{b}}$ ($0<n<m<2n$)} \\
\hline
$n+m = 3k$, $m-n = 3k+1$	& 
$6-2\tilde{d}_{\text{u}} = 2\tilde{d}_{\text{1L}}(3\tilde{h}_1) + 3\tilde{d}_{\text{gb}}N$	& 
$6+2\tilde{d}_{\text{u}} = 2\tilde{d}_{\text{1L}}(3\tilde{h}_2) + 3\tilde{d}_{\text{gb}}N$	&
$2\tilde{d}_{\text{u}} = 3\tilde{d}_{\text{1L}}\tilde{h}_3 + \tilde{d}_{\text{gb}}N$		\\
\hline
$n+m = 3k$, $m-n = 3k+2$	& 
$6+2\tilde{d}_{\text{u}} = 2\tilde{d}_{\text{1L}}(3\tilde{h}_1) + 3\tilde{d}_{\text{gb}}N$	& 
$6-2\tilde{d}_{\text{u}} = 2\tilde{d}_{\text{1L}}(3\tilde{h}_2) + 3\tilde{d}_{\text{gb}}N$	&
$2\tilde{d}_{\text{u}} = 3\tilde{d}_{\text{1L}}\tilde{h}_3 + \tilde{d}_{\text{gb}}N$		\\
\hline
$n+m = 3k+1$ 			& 
$2-2\tilde{d}_{\text{u}} = 2\tilde{d}_{\text{1L}}\tilde{h}_1 + \tilde{d}_{\text{gb}}N$		& 
$2+2\tilde{d}_{\text{u}} = 2\tilde{d}_{\text{1L}}\tilde{h}_2 + \tilde{d}_{\text{gb}}N$		&
$2 = 3\tilde{d}_{\text{1L}}\tilde{h}_3 + \tilde{d}_{\text{gb}}N$							\\
\hline
$n+m = 3k+2$ 			& 
$2-2\tilde{d}_{\text{u}} = 2\tilde{d}_{\text{1L}}\tilde{h}_1 + \tilde{d}_{\text{gb}}N$		& 
$2+2\tilde{d}_{\text{u}} = 2\tilde{d}_{\text{1L}}\tilde{h}_2 + \tilde{d}_{\text{gb}}N$		&
$-2 = 3\tilde{d}_{\text{1L}}\tilde{h}_3 + \tilde{d}_{\text{gb}}N$							\\
\hline
%%%%%%%%%%%%%%%%%%%%%%%%%%%%%%%%%%%%
\multicolumn{4}{|c|}
{$[111]_{(112)}$ $(m,n,-m-n)_{\text{w}} / (n,m,-m-n)_{\text{b}}$ ($0<m<n$)} 	\\
\hline
$n-m = 3k$ & 
$2-6\tilde{d}_{\text{u}} = 2\tilde{d}_{\text{1L}}\tilde{h}_1 + \tilde{d}_{\text{gb}}N$		& 
$2+6\tilde{d}_{\text{u}} = 2\tilde{d}_{\text{1L}}\tilde{h}_2 + \tilde{d}_{\text{gb}}N$		&
$6\tilde{d}_{\text{u}} = 3\tilde{d}_{\text{1L}}\tilde{h}_3 + \tilde{d}_{\text{gb}}N$		\\
\hline
$n-m \ne 3k$ & 
$2-2\tilde{d}_{\text{u}} = 2\tilde{d}_{\text{1L}}\tilde{h}_1 + \tilde{d}_{\text{gb}}N$		& 
$2+2\tilde{d}_{\text{u}} = 2\tilde{d}_{\text{1L}}\tilde{h}_2 + \tilde{d}_{\text{gb}}N$		&
$6 = 3\tilde{d}_{\text{1L}}\tilde{h}_3 + \tilde{d}_{\text{gb}}N$							\\
\hline
\end{tabular}
\end{center}
\end{table}
\end{landscape}

\section{Details of Monte Carlo simulation for grain-boundary roughening}
\label{APPderivation}

The model for MC simulation is shown in Fig.~\ref{GBdescription} and described in Section~\ref{roughening_MC_simulation}. 
First, we derive Eq.~\eqref{tau_i}. 
Based on the coordinate system shown in Fig.~\ref{disconnection}, since periodic boundary condition is applied in the $y$-direction and the period is $L_y$, when there is one dislocation located at $y=0$, we will have an array of identical dislocations located at $y=nL_y$ ($n$ is integer). 
If each dislocation is of edge character and has Burgers vector $\mathbf{b}=b\hat{\mathbf{y}}$, the shear stress due to this array of edge dislocations will be~\cite{HirthLothe}
\begin{align}
\sigma_{yz}(y,z)
&= -\frac{2\mathcal{K}b}{L_y}
\sum_{n=-\infty}^\infty \frac{(Y-n)\left[(Y-n)^2-Z^2\right]}{\left[(Y-n)^2+Z^2\right]^2}
\nonumber \\
&= \sigma_0 \sin(2\pi Y)\left[
\cosh(2\pi Z)-\cos(2\pi Y) - 2\pi Z\sinh(2\pi Z)
\right], 
\end{align}
where
\begin{equation}
\sigma_0
\equiv -\frac{2\pi\mathcal{K}b}{L_y}
\frac{1}{\left[\cosh(2\pi Z)-\cos(2\pi Y)\right]^2}, 
\end{equation}
$Y\equiv y/L_y$ and $Z\equiv z/L_y$. 
Then, the stress at $z=0$ due to an array of edge dislocations located at $y'+nL_y$ is 
\begin{equation}
\sigma_{yz}(y,0)
= \frac{2\pi\mathcal{K}}{L_y} b\cot\left[\frac{\pi}{L_y}(y'-y)\right] \equiv \tau. 
\label{sigma_yz_continuum}
\end{equation}
From Eq.~\eqref{sigma_yz_continuum} and based on the discrete lattice model shown in Fig.~\ref{GBdescription}, the shear stress exerted on the $i'$-th lattice point is 
\begin{equation}
\tau_{i'} 
= \frac{2\pi\mathcal{K}}{L_y}
\sum_{l=1}^N
(u_l-u_{l-1}) \cot\left[\frac{\pi}{N}\left(l-\frac{1}{2}-i'\right)\right], 
\label{app_tau_j}
\end{equation}
where we define $u_0=u_N$, and the summation is over all possible disconnections [the Burgers vector of the disconnection between the $l$- and $(l-1)$-th GB segments $u_l-u_{l-1}$] in one period. 
Now, we derive the expression of $\tau_i$ in Eq.~\eqref{tau_i} for the case where the state of the $i$-th GB segment is changed: $(z_i,u_i)\to (z_i',u_i')=(z_i+h,u_i+b)$. 
Before this state change on the $i$-th lattice point, the stress on any point, e.g. the $i'$-th point, is 
\begin{align}
\tau_{i'}^{(0)}
=\, & \frac{2\pi\mathcal{K}}{L_y}
\left\{
\sum_{l\ne i,i+1} (u_l-u_{l-1})\cot\left[\frac{\pi}{N}\left(l-\frac{1}{2}-i'\right)\right]
\right. 
\nonumber \\
&\left.  + (u_i-u_{i-1})\cot\left[\frac{\pi}{N}\left(i-\frac{1}{2}-i'\right)\right]
+ (u_{i+1}-u_i)\cot\left[\frac{\pi}{N}\left(i+\frac{1}{2}-i'\right)\right]
\right\}, 
\label{app_tau_j_0}
\end{align}
i.e., re-organization of Eq.~\eqref{app_tau_j}. 
Particularly, the stress at the $i$-th point is [i.e., let $i'=i$ in Eq.~\eqref{app_tau_j_0}]
\begin{equation}
\tau_i^{(0)}
= \frac{2\pi\mathcal{K}}{L_y}
\sum_{l=1}^N
(u_l-u_{l-1}) \cot\left[\frac{\pi}{N}\left(l-\frac{1}{2}-i\right)\right], 
\label{app_tau_i_0}
\end{equation}
After the state change on the $i$-th lattice point, the stress becomes 
\begin{align}
\tau_{i'}^{(1)}
=\, & \frac{2\pi\mathcal{K}}{L_y}
\left\{
\sum_{l\ne i,i+1} (u_l-u_{l-1})\cot\left[\frac{\pi}{N}\left(l-\frac{1}{2}-i'\right)\right]
\right. 
\nonumber \\
&\left.  + (u_i'-u_{i-1})\cot\left[\frac{\pi}{N}\left(i-\frac{1}{2}-i'\right)\right]
+ (u_{i+1}-u_i')\cot\left[\frac{\pi}{N}\left(i+\frac{1}{2}-i'\right)\right]
\right\}, 
\label{app_tau_j_1}
\nonumber \\
=\, & \tau_{i'}^{(0)}
+ \frac{2\pi\mathcal{K}}{L_y}
\left\{
\cot\left[\frac{\pi}{N}\left(i-\frac{1}{2}-i'\right)\right]
-\cot\left[\frac{\pi}{N}\left(i+\frac{1}{2}-i'\right)\right]
\right\}. 
\end{align}
Particularly, after the state change on the $i$-th lattice point, the stress at the $i$-th point is 
\begin{equation}
\tau_i^{(1)}
= \tau_i^{(0)}
- \frac{4\pi\mathcal{K}}{L_y} b\cot\left(\frac{\pi}{2N}\right). 
\label{app_tau_i_1}
\end{equation}
For the state change on the $i$-th lattice point, only the stress on this point does work (since relative displacement of both grains occurs only on the GB segment located at this point). 
We determine this stress as an average of $\tau_i^{(0)}$ and $\tau_i^{(1)}$: 
\begin{equation}
\tau_i = \frac{1}{2}
\left(
\tau_i^{(0)} + \tau_i^{(1)}
\right). 
\label{app_tau_i}
\end{equation}
From Eqs.~\eqref{app_tau_i_0} and \eqref{app_tau_i_1}, we will obtain Eq.~\eqref{tau_i}. 
We determine the stress as average of the stresses before and after the state change because, by this approach, the work done by the dislocation itself is automatically canceled out. 

An important point is that there will be phase transition only when the work done by the shear stress is included (i.e., long-range interaction is considered). 
We know that there is no phase transition for 1D Ising model~\cite{chandler1987introduction}. 
If there is no long-range interaction, the energy change is only contributed by the step energy (solid-on-solid model) 
and the dislocation core energy (discrete Gaussian model). 
We consider a particular configuration where 
\begin{equation}
(z_i, u_i) = 
\left\{\begin{array}{ll}
(0,0), & i = 1,\cdots,N/2 \\
(h, b), &  i = N/2+1,\cdots,N \\
\end{array}\right., 
\label{app_configuration}
\end{equation}
where we assume $N$ is even. 
If we take the flat boundary (free of disconnection) as the reference ($E=0$), then 
the energy of this configuration is 
\begin{equation}
\Delta E = 2\mathit{\Gamma}_\text{s}h
+ 2\zeta\mathcal{K}b^2. 
\end{equation}
The configurational entropy of this configuration is $k_\text{B}\ln N$ because we can obtain the energy-degenerate configurations by locating the step between any two lattice points. 
We can find that, as $N\to\infty$, the entropy is much larger than the energy for any finite temperature. 
In other words, at finite temperature, this configuration will be always favored over the flat configuration since this will not influence the energy but increase the entropy (thus decrease the free energy). 
This is the reason why there will be no phase transition for the 1D Ising model if there is no long-range interaction. 
However, there will be phase transition if there is long-range interaction included. 
For the same configuration as Eq.~\eqref{app_configuration}, if the long-range interaction is switched on, the energy of this configuration will become 
\begin{equation}
\Delta E = 2\mathit{\Gamma}_\text{s}h
+ 2\zeta\mathcal{K} b^2
+ 2\mathcal{K} b^2\frac{\pi}{N} \sum_{i=1}^{N/2} \cot\left[\frac{(2i-1)\pi}{2N}\right]. 
\label{DEsum}
\end{equation}
We will approximate the summation by integration. 
\begin{equation}
\sum_{i=1}^{N/2} \cot\left[\frac{(2i-1)\pi}{2N}\right]
= \cot\left(\frac{\pi}{2N}\right)
+ \sum_{i=2}^{N/2} \cot\left[\frac{(2i-1)\pi}{2N}\right]. 
\end{equation}
We evaluate the last summation by Euler-Maclaurin formula~\cite{xu2009derivation}: 
\begin{align}
\sum_{i=2}^{N/2} \cot\left[\frac{(2i-1)\pi}{2N}\right]
=~& \frac{1}{\Delta x}\sum_{i=2}^n \cot(x_i) \Delta x
\nonumber \\
=~& \frac{1}{\Delta x} \int_a^{(2n+1)a} \cot(x)\,\mathrm{d} x
- \frac{1}{2}\left\{ \cot(a) + \cot[(2n+1)a] \right\}
\nonumber \\
&- \frac{\Delta x}{12}\left[ \frac{\mathrm{d} \cot(x)}{\mathrm{d} x}\bigg|_a - \frac{\mathrm{d} \cot(x)}{\mathrm{d} x}\bigg|_{(2n+1)a} \right]
+ O(\Delta x^3), 
\end{align}
where $n = N/2$, $\Delta x = \pi/N$, $a = \pi/2N$ and $x_i = a+(i-1)\Delta x = (2i-1)\pi/2N$. 
Finally, 
\begin{align}
\sum_{i=2}^{N/2} \cot\left[\frac{(2i-1)\pi}{2N}\right]
=~& \frac{N}{\pi} \ln\cot\left(\frac{\pi}{2N}\right)
- \frac{1}{2}\left[ \cot\left(\frac{\pi}{2N}\right) - \tan\left(\frac{\pi}{2N}\right) \right]
\nonumber \\
&- \frac{\pi}{12N}\left[ \cos^{-2}\left(\frac{\pi}{2N}\right) - \sin^{-2}\left(\frac{\pi}{2N}\right) \right]
+ O\left(\frac{1}{N^3}\right)
\end{align}
Equation~\eqref{DEsum} becomes
\begin{align}
\Delta E
= \cdots 
+ 2\mathcal{K}b^2
&\left\{ 
\ln\cot\left(\frac{\pi}{2N}\right)
+ \frac{\pi}{2N} \left[ \cot\left(\frac{\pi}{2N}\right) + \tan\left(\frac{\pi}{2N}\right) \right]  \right.
\nonumber \\
&\left. - \frac{\pi^2}{12N^2} \left[ \cos^{-2}\left(\frac{\pi}{2N}\right) - \sin^{-2}\left(\frac{\pi}{2N}\right) \right]
+ O\left(\frac{1}{N^4}\right)
\right\}. 
\end{align}
As $N\to \infty$, 
\begin{align}
\Delta E
&\to \cdots 
+ 2\mathcal{K}b^2
\left\{
\ln\left(\frac{2N}{\pi}\right)
+ 1
+ \frac{1}{3}
\right\}
\nonumber \\
&= 2\mathit{\Gamma}_\text{s}h
+ 2\mathcal{K}b^2
\left[
\zeta + 4/3 - \ln(\pi/2) + \ln N
\right]. 
\end{align}
Noting again that the configurational entropy is $k_\text{B}\ln N$. 
The energy is as important as the entropy and so the phase transition will occur. 

Finally, we provide the Metropolis MC algorithm:  
\begin{enumerate}
\item[(i)] Initialize the states on each lattice point $\{(z_i, u_i)\}$ ($i = 1,\cdots,N$). 
Calculate and store the shear stress on each lattice point by Eq.~\eqref{app_tau_j}. 

\item[(ii)] Randomly pick one lattice point, e.g. the $i$-th lattice point. 
Randomly pick a change of state $(b,h)$ from the list of disconnection mode $\{(b_n, h_{nj})\}$. 
Change the state of the $i$-th lattice point: 
$(z_i, u_i) \to (z_i', u_i') = (z_i+h, u_i+b)$. 

\item[(iii)] Calculate the energy change due to such change of state according to Eqs.~\eqref{energy_change}, \eqref{energy_change_step}, \eqref{energy_change_core}, \eqref{energy_change_work} and \eqref{tau_i}. 

\item[(iv)] Acceptance or rejection: 
\begin{enumerate}
\item[(a)] If $\Delta E \le 0$, accept the change of state and go to Step (v). 
\item[(b)] If $\Delta E > 0$, calculate the acceptance probability $P = e^{-\Delta EL_x/k_\text{B}T}$ 
and produce a random number $\chi \in [0,1]$. 
If $\chi \le P$, accept the change of state and go to Step (v); 
if $\chi > P$, reject the change of state and go to Step (ii). 
\end{enumerate}

\item[(v)] Update the state on the $i$-th lattice point: $(z_i, u_i) \coloneqq (z_i', u_i')$, 
and re-calculate and store the shear stress on each lattice point according to Eq.~\eqref{app_tau_j_1}. 
Go to Step (ii). 
\end{enumerate}

\section{Derivation of the equation for grain shrinkage coupled with grain rotation}
\label{APPgrainrotation}

Consider the shrinkage of a square grain as shown in Fig.~\ref{grainshrink}a2. 
One possible mechanism to release the stress caused by grain shrinkage and shear-coupled GB migration is grain rotation. 
The grain rotation requires vacancy diffusion from the hollow wedge region to the overlapped wedge region. 

First of all, we will obtain the ration rate based on the GB self-diffusivity. 
The total wedge area is $\mathcal{A}_\text{w}\approx \mathsf{g}_1 r^2 \Delta\theta$, where $r$ is any measure of the grain size (e.g. the size of the edge of the square grain), $\Delta\theta$ is the rotation angle, $\mathsf{g}_1$ is a geometric constant, and the approximation is valid for small $\Delta\theta$. 
The number of the diffusing vacancies (per unit thickness) is $N_\text{v}=\mathcal{A}_\text{w}/\Omega = \mathsf{g}_1r^2\Delta\theta/\Omega$, where $\Omega$ is the atomic volume. 
The change rate of the vacancy number is $\dot{N}_\text{v}=\mathsf{g}_1 r^2 \dot{\theta}/\Omega$. 
On the other hand, the change rate of the vacancy number (per unit thickness) can also be expressed as $\dot{N}_\text{v}=J_\text{v}\lambda$, where $J_\text{v}$ is the vacancy flux and $\lambda$ is the effective GB width. 
Therefore, the rotation rate is 
\begin{equation}
\dot{\theta} 
= \frac{J_\text{v}\lambda\Omega}{\mathsf{g}_1 r^2}.
\label{dtheta}
\end{equation} 
The next problem is to connect $J_\text{v}$ with the GB self-diffusivity $D_\text{gb}$. 

The driving force for grain rotation is
\begin{equation}
\frac{\text{d}E}{\text{d}\theta}
= \frac{\text{d}(\mathsf{g}_2 r \gamma)}{\text{d}\theta}
= \mathsf{g}_2\gamma \frac{\text{d}r}{\text{d}\theta}, 
\label{dEdt}
\end{equation}
where $\mathsf{g}_2$ is a geometric constant and we assume that $\gamma$ is insensitive to the misorientation (this might be true for high-angle GBs). 
If GB migration and GB sliding are coupled by the shear-coupling factor $\beta$, we will have the relation Eq.~\eqref{coupling_relation}. 
Then, Eq.~\eqref{dEdt} becomes 
\begin{equation}
\frac{\text{d}E}{\text{d}\theta}
= - \frac{\mathsf{g}_2\gamma r}{\beta}. 
\end{equation}
The energy change due to the diffusion of one vacancy is 
\begin{equation}
\Delta E_{N_\text{v}=1}
= \Delta \theta_{N_\text{v}=1} \frac{\text{d}E}{\text{d}\theta}
= - \frac{\mathsf{g}_3 \gamma\Omega}{\beta r}. 
\end{equation}
Approximately, the vacancy chemical potential gradient is 
\begin{equation}
\nabla\mu_\text{v}
= \frac{\Delta E_{N_\text{v}=1}}{r/2} = - \frac{2\mathsf{g}_3 \gamma\Omega}{\beta r^2}. 
\end{equation}
Hence, the vacancy flux is 
\begin{equation}
J_\text{v} = -M_\text{v} c_\text{v} \nabla\mu_\text{v}
= -\frac{D_\text{v}}{k_\text{B}T} c_\text{v} \left(- \frac{2\mathsf{g}_3 \gamma\Omega}{\beta r^2}\right)
= \frac{2\mathsf{g}_3\gamma D_\text{gb}}{k_\text{B}T \beta r^2}, 
\label{Jv}
\end{equation}
where $M_\text{v}$ is the vacancy mobility, $c_\text{v}$ is the vacancy concentration, $D_\text{v}$ is the vacancy diffusivity, and $D_\text{gb} = D_\text{v}c_\text{v}\Omega$. 
Substituting Eq.~\eqref{Jv} into Eq.~\eqref{dtheta}, we have the rotation rate 
\begin{equation}
\dot{\theta}
= \frac{\mathsf{g}\lambda D_\text{gb}\gamma\Omega}{k_\text{B}T \beta r^4}. 
\label{dtheta2}
\end{equation}

Now, we connect the rotation rate to the shrinkage rate. 
If we assume that $\beta \approx \theta$, then, from Eq.~\eqref{coupling_relation}, we will have $\text{d}r/\text{d}\theta = -r/\theta \Rightarrow r\theta = r_0\theta_0$. 
Substituting this into Eq.~\eqref{dtheta2}, we will have the equation about $r$, and the solution to this equation will be Eq.~\eqref{r2_2}.

\section{Derivation of the average driving force for grain shrinkage by mode switch}
\label{APPmodeswitch}

For the bicrystal model with two fixed surfaces as shown in Fig.~\ref{coupling_schematic}c, we have had the energy of the bicrystal system $E$ vs. GB migration distance $\bar{z}$ as expressed in Eq.~\eqref{mixed2}. 
In the situation of the shrinkage of a square grain in a polycrystal (see Fig.~\ref{grainshrink}a3), we replace $L_z$ by the size of the edge of the grain $L$ and let $\psi = \gamma/(L - \bar{z}) \approx \gamma/L$ if $\bar{z}\ll L$. 
Then, the energy per unit area in the case of polycrystal is
\begin{equation}
\mathcal{E}(\bar{z})
= -\frac{\gamma}{L} \bar{z} + \frac{\mu \beta_1^2}{L} \bar{z}^2, 
\end{equation}
where the little barriers (i.e., the terms in the first parenthesis of Eq.~\eqref{mixed2}) are ignored, and $\beta_1$ denotes the shear-coupling factor of the first mode activated by the capillary force.  
Stagnation of the first mode occurs when the GB shrinks by the distance $\bar{z}_1 = \gamma/(2\mu\beta_1^2)$ and the energy per unit area reaches $\mathcal{E}_1 = - \gamma^2/(4\mu\beta_1^2 L)$. 

After the GB migration stagnates at the state $(\bar{z}_1, \mathcal{E}_1)$, the GB migration can continue via the operation of the second mode with the shear-coupling factor $\beta_2$, and the energy per unit area is 
\begin{equation}
\mathcal{E}(\bar{z})
= \mathcal{E}_1 - \left[\left(\frac{\gamma}{L} - \frac{\mu \beta_1\beta_2}{L}\bar{z}_1\right)(\bar{z}-\bar{z}_1) - \frac{\mu \beta_2^2}{L}(\bar{z}-\bar{z}_1)^2\right]. 
\end{equation}
Stagnation of the second mode occurs when the GB shrinks by the distance 
\begin{equation}
\bar{z}_2 
= \frac{\gamma}{2\mu}\left(\frac{1}{\beta_1^2} + \frac{1}{\beta_2^2} - \frac{1}{2\beta_1\beta_2}\right), 
\end{equation}
and the energy per unit area reaches 
\begin{equation}
\mathcal{E}_2 
= -\frac{\gamma^2}{4\mu L} \left[\frac{1}{\beta_1^2} + \left(\frac{1}{\beta_2} - \frac{1}{2\beta_1}\right)^2\right]
\end{equation}
The average driving force during this process (activation of the first mode followed by the second mode) is 
\begin{equation}
F = - \frac{\mathcal{E}_2}{\bar{z}_2}
= \mathcal{C}\left(\frac{\beta_2}{\beta_1}\right)\frac{\gamma}{L}, 
\end{equation}
where 
\begin{equation}
\mathcal{C}\left(\frac{\beta_2}{\beta_1}\right)
= \frac{1}{2} \frac{\displaystyle{ \left(\frac{\beta_2}{\beta_1}\right)^2 + \left(1 - \frac{1}{2}\frac{\beta_2}{\beta_1}\right)^2 }}
{\displaystyle{ \left(\frac{\beta_2}{\beta_1}\right)^2 + \left(1 - \frac{1}{2}\frac{\beta_2}{\beta_1}\right) }}
< 1. 
\end{equation}
This suggests that $F < \gamma/L$, implying that the driving force will be lowered if shear-coupled GB migration (via switch between two modes) is taken into account.

\section{The algorithm for the simulation of triple-junction motion and GB migration}
\label{APPtj}

\begin{figure}[t]
\begin{center}
\scalebox{0.5}{\includegraphics{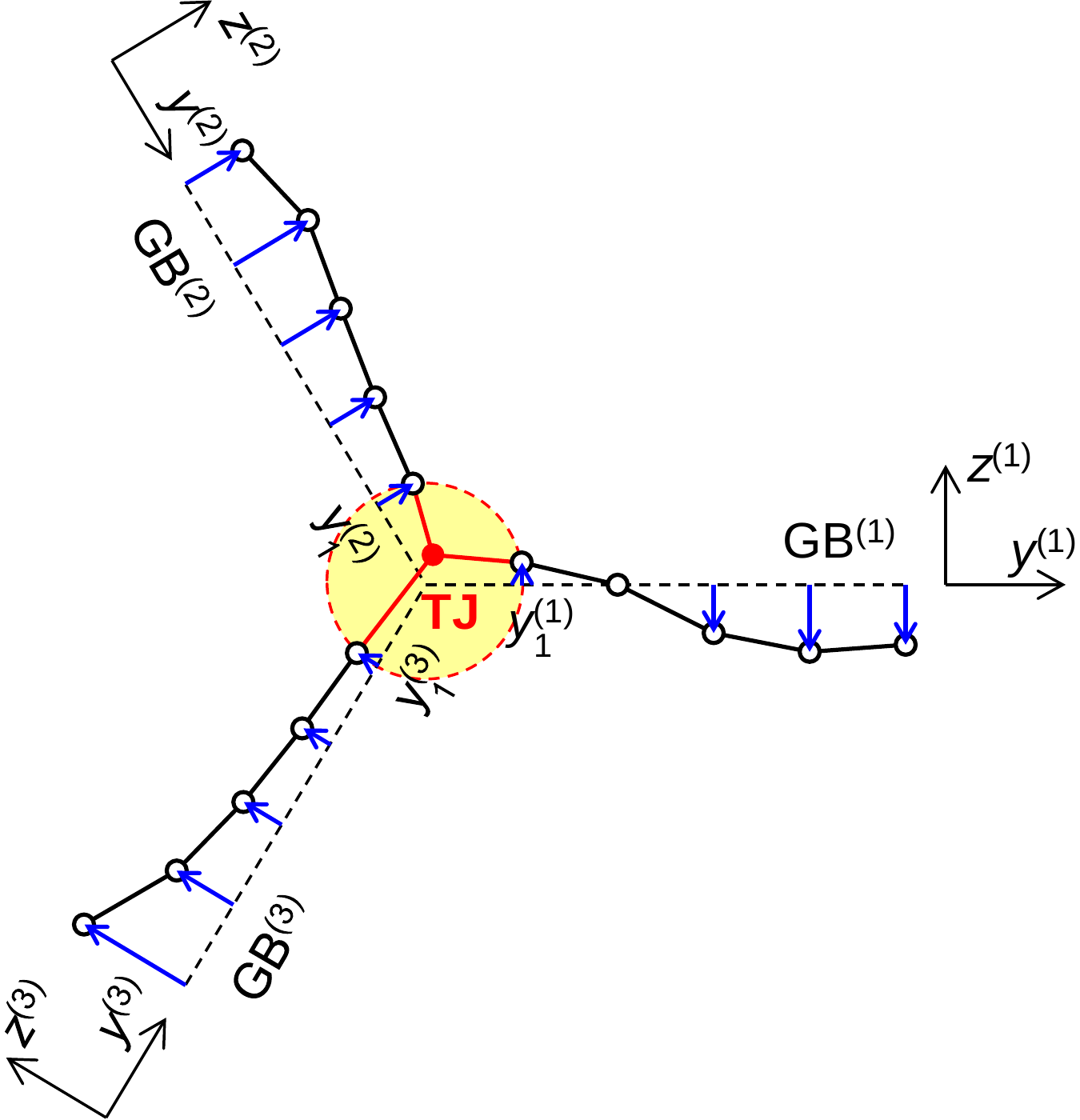}}
\caption{The model for the simulation of TJ motion. 
The black dashed lines represent the reference GB (assumed disconnection free). 
The coordinate system for GB$^{(i)}$ is spanned by the $y^{(i)}$- and $z^{(i)}$-axes. 
The black circles denote the nodes which discretizes the GBs. 
The yellow colored region represents the region for the TJ. 
The node immediately near the TJ point (red point) on GB$^{(i)}$ is located at $y_1^{(i)}$.  } 
\label{TJsimul}
\end{center}
\end{figure}

The model for the simulation of TJ motion is shown in Fig.~\ref{TJsimul}. 
The black dashed lines represent the reference GB (assumed disconnection free). 
The yellow colored region represents the region for the TJ. 
Each GB is discretized for numerical simulation. 
The node immediately near the TJ point (red point) on GB$^{(i)}$ is located at $y_1^{(i)}$. 

The equation of motion for GB$^{(i)}$ is 
\begin{equation}
z_{,t}^{(i)}
= - M_\text{d}^{(i)} \left[
\left(\tau_\text{int}\left[z_{,y}^{(i)}\right] + \tau^{(i)}\right) b^{(i)} + \left(\psi^{(i)} - \mathit{\Gamma}^{(i)} z_{,yy}^{(i)}\right) h^{(i)}
\right]
\left(\left|z_{,y}^{(i)}\right| + \eta^{(i)}\right), 
\label{continuum_eqi}
\end{equation}
for which the coordinate system is attached to GB$^{(i)}$ such that the $y^{(i)}$-axis is parallel to $-\mathbf{t}^{(i)}$, the $z^{(i)}$-axis is parallel to $\mathbf{n}^{(i)}$. 
The algorithm for the simulation of TJ motion along with the migration of three GBs is listed below:  
\begin{itemize}
\item[(i)] Initialize the profile for each GB $z^{(i)}(y^{(i)},t)$. 
Initialize the Burgers vector at the TJ $\mathbf{b}_\text{tj}(t)$, and then calculate the stress state $\tau^{(i)}(t) = \tau^{(i)}(\mathbf{b}_\text{tj}(t))$. 

\item[(ii)] Insert $z^{(i)}(y^{(i)},t)$ and $\tau(t)$ into the right-hand side of Eq.~\eqref{continuum_eqi}, keep the TJ point fixed (this is fixed boundary condition shared by the three GBs), and obtain $z^{(i)}(y^{(i)},t+\Delta t) = z^{(i)}(y^{(i)},t) + z_{,t}^{(i)}(y^{(i)},t)\Delta t$. 

\item[(iii)] Calculate the accumulation rate of Burgers vector at the TJ: $\dot{\mathbf{b}}_\text{tj}(t) = \sum_{i=1}^3 z_{,t}^{(i)}(y_1^{(i)},t) \mathbf{b}^{(i)}/h^{(i)}$.  
Update the Burgers vector at the TJ: $\mathbf{b}_\text{tj}(t+\Delta t) = \mathbf{b}_\text{tj}(t) +\dot{\mathbf{b}}_\text{tj}\Delta t$. 
Update the stress state: $\tau^{(i)}(t+\Delta t) = \tau^{(i)}(\mathbf{b}_\text{tj}(t+\Delta t))$. 

\item[(iv)] Update the position of the TJ point with the nodes located at $y_1^{(i)}$ fixed. The position of the TJ point is found by minimizing the total GB energy in the TJ region (colored yellow). 
If the red GB segment on GB$^{(i)}$ in the TJ region becomes longer than the discretization interval $2\Delta y$, add a node on this segment and denote the coordinate of this node as $y_1^{(i)}$. 
If the red GB segment on GB$^{(i)}$ in the TJ region becomes smaller than the discretization interval $\Delta y$, delete the node at $y_1^{(i)}$ and denote the coordinate of the next node to be $y_1^{(i)}$. 

\item[(vi)] Let $t \coloneqq t+\Delta t$ and return to Step (ii). 
\end{itemize}

\section{Driving force for disconnection climb in a ternary alloy}
\label{APPdriving}

We consider a ternary substitutional alloy composed by A atoms, B atoms and vacancies (denoted by ``A'', ``B'' and ``v'', respectively). 
We focus on the region around a disconnection $(b_\perp, h^\text{L})$; there are $N$ lattice points in this region. 
According to the regular solution model~\cite{meijering1950segregation}, the total free energy of mixing is 
\begin{align}
G(X_\text{v},X_\text{B})
&= Nk_\text{B}T\left[ X_\text{v}\ln X_\text{v} + X_\text{B}\ln X_\text{B} + (1-X_\text{B}-X_\text{v})\ln (1-X_\text{B}-X_\text{v}) \right]
\nonumber \\
&+ N\left[ \omega_\text{vA}X_\text{v}(1-X_\text{B}-X_\text{v}) + \omega_\text{vB}X_\text{v}X_\text{B} + \omega_\text{AB}(1-X_\text{B}-X_\text{v})X_\text{B} \right], 
\label{GXvXB}
\end{align}
where $\omega_\text{vA}$, $\omega_\text{vB}$ and $\omega_\text{AB}$ are interaction parameters; $X_\text{v}$ and $X_\text{B}$ are compositions of vacancies and B atoms (the composition of A atoms is $X_\text{A} = 1-X_\text{B}-X_\text{v}$ due to network constraint). 
The chemical potential of vacancies is 
\begin{equation}
\frac{\partial G}{\partial N_\text{v}}
= \frac{\partial G}{N\partial X_\text{v}}
= k_\text{B}T \ln\frac{X_\text{v}}{1-X_\text{B}-X_\text{v}} + \omega_\text{vA}(1-2X_\text{v}) + (\omega_\text{vB}-\omega_\text{vA}-\omega_\text{AB})X_\text{B}. 
\label{GNv0}
\end{equation} 
The equilibrium compositions of vacancies and B atoms are $X_\text{v}^\text{eq}$ and $X_\text{B}^\text{eq}$ such that $(\partial G/\partial N_\text{v})_{X_\text{v}^\text{eq},X_\text{B}^\text{eq}}=0$, which gives the relationship: 
\begin{equation}
k_\text{B}T \ln\frac{X_\text{v}^\text{eq}}{1-X_\text{B}^\text{eq}-X_\text{v}^\text{eq}} + \omega_\text{vA}(1-2X_\text{v}^\text{eq}) + (\omega_\text{vB}-\omega_\text{vA}-\omega_\text{AB})X_\text{B}^\text{eq}
=0. 
\label{GNv_eq}
\end{equation} 
Obviously, $X_\text{v}^\text{eq}$ and $X_\text{B}^\text{eq}$ are determined by the interaction parameters $\omega_\text{vA}$, $\omega_\text{vB}$ and $\omega_\text{AB}$. 
Note that $X_\text{v}^\text{eq}$ and $X_\text{B}^\text{eq}$ are the equilibrium compositions around a disconnection core rather than in bulk. 
Subtracting Eq.~\eqref{GNv_eq} from Eq.~\eqref{GNv0}, 
\begin{align}
\frac{\partial G}{\partial N_\text{v}}
=~& k_\text{B}T \ln\left(\frac{X_\text{v}}{X_\text{v}^\text{eq}} \frac{1-X_\text{B}^\text{eq}-X_\text{v}^\text{eq}}{1-X_\text{B}-X_\text{v}} \right)
\nonumber \\
&- 2\omega_\text{vA}(X_\text{v}-X_\text{v}^\text{eq}) + (\omega_\text{vB}-\omega_\text{vA}-\omega_\text{AB})(X_\text{B}-X_\text{B}^\text{eq}) 
\nonumber \\
\approx~& \left[\frac{k_\text{B}T(1-X_\text{B}^\text{eq})}{X_\text{v}^\text{eq}(1-X_\text{B}^\text{eq}-X_\text{v}^\text{eq})} - 2\omega_\text{vA} \right](X_\text{v}-X_\text{v}^\text{eq})
\nonumber \\
&+ \left[\frac{k_\text{B}T}{1-X_\text{B}^\text{eq}-X_\text{v}^\text{eq}} + (\omega_\text{vB}-\omega_\text{vA}-\omega_\text{AB})\right](X_\text{B}-X_\text{B}^\text{eq})
\nonumber \\
\equiv~& \mathcal{E}_\text{vv}(X_\text{v}-X_\text{v}^\text{eq}) + \mathcal{E}_\text{vB}(X_\text{B}-X_\text{B}^\text{eq}), 
\label{GNv}
\end{align} 
where Taylor expansion has been done about $X_\text{v}=X_\text{v}^\text{eq}$ and $X_\text{B}=X_\text{B}^\text{eq}$ to the first order. 
Following the similar derivation and approximation, the chemical potential of B atoms is 
\begin{align}
\frac{\partial G}{\partial N_\text{B}}
\approx~& \left[\frac{k_\text{B}T}{1-X_\text{B}^\text{eq}-X_\text{v}^\text{eq}} + (\omega_\text{vB}-\omega_\text{vA}-\omega_\text{AB})\right](X_\text{v}-X_\text{v}^\text{eq})
\nonumber \\
&+ \left[\frac{k_\text{B}T(1-X_\text{v}^\text{eq})}{X_\text{B}^\text{eq}(1-X_\text{B}^\text{eq}-X_\text{v}^\text{eq})} - 2\omega_\text{AB} \right](X_\text{B}-X_\text{B}^\text{eq})
\nonumber \\
\equiv~& \mathcal{E}_\text{Bv}(X_\text{v}-X_\text{v}^\text{eq}) + \mathcal{E}_\text{BB}(X_\text{B}-X_\text{B}^\text{eq}). 
\label{GNB}
\end{align}
Note that the coefficient $\mathcal{E}_\text{IJ} = \mathcal{E}_\text{IJ}(\omega_\text{vA},\omega_\text{vB},\omega_\text{AB};T)$, where $\text{I},\text{J} = \text{B}, \text{v}$, and has the dimension of energy. 

We first think of the case of homophase interface, for which $\bar{X}_\text{B} = X_\text{B}$ and $\Delta X_\text{B} = 0$. 
From Eqs.~\eqref{IB} and \eqref{Iv}, 
\begin{equation}
\frac{\mathrm{d}N_\text{B}}{L_x\mathrm{d}y} = X_\text{B}\frac{b_\perp}{\Omega}
\text{ and }
\frac{\mathrm{d}N_\text{v}}{L_x\mathrm{d}y} = -\frac{b_\perp}{\Omega}. 
\label{dNdy}
\end{equation}
Then, from Eqs.~\eqref{GNv}, \eqref{GNB} and \eqref{dNdy}, the chemical driving force applied on a disconnection is 
\begin{align}
\frac{\mathrm{d}G}{L_x\mathrm{d}y}
&= \frac{\partial G}{\partial N_\text{v}} \frac{\mathrm{d}N_\text{v}}{L_x\mathrm{d}y}
+ \frac{\partial G}{\partial N_\text{B}} \frac{\mathrm{d}N_\text{B}}{L_x\mathrm{d}y}
\nonumber \\
&\approx -\frac{b_\perp}{\Omega} \left[
(\mathcal{E}_\text{vv}-\mathcal{E}_\text{Bv}X_\text{B}^\text{eq})(X_\text{v}-X_\text{v}^\text{eq})
+(\mathcal{E}_\text{vB}-\mathcal{E}_\text{BB}X_\text{B}^\text{eq})(X_\text{B}-X_\text{B}^\text{eq})
\right]
\nonumber \\
&\equiv -\frac{b_\perp}{\Omega} \left[\mathcal{E}_\text{v}(X_\text{v}-X_\text{v}^\text{eq})
+\mathcal{E}_\text{B}(X_\text{B}-X_\text{B}^\text{eq})\right], 
\end{align}
where we only keep the first-order term.  
Note that $\mathcal{E}_\text{v/B}=\mathcal{E}_\text{v/B}(\omega_\text{vA},\omega_\text{vB},\omega_\text{AB};T)$; and as $X_\text{v}^\text{eq}\ll 1$, we can easily show that $\mathcal{E}_\text{v} \approx k_\text{B}T/X_\text{v}^\text{eq}>0$. 
Aside from the chemical driving force, there may be another type of driving force, i.e., Peach-Koehler force~\cite{HirthLothe}: $f_\text{PK} = \boldsymbol{\sigma}\mathbf{b}\times \mathbf{o}\cdot\mathbf{p} = \sigma_{nn} b_\perp$ (see Fig.~\ref{energetics} for the coordinate system), where $\sigma_{nn} = \mathbf{n}\cdot\boldsymbol{\sigma}\mathbf{n}$.  
Driven by both composition difference and Peach-Koehler force, the disconnection climb velocity is 
\begin{align}
v_\text{cl} 
&= M_\text{d}F
= M_\text{d}\left(\frac{\mathrm{d}G}{L_x\mathrm{d}y} + f_\text{PK}\right)
\nonumber \\
&= -M_\text{d}\frac{b_\perp}{\Omega} \left[\mathcal{E}_\text{v}(X_\text{v}-X_\text{v}^\text{eq})
+\mathcal{E}_\text{B}(X_\text{B}-X_\text{B}^\text{eq}) - \sigma_{nn}\Omega\right]. 
\label{vclMdF}
\end{align}
The vacancy flux into the interface required by disconnection climb is 
\begin{equation}
\varrho I_\text{v}
= -\varrho\frac{b_\perp}{\Omega} v_\text{cl}
= M_\text{d}\varrho\left(\frac{b_\perp}{\Omega}\right)^2 
\left[\mathcal{E}_\text{v}(X_\text{v}-X_\text{v}^\text{eq})
+\mathcal{E}_\text{B}(X_\text{B}-X_\text{B}^\text{eq}) - \sigma_{nn}\Omega\right].
\end{equation}

We now think of the case of a heterophase interface between a stoichiometric compound (with fixed composition $X_\text{B}^\text{w}$) and a solid solution. 
Equations~\eqref{GNv} and \eqref{GNB} still applies at the side of solid solution. 
From Eqs.~\eqref{IB} and \eqref{Iv}, 
\begin{equation}
\frac{\mathrm{d}N_\text{B}}{L_x\mathrm{d}y} 
= \left(\bar{X}_\text{B} b_\perp - \Delta X_\text{B} h^\text{L}\right)\frac{1}{\Omega}
\text{ and }
\frac{\mathrm{d}N_\text{v}}{L_x\mathrm{d}y} = -\frac{b_\perp}{\Omega}. 
\label{dNdy2}
\end{equation}
Then, from Eqs.~\eqref{GNv}, \eqref{GNB} and \eqref{dNdy2}, the chemical driving force applied on a disconnection is 
\begin{align}
\frac{\mathrm{d}G}{L_x\mathrm{d}y}
\approx~& -\frac{h^\text{L}}{\Omega} \left\{
\left[\mathcal{E}_\text{vv}\beta_\perp-\mathcal{E}_\text{Bv}(\bar{X}_\text{B}^\text{eq}\beta_\perp-\Delta X_\text{B}^\text{eq})\right](X_\text{v}-X_\text{v}^\text{eq})
\right.
\nonumber \\
&\left.+\left[\mathcal{E}_\text{vB}\beta_\perp-\mathcal{E}_\text{BB}(\bar{X}_\text{B}^\text{eq}\beta_\perp-\Delta X_\text{B}^\text{eq})\right](X_\text{B}-X_\text{B}^\text{eq})
\right\}
\nonumber \\
\equiv~& -\frac{h^\text{L}}{\Omega} \left[\mathcal{E}''_\text{v}(X_\text{v}-X_\text{v}^\text{eq})
+ \mathcal{E}''_\text{B}(X_\text{B}-X_\text{B}^\text{eq})\right], 
\label{dGLxdy}
\end{align}
where $\bar{X}_\text{B}^\text{eq}\equiv (X_\text{B}^\text{eq}+X_\text{B}^\text{w})/2$, $\Delta X_\text{B}^\text{eq}\equiv X_\text{B}^\text{eq}-X_\text{B}^\text{w}$ and $\beta_\perp \equiv b_\perp/h^\text{L}$. 
Note that $\mathcal{E}''_\text{v/B}=\mathcal{E}''_\text{v/B}(\omega_\text{vA},\omega_\text{vB},\omega_\text{AB},X_\text{B}^\text{w},\beta_\perp;T)$. 
Together with Peach-Koehler force, the disconnection climb velocity is 
\begin{equation}
v_\text{cl} 
= -M_\text{d}\frac{h^\text{L}}{\Omega} \left[\mathcal{E}''_\text{v}(X_\text{v}-X_\text{v}^\text{eq})
+\mathcal{E}''_\text{B}(X_\text{B}-X_\text{B}^\text{eq}) - \sigma_{nn}\beta_\perp\Omega\right]. 
\end{equation}
The flux of B atoms into the interface required by the motion of one disconnection is 
\begin{align}
\varrho I_\text{B}
=~& \varrho \left[(X_\text{B}+X_\text{B}^\text{w})\beta_\perp/2-(X_\text{B}-X_\text{B}^\text{w})\right] \frac{h^\text{L}}{\Omega}v_\text{cl}
\nonumber \\
=~& M_\text{d}\varrho \left(\frac{h^\text{L}}{\Omega}\right)^2
\left[(X_\text{B}-X_\text{B}^\text{w}) - (X_\text{B}+X_\text{B}^\text{w})\beta_\perp/2\right]
\nonumber \\
&\times \left[
\mathcal{E}''_\text{v}(X_\text{v}-X_\text{v}^\text{eq})
+\mathcal{E}''_\text{B}(X_\text{B}-X_\text{B}^\text{eq})
- \sigma_{nn}\beta_\perp\Omega
\right]
\nonumber \\
\approx~&  M_\text{d}\varrho \left(\frac{h^\text{L}}{\Omega}\right)^2
\left[ \mathcal{E}'_\text{v}(X_\text{v}-X_\text{v}^\text{eq}) + \mathcal{E}'_\text{B}(X_\text{B}-X_\text{B}^\text{eq}) - \mathcal{E}'_\text{s} \sigma_{nn}\beta_\perp\Omega \right].
\end{align}
In particular, when $\sigma_{nn}=0$, 
\begin{equation}
\mathcal{E}'_\text{B}
= (\Delta X_\text{B}^\text{eq}-\bar{X}_\text{B}^\text{eq} \beta_\perp)
\left[
\mathcal{E}_\text{vB}\beta_\perp + \mathcal{E}_\text{BB}(\Delta X_\text{B}^\text{eq}-\bar{X}_\text{B}^\text{eq} \beta_\perp)
\right]. 
\end{equation}

For a special disconnection mode with $b_\perp =0$, i.e., pure step mode (for which $\beta_\perp \to 0$), Eq.~\eqref{dGLxdy} becomes 
\begin{equation}
\frac{\mathrm{d}G}{L_x\mathrm{d}y}
\approx -\frac{h^\text{L}}{\Omega} \Delta X_\text{B}^\text{eq} \left[
\mathcal{E}_\text{Bv}(X_\text{v}-X_\text{v}^\text{eq})
+\mathcal{E}_\text{BB}(X_\text{B}-X_\text{B}^\text{eq})
\right]. 
\end{equation}
Since $b_\perp =0$, there is no Peach-Koehler force. 
The disconnection climb velocity, in the special case of pure step mode, is 
\begin{equation}
v_\text{cl}
= -M_\text{d}\frac{h^\text{L}}{\Omega} \Delta X_\text{B}^\text{eq} \left[
\mathcal{E}_\text{Bv}(X_\text{v}-X_\text{v}^\text{eq})
+\mathcal{E}_\text{BB}(X_\text{B}-X_\text{B}^\text{eq})
\right]. 
\end{equation}
The flux of B atoms into the interface required by the motion of one disconnection (pure step) is 
\begin{align}
\varrho I_\text{B}
&= -\varrho(X_\text{B}-X_\text{B}^\text{w})\frac{h^\text{L}}{\Omega}v_\text{cl}
\nonumber \\
&\approx M_\text{d}\varrho \left(\frac{h^\text{L}}{\Omega}\right)^2
(\Delta X_\text{B}^\text{eq})^2 \left[
\mathcal{E}_\text{Bv}(X_\text{v}-X_\text{v}^\text{eq})
+\mathcal{E}_\text{BB}(X_\text{B}-X_\text{B}^\text{eq})
\right]. 
\end{align}

\section{Criteria of dislocation transmission across a grain boundary}
\label{APPcriteria}

\begin{figure}[!t]
\begin{center}
\scalebox{0.45}{\includegraphics{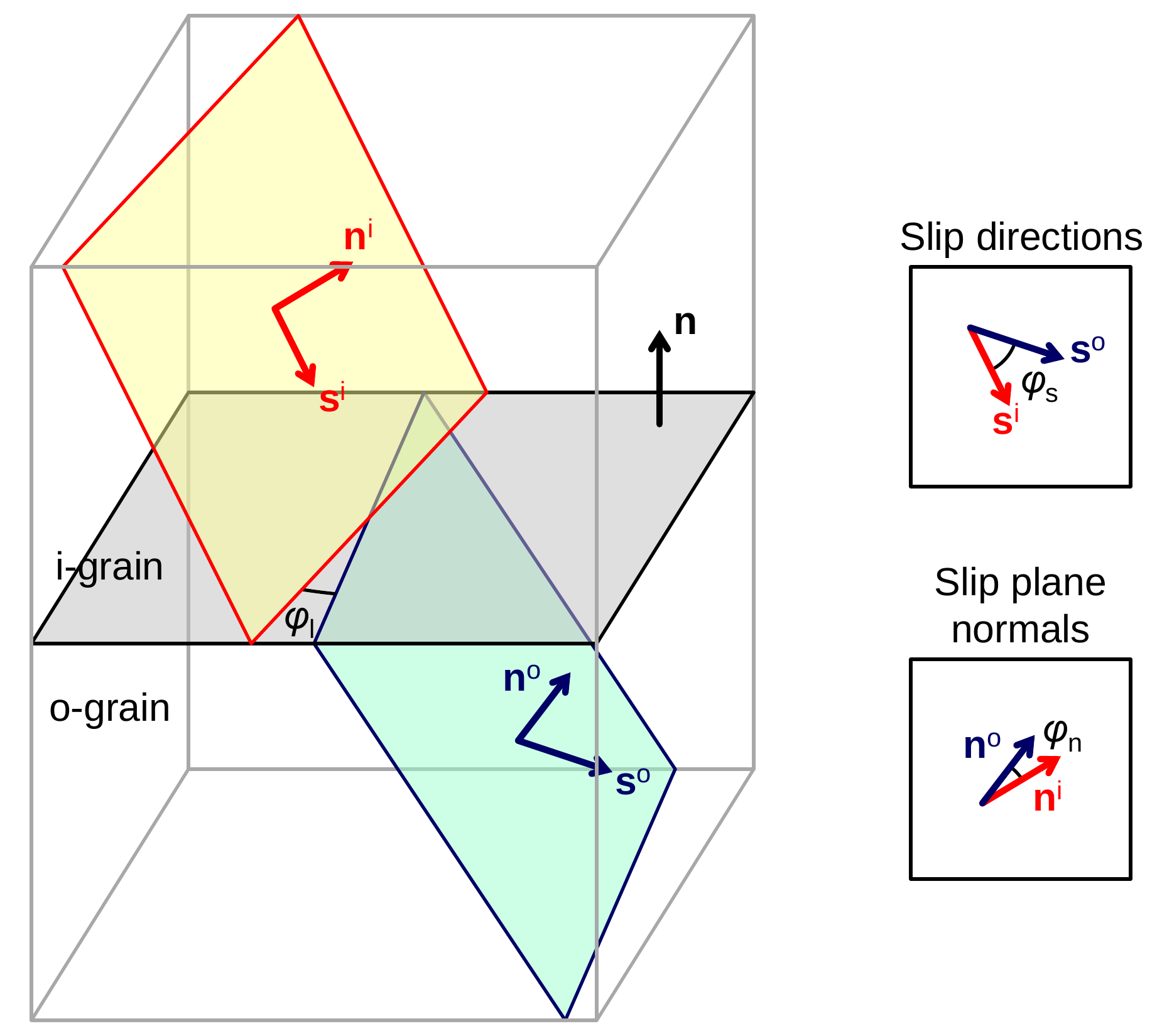}}
\caption{Geometry for describing dislocation transmission across a GB in a bicrystal. 
The lattice dislocation is assumed to come into the GB from the upper grain (denoted ``i-grain'') and go out of the GB into the lower grain (denoted ``o-grain''). 
The slip systems of the incoming and outgoing dislocations are, respectively, $(\mathbf{n}^\text{i}, \mathbf{s}^\text{i})$ and $(\mathbf{n}^\text{o}, \mathbf{s}^\text{o})$. 
See text in \ref{APPcriteria} for definition of the angles $\varphi_\text{l}$, $\varphi_\text{s}$ and $\varphi_\text{n}$. } 
\label{DislTransCriteria}
\end{center}
\end{figure}

There is literature that provides reviews of the criteria for dislocation transmission across a GB~\cite{bieler2014grain,spearot2014insights,zhang2016review,stabix}. 
Here, we will summarize these criteria.  
The description is based on the bicrystal geometry, slip systems and notation shown in Fig.~\ref{DislTransCriteria}. 
The lattice dislocation is assumed to come {\it into} the GB from the upper grain and go {\it out of} the GB into the lower grain; the upper and lower grains are denoted as ``i-grain'' and ``o-grain'', respectively. 
$\mathbf{n}^\text{i}$ and $\mathbf{n}^\text{o}$ are the slip plane normals of the incoming and outgoing dislocations, respectively. 
$\mathbf{s}^\text{i}$ and $\mathbf{s}^\text{o}$ are the slip directions of the incoming and outgoing dislocations, respectively. 
$\varphi_\text{l}$ is the angle between the intersection line of the incoming slip plane with the GB and that of the outgoing slip plane with the GB. 
$\varphi_\text{s}$ and $\varphi_\text{n}$ are the angles between the slip directions and slip plane normals, respectively, of the incoming and outgoing slip systems.

(i) The slip systems involved in slip transfer across a GB are those which can minimize the magnitude of residual Burgers vector~\cite{lim1985continuity,lee1989prediction,abuzaid2012slip}.  
The residual Burgers vector (i.e., the difference between the Burgers vector of the incoming dislocation and that of the outgoing dislocation) is 
\begin{equation}
\mathbf{b}_\text{r} = \mathbf{b}^\text{i}-\mathbf{b}^\text{o}, 
\end{equation}
where $\mathbf{b}^\text{i}$ and $\mathbf{b}^\text{o}$ are the Burgers vectors of the incoming and outgoing lattice dislocations, respectively. 
If slip systems in i- and o-grains are same (this is not necessarily true; e.g. anomalous slip was observed in experiments~\cite{lee1989anomalous}), $|\mathbf{b}^\text{i}|=|\mathbf{b}^\text{o}| \equiv b^\text{L}$. 
In this case, the difficulty of slip transfer may be measured by the quantity $|\mathbf{b}_\text{r}|=2 b^\text{L} \sin(\varphi_\text{s}/2) \approx b^\text{L}\varphi_\text{s}$ (for $\varphi_\text{s}\ll 1$). 
Two factors are considered in this criterion: (1) $\varphi_\text{s}$, which is directly related to the misorientation of two grains and (2) the Burgers vector of lattice dislocations $b^\text{L}$. 

(ii) The slip systems involved in slip transfer across a GB are those which can maximize the Schmid factor for slip in the i- and/or o-grain. 
The Schmid tensor of the slip system in the i- or o-grain is 
\begin{equation}
\mathbf{S}^\text{i/o} 
= \mathbf{s}^\text{i/o} \otimes \mathbf{n}^\text{i/o}. 
\end{equation}
Under general stress $\boldsymbol{\sigma}$, the resolved shear stress on the slip system in the i- or o-grain  is 
\begin{equation}
\tau^\text{i/o} = \mathbf{S}^\text{i/o} : \boldsymbol{\sigma}. 
\end{equation}
If uniaxial tension $\boldsymbol{\sigma} = \sigma \mathbf{e}\otimes\mathbf{e}$ ($\mathbf{e}$ is the direction of tension) is applied, the Schmid factor is $\mathscr{S}^\text{i/o} = \tau^\text{i/o}/\sigma$. 
Abuzaid et al. suggest using the combined Schmid factor as the quantity to maximize~\cite{abuzaid2012slip}: 
\begin{equation}
\mathfrak{N}_\sigma
= \mathscr{S}^\text{i} + \mathscr{S}^\text{o}
= (\mathbf{s}^\text{i}\otimes\mathbf{n}^\text{i} + \mathbf{s}^\text{o}\otimes\mathbf{n}^\text{o}) : (\mathbf{e}\otimes\mathbf{e})
= (\mathbf{s}^\text{i}\cdot\mathbf{e}) (\mathbf{n}^\text{i}\cdot\mathbf{e}) + (\mathbf{s}^\text{o}\cdot\mathbf{e}) (\mathbf{n}^\text{o}\cdot\mathbf{e}). 
\label{Nsigma}
\end{equation}
If pure shear stress $\boldsymbol{\sigma} = \tau^\text{i} (\mathbf{s}^\text{i}\otimes\mathbf{n}^\text{i} + \mathbf{n}^\text{i}\otimes\mathbf{s}^\text{i})$ is applied on the incoming slip system, the Schmid factor of the outgoing slip system $(\mathbf{s}^\text{o}, \mathbf{n}^\text{o})$ in the o-grain is 
\begin{equation}
\mathfrak{N}_\tau
\equiv \tau^\text{o}/\tau^\text{i}
= (\mathbf{s}^\text{o}\otimes\mathbf{n}^\text{o}) : (\mathbf{s}^\text{i}\otimes\mathbf{n}^\text{i}+\mathbf{n}^\text{i}\otimes\mathbf{s}^\text{i})
= (\mathbf{n}^\text{i}\cdot\mathbf{n}^\text{o})(\mathbf{s}^\text{i}\cdot\mathbf{s}^\text{o}) + (\mathbf{n}^\text{i}\cdot\mathbf{s}^\text{o})(\mathbf{s}^\text{i}\cdot\mathbf{n}^\text{o}). 
\end{equation}
Livingston and Chalmers suggested that, for a particular slip system in the i-grain $(\mathbf{s}^\text{i}, \mathbf{n}^\text{i})$, the operative slip system in the o-grain can be predicted by maximizing the value of $\mathfrak{N}$ with respect to all possible slip systems $(\mathbf{s}^\text{o}, \mathbf{n}^\text{o})$~\cite{livingston1957multiple}. 

(iii) The slip systems involved in slip transfer across a GB are those which can maximize the geometric quantity: 
\begin{equation}
\mathfrak{M}
\equiv \cos\varphi_\text{l} \cos\varphi_\text{s}
= \left[(\mathbf{n}^\text{i}\times\mathbf{n})\cdot(\mathbf{n}^\text{o}\times\mathbf{n})\right]
\left(\mathbf{s}^\text{i}\cdot\mathbf{s}^\text{o}\right). 
\label{M}
\end{equation}
Obviously, the GB would be completely transparent for slip transfer if $\varphi_\text{l} = \varphi_\text{s}=0$, i.e., the incoming and outgoing slip systems are perfectly coincident; this would lead to the theoretically maximum value $\mathfrak{M}=1$. 
This criterion is originally proposed by Shen, Wagoner and Clark~\cite{shen1986dislocation,shen1988dislocation}. 
It is also suggested to modify the quantity $\mathfrak{M}$ to be 
\begin{equation}
\mathfrak{M}_\text{c}
\equiv \left\{\begin{array}{ll}
\cos\left(\dfrac{\pi}{2}\dfrac{\varphi_\text{l}}{\varphi_\text{l}^*}\right) \cos\left(\dfrac{\pi}{2}\dfrac{\varphi_\text{s}}{\varphi_\text{s}^*}\right),
& |\varphi_\text{l}|\le\varphi_\text{l}^* \text{ and } |\varphi_\text{s}|\le\varphi_\text{s}^* \\
0, & \text{otherwise}
\end{array}\right., 
\label{Mc}
\end{equation}
where $\varphi_\text{l}^*$ and $\varphi_\text{s}^*$ are critical values such that slip transfer is prohibited when either angle is larger than the critical value~\cite{beyerlein2012structure}.

(iv) The slip systems involved in slip transfer across a GB are those which can maximize the degree of coplanarity: 
\begin{equation}
\mathfrak{M}'
\equiv \cos\varphi_\text{n} \cos\varphi_\text{s}
= \left(\mathbf{n}^\text{i} \cdot \mathbf{n}^\text{o}\right)
\left(\mathbf{s}^\text{i}\cdot\mathbf{s}^\text{o}\right). 
\label{Mp}
\end{equation}
In comparison with Eq.~\eqref{M}, we find that the quantity $\mathfrak{M}'$ is just the quantity $\mathfrak{M}$ with the angle $\varphi_\text{l}$ replaced by $\varphi_\text{n}$. 
Since the acquirement of $\varphi_\text{l}$ needs measurement of the GB inclination, which is usually hard in experiment, $\mathfrak{M}'$ is easier to evaluate than $\mathfrak{M}$. 
The replacement of $\varphi_\text{l}$ by $\varphi_\text{n}$ was suggested by Luster and Morris~\cite{luster1995compatibility}
Similar to Eq.~\eqref{Mc}, Werner and Prantl suggested to set critical values for $\varphi_\text{n}$ and $\varphi_\text{s}$~\cite{werner1990slip}: 
\begin{equation}
\mathfrak{M}'_\text{c}
\equiv \left\{\begin{array}{ll}
\cos\left(\dfrac{\pi}{2}\dfrac{\varphi_\text{n}}{\varphi_\text{n}^*}\right) \cos\left(\dfrac{\pi}{2}\dfrac{\varphi_\text{s}}{\varphi_\text{s}^*}\right),
& |\varphi_\text{n}|\le\varphi_\text{n}^* \text{ and } |\varphi_\text{s}|\le\varphi_\text{s}^* \\
0, & \text{otherwise}
\end{array}\right.
\end{equation}

(v) The slip systems involved in slip transfer across a GB are those which can minimize the misorientation angle for STGBs or TwGBs~\cite{chalmers1937influence,aust1954effect,clark1954mechanical}. 
(This criterion is the earliest one proposed.) 
The misorientation angle is 
\begin{equation}
\theta
= \arccos\left[(\mathrm{Tr}\mathbf{R}-1)/2\right], 
\label{theta}
\end{equation}
where $\mathbf{R}$ is the rotation matrix which relates the orientation of both grains. 
This criterion does not require information about slip systems. 

(vi) A combination of some of the above criteria~\cite{clark1992criteria,tsuru2016predictive}. 

If we want to measure the dislocation transferability across a GB only based on the GB geometry without knowledge of the incoming slip system and applied stress, only Criterion (v) can be used. 
However, we can construct other useful criteria from Criterion (ii), (iii) or (iv). 
One approach is to go through all possible slip systems in the i- and o-grains and find the pair corresponding to the maximum value of $\mathfrak{N}$ ($\mathfrak{N}\equiv \mathfrak{N}_\sigma$, $\mathfrak{N}_\tau$, $\mathfrak{M}$, $\mathfrak{M}_\text{c}$, $\mathfrak{M}'$ or $\mathfrak{M}'_\text{c}$). 
This criterion is to maximize the quantity~\cite{shen1986dislocation,shen1988dislocation}: 
\begin{equation}
\mathfrak{N}_\text{m}
= \max_{\mathbf{s}^\text{i},\mathbf{n}^\text{i},\mathbf{s}^\text{o},\mathbf{n}^\text{o}}\mathfrak{N}. 
\label{Nm}
\end{equation}
Another approach is to go through all possible slip systems in the i- and o-grains and consider the average value of $\mathfrak{N}$. 
This criterion is to maximize the quantity~\cite{werner1990slip,bieler2014grain}: 
\begin{equation}
\bar{\mathfrak{N}}
= \frac{1}{N^\text{i}N^\text{o}} \sum_{\mathbf{s}^\text{i},\mathbf{n}^\text{i},\mathbf{s}^\text{o},\mathbf{n}^\text{o}}\mathfrak{N},  
\label{Nbar}
\end{equation}
where $N^\text{i}$ and $N^\text{o}$ are the numbers of potential slip systems in the i- and o-grains, respectively. 
The quantities $\theta$ [Eq.~\eqref{theta}], $\mathfrak{N}_\text{m}$ [Eq.~\eqref{Nm}] and $\bar{\mathfrak{N}}$ [Eq.~\eqref{Nbar}] can all be used to define the dislocation transferability across a GB. 
Note that transferability is only the property of the GB of concern.

\end{document}